\documentclass[twocolumn,traditabstract,longauth]{aa}

\usepackage{amsmath}
\usepackage{fixltx2e}
\usepackage[english]{babel}
\usepackage{graphicx}
\usepackage{epstopdf}
\usepackage{epsf,color}
\usepackage[mathscr]{eucal}
\usepackage{amsmath}
\usepackage{amssymb,amsfonts}
\usepackage{natbib}
\usepackage{graphicx}
\usepackage{txfonts}
\usepackage{dsfont}
\definecolor{Mygreen}{rgb}{0.75, 0.0, 0.0}
\definecolor{Mypink}{rgb}{1.0, 0.0, 0.5}
\definecolor{Myred}{rgb}{0.7, 0.0, 0.0}
\usepackage[breaklinks, citecolor=blue, linkcolor=Myred, urlcolor=Myred, colorlinks=true, debug, baseurl=' ']{hyperref}
\usepackage{float}
\usepackage{color}
\usepackage{ulem}
\usepackage{subcaption}
\usepackage{wasysym}

\bibpunct{(}{)}{;}{a}{}{,}
\begin{document}
\title{\textit{Euclid} preparation III. Galaxy cluster detection in the wide photometric survey, performance and algorithm selection}
\author{
Euclid Collaboration
\and R.~Adam\inst{\ref{i1},\ref{i2},\ref{i2b}}\thanks{Corresponding author: R\'emi Adam}
\and M.~Vannier\inst{\ref{i2}}
\and S.~Maurogordato\inst{\ref{i2}}
\and A.~Biviano\inst{\ref{i3}}
\and C.~Adami\inst{\ref{i4}}
\and B.~Ascaso\inst{\ref{i5}}
\and F.~Bellagamba\inst{\ref{i6},\ref{i7}}
\and C.~Benoist\inst{\ref{i2}}
\and A.~Cappi\inst{\ref{i2},\ref{i7}}
\and A.~D\'{\i}az-S\'anchez\inst{\ref{i8}}
\and F.~Durret\inst{\ref{i9}}
\and S.~Farrens\inst{\ref{i10}}
\and A.H.~Gonzalez\inst{\ref{i11}}
\and A.~Iovino\inst{\ref{i12}}
\and R.~Licitra\inst{\ref{i5},\ref{i13}}
\and M.~Maturi\inst{\ref{i14}}
\and S.~Mei\inst{\ref{i5},\ref{i13},\ref{i15}}
\and A.~Merson\inst{\ref{i15},\ref{i16}}
\and E.~Munari\inst{\ref{i3},\ref{i17},\ref{i18}}
\and R.~Pell\'{o}\inst{\ref{i19}}
\and M.~Ricci\inst{\ref{i2}}
\and P.F.~Rocci\inst{\ref{i2}}
\and M.~Roncarelli\inst{\ref{i6},\ref{i7}}
\and F.~Sarron\inst{\ref{i9}}
\and Y.~Amoura\inst{\ref{i9}}
\and S.~Andreon\inst{\ref{i12}}
\and N.~Apostolakos\inst{\ref{i20}}
\and M.~Arnaud\inst{\ref{i10},\ref{i21}}
\and S.~Bardelli\inst{\ref{i7}}
\and J.~Bartlett\inst{\ref{i22}}
\and C.M.~Baugh\inst{\ref{i23}}
\and S.~Borgani\inst{\ref{i3},\ref{i17},\ref{i24}}
\and M.~Brodwin\inst{\ref{i24b}}
\and F.~Castander\inst{\ref{i25},\ref{i26}}
\and G.~Castignani\inst{\ref{i5}, \ref{i27}, \ref{i2}, \ref{ext66}}
\and O.~Cucciati\inst{\ref{i7}}
\and G.~De Lucia\inst{\ref{i3}}
\and P.~Dubath\inst{\ref{i20}}
\and P.~Fosalba\inst{\ref{i25},\ref{i26}}
\and C.~Giocoli\inst{\ref{i6},\ref{i7},\ref{i28}}
\and H.~Hoekstra\inst{\ref{i29}}
\and G.A.~Mamon\inst{\ref{i9}}
\and J.B.~Melin\inst{\ref{i10}}
\and L.~Moscardini\inst{\ref{i6},\ref{i7},\ref{i28}}
\and S.~Paltani\inst{\ref{i20}}
\and M.~Radovich\inst{\ref{i30}}
\and B.~Sartoris\inst{\ref{i17}}
\and M.~Schultheis\inst{\ref{i2}}
\and M.~Sereno\inst{\ref{i7},\ref{i6}}
\and J.~Weller\inst{\ref{i31},\ref{i32},\ref{i33}}
\and C.~Burigana\inst{\ref{ext00},\ref{ext01},\ref{ext02}}
\and C.~S.~Carvalho\inst{\ref{ext10}}
\and L.~Corcione\inst{\ref{ext15}}
\and H.~Kurki-Suonio\inst{\ref{ext20}}
\and P.~B.~Lilje\inst{\ref{ext30}}
\and G.~Sirri\inst{\ref{i28}}
\and R.~Toledo-Moreo\inst{\ref{ext50}}
\and G.~Zamorani\inst{\ref{i7}}
}

\institute{\tiny
Centro de Estudios de F\'isica del Cosmos de Arag\'on (CEFCA), Plaza San Juan, 1, planta 2, E-44001, Teruel, Spain\goodbreak
\label{i1}
\and
Universit\'e C\^ote d'Azur, Observatoire de la C\^ote d'Azur, CNRS, Laboratoire Lagrange, France\goodbreak
\label{i2}
\and
Laboratoire Leprince-Ringuet, Ecole Polytechnique, CNRS/IN2P3, 91128 Palaiseau, France\goodbreak
\label{i2b}
\and
INAF-Osservatorio Astronomico di Trieste, via G. B. Tiepolo 11, 34143, Trieste, Italy\goodbreak
\label{i3}
\and 
Aix Marseille Univ, CNRS, CNES, LAM, Marseille, France\goodbreak
\label{i4}
\and
Sorbonne Universit\'{e}, Observatoire de Paris, Universit\'{e} PSL, CNRS, LERMA, F-75014, Paris, France\goodbreak
\label{i5}
\and
Dipartimento di Fisica e Astronomia, Universit\`a di Bologna, via Gobetti 93/2, I-40129 Bologna, Italy\goodbreak
\label{i6}
\and
Istituto Nazionale di Astrofisica (INAF) - Osservatorio di Astrofisica e Scienza dello Spazio (OAS), via Gobetti 93/3, I-40127 Bologna, Italy\goodbreak
\label{i7}
\and
Departamento F\'isica Aplicada, Universidad Polit\'ecnica de Cartagena, Campus Muralla del Mar, 30202 Cartagena, Murcia, Spain\goodbreak
\label{i8}
\and
Institut d'Astrophysique de Paris (UMR 7095: CNRS \& Sorbonne Universit\'e), 98 bis Bd Arago, 75014 Paris, France\goodbreak
\label{i9}
\and
IRFU, CEA, Universit\'e Paris-Saclay, F-91191 Gif Sur Yvette, France\goodbreak
\label{i10}
\and 
Department of Astronomy, University of Florida, Bryant Space Science Center, Gainesville, FL 32611, USA\goodbreak
\label{i11}
\and
INAF - Osservatorio Astronomico di Brera, via Brera 28, 20122 Milano, via. E. Bianchi 46, 23807 Merate, Italy\goodbreak
\label{i12}
\and
University of Paris Denis Diderot, University of Paris Sorbonne Cit\'e (PSC), 75205 Paris Cedex 13, France\goodbreak
\label{i13}
\and
Zentrum f\"ur Astronomie, Universit\"at Heidelberg, Philosophenweg 12, D-69120 Heidelberg, Germany\goodbreak
\label{i14}
\and
Jet Propulsion Laboratory, Cahill Center for Astronomy \& Astrophysics, California Institute of Technology, 4800 Oak Grove Drive, Pasadena, California, USA\goodbreak
\label{i15}
\and
Infrared Processing and Analysis Center, California Institute of Technology, Pasadena, CA 91125, USA\goodbreak
\label{i16}
\and
Dipartimento di Fisica - Sezione di Astronomia, Universit\`a di Trieste, via Tiepolo 11, I-34131 Trieste - Italy\goodbreak
\label{i17}
\and
Dark Cosmology Centre, Niels Bohr Institute, University of Copenhagen, Juliane Maries Vej 30, DK-2100 Copenhagen,Denmark\goodbreak
\label{i18}
\and
Institut de Recherche en Astrophysique et Plan\'etologie (IRAP), Universit\'e de Toulouse, CNRS, UPS, CNES, 14 Av. Edouard Belin, F-31400 Toulouse, France\goodbreak
\label{i19}
\and
Department of Astronomy, University of Geneva, ch. d’\'Ecogia 16, CH-1290 Versoix, Switzerland\goodbreak
\label{i20}
\and 
Universit\'e Paris Diderot, AIM, Sorbonne Paris Cit\'e, CEA, CNRS, F-91191 Gif-sur-Yvette, France\goodbreak
\label{i21}
\and
APC, AstroParticule et Cosmologie, Universit\'e Paris Diderot, CNRS/IN2P3, CEA/lrfu, Observatoire de Paris, Sorbonne Paris Cit\'e, 10 rue Alice Domon et L\'eonie Duquet, 75205, Paris Cedex 13, France\goodbreak
\label{i22}
\and
Institute for Computational Cosmology, Department of Physics, Durham University, South Road, Durham, DH1 3LE, UK\goodbreak
\label{i23}
\and
INFN, Instituto Nazionale di Fisica Nucleare, Trieste, Italy\goodbreak
\label{i24}
\and
Department of Physics and Astronomy, University of Missouri, 5110 Rockhill Road, Kansas City, MO 64110\goodbreak
\label{i24b}
\and 
Institute of Space Sciences (ICE, CSIC), Campus UAB, Carrer de Can Magrans, s/n, 08193 Barcelona, Spain\goodbreak
\label{i25}
\and 
Institut d'Estudis Espacials de Catalunya (IEEC), 08193 Barcelona, Spain\goodbreak
\label{i26}
\and
Coll\`{e}ge de France, 11 Place Marcelin Berthelot, 75231 Paris, France\goodbreak
\label{i27}
\and
INFN - Sezione di Bologna, viale Berti-Pichat 6/2, I-40127 Bologna, Italy\goodbreak
\label{i28}
\and
Leiden Observatory, Leiden University, Niels Bohrweg 2, 2333 CA Leiden, The Netherlands\goodbreak
\label{i29}
\and
INAF - Osservatorio Astronomico di Padova, Vicolo Osservatorio 5, 35122 - Padova, Italy\goodbreak
\label{i30}
\and
Universit\"ats-Sternwarte M\"unchen, Fakult\"at f\"ur Physik, Ludwig-Maximilians-Universit\"at M\"unchen, Scheinerstrasse 1, 81679 M\"unchen, Germany\goodbreak
\label{i31}
\and
Excellence Cluster Universe, Boltzmannstr. 2, D-85748 Garching, Germany\goodbreak
\label{i32}
\and
Max Planck Institute for Extraterrestrial Physics, Giessenbachstr. 1, D-85748 Garching, Germany\goodbreak
\label{i33}
\and 
INAF, Istituto di Radioastronomia, Via Piero Gobetti 101, I-40129 Bologna, Italy\goodbreak
\label{ext00}
\and
Dipartimento di Fisica e Scienze della Terra, Universit\'a di Ferrara, Via Giuseppe Saragat 1, I-44122 Ferrara, Italy\goodbreak
\label{ext01}
\and
Istituto Nazionale di Fisica Nucleare, Sezione di Bologna, Via Irnerio 46, I-40126 Bologna, Italy\goodbreak
\label{ext02}
\and
Instituto de Astrof\'isica e Ci\^encias do Espa\c{c}o, Faculdade de Ci\^encias, Universidade de Lisboa, Tapada da Ajuda, PT-1349-018 Lisboa, Portugal\goodbreak
\label{ext10}
\and
INAF - Osservatorio Astrofisico di Torino, via Osservatorio 20, 10025 Pino Torinese (TO), Italy\goodbreak
\label{ext15}
\and
Department of Physics and Helsinki Institute of Physics, Gustaf H\"allstr\"omin katu 2, 00014 University of Helsinki, Finland\goodbreak
\label{ext20}
\and
Institute of Theoretical Astrophysics, University of Oslo, P.O. Box 1029 Blindern, N-0315 Oslo, Norway\goodbreak
\label{ext30}
\and
Universidad Politécnica de Cartagena, Departamento de Electrónica y Tecnología de Computadoras. 30202, Cartagena, Spain\goodbreak
\label{ext50}
\and
Laboratoire d'astrophysique, \'{E}cole Polytechnique F\'{e}d\'{e}rale de Lausanne (EPFL), Observatoire de Sauverny, 1290 Versoix, Switzerland\goodbreak
\label{ext66}
}

\date{Received \today \ / Accepted --}
\abstract {
Galaxy cluster counts in bins of mass and redshift have been shown to be a competitive probe to test cosmological models. This method requires an efficient blind detection of clusters from surveys with a well-known selection function and robust mass estimates, which is particularly challenging at high redshift.
The \textit{Euclid} wide survey will cover 15000 deg$^2$ of the sky, avoiding contamination by light from our Galaxy and our Solar System in the optical and near-infrared bands, down to magnitude 24 in the $H$-band. The resulting data will make it possible to detect a large number of galaxy clusters spanning a wide-range of masses up to redshift $\sim 2$ and possibly higher.
This paper presents the final results of the \textit{Euclid} Cluster Finder Challenge (CFC), fourth in a series of similar challenges. The objective of these challenges was to select the cluster detection algorithms that best meet the requirements of the \textit{Euclid} mission. The final CFC included six independent detection algorithms, based on different techniques, such as photometric redshift tomography, optimal filtering, hierarchical approach, wavelet and friend-of-friends algorithms. These algorithms were blindly applied to a mock galaxy catalog with representative \textit{Euclid}-like properties. The relative performance of the algorithms was assessed by matching the resulting detections to known clusters in the simulations down to masses of $M_{200} \sim 10^{13.25}$ M$_{\odot}$.
Several matching procedures were tested, thus making it possible to estimate the associated systematic effects on completeness to $<3$\%.
All the tested algorithms are very competitive in terms of performance, with three of them reaching $>80$\% completeness for a mean purity of 80\% down to masses of $10^{14}$ M$_{\odot}$ and up to redshift $z=2$.
Based on these results, two algorithms were selected to be implemented in the \textit{Euclid} pipeline, the Adaptive Matched Identifier of Clustered Objects (AMICO) code, based on matched filtering, and the PZWav code, based on an adaptive wavelet approach.}
\titlerunning{Galaxy cluster detection with \textit{Euclid}}
\authorrunning{Euclid Collaboration}
\maketitle

\section{Introduction}\label{sec:Introduction}
Galaxy clusters are good tracers of the matter density peaks in the cosmic web. They additionally provide efficient tests for cosmological models as they form via gravitational collapse in the expanding Universe \citep[for a review, see][]{Allen2011}. In particular, the number density of galaxy clusters as a function of mass and redshift enables us to constrain cosmological parameters primarily through the linear growth rate of perturbations. This has been proven to be very competitive and complementary to other probes \citep[e.g.,][]{Vikhlinin2009,Rozo2010,Planck2013XX,Bohringer2014,Mantz2015,Planck2015XXIV,deHaan2016}. The spatial distribution of clusters can provide additional information to help constrain cosmological parameters via the measurement of the cluster-cluster two-point correlation function \citep[e.g.,][]{Majumdar2004,Mana2013,Veropalumbo2014,Sridhar2017}. In particular, clusters probe a redshift range that is sensitive to dark energy and hence they can be used to constrain extensions of the standard model. However, any cosmological inference using cluster counts or spatial distribution requires accurate calibration of the halo mass function, an accurate knowledge of the cluster sample selection function, and primary observables that tightly correlate to cluster masses via scaling relations (including an understanding of the intrinsic scatter in the scaling relations). The calibration of the proper mass scale is also fundamental for cluster physics studies.

Galaxy clusters can be detected through their hot gas content, either from their X-ray emission \citep[see e.g.,][]{Bohringer2001,Pacaud2016}, or using their imprint in the Cosmic Microwave Background (CMB) via the thermal Sunyaev--Zel'dovich effect \citep[tSZ,][]{Sunyaev1972} at millimeter wavelengths \citep[e.g.,][]{Hasselfield2013,Bleem2014,Planck2015XXVII}. In the optical \citep[e.g.,][]{Kepner1999,Rykoff2014} or near-infrared \citep[NIR; e.g.,][]{Eisenhardt2008,Wylezalek2013,Rettura2014} clusters can be identified using galaxy overdensities. Additionally, optical imaging and analysis methods have now reached the maturity to construct convergence maps via the weak lensing (WL) of background galaxies, where massive clusters appear as peaks \citep[e.g.,][]{Gavazzi2007,Shan2012,Jeffrey2018}. In a cosmological context, the quest for a well-characterized cluster sample, preferably as complete and as pure as possible, is important in quantifying the likelihood of cluster detections for a given set of cosmological parameters.

The properties of galaxy groups and clusters are also essential for understanding galaxy formation because they constitute the local environment in which a significant fraction of galaxies evolve \citep[see, e.g.,][]{DeLucia2012,Raichoor2012}. Observations show that, at fixed stellar mass, cluster core galaxies present specific properties compared to field galaxies such as lower star formation rates, early-type morphologies and a tight red sequence up to redshift $z \sim 1$ \citep[e.g.,][]{Mei2009,George2011,Wetzel2013}. At higher redshifts, higher star formation rates are observed in cluster cores as well as more disturbed morphologies \citep[e.g.,][]{Brodwin2013,Alberts2016,Noirot2016}. A deeper understanding of the mechanisms that trigger such properties and their evolution will be achievable with future large-scale optical or NIR surveys such as \textit{Euclid} \citep{Laureijs2011}, the Large Synoptic Survey Telescope \citep[LSST,][]{LSST2009}, the Javalambre-Physics of the Accelerated Universe Astrophysical Survey \citep[J-PAS,][]{Benitez2014}, and the Wide Field Infrared Survey Telescope \citep[\textit{WFIRST,}][]{Spergel2015}, which will reach cluster masses down to a few $10^{14}$ M$_{\odot}$ up to $z \sim 2$ \citep{Sartoris2016,Ascaso2017}. Optical or NIR observation can also potentially select the most massive clusters at high redshifts \citep[see e.g.,][]{Andreon2009,Brodwin2012}, and those are likely the place where the first massive galaxies form.

\textit{Euclid} is a European Space Agency (ESA) mission planned for launch in 2021 that aims at providing a better understanding of the origin of the accelerated expansion of the Universe, particularly the nature of dark energy, dark matter, and gravity \citep{Laureijs2011,Amendola2013}. Through its dedicated wide survey, \textit{Euclid} will observe 15000 deg$^2$, that is a large fraction of the sky (outside of the Galactic plane), in a wide optical band (VIS, down to magnitude 24.5 for a 10$\sigma$ extended object) and three near-infrared bands ($Y$, $J$, $H$, down to magnitude 24 for a 5$\sigma$ point-source). Deep surveys will cover about 40 deg$^2$, which is two magnitudes deeper. Using the Near Infrared Spectrometer and Photometer (NISP) slitless spectrograph, photometric data will be complemented by spectroscopy, which is expected to release redshifts for several tens of millions of galaxies. Photometric redshifts that will be obtained by combination with ground based photometric surveys \citep[such as the LSST, J-PAS or the Dark Energy Survey, DES,][]{Abbott2018} will enable \textit{Euclid} to detect galaxy clusters over a large range of masses and up to redshift $\sim 2$. As an optical and NIR survey, the rest-frame optical richness of clusters will be the natural mass proxy, for which \textit{Euclid} will be able to provide an internal calibration using WL mass estimates and velocity dispersion from spectroscopy using stacking techniques. A recent assessment of \textit{Euclid} performance in terms of weak lensing mass estimates of ensemble clusters \citep{Kohlinger2015} has shown that statistical uncertainties are expected to reach a very low level, and that usually predominant systematic errors such as multiplicative bias and additive bias are expected to be negligible. The richness estimates will also be complemented by other multiwavelength (X-ray, tSZ) mass proxies to reduce systematic uncertainties in the calibration. The combination of these properties should allow \textit{Euclid} to push cluster cosmology to an unprecedented level \citep[e.g., constraints of the order of a few percent on the dynamical evolution of dark energy or the growth factor parameter $\gamma$,][]{Sartoris2016}.

In order to reach these goals, several cluster finders have been developed within the \textit{Euclid} consortium. It was then necessary to develop a work frame to test and evaluate the performance of these different algorithms in the context of \textit{Euclid}. Two main methodologies are generally used in the literature, both presenting advantages and limitations: 1) the use of end-to-end simulated data, aiming at matching the expected properties of the real data \citep[e.g.,][]{Koester2007,Knobel2009, Adami2010,Old2015}, or 2) the injection of simulated clusters in a given existing data set \citep[e.g.,][]{Adami2000,Goto2002,Kim2002,Rykoff2014,Planck2015XXVII}. Given the rise of multiwavelength data-sets, the comparison of the cluster detections based on different tracers is also now a powerful way to cross-validate the selection functions \citep[e.g.,][]{Saro2015}. On one hand, the first method includes realistic projection effects associated with the spatial correlation between structures, while they are difficult to reproduce using the second method. This is particularly relevant in the case of cluster detection based on the galaxy distribution because the background is expected to be correlated with the targeted objects. On the other hand, the first method relies on the implementation of complex recipes to model the data, while the second method by construction is based on data. The second method is also more flexible regarding the modeling of the simulated cluster. Finally, arbitrary large volumes may in principle be created using the first method, while the second approach requires having in-hand data that are representative of the given survey under consideration, and large volumes to test the detection with sufficient statistics. Recently, the joint use of data and mocks has been shown to be extremely successful to fully account for correlated and uncorrelated background in the determination of richness \citep{Costanzi2019}, demonstrating the benefits of both approaches.

For the purpose of this paper, we use mocks to evaluate and compare the performance of cluster finders. This choice was motivated by several factors: i) mocks allow us to probe the whole redshift range that will be covered by \textit{Euclid} on a wide-range of richnesses and masses ; ii) they provide the distribution of halos of a given mass and redshift, which can be used as a truth table ; and iii) they preserve the effect of the correlated background. We stress that the main limitation of this approach is the fact that simulations may not fully reproduce all the cluster properties, and the absolute performance derived may therefore be taken with caution. However, we found it the most operational way to compare the relative performance of the different algorithms on a common ground. The full methodology currently developed to determine the selection function and the related mass proxy will be addressed in future work.

The performance of the cluster finder algorithms has been tested and compared in a series of four Cluster Finder Challenges (CFC) between 2013 and 2017. The codes were tested on \textit{Euclid} survey-like mock catalogs based on semianalytic models \citep{Merson2013,Gonzalez-Perez2014} and halo occupation modeling \citep{Carretero2015}. The positions of the mock clusters were unknown to the participants of the challenges. Through the years, the mock catalogs were refined to better represent the properties of galaxies within clusters. In particular, photometric redshifts were assigned to galaxies in order to run the codes as they would be run in the \textit{Euclid} context. In the first challenges, photometric redshifts were assigned following a simplistic Gaussian distribution, while in the later ones, photometric redshift codes were used. At the end of the third cluster finder challenge, the methodology and analysis pipeline were sufficiently mature for a first assessment of the relative performance of the different codes. While eight cluster finder codes in total were tested in the three preliminary challenges, only six of them took part in the final challenge described in this paper.

In this article, we present the methodology used to assess the performance of the codes and the results obtained from the final cluster finder challenge. The detection codes were applied blindly to a realistic galaxy mock, built using {\tt PhotReal} \citep{Ascaso2015} on the \textit{Euclid} wide light-cone \citep{Merson2013}, which was considered to be the best compromise available in terms of angular size (300 deg$^2$), depth ($z>2.5$), and realistic modeling of galaxy properties. We present the main assumptions and methodology of each of the competing codes and discuss the main properties of the simulated mock in the context of cluster detection. The code detections were matched to the true mock clusters and this information was used to evaluate the performance of the algorithms. Special care was given to the matching procedure by using several methods, allowing us to estimate the associated systematic uncertainties. In light of the mock properties, the performance comparison of the different algorithms participating in the challenge guided our selection of those now being validated and implemented in the \textit{Euclid} pipeline. At this stage, we stress that the goal of this paper is not yet to compute a robust selection function and robust mass proxies, but instead, to compare the relative performance of different algorithms and to test different methodologies. The definition and assessment of the selection function and the best mass proxies will be addressed in future publications.

This paper is organized as follows. In Section~\ref{sec:Algorithms}, we present the competing algorithms. Section~\ref{sec:Simulations} describes the characterization of the simulations that are used. The matching procedure, of associating the detected clusters to the mock clusters, is detailed in Section~\ref{sec:Matching}, and the performance of the algorithms is given in Section~\ref{sec:Performance}. We discuss the results and the \textit{Euclid} algorithm selection in Section~\ref{sec:Discussions}. Conclusions are given in Section~\ref{sec:Conclusions}. A brief summary of the previous challenges, as well as the description of the previously employed codes are given in the Appendix. Throughout this paper, we assume a flat $\Lambda$CDM cosmology according to that used in the mock, with $H_0 = 73$ km s$^{-1}$ Mpc$^{-1}$, $h = H_0/100$ km s$^{-1}$ Mpc$^{-1}$, $\Omega_{\rm m} = 0.25$, $\Omega_{\Lambda} = 0.75$, and $\sigma_8 = 0.9$. All logarithmic quantities shown in this paper are defined using base 10. All the magnitudes in the paper are given in the $\it{AB}$ system.

\section{Galaxy cluster detection algorithms}\label{sec:Algorithms}
The detection of galaxy clusters from photometric (or spectroscopic) surveys at optical and NIR wavelengths is a longstanding issue \citep[see e.g., the pioneering work by][]{Abell1958}. Several techniques have been developed, using different kinds of information. Some algorithms are based on the geometrical distribution of galaxies, both in projected coordinates and in photometric redshift space, while others also focus on known properties of cluster galaxies, such as colors, luminosities, and density profiles. Cluster finders are generally classified by methodology (or a combination of methodologies), of which a large variety exists in the literature. Some common examples include the use of the cluster red sequence \citep[e.g.,][]{Gladders2000,Rykoff2014}, the presence of brightest cluster galaxies \citep[BCG; e.g.,][]{Koester2007}, percolation algorithms \citep[e.g.,][]{Dalton1997}, matched filtering \citep[e.g.,][]{Postman1996,Olsen2007}, Voronoi tessellation methods \citep[e.g.,][]{Ramella2001}, friends-of-friends \citep[FoF; e.g.,][]{Wen2012}, the use of smoothing kernel techniques \citep[e.g.,][]{Gal2003,Mazure2007}, or wavelet filtering techniques \citep[see e.g., the pioneering work of][]{Eisenhardt2008}. These techniques have been extensively used to build large samples of clusters \citep[e.g.,][]{Gilbank2011} and have also led to the discovery of some massive clusters at high redshifts \citep[e.g.,][]{Stanford2012}. All detection techniques present advantages and drawbacks regarding selection effects, however different techniques are often complementary to one another. For instance, searching for the presence of a red sequence can be an efficient way to detect clusters at low and intermediate redshifts. This property, however, is expected to fade at higher redshifts \citep[e.g.,][and references therein]{Strazzullo2016} making it less effective for detecting distant clusters. For a review on cluster detection, see for example \cite{Gal2006}, or for a detailed discussion about the necessary features of galaxy cluster finders in the context of large photometric surveys, see for example \cite{Rykoff2014}.

The detection of galaxy clusters in the \textit{Euclid} survey will be largely driven by photometric data. Indeed, analytical estimates \citep{Sartoris2016} have shown that the mass detection limits obtained using spectroscopic redshifts are significantly higher than those obtained with photometry. Spectroscopic redshifts may also be used to improve the detection procedure, nevertheless this has not been taken into consideration for this work and is left for future studies. Spectroscopic data will, however, be used to confirm and refine the redshifts of the clusters detected by photometry.

Six algorithms participated in the final CFC. They were all blindly applied to a simulated mock catalog (see Section~\ref{sec:Simulations}) to provide a cluster catalog with the coordinates of the objects (sky coordinates: right ascension, RA, and declination, Dec, and redshift), a mass proxy (typically the richness) and a ranking of the likeliest true detections (mainly by signal-to-noise ratio, S/N). Four algorithms also provided the probability of the cluster member galaxies associated with each detected cluster. The names of the cluster finders, as used hereafter, and their main detection principles are provided in Table~\ref{tab:cluster_finder_summary}. The following subsections provide an overview of the methodology and the assumptions used by each code.

\begin{table*}[h]
\caption{\footnotesize{Summary of properties and names of eight cluster finder algorithms that participated in CFC. The properties listed here correspond to those of the final CFC. All algorithms performed redshift slicing or made use of a grid, and all rely on the $H$-band in the case of the final CFC. RedGOLD and Voronoi did not participate in the last challenge for reasons not related to their performance in the earlier ones.}}
\begin{center}
\resizebox{1\textwidth}{!} {
\begin{tabular}{c|c|c|c|c|c|c}
\hline
\hline
Name & CFC participation & Detection principle & Main reference & Cluster properties assumptions & Use of calibration field & Membership \\
\hline
AMASCFI & 1,2,3,4 & Adaptive kernel & \cite{Adami1999} & Typical size and $m_{H}^{\star}$ calibration & Yes & \text{\sffamily X} \\
AMICO & 1,2,3,4 & Optimal filtering & \cite{Bellagamba2018} & LF and profile & No & \checked \\
HCFA & 3,4 & Hierarchical finder & \textcolor{blue}{D\'{\i}az-S\'anchez (in prep.)} & Typical size only & No & \checked \\
PZWav & 1,2,3,4 & Wavelet adaptive & \cite{Gonzalez2014} & Typical size and $m_{H}^{\star}$ evolution & No & \text{\sffamily X}\\
sFoF & 1,2,3,4 & Friends-of-friends & \cite{Farrens2011} & None & Yes & \checked \\
WaZP & 1,2,3,4 & Wavelet & \cite{Benoist2014} & Typical size and $m_{H}^{\star}$ evolution & Yes & \checked \\
\hline
RedGOLD & 1,2 & Red sequence & \cite{Licitra2016a} & -- & -- & -- \\
Voronoi & 1 & Voronoi tessellation & \textcolor{blue}{Iovino (in prep.)} & -- & -- & -- \\
\hline
\end{tabular}
}
\end{center}
\label{tab:cluster_finder_summary}
\end{table*}

\subsection{AMASCFI: Adami, Mazure \& Sarron cluster finder}
The Adami, Mazure \& Sarron cluster finder (AMASCFI) algorithm \citep{Sarron2018} searches for clusters in large multi-band imaging surveys using photometric redshift ($z_{\rm phot}$) tomography. As an input, the AMASCFI algorithm requires a galaxy catalog with sky positions (RA, Dec) and photometric redshifts. The photometric redshift catalog is first divided in redshift slices of variable width according to the evolution of the photometric redshift error, $\sigma_{\rm z_{\rm phot}}(z_{\rm spec})$, which is estimated using spectroscopic redshifts from the calibration field (see Section~\ref{sec:Simulations}). All slices overlap by 0.05 in redshift, taken as a constant so that the cluster photometric redshifts are sampled with the same resolution whatever the redshift. Galaxy density maps are built for each redshift slice, based on an adaptive kernel technique, with an initial kernel size (diameter) fixed at 1.5 Mpc. This way the adaptive kernel size in the densest region (corresponding to galaxy clusters) is about 1 Mpc (i.e., the typical size of cluster cores). Structures in these density maps are detected using the source extraction software, SExtractor \citep{Bertin1996}, in the different redshift bins with a detection threshold set to a given number of galaxies per Mpc$^{2}$. The initial structures are then assembled into larger structures using a minimal spanning tree FoF algorithm \citep[see][]{Adami1999}. Any two detections less than 1 Mpc apart and with $\Delta z \le 0.05$ are merged. A detailed description of each step of the algorithm, as well as a discussion of the influence of the choice of parameters can be found in \cite{Sarron2018}.

The sky coordinates (RA, Dec) and redshift of each candidate cluster are taken to be the mean of each of its individual merged detections weighted by its galaxy number density. For each redshift slice, the S/N of detected peaks is computed from the 2D density map as $\left(\langle n_{\rm cluster} \rangle A - \langle n_{\rm field} \rangle A\right)/\sqrt{\langle n_{\rm field} \rangle A}$, where $\left\langle n_{\rm cluster}\right\rangle$ and $\left\langle n_{\rm field}\right\rangle$ correspond to the average number density of galaxies per Mpc$^{2}$ in a slice of width $\Delta z$ for cluster and field area, respectively, and $A$ is the cluster area (taken to 500 kpc radius) projected on the sky. For each cluster candidate, the final S/N is taken as the maximum S/N of its individual merged detections. The richness $\lambda_{\rm det}$ is computed from a modified version of the \cite{Licitra2016a} estimator. AMASCFI first counts the number of galaxies with $m_{H} < m_{H}^{\star}+2.5$ in a cylinder of radius $R_{\rm det} = 1$ Mpc $h^{-1}$ and length $\pm 2 \sigma_{z_{\rm phot}}$ around the cluster center, and removes the galaxy background contribution. The knee magnitude of the luminosity function (LF), $m_{H}^{\star}$, was calibrated using the value measured for the Coma Cluster obtained by \cite{dePropris1998}. It then iteratively rescales the detection radius as $R_{\rm det} = \left(\lambda_{\rm det}\left(<R_{\rm det}\right)/100\right)^{0.2}$ until convergence. For the last CFC, the rank was determined by sorting the S/N values. The richness was used to establish the relative rank for objects with identical S/N values. AMASCFI was applied to the CFHTLS in \cite{Sarron2018} and the previous version of the AMASCFI algorithm (AMACFI) was used to search for clusters in the CFHTLS \citep{Mazure2007,Adami2010,Durret2011} and in the SDSS Stripe 82 data \citep{Durret2015}.

\subsection{AMICO: Adaptive Matched Identifier of Clustered Objects}
The Adaptive Matched Identifier of Clustered Objects (AMICO) algorithm \citep{Bellagamba2011,Bellagamba2018} is an enhanced matched filter algorithm that looks for cluster candidates by convolving the 3D galaxy distribution with a redshift-dependent filter. The input of the algorithm is a galaxy catalog that includes sky coordinates (RA, Dec), photometric redshifts and magnitudes. The filter is defined on the basis of a cluster and noise model that has the purpose of amplifying the contrast between the two components. Originally this filtering method was used to detect galaxy clusters in weak lensing data \citep{Maturi2005}. The noise is modeled by assuming a spatially uniform LF, while the cluster model is the combination of a cluster galaxy LF and a galaxy density profile. In the CFC, AMICO considered only the $H$-band for detection, but it can use any other magnitude or a combination of two or more. It also accounts for the full shape of the photometric redshift probability distribution function (PDF), $P(z)$, provided by the mock. The convolution of the galaxy distribution with the AMICO filter generates a 3D amplitude map, whose peaks represent the detections. In addition to standard matched filter algorithms, AMICO defines a membership probability for each galaxy to belong to a given detection. It uses this information to remove signals in the original amplitude map in order to search for further detections, which might be blended with other structures, without any further assumptions. This has proven to be an efficient method to disentangle close-by objects.

The output sky coordinates (RA, Dec) and redshift of the candidate clusters are given by the position of the peaks in the likelihood on the 3D grid. The uncertainty on the amplitude is derived from the expected variance in the measurement, due to the background fluctuations and the shot-noise in the cluster galaxy distribution. The S/N associated to the candidate clusters is then the ratio of the amplitude over its uncertainty. The mass proxy provided by AMICO is the amplitude, a measure of the cluster galaxy abundance in units of the cluster model. Detections are ranked according to their S/N. We note that AMICO can provide another mass proxy, given by the sum of the membership probabilities for each detection \citep[a measurement of the richness, see][]{Bellagamba2018b}, but this quantity was not used in this work. AMICO was recently used to identify galaxy clusters in the Kilo Degree Survey \citep[KiDS,][]{Radovich2017,Maturi2018}.

\subsection{HCFA: Hierarchical Cluster Finder Algorithm}
The Hierarchical Cluster Finder Algorithm (HCFA) algorithm (\textcolor{blue}{D\'{\i}az-S\'anchez, in prep.}) searches for overdensities of galaxies using different angular scales in a hierarchical approach. The HCFA algorithm requires only the position and the photometric redshift of the galaxies as inputs. It first uses overlapping redshift bins of size $\Delta z = 0.05$ (as for AMASCFI) to identify the galaxies that are in local overdensity regions. Each galaxy is then labeled with its local density, $n_{\rm g}$, according to the galaxies in its neighborhood. HCFA uses a primary angular scale of 0.2 Mpc for this purpose. A critical density $n_{\rm g \it c}$ is defined as $3 \sigma_{n_{\rm g}}$ above the mean local density $\left\langle n_{\rm g}\right\rangle$, $n_{\rm g \it c} = 3 \sigma_{n_{\rm g}} + \left\langle n_{\rm g}\right\rangle$, where $\sigma_{n_{\rm g}}$ is the standard deviation of the local galaxy density field, and galaxies labeled with lower densities are removed from the sample. The remaining galaxies are merged using a FoF algorithm with an angular linking scale equal to the primary one. The overdensity factor is calculated for each resulting group of galaxies. Membership probabilities are defined according to the local density of each galaxy. Groups with densities lower than the critical density are removed and the FoF algorithm is repeated iteratively, increasing the angular linking scale, until groups do not merge any more or the linking scale reaches 0.6 Mpc. In this way, HCFA identifies galaxy clusters composed of hierarchical overdensities. The algorithm uses a sky tiling of 36 arcmin$^2$ (chosen for convenience) and tiles are processed in parallel.

The cluster candidate centroids are calculated taking into account all the galaxies in the cluster, while the redshift is given by the mean redshift of the galaxies. A S/N is defined for each galaxy as $\left(n_{\rm g} - \left\langle n_{\rm g}\right\rangle\right) / \sigma_{n_{\rm g}} $. From this definition, the S/N of the candidate clusters are set to the mean S/N of the five galaxies with the highest S/N values in the cluster. A minimum of five galaxies are required in order to define a candidate cluster. The richness is given by the total over-density factor of the cluster, i.e., the number of galaxies in the cluster multiplied by the S/N of each galaxy. The candidate clusters are ranked according to the S/N. The HCFA algorithm has not yet been applied to real data.

\subsection{PZWav}
The cluster finding algorithm PZWav \citep{Gonzalez2014} is a wavelet-style algorithm that searches for overdensities on fixed physical scales. PZWav requires a galaxy catalog with sky coordinates, photometric redshifts, and magnitudes. It uses a difference-of-Gaussian smoothing kernel and incorporates for each galaxy the full probability distribution associated with the photometric redshift, $P(z)$. As a preprocessing step, the galaxy catalog is culled to contain only galaxies brighter than a given limit, taken as $m_{H} < m_{H}^{\star}+2$ in $H$-band, so that galaxies out to $z=1.5$ are selected down to the same limit, as traced by any model of galaxy evolution. This preprocessing step minimizes the redshift dependence of the mass threshold for cluster detection. After this preprocessing is complete, the algorithm first constructs a series of redshift slices spanning the redshift range of interest, and then inserts each galaxy into these redshift slices, weighted by the probability that the galaxy lies at a given redshift. These density maps are next convolved with a difference-of-Gaussians smoothing kernel of a fixed physical size, which is approximately matched to the physical size of cluster cores. A second set of density maps is also constructed for which the redshift probability distributions have been randomly shuffled relative to the positional information. These random density maps are used for bootstrap simulations to calculate a uniform noise threshold as a function of redshift that is independent of the mean galaxy density. Galaxy cluster candidates are next identified in each redshift slice, and these detections are merged across the redshift slices. All detections that lie near the edge of the survey field are rejected, and redshift estimates are refined for each cluster using a secondary code that sums the probability distributions of all galaxies within a fixed radius of the cluster detection.

The cluster centroids come directly from the smoothed density maps, corresponding to the peak location of each detected overdensity. Cluster redshifts are derived by computing the $\sigma-$clipped median photometric redshift from all galaxies that lie within 30$^{\prime\prime}$ of the centroid and lie within $\Delta z=0.12$ of the redshift slice in which a cluster is detected. The direct observable from this search is the peak amplitude of each detected overdensity, which can be taken as a proxy for richness. Candidates are ranked by this peak amplitude. The version of PZWav used for the challenges did not calculate the S/N, reporting only the peak amplitude. The current version of the code calculates the S/N based upon the fluctuations in the random maps. This algorithm is based upon the approach initially developed for the IRAC Shallow Cluster Survey \citep{Elston2006,Eisenhardt2008}, also used in the work of \cite{Stanford2012}, but has been optimized and refined to work efficiently with \textit{Euclid}-like data.

\subsection{sFoF: Friends-of-friends}
The sFoF algorithm is a friends-of-friends galaxy cluster detection algorithm \citep{Farrens2011} that follows the principles established by \cite{Huchra1982} and later modifications implemented by \cite{Botzler2004}. The algorithm operates using an input galaxy catalog with either spectroscopic redshifts (3D: using sky coordinates and redshifts) or photometric redshifts (2+1D, as in the present case), using sky coordinates stacked in bins of photometric redshift. All of the internal operations are performed in angular space and no assumptions are made about the nature of clusters of galaxies (e.g., size, color, shape). Two primary free parameters, the transverse linking and the line-of-sight linking lengths, determine the total number of cluster candidates and their corresponding properties. These linking parameters change as a function of redshift to account for selection effects, which in turn provides a redshift independent richness estimate for each cluster candidate. The parameters were optimized using the calibration field provided with the mock (see Section~\ref{sec:Simulations}). Each FoF group galaxy is marked as a cluster member and its membership probability is set to unity, while non cluster members have a membership probability that is set to zero. The code implements k-dimensional tree and Open Multi-Processing routines to improve the performance of a single run.

The cluster candidate coordinates (RA, Dec and redshift) are obtained from the median of the member positions. The S/N is computed as $\left(\lambda_{\rm det} - A \, n_{\rm field}\right) / \sqrt{A \, n_{\rm field}}$, where $\lambda_{\rm det}$ is the estimated richness, $A$ is the cluster area projected on the sky, and $n_{\rm field}$ is the galaxy background level at the cluster redshift. The richness is given by the number of FoF objects found for a given cluster, which is also the sum of the membership probabilities. Because the linking parameters change as a function of redshift, this roughly gives a redshift independent estimate. Candidate clusters were ranked according to the richness. The sFoF algorithm was applied to the 2SLAQ spectroscopic survey \citep{Cannon2006} of potential luminous red galaxies in \cite{Farrens2011}.

\subsection{WaZP: Wavelet Z-Photometric cluster finder}
The Wavelet Z-Photometric cluster finder (WaZP) algorithm \citep{Benoist2014,Dietrich2014} is an optical cluster finder based on the identification of galaxy overdensities in (RA, Dec, $z_{\rm phot}$) space. WaZP requires a galaxy catalog with sky coordinates (RA, Dec), photometric redshifts and magnitudes. The detection process makes no assumptions on the LF of cluster galaxies nor on the galaxy density profile. From an operational point of view the WaZP algorithm goes through the sequence described below. The galaxy catalog is sliced along the photometric redshift axis in overlapping redshift bins of variable widths controlled by the scatter of $P(z)$. In each slice, galaxies are weighted by the fraction of their PDF intersecting the slice. In addition, in the context of this work, detection was performed only using galaxies with $m_{H} \le  m_{H}^{\star}+1$. The resulting projected galaxy distribution is then pixelated on a grid with a physical step size of 1/16$^{\rm th}$ of a Mpc. The pixelated galaxy catalog is filtered using the wavelet task {\texttt MR\_FILTER} from the multiresolution package MR/1 \citep[][]{Starck1998}. This task incorporates a statistically rigorous treatment of the Poisson noise, which makes it possible to keep significant structures in an appropriate scale range. Here structures with scales up to 1 Mpc are selected and a $3\, \sigma$ iterative multiresolution thresholding with a B-spline wavelet transform is applied. From each wavelet map, peaks are extracted and merged with peaks from consecutive slices to produce a final cluster list.

Each peak detected in the projected filtered maps is characterized by i) a position defined as the mode of the peak, ii) a radius $R_{\rm det}$ defined as the mean extent of the peak, iii) a redshift defined as the median redshift of the photometric redshifts selected within a projected distance $\le R_{\rm det}$ from the center and within $\pm 3 \sigma_{z_{\rm phot}}$ around the mean redshift of the map, and iv) a S/N defined as $\left(n - \left\langle n\right\rangle\right) / \sigma_{\rm bg}$ where $n$ and $\left\langle n\right\rangle$ are the galaxy density within 300 kpc from the peak center and the galaxy local background density respectively. The quantity $\sigma_{\rm bg}$ is given by the second order moments of galaxy counts in cells. When a cluster is detected in several consecutive slices, it is associated to the peak with the largest S/N. For each cluster, membership probabilities are computed following the prescription given in \cite{Castignani2016}, based here on a local background density modeling. Membership probabilities are computed up to a radius corresponding to a given galaxy density contrast. Finally each cluster is characterized by a richness defined as the sum of the membership probabilities for galaxies with a magnitude $m_{H} \le m_{H}^{\star} +1$. Clusters are ranked according to their S/N. The WaZP algorithm was applied to N-body simulations in \cite{Dietrich2014} and to the CFHTLS data to search for optical counterparts to the XXL survey \citep{Pierre2016} X-ray clusters (\textcolor{blue}{Benoist et al., in prep.}).

\section{\textit{Euclid} mock galaxy catalog}\label{sec:Simulations}
The final \textit{Euclid} CFC made use of a main mock galaxy catalog \citep{Ascaso2015} in order to test the behavior of the detection algorithms on \textit{Euclid}-like data. This mock includes photometric redshifts, $z_{\rm phot}$, and their errors. It was limited to $H$-band magnitudes brighter than $H_{\rm AB}=24$ to mimic the context of the \textit{Euclid} wide survey ($H_{\rm AB}=24$ for $5\sigma$ point-source). A 20 deg$^2$ region including both photometric and spectroscopic redshifts was also provided as a calibration field for the photometric redshifts or for the detection code parameters. While it is not the purpose of this paper to make an assessment of the validity of the semianalytic models on which the mock is based, we do aim to verify the reliability of the model predictions. This is done in order to quantify how realistic the performance of the cluster finders are when applied to the mock. We discuss
the construction of the mock in Section~\ref{sec:Mock_galaxy_catalog}.

\subsection{Construction of the mock galaxy catalogs}\label{sec:Mock_galaxy_catalog}
We placed some constraints on the properties of the mock as we aimed to test the performance of the cluster finders at high redshift (up to about 2) and high mass (larger than about $10^{14}$ M$_{\odot}$) in the \textit{Euclid} regime. In order to satisfy these requirements, the mock has to be complete in magnitude to at least $H_{\rm AB} = 24$, to cover a redshift range up to $z \gtrsim 2$, and to have a reasonable sky coverage in order to get enough statistics on the high mass and high redshift clusters. We therefore chose a parent sample of 500 deg$^2$ from which we extracted a 300 deg$^2$ mock. Finally, this mock was blinded by applying a rotation and translation.

\subsubsection{Galaxy catalog}\label{sec:Galaxy _catalog}
The galaxy catalog was extracted from the \cite{Ascaso2015} mock, which was based on the $H$-band wide light-cone from \cite{Merson2013}. The light-cone was generated from the Millennium simulation \citep{Springel2005} using semianalytical modeling of galaxy formation with the {\tt GALFORM} model \citep{Lagos2012}. The mock was reprocessed with the software {\tt PhotReal} \citep{Ascaso2015} to obtain realistic galaxy photometry compliant with \textit{Euclid} depth in $YJH$ (down to magnitude 24 at $5\, \sigma$, point sources) and $grizY$ (down to magnitudes 25.2, 24.8, 24.0, 23.4 and 21.7 at $10\, \sigma$, extended sources), assuming complementary ground-based DES data \citep{Mohr2012}. This corresponds to the pessimistic case in \cite{Ascaso2015}, as opposed to the combination of the \textit{Euclid} observations with deeper ground-based photometry from LSST \citep[the optimistic case in][]{Ascaso2015}. In this sense the performance derived hereafter is expected to be conservative.

The photometry was also modified by {\tt PhotReal} using a set of empirical templates to fit observed spectral distributions and make the galaxy colors, luminosity and mass functions more consistent with current observations \citep[see][for more details]{Ascaso2015}. Photometric redshifts were estimated using the Bayesian Photometric Redshifts software \citep[BPZ,][]{Benitez2000,Benitez2004,Coe2006} applied to the {\tt PhotReal} photometry. The most likely redshifts (PDF peaks) were derived, as well as their probability distribution functions.

We note that the magnitude cut applied to the mock used in the present paper introduces and extra idealization. Indeed, in practice the \textit{Euclid} catalog will extend to fainter magnitudes (albeit being incomplete), which may benefit to the detection codes, in particular for the detection of high redshift clusters. In this sense, the results presented in this paper are conservative in terms of performance, as the magnitude cut applied limits the sampling of the luminosity function at high redshift (however still reaching $m^{\star}+1.5$ at redshift 2). In addition, accurate photometry in crowded cluster fields, with the intra cluster light also contributing to the background, is a real challenge as shown in recent studies based on Hubble Space Telescope observations \citep[e.g.,][]{Molino2017}. Such effects, which are not included in the mock used in this paper, may boost the photometric redshifts uncertainties of the corresponding galaxies, and we leave their detailed investigation for future work, when the end-to-end \textit{Euclid} simulations including all observational effects, the final pattern of ground-based complementary observations, and the estimation of photometric redshifts performed with the \textit{Euclid} code, will be available.

\subsubsection{Mock cluster catalogs}\label{sec:Halo_catalogs}
Dark matter halos were identified in the simulation using the algorithm defined in \cite{Jiang2014}, such that galaxies were given a group identifier and the central galaxies were marked. A cluster catalog was thus constructed by grouping galaxies that belonged to the same halo, using their unique identifiers. The coordinates of each cluster were taken to be those of the central galaxy, both in sky coordinates and redshift. We also observed that defining the mock cluster center using the barycenter of the member galaxies marginally impacts the results presented in this paper and differences are discussed hereafter whenever relevant. For each mock cluster we calculated the quantities RA$_{\rm min}$, RA$_{\rm max}$, Dec$_{\rm min}$, Dec$_{\rm max}$, i.e., the minimum and maximum right ascension and declination of the members. This defines a rectangular area that includes all the galaxies belonging to a given mock cluster.

The mock cluster masses, \textit{Dhalo} ($M_{\rm DH}$), were also defined according to \cite{Jiang2014}. The $M_{\rm DH}$ values are related to the masses that are generally used in observations, such as $M_{200}$\footnote{The mass $M_{200}$ corresponds to the mass enclosed within a radius $R_{200}$, within which the mean density of the cluster is equal to 200 times the critical density of the Universe at the cluster redshift.}. The median ratio between $M_{\rm DH}$ and $M_{200}$ is equal to about 1.25 and the distribution remains confined between $\gtrsim 1$ and $\lesssim 1.5$ at 90\% C.L., being fairly flat \citep{Jiang2014}. We note that in \cite{Jiang2014}, the mass ratio is well characterized up to $M_{\rm DH} \simeq 10^{14}$ $h^{-1}$ M$_{\odot}$. Given the smooth evolution of the ratio with mass over several orders of magnitude, we assume that extrapolation is accurate up to the high mass tail considered here, $M \sim 10^{15.5}$ M$_{\odot}$. The final mock cluster catalogs were constructed by selecting all clusters down to masses of $10^{13.25}$ M$_{\odot}$. The implications of this limit on our results is further discussed in sections~\ref{sec:Performance}~and~\ref{sec:Discussions}. Hereafter, the masses are referred as $M$.

The characteristic radius was estimated as $\tilde{R}_{200} \equiv \left[M / \left(\frac{4}{3} \pi \, 200 \, \rho_c\right)\right]^{1/3}$. This quantity is related to the mass of each mock cluster and uses the critical density at the cluster redshift, $\rho_c$, as computed from the mock cosmological parameters in the flat $\Lambda$CDM model. Because the masses we used are not defined as $M_{200}$, our estimates of $R_{200}$ are biased high by around 8\% for the median of the cluster population, and remain less than 17\% larger at 95\% C.L. It should be noted, however, that these $\tilde{R}_{200}$ values were only used to associate detected clusters to mock clusters and hence this does not significantly affect our results, as discussed further in Section~\ref{sec:Matching}.

\subsection{Properties of galaxies and galaxy clusters in the mocks}
To facilitate the interpretation of the results of the final CFC and to validate the simulations for our purposes, we explore the properties of the mock in terms of photometric redshift reconstruction, mass-richness relation, cluster galaxy density profiles and galaxy cluster LF. An analysis of the galaxy properties in the mock is provided in \cite{Ascaso2015}. In the following subsections we complement this analysis, particularly with regards to cluster environment.

\subsubsection{Photometric redshift properties}\label{sec:Photometric_redshift_properties}
The precision of the photometric redshift estimates is expected to have a significant impact on cluster finder performance. Clusters appear as overdensities not only in projected space, but also in redshift space, information that is used by the detection algorithms via the photometric redshifts. \cite{Ascaso2015} validated BPZ photometric redshifts comparing them to spectroscopic redshifts and assessing their performance in terms of resolution and outliers (see Section 5 of their paper and Tables 1 and 2). We briefly summarize their results and present an internal validation performed in the context of the CFC. \cite{Ascaso2015} showed that for the \textit{Euclid pessimistic} case $\sigma_{\rm NMAD} \leq 0.03$ for galaxy $m_{H} < 22.5$ and increases up to $\sigma_{\rm NMAD} \sim 0.08$ at $m_{H} \sim 24$, using the normalized median absolute deviation (NMAD)\footnote{The NMAD associated to the variable $X$ is defined as $\sigma_{\rm NMAD}(X) = 1.48 \ {\rm median} \left| X - {\rm median} \left(X\right)\right|$.}. When considering all magnitudes up to $m_{H} = 24$, $\sigma_{\rm NMAD} \leq 0.045$ for redshift $z<1.5$ and $\sigma_{\rm NMAD} \sim 0.06$ at $1.5<z<3$. These limits increase when using the {\it odds} parameter in BPZ (not used in the CFC). In terms of outliers, the {\it Euclid pessimistic} case shows a rate of outliers in the range 10-20\%, with the highest fractions in the redshift ranges $0.5<z<1$ and $2<z<3$. These results are shown in \cite{Ascaso2015} Tables 1 and 2 as a function of galaxy magnitude and redshift, and in Figures 17 to 22. As a general comment, the photometric redshift resolution of the {\it Euclid optimistic case} is a factor of two to five better than the pessimistic case both in terms of photometric redshift accuracy and bias.

We hereafter present the internal challenge validation of the photometric redshift quality in the simulation. For this, we follow \cite{Ricci2018}, adapted from \cite{Ilbert2006}\footnote{See also the photometric redshift release explanatory document \url{http://cesam.lam.fr/cfhtls-zphots/files/cfhtls_wide_T007_v1.2_Oct2012.pdf}}. For each redshift bin, we compute the difference $z_{\rm phot} - z_{\rm true}$, and use the resulting distributions to extract the bias, the catastrophic failure fraction and the dispersion. Here, $z_{\rm true}$ refers to the true spectroscopic redshifts. These values account for peculiar velocities, which are known for all the galaxies in the simulation and are not affected by selection effects. The bias is computed as the median of the distribution. The outlier fraction is given by the fraction of objects satisfying $\left| z_{\rm phot} - z_{\rm true} - {\rm bias} \right| > 0.15 \left(1+z_{\rm true}\right)$. The dispersion is computed both using NMAD as in \cite{Ascaso2015}, and percentiles by integrating the distributions up to a 68.2\% confidence level on the positive and negative parts. We also reproduce this analysis after removing galaxies with $H$-band $m_{H} > 23$ magnitude to highlight the effects of contamination from low S/N objects. We note that below this limit, the distribution remains fairly stable. Similarly, we reproduce this analysis by selecting cluster member galaxies above a given halo mass, to investigate potential environmental effects.

\begin{figure}[h]
\centering
\includegraphics[width=0.5\textwidth]{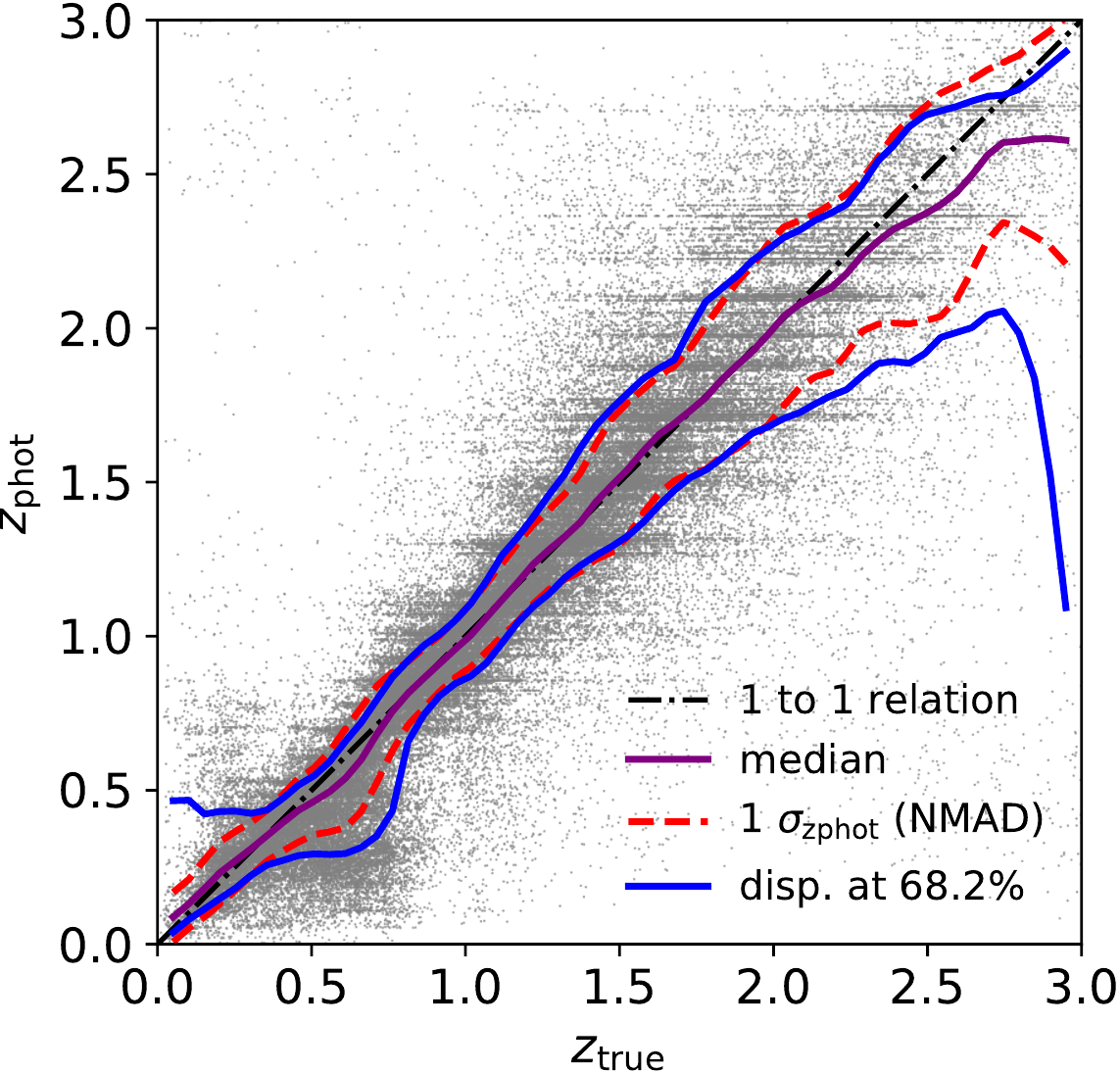}
\caption{\footnotesize{Comparison between photometric redshift, $z_{\rm phot}$, and true spectroscopic redshifts, $z_{\rm true}$. The bias is shown by the purple solid line, the NMAD is shown as the red dashed line, and the dispersion computed as percentiles is shown by the blue solid line. The black dashed-doted line provides the one-to-one relation for reference.}}
\label{fig:cluster_redshift_distribution}
\end{figure}

\begin{figure*}[h]
\centering
\includegraphics[width=0.49\textwidth]{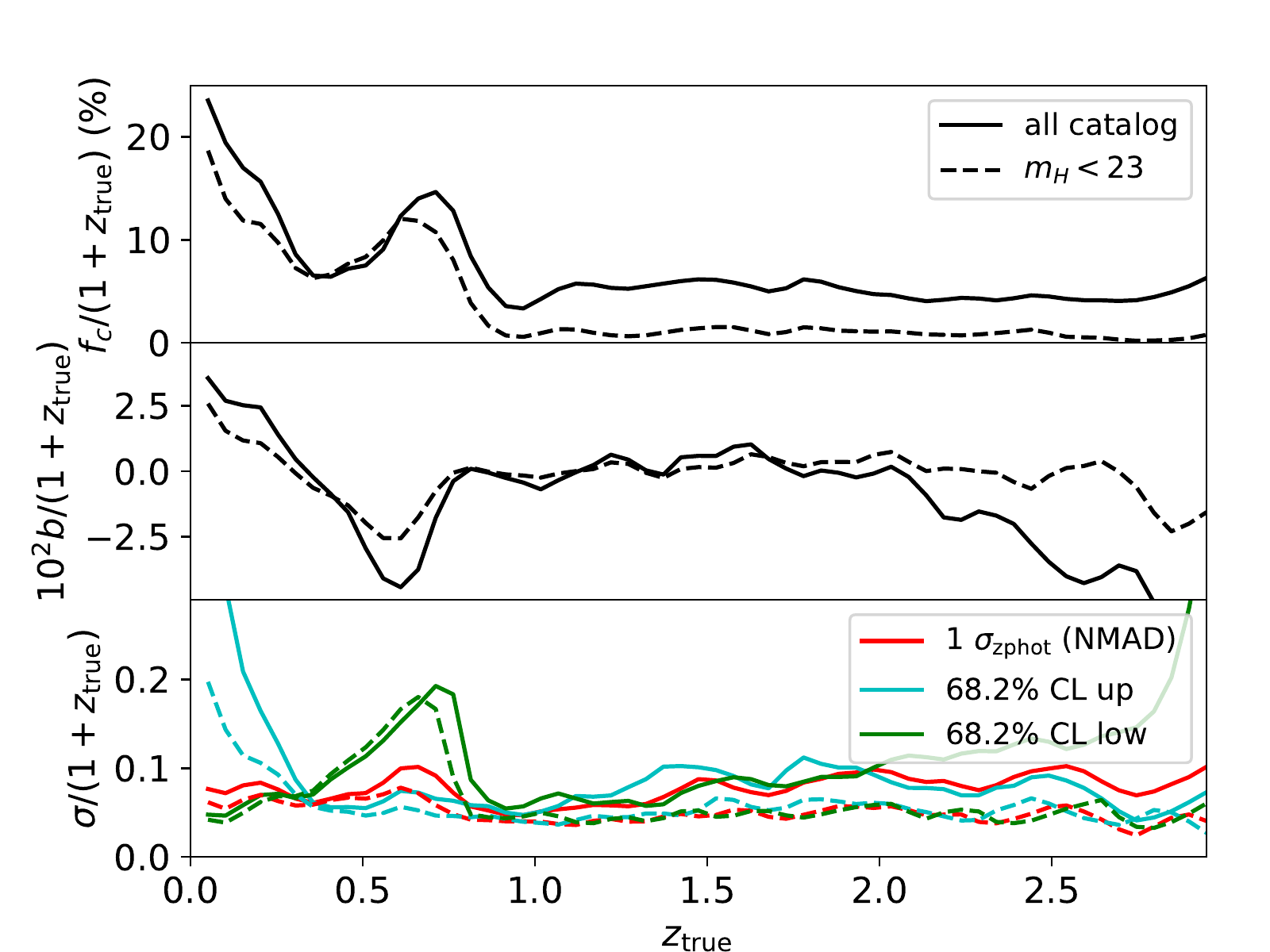}
\includegraphics[width=0.49\textwidth]{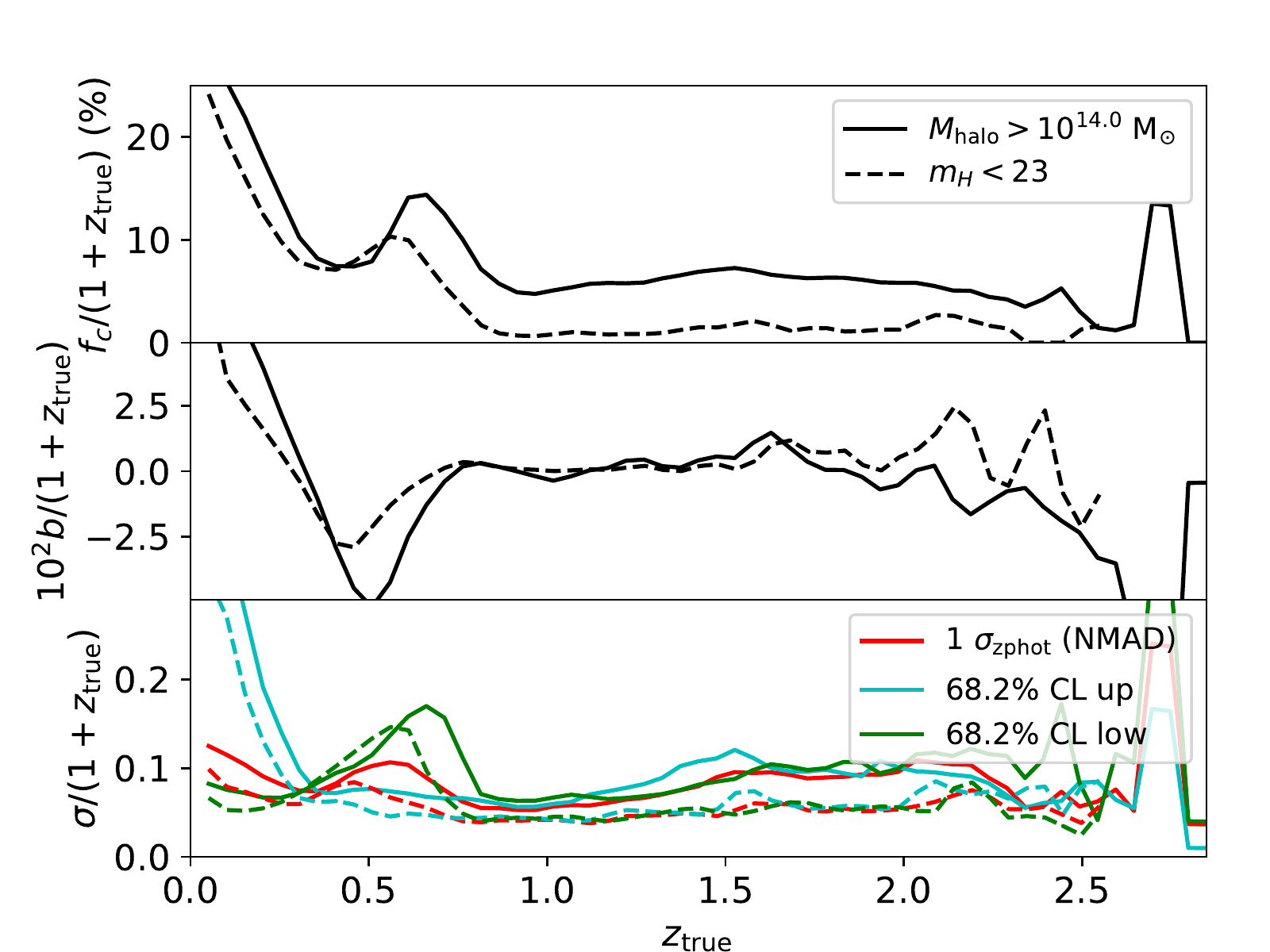}
\caption{\footnotesize{Redshift evolution of the catastrophic outlier fraction ($f_c$, upper panel), the bias ($b$, middle panel), and different estimates of the dispersions ($\sigma$, lower panel) as a function of spectroscopic redshift. The solid lines correspond to the full catalog, while the dashed lines correspond to the catalog once objects fainter than magnitude $m_{H} = 23$ are removed. Upper and lower values of the dispersion computed using percentiles with respect to the de-biased distributions are shown according to the legend. The left panel provides the distributions for the field plus cluster member galaxies and the right panel focuses on cluster member galaxies, i.e., those within haloes more massive than $10^{14}$ M$_{\odot}$.}}
\label{fig:zphot_vs_zcos_mock}
\end{figure*}

Figure~\ref{fig:cluster_redshift_distribution} shows the comparison between the true spectroscopic redshift, $z_{\rm true}$, and the photometric redshifts $z_{\rm phot}$, for a randomly selected subsample of galaxies from the mock ($\sim 10^5$ galaxies are shown). This figure also provides the bias and the two estimates of the dispersion. Figure~\ref{fig:zphot_vs_zcos_mock} shows the redshift evolution of the catastrophic outlier fraction (top panel), the bias (central panel) and the different estimates of the dispersion (bottom panel) for the full mock and after removing objects with $m_{H} > 23$. The left panel includes cluster and field galaxies while the right panel focuses on cluster member galaxies, belonging to haloes of mass larger than $10^{14}$ M$_{\odot}$. We measure the overall mean photometric uncertainty to $\sigma_{z_{\rm phot}} = 0.050 \left(1 + z_{\rm true}\right)$. The dispersion increases by a factor of $\sim 2$ and becomes very asymmetric at $z_{\rm true} \sim 0.5-0.6$. It also increases by a similar amount at redshifts below 0.2 and above 2.5 for the full catalog, but remains relatively flat for the high S/N catalog ($m_{H}<23$). The bias becomes large where the photometric uncertainties are large, even for the $m_{H}<23$ catalog. The fraction of catastrophic redshifts is small at redshifts above 0.8 ($\lesssim 0.05$ even for the full catalog, and about 0.01 for the high S/N catalog). However, it becomes large at lower redshifts, reaching up to 20\% for the full catalog and 15\% for the $m_{H}<23$ catalog. The distribution remains very similar in the case where cluster member galaxies are selected, independently of the exact value adopted for the mass cut. We note that the overall quality of the photometric redshifts measured corresponds to the pessimistic case, as expected from the catalog used. In the context of \textit{Euclid}, the standard deviation of the photometric redshifts with respect to the true redshifts is required to be $\sigma_z/(1+z) < 0.05$, keeping as a goal $\sigma_z/(1+z) < 0.03$ \citep{Laureijs2011}. Similarly, the catastrophic failures requirement is less than 10\% beyond $0.15 (1+z_{\rm true})$, while the goal is to keep this less than 5\% beyond $0.15 (1+z_{\rm true})$. Our internal validation is consistent with the mock validation performed in \cite{Ascaso2015} where a more optimistic case is also presented in addition to the pessimistic one used here. We note that the large number of outliers, the large bias and the large dispersion at redshifts below 0.3, above 2.3 or near 0.6 are largely due to the fact that no $u$-band is used in the pessimistic case, while it would be available in the optimistic case. 

Based on the photometric redshift properties of the catalog, we expect cluster finder detection properties to be altered in the redshift range in which the catastrophic outlier fraction is large ($z_{\rm true} \sim 0.5-0.6$, and $z_{\rm true} \lesssim 0.2$). This is even more true for clusters with fewer member galaxies (i.e., at lower masses). This alteration might show up as an increased number of false detections or larger uncertainties in the redshift recovery of the clusters, depending on how the photometric redshifts are used by the finders. The bias can also affect the matching performed to associate the detections to the true clusters (see Section~\ref{sec:Matching}). At redshifts $0.8<z_{\rm true}<2$, the photometric redshift distribution is nearly Gaussian (with small bias and a small catastrophic outlier fraction). Therefore, the cluster finders are expected to behave well despite the fact that the larger photometric errors and the lower number of galaxies, reduced by redshift dimming, should impact the completeness.

\subsubsection{Mass-richness relation}\label{sec:mock_mass_lambda_relation}
The richness of galaxy clusters is a fundamental quantity derived from optical or NIR surveys. It generally serves as the primary mass proxy and its normalization is tightly related to the detection performance at a given mass. In the context of the CFC, it was necessary to characterize the mass-richness relation of the mock itself in order to estimate the scatter introduced to richness measurements (see Section~\ref{sec:Performance}). See also the work by \cite{Ascaso2017} for the characterization of the cluster total stellar mass as a cluster mass proxy, using the same mock.

For each mock cluster, we compute an estimate of the richness as the number of galaxies associated to the halo as
\begin{equation}
	\lambda_{\rm mock} = N_{\rm gal} \left( m_{H} < m_{H, \ {\rm ref}}^{\star}(z_{\rm true}) + 2 \right).
\label{eq:richness_def}
\end{equation}
In order to account for a redshift dependence of the richness definition, through the magnitude evolution, we exclude galaxies with $m_{H}$ larger than $m_{H, \ {\rm ref}}^{\star} + 2$. This allows us to have a complete sample up to $m_{H} = 24$ at redshift 2.5 (see also the discussion on the LF in Section~\ref{sec:Cluster_galaxies_luminosity_function}). The reference magnitude $m_{H, \ {\rm ref}}^{\star}$ is derived from the passive evolution of a starburst galaxy with a formation redshift $z_{\rm form} = 3$ taken from the {\tt PEGASE2} library \citep[{\tt burst\_sc86\_zo.sed},][]{Fioc1997}. It is calibrated using the value of $K^{\star}$ at redshift 0.25 derived by \cite{Lin2006} from an observed cluster sample. The validity of this evolution is addressed in Section~\ref{sec:Cluster_galaxies_luminosity_function} (see also Figure~\ref{fig:luminosity_function_mock}, right panel) and the exact $m_{H, \ {\rm ref}}^{\star}$ model used to compute $\lambda_{\rm mock}$ has a negligible impact on our results, especially given that it reproduces well the trend seen in the mock at the relevant redshifts.

\begin{figure}[h]
\centering
\includegraphics[width=0.5\textwidth]{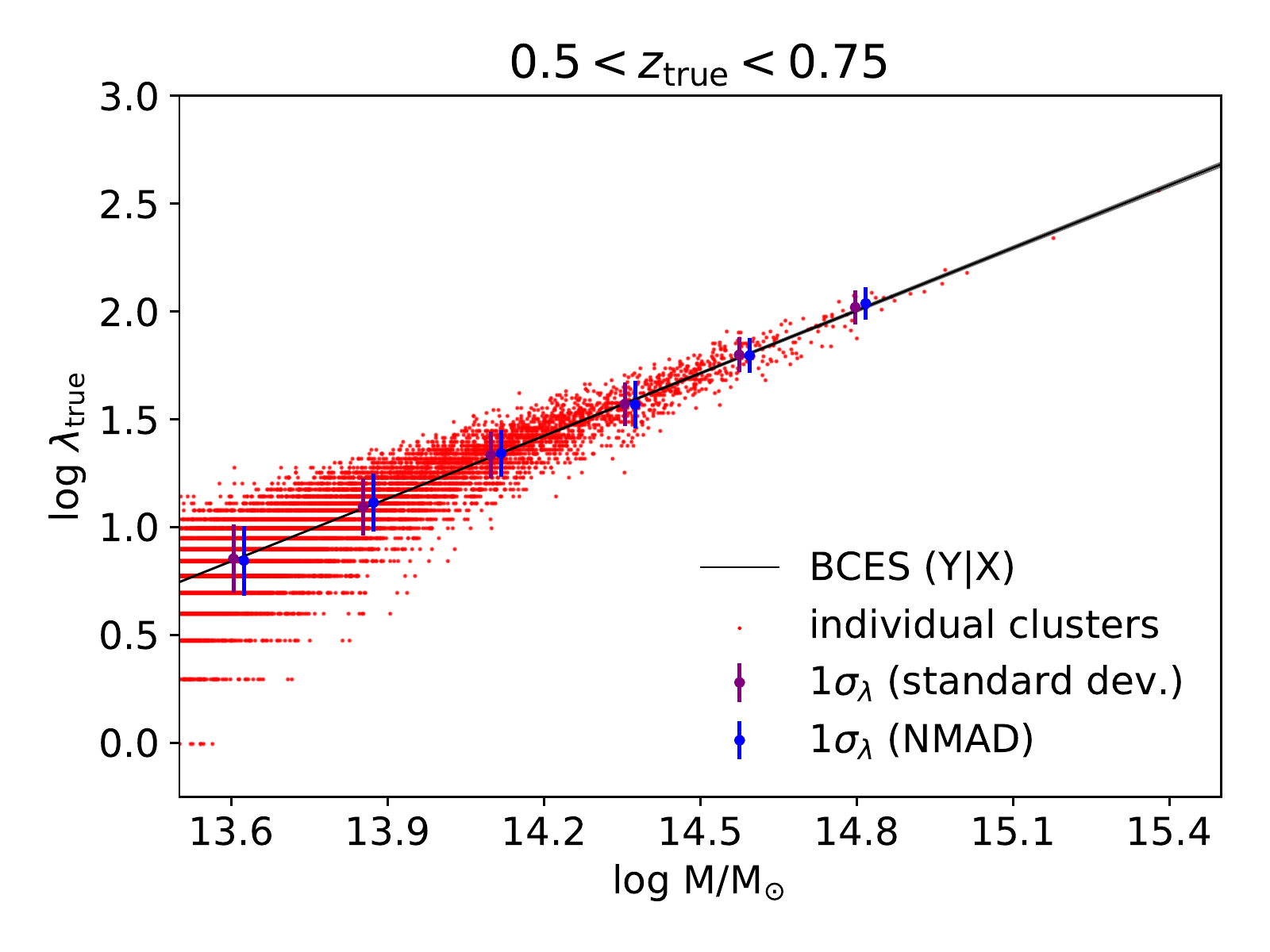}
\caption{\footnotesize{Example of the mass-richness scaling, for the redshift range $z_{\rm true} = [0.5, 0.75]$. The red dots show the cluster population. The blue points with error bars represent the median richness and scatter computed as the normalized median absolute deviation, while the purple points correspond to the mean richness and the scatter computed as the standard deviation within each bin.}}
\label{fig:mass_lambda_mocks}
\end{figure}

In Figure~\ref{fig:mass_lambda_mocks}, we provide an example of the scaling between the mass and the richness, computed for all clusters in the redshift range $[0.5-0.75]$. The mass-richness relation is modeled by power law and fitted using the bivariate correlated errors and intrinsic scatter \citep[BCES,][]{Akritas1996} method. The best-fit model is subtracted from the data and the residual is used to compute the scatter in the richness at fixed mass. The blue and purple dots provide the median and mean richness of the corresponding mass bin, while the error bars represent the scatter computed as the NMAD and the standard deviation, respectively. While the standard deviation is accurate for lognormal scatter, the NMAD is more robust to outliers and we use it as the baseline. The differences between the two methods are insignificant. The slope is consistent with unity within a few percent at all redshifts. The scatter does not significantly evolve with redshift (not shown), but it does decrease linearly with $\log M$ ($\sigma_{{\rm log} \lambda} \simeq 0.1$ at $M = 10^{13.5}$ M$_{\odot}$ and $\sigma_{\log \lambda} \simeq 0.05$ at $M = 10^{14.5}$ M$_{\odot}$). This intrinsic scatter will be later used when quantifying the scatter introduced by the detection algorithms in Section~\ref{sec:Performance}. We observe outliers at low richness in the scaling relation when using the mock cluster catalog based on the barycenter of cluster galaxies (not shown). They correspond to clusters that are on the edge of the footprints since their number of member galaxies is generally truncated while their mass remains the same. In principle, these clusters also affect the detections, but we have observed that they have a negligible impact on the global performance presented in this paper. In practice, the \textit{Euclid} survey will be affected by masks, or varying depth, but at this stage not all the algorithms are able to handle such effects and we leave the investigation of their impact on the detection of galaxy clusters for future work.

\subsubsection{Cluster galaxy density profile}
The radial structure of galaxy clusters is a key property, which may affect any cluster finder. Since the CFC detection algorithms are driven by photometric data with photometric redshift uncertainties that are much larger than the cluster extent along the line of sight, they are mainly sensitive to the projected galaxy radial number density distribution of the clusters, $\Sigma (R)$. We therefore investigate the projected radial profiles of the mock clusters by stacking galaxies belonging to clusters in mass and redshift bins. Prior to the stacking, we normalize the projected clustercentric distances by the characteristic radius $R_{200}$.

In order to study in a quantitative way the mass and redshift evolution of the profiles and compare it to observations from the literature, we use the following approach. We model the profiles by a Navarro, Frenk and White \citep[NFW,][]{Navarro1996} distribution, as expected from observations \citep[e.g.,][]{Carlberg1997,Lin2004}. However, as we observe a deficit of galaxies in the outskirts of the profiles, we also include a truncation radius, $r_{\rm max}$, above which the number density of galaxies is set to zero. The 3D profile of the cluster galaxy space density, $n$, can thus be written as
\begin{equation}
	n(r/R_{\rm 200}) =  \frac{n_0}{\left(c\, r/R_{200} \right) \left( c \, r / R_{200} + 1\right)^2} \mathcal{H}\left(r_{\rm max} - r\right),
\label{eq:tNFW}
\end{equation}
where $\mathcal{H}$ is the Heaviside step function, $n_0$ the normalization, $c = R_{\rm 200} /r_c$ the concentration, with $r_c$ a characteristic radius, and $r_{\rm max}$ a truncation radius. We fit the stacked normalized number surface density profiles, $\Sigma(R)$, as described by equation~(\ref{eq:tNFW}), using the analytical projection given in \cite{Mamon2010}. The parameter space (normalization $n_0$, number concentration $c$, and truncation radius $r_{\rm max}$) are sampled using a Markov Chain Monte Carlo method, using the algorithm described in \cite{Adam2015}.

\begin{figure*}[h]
\centering
\includegraphics[height=6.0cm]{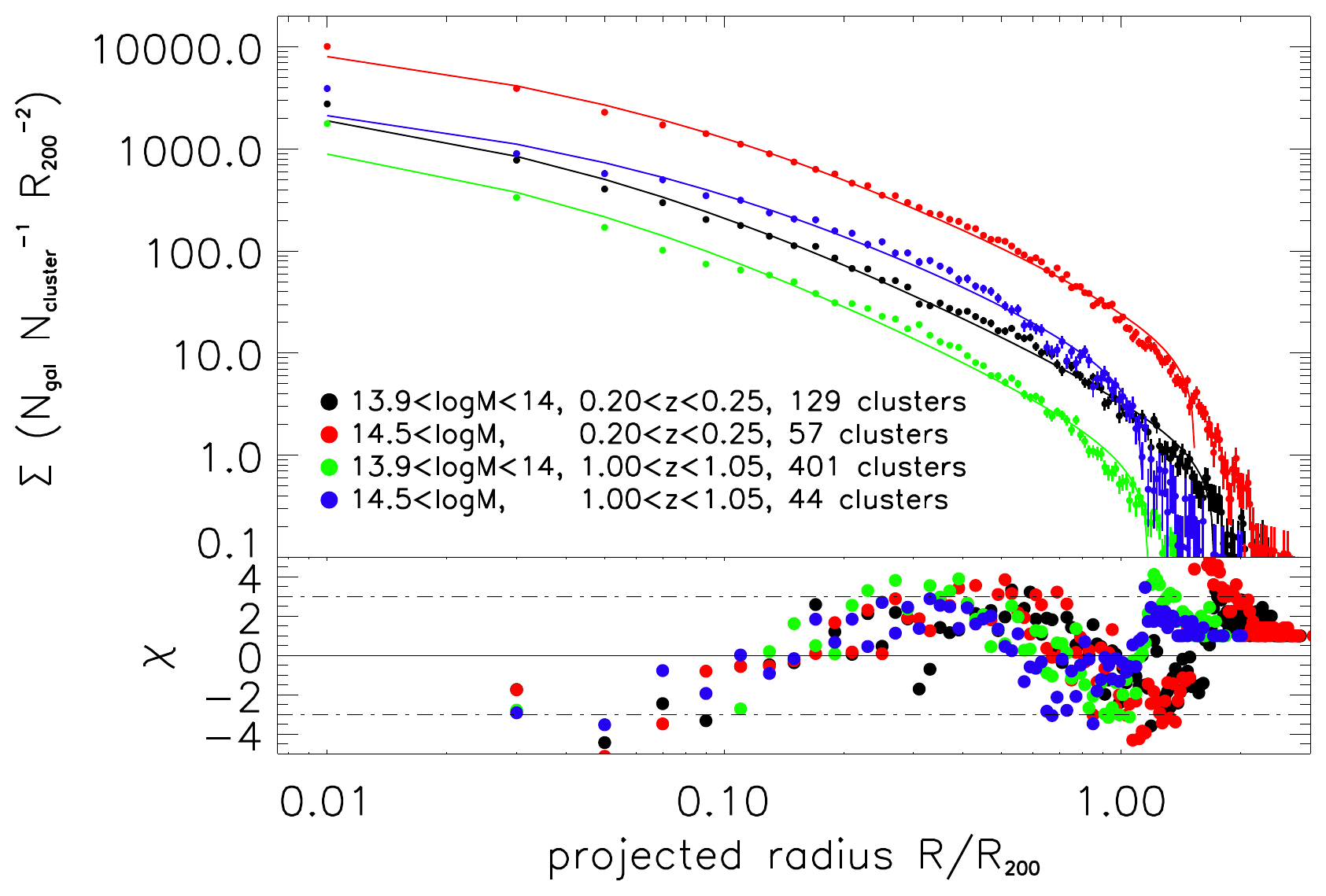}
\includegraphics[height=6.0cm]{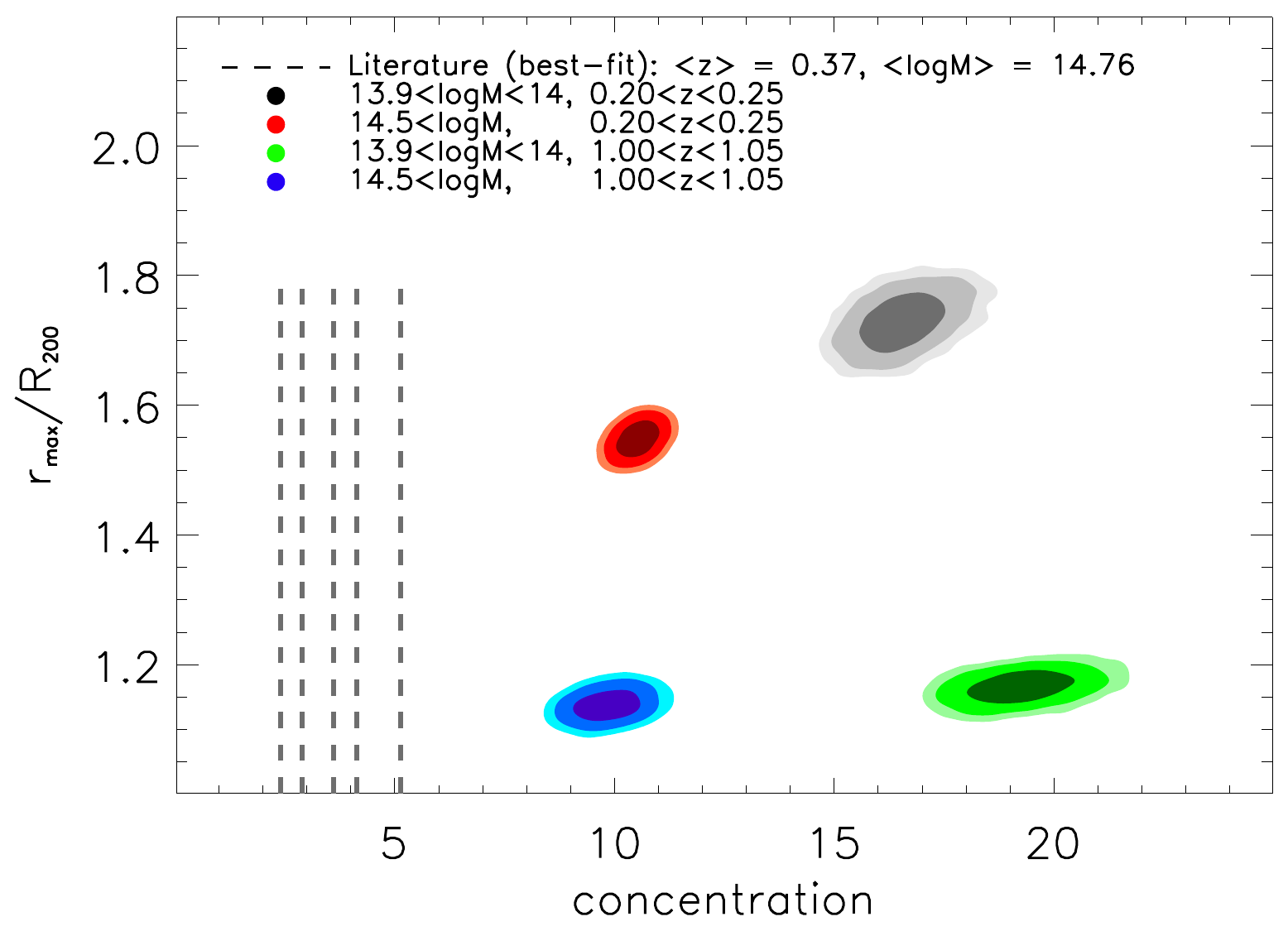}
\caption{\footnotesize{{\bf Left:} stacked surface density (projected) profile of cluster galaxies. The different colors indicate different mass and redshift bins, as indicated in the legend. The solid lines provide the best-fit models of equation~(\ref{eq:tNFW}) in each case. The residual normalized by the error, $\chi$, is also provided. {\bf Right:} posterior likelihood on the model truncation radius parameters $r_{\rm max}$ and concentration $c$ for each bin, providing the 68\%, 95\% and 99\% C.L. The vertical dashed lines represent the best-fit number concentration cluster observational data from the literature, namely: $c = 2.90 \pm 0.22$ \citep[at a median redshift ${\rm med}(z) = 0.04$, and median mass ${\rm med}(M_{200}) = 4 \times 10^{14}$ M$_{\odot}$,][]{Lin2004}, $c = 4.13 \pm 0.57$ \citep[at ${\rm med}(z) = 0.31$, ${\rm med}(M_{200}) = 3 \times 10^{14}$ M$_{\odot}$,][]{Muzzin2007}, $c = 5.14^{+0.54}_{-0.63}$ \citep[at ${\rm med}(z) = 1.00$, ${\rm med}(M_{200}) = 2 \times 10^{14}$ M$_{\odot}$,][]{vdb2014}, $1/c = 0.278 \pm 0.065$ \citep[at ${\rm med}(z) = 0.06$, ${\rm med}(M_{200}) = 6 \times 10^{14}$ M$_{\odot}$,][]{vdb2014}, $c = 2.40 \pm 0.30$ \citep[at ${\rm med}(z) = 0.44$, ${\rm med}(M_{200}) = 14 \times 10^{14}$ M$_{\odot}$,][]{Annunziatella2014}.}}
\label{fig:profile}
\end{figure*}

The left panel of Figure~\ref{fig:profile} provides the stacked projected profiles of clusters in four redshift and mass bins together with the best-fit models. Overall, the clusters are relatively well described by a truncated NFW model. However, some excess is seen above the best-fit truncation radius, probably due to the fact that each cluster may present a slightly different $r_{\rm max}$ value, while we are introducing blurring in the profile when stacking and only fitting for a unique $r_{\rm max}/R_{200}$. In addition, the mock clusters present a significantly shallower slope in the center. The best fits are thus slightly biased high in the center, and biased low in the intermediate regions, as seen in the residual. The right panel of Figure~\ref{fig:profile} gives the marginalized posterior likelihood for the parameter $r_{\rm max}$ versus $c$. The truncation radius decreases with redshift, being $r_{\rm max}/R_{200} \sim 1.1-1.8$. Such a trend could be due to the fact that $r_{\rm max}$ measures more closely the virial radius, which is defined at higher densities at higher redshifts, leading to radii that will be smaller. However the size of the effect we find is larger than expected. This truncation is not expected from observations, which indicate that the intrinsic cluster number density profile (not counting galaxies in other groups for clusters) extends to over ten virial radii \citep{Trevisan2017}. 

The number concentration parameter, is $c \sim 10$ at high masses ($>10^{14.5}$ M$_{\odot}$) and increases up to 20 at lower masses (about $10^{13.9}$ M$_{\odot}$). We note that the values of number concentrations found in these simulations are higher with respect to those estimated fitting radial number density and stellar mass density profiles of satellite galaxies in observed massive clusters by a factor of about two or more, depending on mass and redshifts \citep{Carlberg1997,Lin2004,Collister2005,Muzzin2007,vdb2014,vdb2015,Cava2017}. Some of these observational values of concentration and truncation radius (normalized to $R_{200}$) are reported in the right panel of Figure~\ref{fig:profile}, showing a significant offset with respect to the values estimated from the mock. This discrepancy is similar to that found by \cite{Budzynski2012} comparing number density profiles estimated from SDSS DR7 groups and clusters and predicted profiles from semianalytical modeling of galaxy formation. Indeed, the treatment of the galaxy mergers in the model is shown to impact on the profile shape. When a galaxy becomes a satellite, an analytic estimate of the merger time is made and the galaxy merges regardless of whether or not its host sub-halo can still be resolved. According to the way this merger dynamical timescale is calculated may lead to steeper inner satellite number density profiles in the case of semianalytic models as compared to observed ones. Our main concern here is if this difference could hamper our performance estimation of cluster finders.

As highly concentrated clusters are expected to be more easily identified by cluster finders, this high concentration potentially affects the detections. This may boost high the absolute estimate of the performance, in particular for low S/N objects. However, all the cluster finders are density-based, so their relative performance should not be affected by the higher concentrations. In addition, the truncation of the simulated clusters at 1 to 2 $R_{200}$ facilitates the distinction of the cluster with the background galaxy density, helping the cluster finders limit the dimensions of the clusters on the sky. However, this effect is likely to have a minor impact on the results because the truncation happens at large radii and only marginal effects are visible in the inner part of the clusters once projected along the line of sight.

\subsubsection{Cluster galaxy luminosity function}\label{sec:Cluster_galaxies_luminosity_function}
Another important cluster property which can {\it a priori} affect its detection is its luminosity function. If the galaxy luminosity function in the mock clusters was significantly different from that of the real data, this may impact on the estimate of the absolute performances of the cluster finders. We note however that this is not of major importance for this analysis, in which we are mostly interested to the relative properties of the cluster finders.

We follow the same approach as for the profiles in order to investigate the LF of galaxy clusters within the simulations \citep[see also][where the mock galaxy luminosity and mass functions were shown to be in good agreement with observations]{Ascaso2015}. We count the number of cluster galaxies in bins of magnitude (in the $H$-band prior to introducing any noise on the galaxy fluxes), within a projected radius of $R_{200}$ and per Mpc$^2$. This is done after selecting clusters within bins of mass and redshift. The LF is then fitted by a Schechter function \citep{Schechter1976}, given by \citep[see e.g.,][]{Driver1994}
\begin{equation}
	\Phi(m) =  0.4 \ {\rm log}\left(10\right) \phi^{\star} 10^{0.4 (m^{\star}-m) (\alpha+1)} {\rm exp}\left(-10^{0.4 (m^{\star}-m)}\right).
\label{eq:shechter}
\end{equation}
As in the case of the galaxy density profile, we fit for the parameters $\phi^{\star}$, $m^{\star}$, and $\alpha$, which set the normalization, the characteristic magnitude, and the faint-end slope of the population, respectively. Several observational estimates of the cluster LF have shown that a single Schechter function may not reproduce well both the bright and the faint part of the LF, for various reasons \citep[e.g.,][]{Popesso2005,Barkhouse2007,Yang2008,Trevisan2017b}. However it can be used successfully to model its bright part. Here, the Schechter function is not able to describe the mock LF in the faint part (typically $m>m^{\star} + 3$), where a more sophisticated modeling would be necessary. Therefore, we first focus on the bright end of the LF studying the evolution of the parameter $m^{\star}$. To do so, we perform the fit of equation~(\ref{eq:shechter}) in the magnitude range limited to $m_{\rm brightest}+3$, where $m_{\rm brightest}$ is the magnitude of the brightest galaxy in the bin we consider. This ensures good modeling of the mock LF in this regime. We check that our best-fit is not sensitive to this magnitude limit. The faint end properties of the LF are addressed as a function of redshift without relying on a model.

\begin{figure*}[h]
\centering
\includegraphics[height=4.9cm]{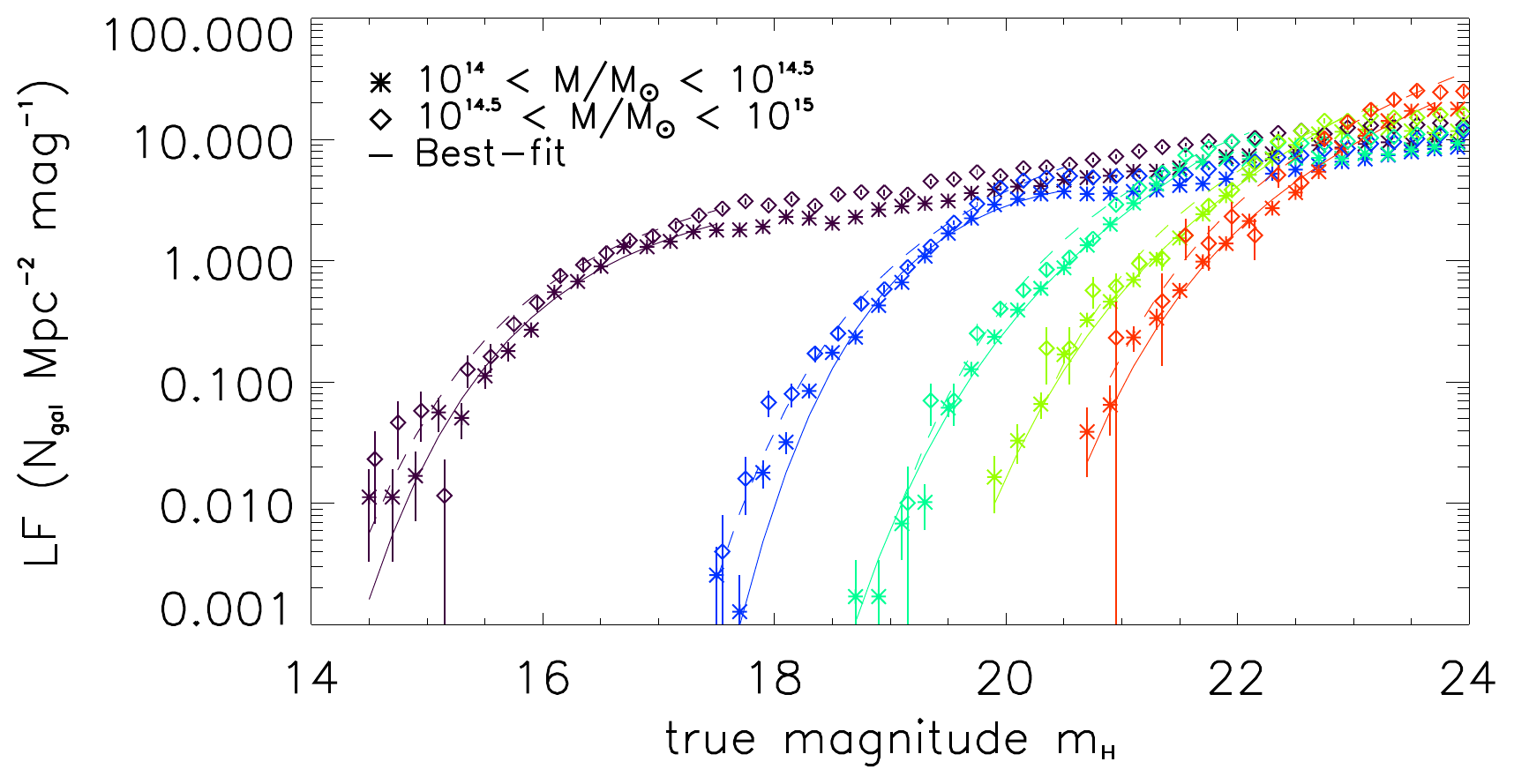}
\includegraphics[height=4.9cm]{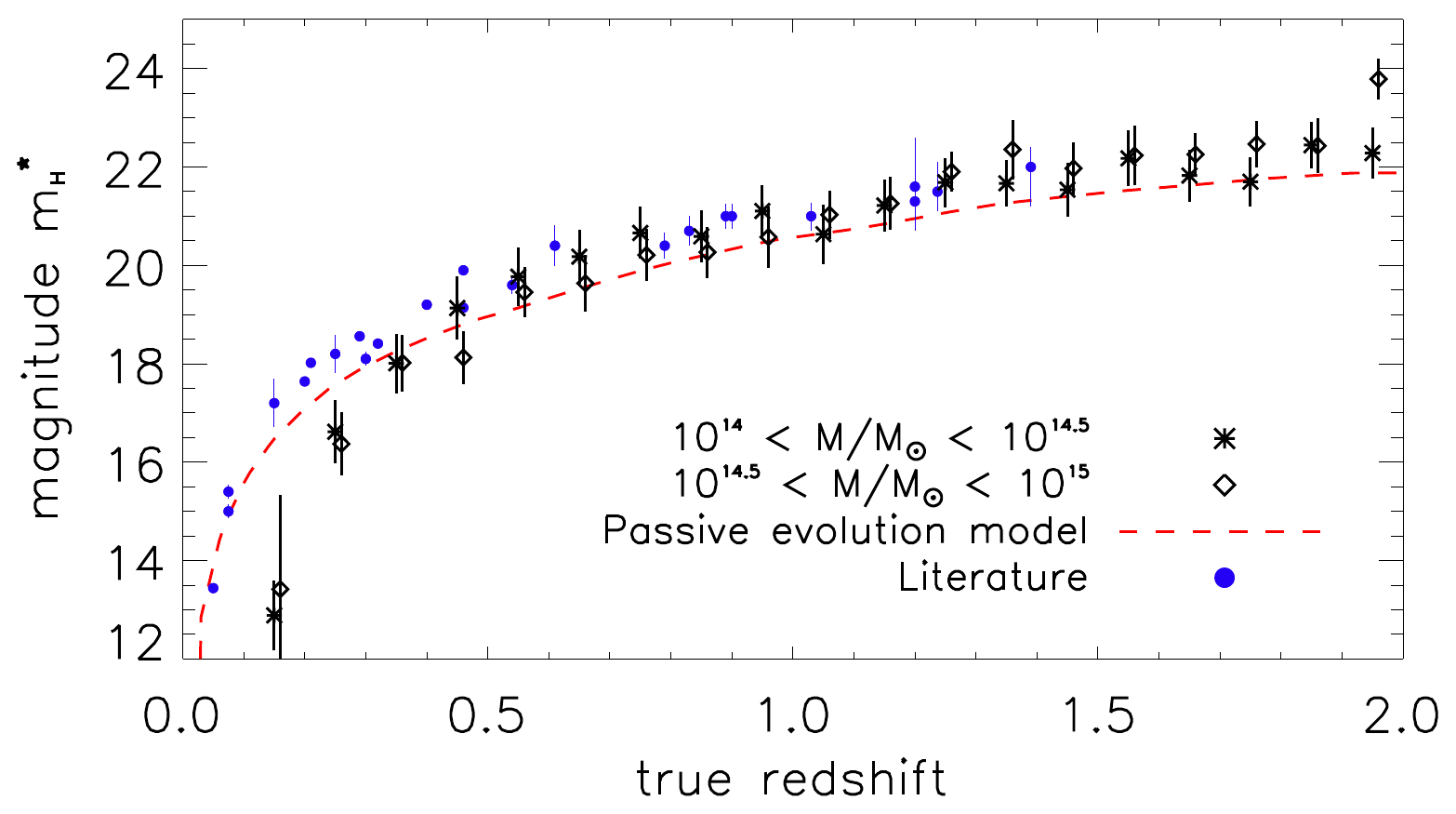}
\caption{\footnotesize{{\bf Left:} stacked LF of cluster galaxies. The different colors indicate different redshift bins, of width 0.1, used to compute them. Only redshift bins centered on 0.15 (purple), 0.55 (blue), 0.95 (cyan), 1.35 (green) and 1.75 (red) are shown for clarity. As indicated in the legend, the star and diamond symbols correspond to the two mass bins, in the range $10^{14}-10^{14.5}$ M$_{\odot}$ and $10^{14.5}-10^{15}$ M$_{\odot}$, respectively. The solid and dashed lines provide the best-fit models of equation~(\ref{eq:shechter}) in the bright magnitude regime, in the low and high mass bins, respectively. We note that in the high mass bin, the number of clusters per bin may be less than 10 at redshifts larger than 1.5, and reaches 2 in the last bin. {\bf Right:} redshift evolution of the parameter $m^{\star}_{H}$, for each mass bin using similar symbols, and comparison to the passive evolution model, as the red dashed line, from \cite{Fioc1997}, and calibrated using the work by \cite{Lin2006}. The blue points indicate $m_H^{\star}$ values from the literature (see text). The error bars provide the standard deviation of the posterior distribution of the parameters $m^{\star}_{H}$, but we stress that the distributions are generally non gaussian and non symmetric \citep[see][for a detailed discussion on this topic]{Ricci2018}.}}
\label{fig:luminosity_function_mock}
\end{figure*}

The left panel of Figure~\ref{fig:luminosity_function_mock} provides the cluster galaxy LF in two bins of mass (above $10^{14}$ M$_{\odot}$) and five redshift bins (among the twenty considered, from $z=0$ to $2$). We observe that the mock LF are well described by the Schechter function in the bright regime, but that the faint part may require more sophisticated modeling. The right panel of Figure~\ref{fig:luminosity_function_mock} compares the evolution of the best-fit $m^{\star}$ parameter to a passive evolution model derived from \cite{Fioc1997}, as well as from data taken from the literature. The blue points indicate $m_{H}^{\star}$ values from \cite{DePropris1999,Nakata2001,Ellis2004,Lin2004,Toft2004,Andreon2005,Strazzullo2006,Muzzin2007,Strazzullo2010,DePropris2017}. They were obtained from studies of K-band cluster luminosity functions at different redshifts. We converted the $m_{K}^{\star}$ values to the H-band using the early-type k-corrections of \cite{Mannucci2001} and the mean rest-frame color for cluster galaxies, $m_{H} - m_{K} = 0.26$, obtained as an average of the values provided by \cite{Boselli1997,dePropris1998b,Ramella2004}, and adopting when needed the transformation to the AB-system $m_{H_{\rm AB}} = m_{H} + 1.37$ \citep{Ciliegi2005}. The evolution of the mock is relatively well described by the model and matches well the literature data at redshift larger than 0.3, for the two mass bins considered, but the value of $m^{\star}$ is overall lower by about 0.5 magnitude for the passive evolution model. At lower redshifts, the evolution is stronger with redshift and the mock $m^{\star}$ values are lower than the model and the literature values.

We have also investigated if the performance of the cluster finders could be affected differently according to the way the luminosity function is used in the detection process. While sFoF and HCFA algorithms do not make use of the luminosity function, AMASCFI, AMICO, PZWav and WaZP do. In the case of AMICO, the procedure adopted is fully general and treats the mock as real data. The procedure starts from an initial simple model (built in a blind way) with a luminosity function extracted from all galaxies in the catalog. AMICO is run to define a first set of detections that have been used to refined the cluster model, now introducing a different LF for clusters and field. Finally,  AMICO is run with this refined model to derive the final catalog. In the case of PZWav and WaZP, a value of $m^{\star}$ derived from passive evolution model is used to define a constant stellar mass threshold with redshift for detection. However, the dependance of the performance on the $m^{\star}$ cut was tested and found to be negligible. AMASCFI, PZWav and WAZP also use $m^{\star}$ parametrization for richness estimation, but here again richnesses are only used as relative quantities. Therefore, the impact of different uses of the luminosity function by the cluster finders is expected to be negligible on their relative performance.

In addition to the LF itself, we have checked the luminosity differences between the BCG and the central galaxies (i.e., the one coincident with the dark matter halo center in the mocks). The BCG is coincident with the central galaxy in about 70\% of the clusters. This number increases with mass, reaching nearly 100\% for the most massive clusters. When the BCG is not the central galaxy, the distance from the BCG to the cluster center (either defined as the central galaxy or the barycenter), is about 0.45 $R_{200}$, decreasing by a few percent as mass increases. However, the distribution extends up to around 2$R_{200}$ in the low mass clusters. Even when it is not the BCG, the central galaxy is among the brightest members and the magnitude difference with the BCG does not exceed $\Delta m_{H} \sim 2$, or $\Delta m_{H} \sim 0.5$ at high mass. The differences between the BCG and the central galaxy can affect the cluster finders to some extent, but we note that no finder relies on the BCG directly. As discussed in Section~\ref{sec:Matching}, the associations between the detection based on the BCG and the clusters in the mock could even be missed in a small fraction of the cases, but we have verified that this does not significantly impact the results. We have also checked that the distribution of halo BCG magnitudes in the mocks was in good agreement with observations.

Another important property of the galaxy distribution to be fiducially reproduced by the mocks is the color distribution. We do not focus on that point in this paper since none of the cluster finders participating in the last CFC was relying on galaxy colors. We refer to the work by \cite{Ascaso2015} who found a good agreement in the red sequence properties and the blue valley location between mocks and observation in the redshift range $[0.3, 1.65]$.

\section{Mock cluster to detected cluster associations}\label{sec:Matching}
The assessment of the performance of an algorithm requires associating the candidate clusters and the mock clusters, which are known from the simulation \citep[see e.g.,][and in particular their Figure 3]{Knobel2009}. In this section, we present the methodology developed to perform this association as well as an estimation of the corresponding systematic effects.

\subsection{Matching procedures}
The association between candidate clusters and mock clusters, or any pairs between cluster catalogs, is a non-trivial task. In order to validate our methodology and test for systematic effects, we have developed three different matching methods. They are hereafter referred to as {\tt geometrical}, {\tt ranking}, and {\tt membership} matching. The matching can generally be performed in two ways, starting from the mock clusters and searching for associated detections, or starting from the candidate clusters and searching for counterparts in the mock. We define the one-way associations as the clusters for which the association has been made in one direction, but not the other one. Similarly, we define the two-way associations as the ones for which the associations are bijective.

\begin{figure}[h]
\centering
\includegraphics[width=0.4\textwidth]{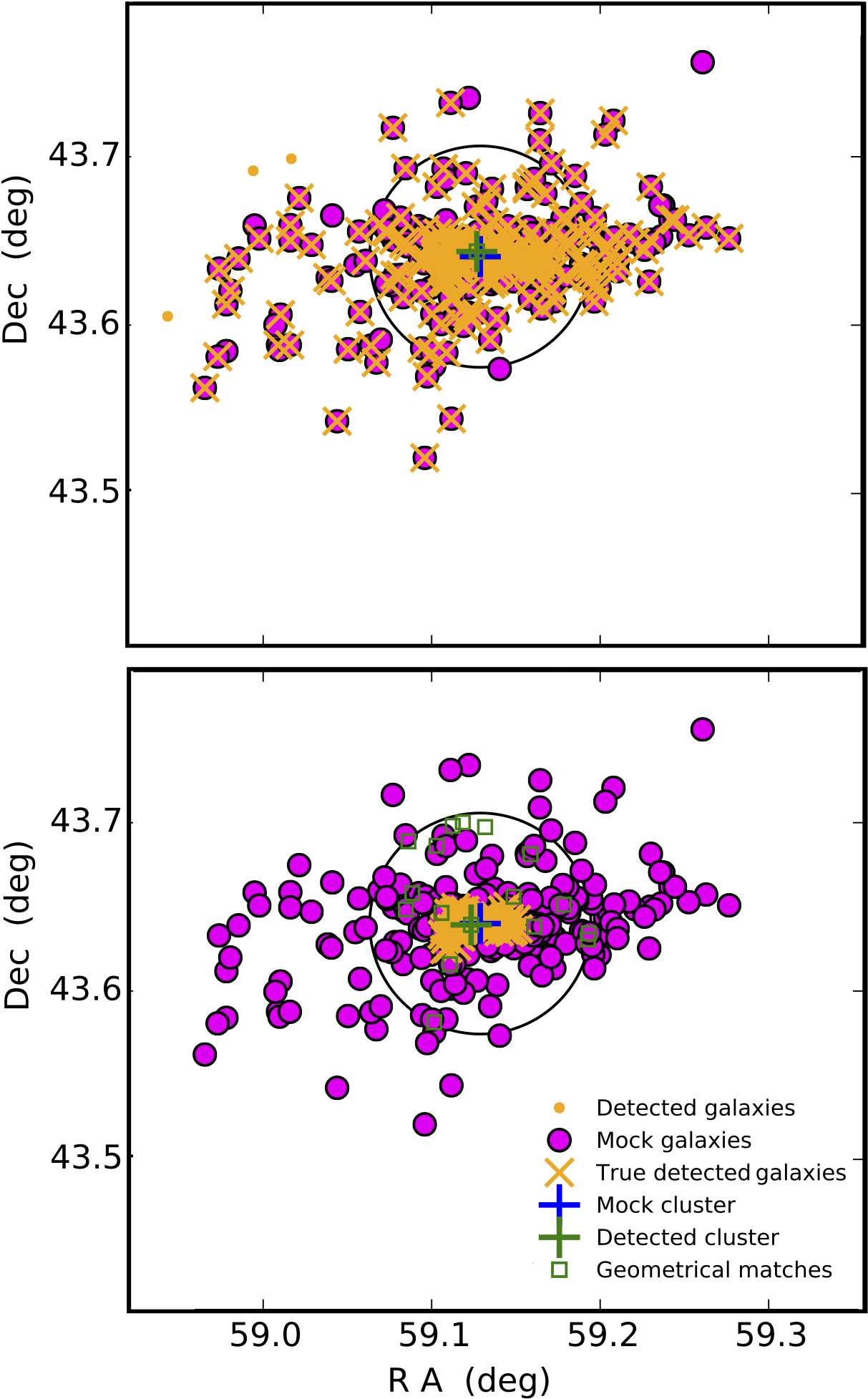} \put(-155,165){\makebox(0,0){\rotatebox{0}{\large sFoF}}} \put(-155,325){\makebox(0,0){\rotatebox{0}{\large AMICO}}}
\caption{\footnotesize{Illustration of the matching procedure performance in the case of a test cluster ($z=0.285$, $M = 10^{14.04}$ M$_{\odot}$), for two algorithms for which the galaxy membership probability was available (AMICO on the top and sFoF on the bottom panel). The true cluster galaxies, known from the mock, are given in purple. The cluster galaxies identified as members by the cluster finders are given in yellow (only galaxies with $P_i > 0.25$ are shown for clarity), with a cross on top in case they are true cluster members. The target mock cluster is given as a blue cross. All the candidate clusters that lie within the association volume are given as green squares, with a cross on top of the one that is the best match. The black circle represents $\theta_{200}$. In the case of sFoF, 17 one-way {\tt geometrical} associations would be possible (number of green squares), highlighting a large fragmentation rate.}}
\label{fig:matching_illustration}
\end{figure}

\subsubsection{Geometrical matching}
The {\tt geometrical} matching method is implemented via the following steps.
\begin{enumerate}
\item For each mock cluster, we search for detection counterparts within a volume around the mock cluster. The volume depth along the redshift axis is controlled by the parameter $\Delta z_{\rm match} = k \ \sigma_0 (1+z)$ where $\sigma_0 = 0.05$ (see Section~\ref{sec:Photometric_redshift_properties}). We use $k=4$, i.e., a width of four times the typical photometric error at the given true redshift to ensure avoiding missing matches and minimize false associations. We do not consider any photometric redshift bias or dependence of scatter with redshift, as given in Figure~\ref{fig:zphot_vs_zcos_mock}. As a result of the photometric redshift uncertainty and inaccuracy, cluster finders for which redshifts are inaccurately calibrated might loose detections that will be considered as impurities, lowering the completeness of the sample. The footprint of the volume, in terms of sky coordinates, is first restricted to the extent of the galaxies belonging to the mock cluster: RA$_{\rm min}$, RA$_{\rm max}$, Dec$_{\rm min}$ and Dec$_{\rm max}$. This ensures that the galaxies that are driving the detection are true cluster members and not nearby line-of-sight projected structures. In addition, the volume footprint is restricted to be within $\theta_{200}$ of the mock cluster, the angular radius corresponding to $R_{200}$, given the mock cosmological parameters. In the case of massive and nearby clusters, this last condition is more restrictive than the first one. However, as redshift increases and mass decreases, the number of cluster galaxies remaining above the mock flux limit drops, and for a given cluster, all the mock cluster galaxies are eventually enclosed within $\theta_{200}$. In this case, this secondary constraint becomes ineffective with respect to the first one.
\item In the case of multiple counterparts within the volume, we define the matched cluster as the one which is the closest (projected on the sky) to the mock cluster. Nevertheless, we record the total number of possible matches for all mock clusters, as they correspond to fragmented detections.
\item We repeat the first step (search for counterparts in the volume around the cluster), using the candidate clusters as the reference and searching for mock counterparts. While the redshift criterium is symmetric and remains the same, it is not the case for the projected area because the detection algorithms do not provide a characteristic radius of the detected objects. Therefore, a mock cluster is associated with the detection if it is at a projected distance that is lower than its own $\theta_{200}$.
\item We repeat step 2 with candidate clusters as the reference.
\item By comparing mock and cluster detection counterparts, we identify mock clusters and detected clusters for which the association is identical both ways.
\end{enumerate}
This method allows us to define both the one-way and the two-way associations. In case a mock cluster is associated with a multiple number of detections, this indicates that fragmentation has occurred and this is an important quality assessment of a cluster finder. Similarly, detected clusters that are matched to multiple mock clusters correspond to over-merging events. The two-way {\tt geometrical} matching is taken as the baseline method in the present paper.

\subsubsection{Ranking matching}
The {\tt ranking} matching method follows the same initial condition as the {\tt geometrical} matching (first step: search for counterparts in the volume around the cluster). However, instead of performing the matching both ways, it associates candidate clusters to mock clusters after ranking them by decreasing mass, as provided from the mock catalogs, and richness, as provided by the cluster finders, respectively. The richest detected clusters are then matched to the most massive mock clusters, and subsequently removed from the list. If two or more clusters have the same richness within the association volume, the nearest one to the mock cluster center is selected. Because detected and mock clusters are subtracted from the cluster list as they are matched to one another, this matching procedure is bijective by construction and thus corresponds to a two-way matching. It cannot, therefore, be used to address fragmentation and over-merging issues. The {\tt ranking} matching follows the idea that the most massive clusters, i.e., the richest ones, are the first ones detected.

\subsubsection{Membership matching}
The {\tt geometrical} matching and the {\tt ranking} matching do not directly rely on the cluster member galaxies. In contrast, the third method we developed, hereafter {\tt membership} matching, consists in defining the associations using the galaxies that are detected as cluster members by the algorithms. Because not all the detection algorithms provide the galaxy membership information, this method is only used as a crosscheck (see Table~\ref{tab:cluster_finder_summary} for the algorithms that provide the membership). The main steps of the procedure are summarized as follows.
\begin{enumerate}
\item For each mock cluster, we search for detection counterparts within the volume as defined in the case of the {\tt geometrical} matching.
\item If matches are found, we define the fraction of common galaxies between the candidate cluster and the mock cluster, with respect to the mock cluster (i.e., the success rate) as
\begin{equation}
	f_{\rm com, \ mock} = \frac{\sum_i P_i^{\rm match}}{N_{\rm gal, \ mock}},
\label{eq:common_fraction1}
\end{equation}
where $P_i^{\rm match}$ are the cluster membership probabilities of the galaxies which are indeed true cluster members and $N_{\rm gal, \ mock}$ is the number of galaxies that belong to the cluster according to the mock catalog. The matches are then accepted if $f_{\rm com, \ mock} > f_{\rm cut}$, with $f_{\rm cut}$ a free parameter of the matching. In the case of possible multiple matches, the candidate cluster with the largest $f_{\rm com, \ mock}$ is defined as the best match, but we still record the total number of possible matches for fragmentation estimates. Since the definition of membership probability is different for the different detection algorithms (e.g., a large number of galaxies dominated by low probability objects due to large photometric redshift error, or few galaxies with binary membership probabilities), the minimum fraction of common galaxies was taken to be $f_{\rm cut} = 0$, i.e., any cluster with non-zero $f_{\rm com, \ mock}$ was considered as a possible match.
\item We repeat steps 1 and 2, using candidate clusters as the reference and searching for mock counterparts. This time, the fraction of common galaxies is defined with respect to the candidate clusters as
\begin{equation}
	f_{\rm com, \ det} = \frac{\sum_i P_i^{\rm match}}{\sum_i P_i^{\rm all}}.
\label{eq:common_fraction2}
\end{equation}
The minimum fraction of common galaxies was also taken to be $f_{\rm cut} = 0$ in this direction.
\item By comparing mock and cluster detection counterparts, we identify clusters for which the association is bijective.
\end{enumerate}
Because of the low value of $f_{\rm cut}$, the one-way associations are redundant with the one-way {\tt geometrical} matches. On the other hand the two-way associations rely on bijective galaxy associations instead of distance.

In Figure~\ref{fig:matching_illustration}, we illustrate the different matching procedures using one test cluster at $z=0.285$ and $M=10^{14.04}$ M$_{\odot}$, in the case of two cluster finders that provide the membership probabilities (AMICO and sFoF). While all the matching methods agree on the detection of this mock cluster for both algorithms, the {\tt geometrical} and {\tt membership} matching methods show that fragmentation is important for sFoF in this test case.

\begin{figure*}[h]
\centering
\includegraphics[width=0.45\textwidth]{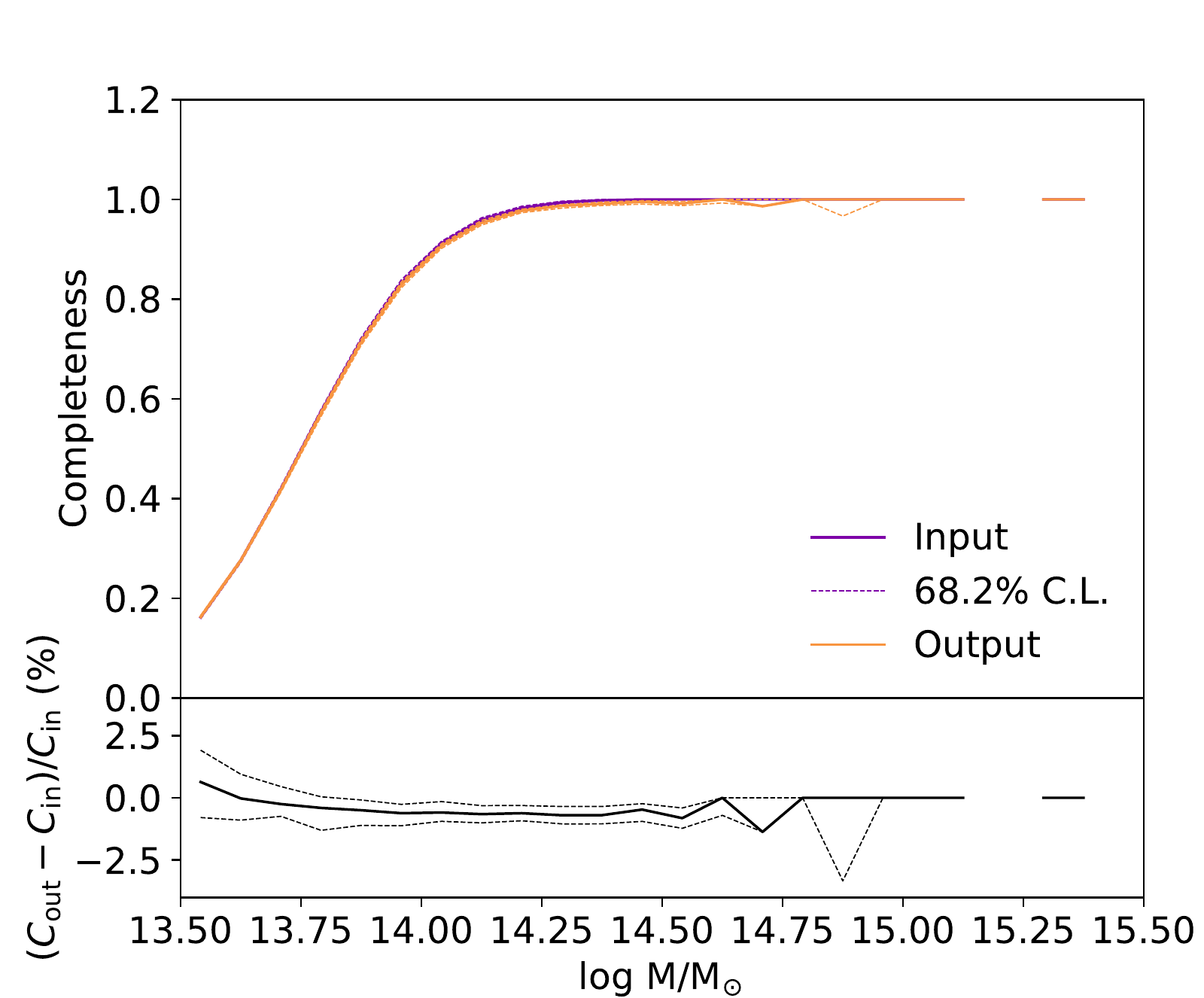}
\includegraphics[width=0.45\textwidth]{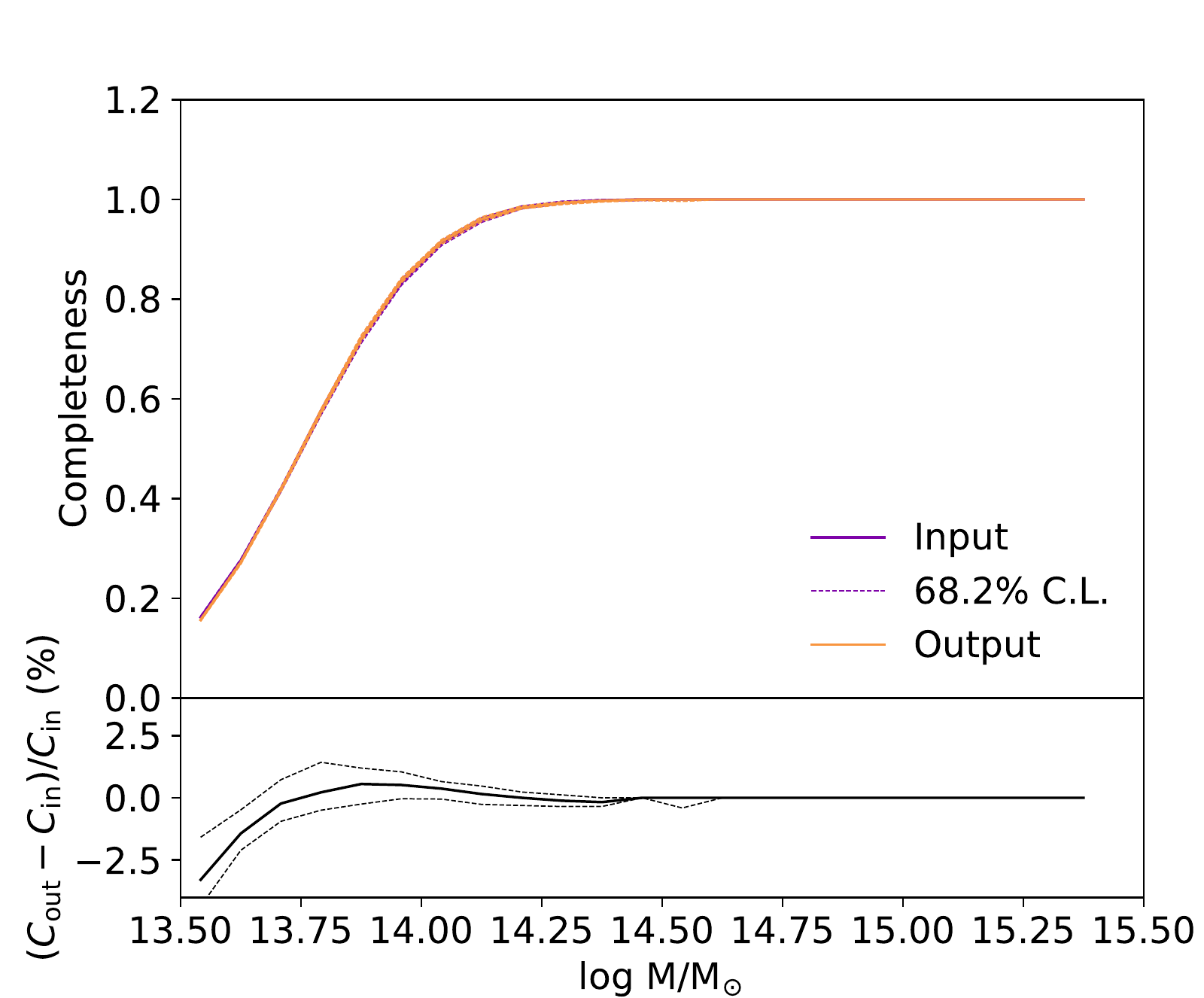}
\caption{\footnotesize{Comparison between the input completeness, as a function of mass, and the recovered completeness for the {\tt geometrical} matching (left) and the {\tt ranking} matching (right). The normalized residual is provided as the bottom plot. The dashed line provide the statistical uncertainty on the bias, computed as the 68.2\% statistical limit over all the Monte Carlo realizations.}}
\label{fig:matching_systematics}
\end{figure*}

\subsection{Purity, completeness, fragmentation and over-merging}\label{sec:Purity_completeness_fragmentation_and_overmerging}
The performance of an algorithm is related to the quality of the cluster catalog that it produces. It is generally quantified in terms of completeness (i.e., the number of detected clusters normalized by the number of clusters in the simulation) and purity (i.e., the number of true detections normalized by the overall detection number), which can be expressed as a function of cluster properties such as redshift, mass or richness. With the matching results in-hand, it is straightforward to define the one-way purity and completeness as
\begin{align}
    \begin{cases}
    	\vspace{0.2cm}
        \displaystyle P_1(\Delta \overrightarrow{\mu}) = \frac{\mathcal{A}_1\left[N^{\rm det}(\Delta \overrightarrow{\mu}) \longrightarrow N^{\rm true}\right]}{N^{\rm det}(\Delta \overrightarrow{\mu})} \\
        \displaystyle C_1(\Delta \overrightarrow{\mu}) = \frac{\mathcal{A}_1\left[N^{\rm true}(\Delta \overrightarrow{\mu}) \longrightarrow N^{\rm det}\right]}{N^{\rm true}(\Delta \overrightarrow{\mu})},
    \end{cases}
\label{eq:P1C1}
\end{align}
where $N^{\rm true}$ is the number of mock clusters and $N^{\rm det}$ is the number of detected clusters. The quantity $\Delta \overrightarrow{\mu}$ stands for a bin in terms of cluster parameters such as $\overrightarrow{\mu} \equiv \left(M, z, \lambda\right)$, so that $\mathcal{A}_1\left[N^{\rm det}(\Delta \overrightarrow{\mu}) \longrightarrow N^{\rm true}\right]$ is the number of mock clusters in the bin $\Delta \overrightarrow{\mu}$ that are associated with a detection. We note that while the completeness can be estimated as a function of true mass, redshift or richness, purity cannot be expressed in a given mass bin unless the measured mass-richness relation is used. Similarly, the two-way purity and completeness (available for the {\tt geometrical} and {\tt membership} matching procedures), are defined as
\begin{align}
    \begin{cases}
    	\vspace{0.2cm}
        \displaystyle P_2(\Delta \overrightarrow{\mu}) = \frac{\mathcal{A}_2\left[N^{\rm det}(\Delta \overrightarrow{\mu}) \longleftrightarrow N^{\rm true}\right]}{N^{\rm det}(\Delta \overrightarrow{\mu})} \\
        \displaystyle C_2(\Delta \overrightarrow{\mu}) = \frac{\mathcal{A}_2\left[N^{\rm true}(\Delta \overrightarrow{\mu}) \longleftrightarrow N^{\rm det}\right]}{N^{\rm true}(\Delta \overrightarrow{\mu})},
    \end{cases}
\label{eq:P2C2}
\end{align}
in which case we impose that the associations be bijective.

We note that impurities do not only correspond to spurious objects, such as improperly identified structures along the line-of-sight, or unmatched fragments of larger clusters. They may also correspond to clusters for which the mass is below the threshold of the mock cluster catalog constructed as defined in Section~\ref{sec:Halo_catalogs}. This point is discussed further in sections~\ref{sec:Performance} and~\ref{sec:Discussions}.

We define the N-fragmentation rate as the fraction of mock clusters for which more than N-associations to detected clusters are possible. Similarly to completeness and purity, it can be expressed as a function of mass, redshift and richness. The N-over-merging rate is given by the fraction of detected clusters for which more than N-associations to mock clusters are possible. Having $C_1 \simeq C_2$ and $P_1 \simeq P_2$ is also an indication of low fragmentation and low over-merging.

For each detected cluster that is matched to a mock cluster, we compute the redshift difference between the true cluster redshift and the recovered one. We additionally compute the projected sky coordinate offsets (see Section~\ref{sec:centroid_and_redshift_differences} for further discussions and the use of these quantities).

\subsection{Systematic effects}
The matching procedure is ambiguous in the sense that the detection of a cluster has to be addressed based on somewhat arbitrary criteria. In addition, galaxy clusters are extended objects, with internal structure that varies among the cluster population. As a consequence, the matching itself is not immune to artifacts that are reflected in the selection function.

In order to estimate the systematic effects associated with the matching, we use the following procedure. We construct a new mock cluster catalog by removing clusters randomly, following a realistic input completeness that we define, $C_{\rm in}$, based on the results described in Section~\ref{sec:Performance} (typically using the overall results of the different algorithms). As shown in Figure~\ref{fig:matching_systematics}, the completeness is described by an error function (erf) with characteristic mass of $10^{13.75}$ M$_{\odot}$ and with $0.21$ dex width. We also introduce noise in the sky and redshift coordinates of the clusters, modeled by a Gaussian distribution, representative of the results of Section~\ref{sec:centroid_and_redshift_differences}, as a function of mass and redshift. The redshift standard deviation is set to $0.03 (1+z_{\rm true})$ at $M = 10^{13.5}$ M$_{\odot}$, and evolves as $1/\sqrt{M}$. The position standard deviation is fixed to 0.25 arcmin. Since the {\tt ranking} matching also relies on the richness estimates, we introduce an extra scatter of 0.2 dex in the richness, based on the statistical scatter as measured in Section~\ref{sec:Mass_richness_scaling_relation}. The new constructed cluster catalog mimics the properties of the catalog produced by the detection algorithms and it is matched to the original catalog to estimate the biases induced by the matching procedure on the completeness. As each newly generated mock catalog is only a noisy representation of the mean catalog, this procedure is repeated by generating 100 Monte Carlo realizations of such a catalog. The mean recovered completeness, $C_{\rm out}$, allows us to measure biases and the dispersion from the Monte Carlo realization used to estimate statistical errors. This method is applied only to the {\tt ranking} matching and the {\tt geometrical} two-way matching, because it cannot handle the galaxy membership required by the {\tt membership} matching procedure. It is therefore only used as a cross check.

Figure~\ref{fig:matching_systematics} provides the comparison between the input completeness and the recovered one, for the {\tt geometrical} two-way and the {\tt ranking} matching, as a function of mass. For the {\tt ranking} matching, we observe a mass dependent bias, reaching up to 3\% at $10^{13.5}$ M$_{\odot}$. This is due to the fact that low mass clusters can be missed when there is a more rich, competing cluster, in their surrounding. The overall systematic effects due to the matching procedure remain below 3\%.

\section{Results and performance}\label{sec:Performance}
The blind detection of galaxy clusters using the algorithms presented in Section~\ref{sec:Algorithms} on the mock catalog (discussed in Section~\ref{sec:Simulations}) was followed by the association between detections and mock clusters (Section~\ref{sec:Matching}). In this section we address the detection performance expected for \textit{Euclid}. The overall performance is summarized in Table~\ref{tab:perf_summary}.

\subsection{Comparison criteria}\label{sec:Comparison_criteria}
\begin{figure}[h]
\centering
\includegraphics[width=0.43\textwidth]{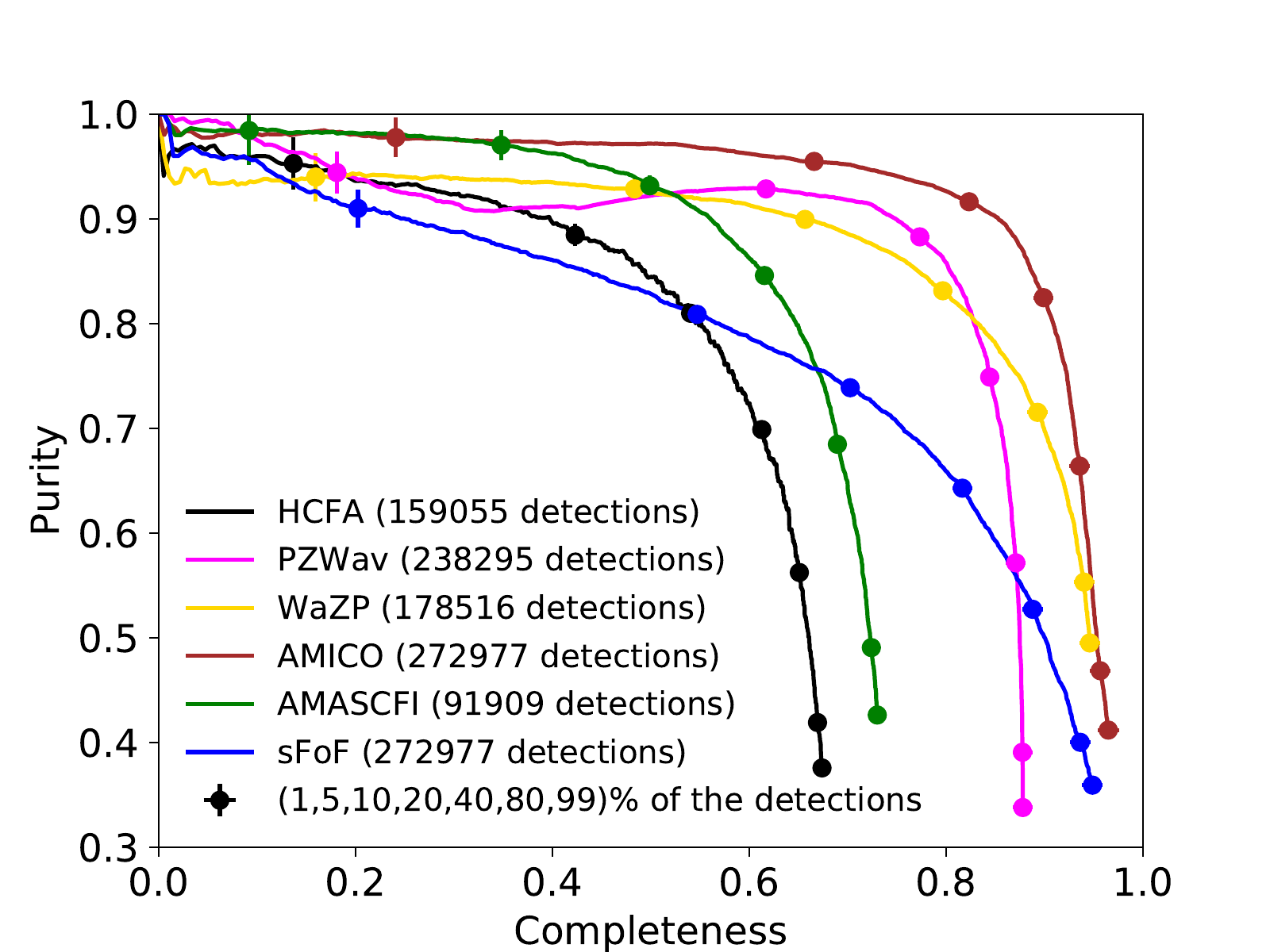} \put(-110,155){\makebox(0,0){\rotatebox{0}{\large $M>10^{14}$ M$_{\odot}$, $z_{\rm true}<2$}}}\\
\includegraphics[width=0.43\textwidth]{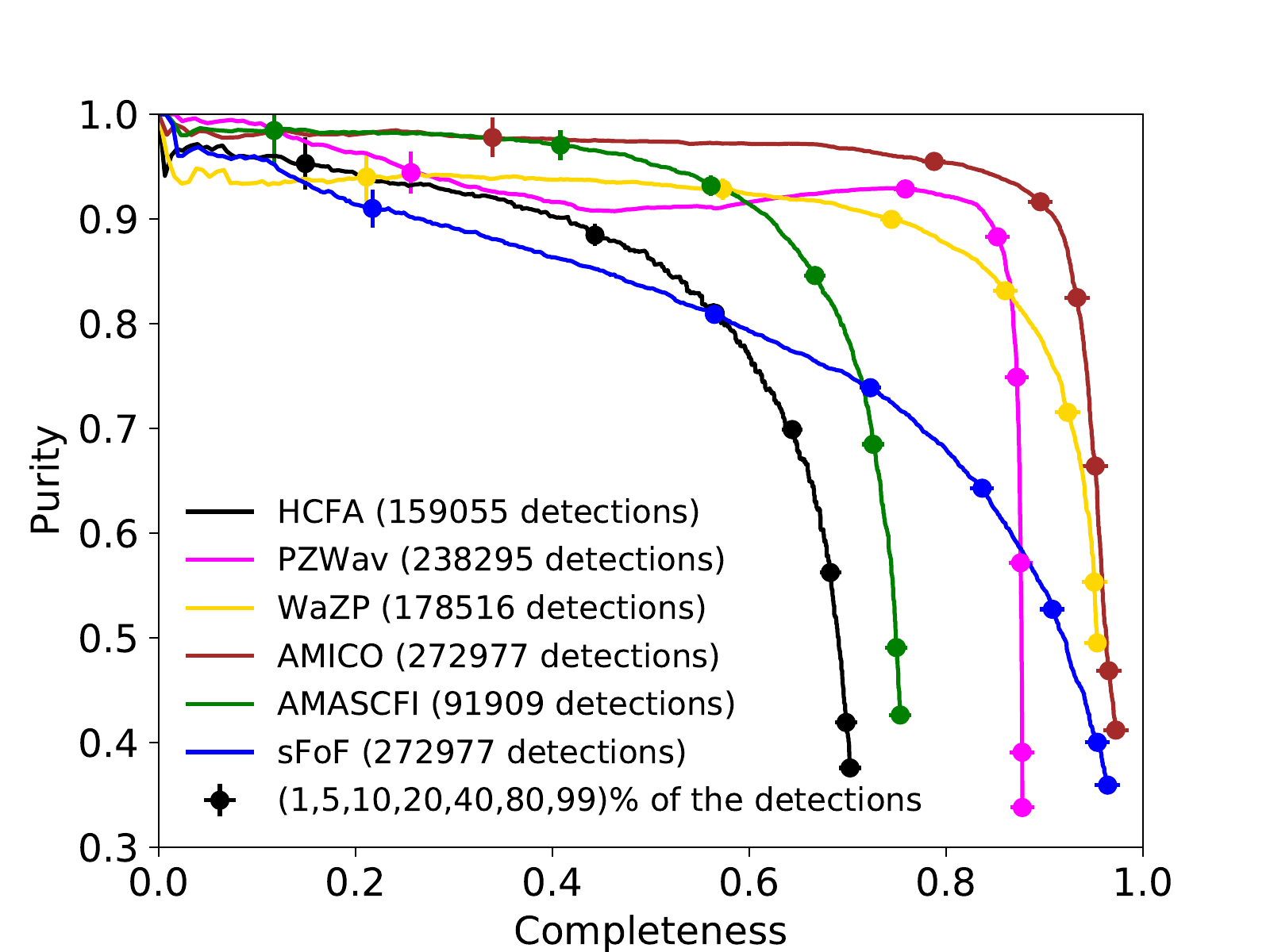} \put(-110,155){\makebox(0,0){\rotatebox{0}{\large $M>10^{14}$ M$_{\odot}$, $z_{\rm true}<1$}}}\\
\includegraphics[width=0.43\textwidth]{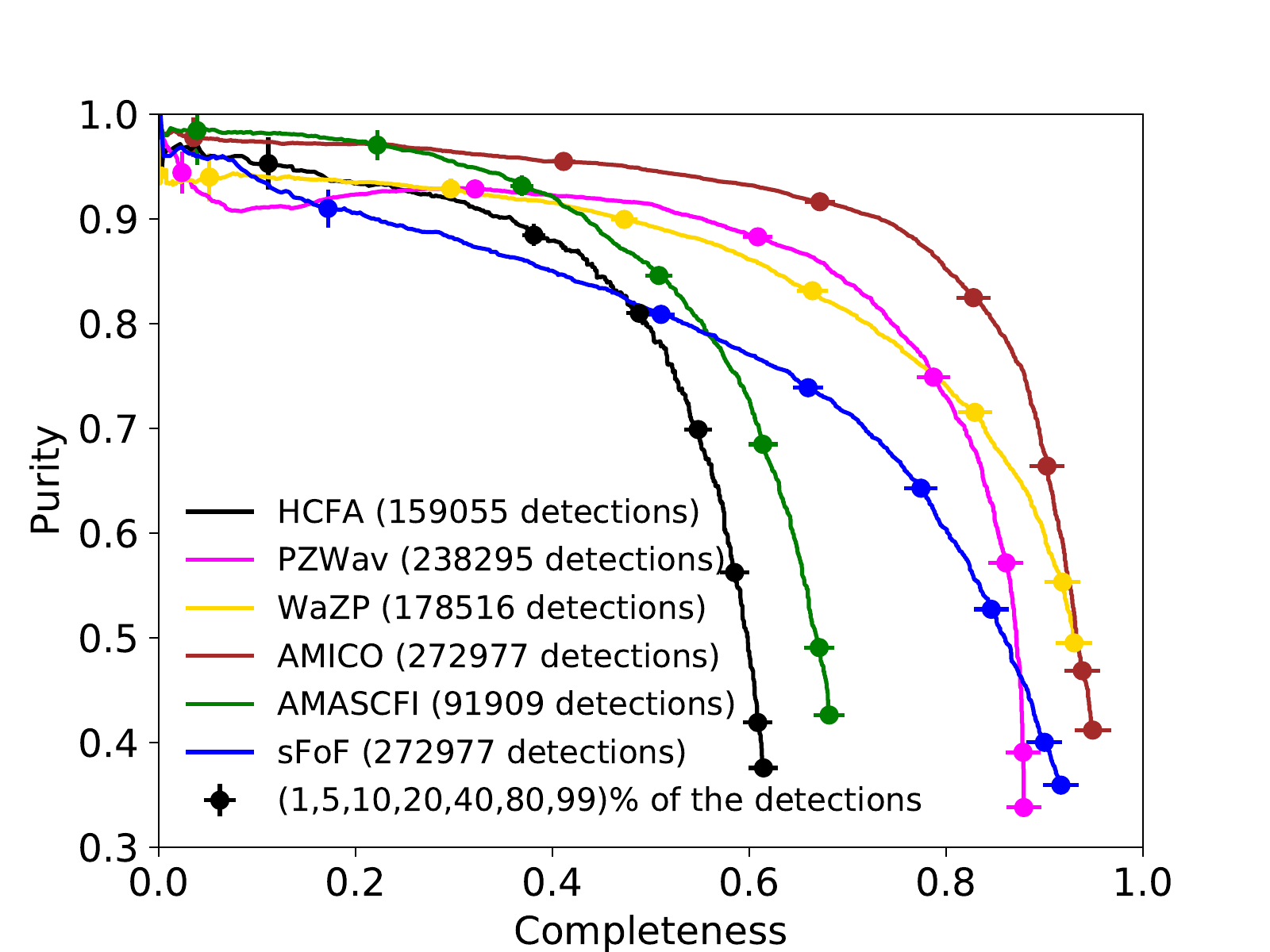} \put(-110,155){\makebox(0,0){\rotatebox{0}{\large $M>10^{14}$ M$_{\odot}$, $1<z_{\rm true}<2$}}}\\
\caption{\footnotesize{Purity as a function of completeness, given for all cluster finder catalogs as a function of the ranking of the detections. As the number of considered detections above a given rank increases, the detection properties evolve from the high purity low completeness regime to the low purity high completeness regime. The dots provide the percentage of considered detections at a given coordinate on the curve, the total number of detections being given in the legend for each finder. The top panel only accounts for mock clusters in the range $z_{\rm true}<2$ and $M>10^{14}$ M$_{\odot}$ when computing the completeness. For illustration, we also provide the same figure in the range $z_{\rm true}<1$ and $1<z_{\rm true}<2$ in the bottom panels.}}
\label{fig:roc_curve}
\end{figure}

\begin{table*}[]
\caption{\footnotesize{Summary of the performance of the cluster finders, as applied on the mock. Note that in the case of WaZP, the median centroid offset is zero because most detections have been assigned to the true mock cluster central galaxies.}}
\begin{center}
\resizebox{\textwidth}{!} {
\begin{tabular}{l||c|c|c|c|c}
\hline
\hline
Finder & Mean $C$ for a $P=0.8$ & Mean $P$ for a $C=0.8$ & Mean statistical $M$--$\lambda_{\rm det}$ scatter & Redshift RMS / $(1+z_{\rm true})$ & Centroid offset (mean, median, RMS) \\
-- & (at $M>10^{14}$ M$_{\odot}$, $z_{\rm true}<2$) & (at $M>10^{14}$ M$_{\odot}$, $z_{\rm true}<2$) & (at $M>10^{14}$ M$_{\odot}$, $z_{\rm true}<2$, in dex) & (at $M>10^{14}$ M$_{\odot}$) & (at $M>10^{14}$ M$_{\odot}$, arcmin) \\
\hline
AMASCFI  & 65\% & Not reached & 0.27 & 0.025 & (0.44, 0.36, 0.33) \\
AMICO    & 91\% & 93\%        & 0.19 & 0.015 & (0.22, 0.19, 0.19) \\
HCFA     & 55\% & Not reached & 0.15 & 0.021 & (0.46, 0.33, 0.45) \\
PZWav    & 83\% & 86\%        & 0.16 & 0.020 & (0.27, 0.21, 0.45) \\
sFoF     & 57\% & 66\%        & 0.17 & 0.017 & (0.16, 0.11, 0.15) \\
WaZP     & 83\% & 83\%        & 0.18 & 0.016 & (0.10, 0.00, 0.26) \\
\hline
\end{tabular}
}
\end{center}
\label{tab:perf_summary}
\end{table*}

The competing cluster finders do not provide cluster catalogs down to the same detection limit. Additionally, the S/N of the detections are only available for a subset of the cluster finders. Nevertheless, a ranking of the detections, according to their reliability, is provided for all of them. In most cases, the ranking is performed according to S/N but for cluster finders not providing S/N ranking is evaluated according to richness. In order to compare the global performance of the algorithms, we impose a minimum purity on the cluster detection catalogs. This is done by removing the detection with lowest ranking, until the minimal purity is reached for the overall sample. This is illustrated in Figure~\ref{fig:roc_curve}, where we compute the purity as a function of completeness by limiting the detection catalog to the most reliable objects up to a given rank, which is varied from unity (i.e., only the best detection) to the total number of detections in the catalogs (i.e., until the least reliable detections are included). To illustrate the expected \textit{Euclid} data usage in a cosmological context, the target mock catalog is restricted to redshifts $z_{\rm true}<2$ and masses $M>10^{14}$ M$_{\odot}$. The completeness is thus computed using only the objects satisfying these limits, while the purity reflects the content of the full detection catalogs. The completeness will be discussed in more detail in Section~\ref{sec:Completeness_and_purity}.

\begin{figure}[h]
\centering
\includegraphics[width=0.5\textwidth]{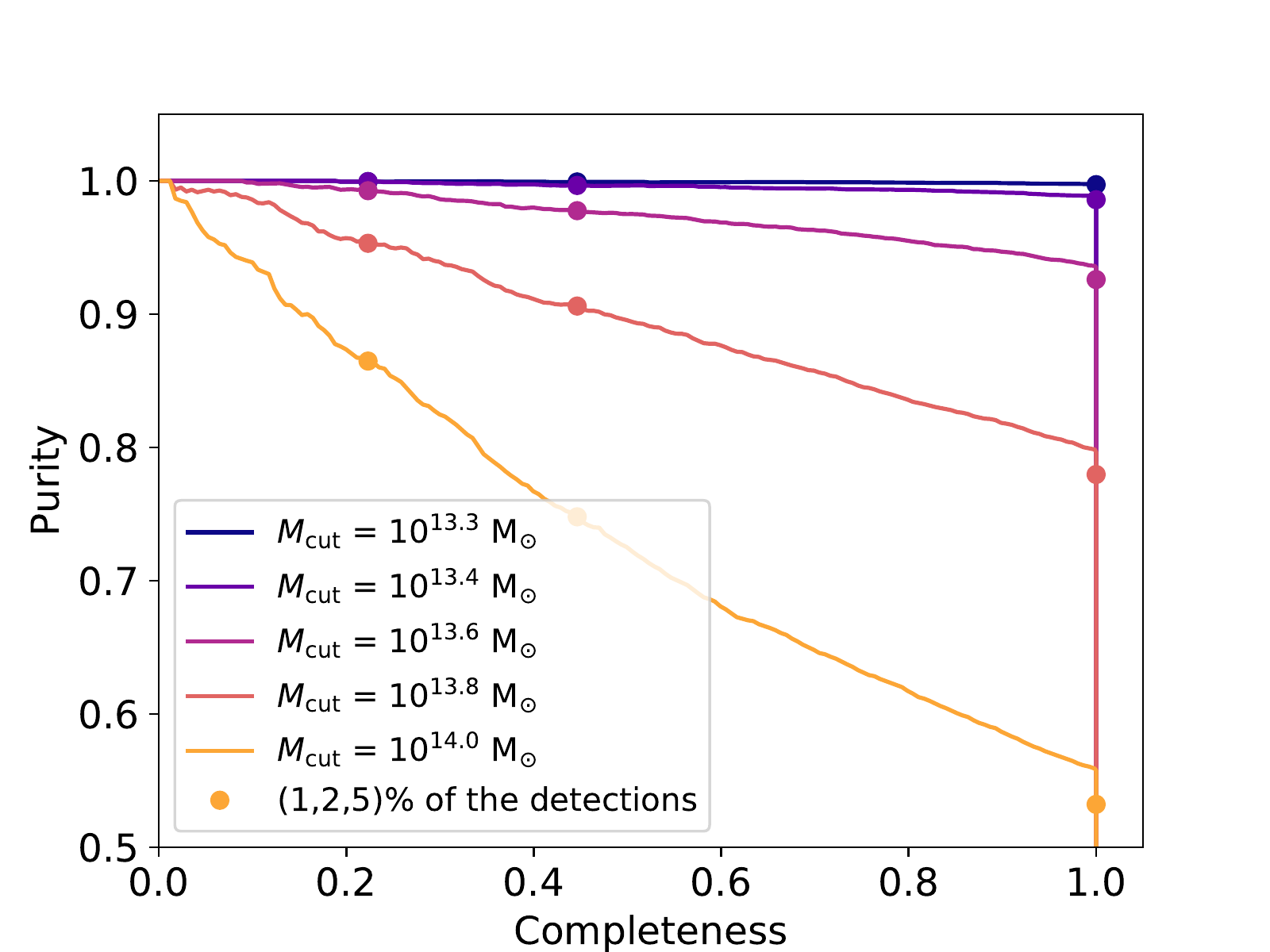}
\caption{\footnotesize{Purity versus completeness constructed for an ideal detection catalog. The different curves correspond to different mass cuts assumed for the mock catalog. The redshift and mass ranges considered are the same as in the top panel of Figure~\ref{fig:roc_curve}.}}
\label{fig:roc_curve_valid}
\end{figure}

Since the full mock cluster catalog only includes objects with masses larger than $10^{13.25}$ M$_{\odot}$, detections at masses below this limit will appear as impurities even though they can correspond to real groups. To estimate the importance of this effect, considering the mass range we adopted, we produce a perfect detection-like catalog using the true mock catalog. The clusters are ranked by richness, accounting for redshift evolution as detailed in Section~\ref{sec:mock_mass_lambda_relation} and including an extra scatter of 0.2 dex that mixes the order of the detections as expected for real cluster finders. We then consider the detected clusters to be true detections (i.e., they have been matched to mock clusters) only if their mass is above a given threshold, $M_{\rm cut}$. This threshold thus mimics the mass limit that we define when constructing the mock catalog ($10^{13.25}$ M$_{\odot}$ for our baseline). We reproduce Figure~\ref{fig:roc_curve}, in the same range of mass and redshift for completeness computation, but for different values of the threshold $M_{\rm cut}$. The results are shown in Figure~\ref{fig:roc_curve_valid}, where we can see that while the mock catalog mass threshold remains close enough to the detection limit, the loss of purity is small. A bias of $\lesssim 5$\% is observed up to mass cut as large as $M_{\rm cut} = 10^{13.6}$ M$_{\odot}$ (compared to the limit of $10^{13.25}$ M$_{\odot}$ we have used). Based on the detection limit of the cluster finders (see Section~\ref{sec:Completeness_and_purity}), the bias is expected to be about $1-2$\% at most. In addition to this crosscheck, the calculations of Figure~\ref{fig:roc_curve} were repetead for a catalog mass cut of $10^{13}$ M$_{\odot}$ (instead of $10^{13.25}$ M$_{\odot}$), showing no significant differences with the baseline choice.

Prior to further post-detection analysis, it was necessary to restrict the cluster detection catalogs to a common detection significance. This was done in order to compare the performance of the different algorithms, particularly in terms of completeness, given the heterogeneous nature of the detection catalogs provided for the CFC. Cluster cosmology requires a well-defined cluster catalog and we thus restrict the cluster finder samples to ranks good enough to ensure that the mean sample purity is equal to 80\% over the mass and redshift range given by $M > 10^{14}$ M$_{\odot}$ and $z_{\rm true}<2$. This corresponds to defining a threshold at 80\% purity in Figure~\ref{fig:roc_curve} and excluding clusters with ranks beyond the corresponding limit. This cut at $P=0.8$ also corresponds to the limit for which the purity starts to drop rapidly with limited improvement for the completeness. The mean completeness of the corresponding catalog is given in Table~\ref{tab:perf_summary}, for this purity threshold, as well as the purity for a completeness threshold of 80\%. This baseline limit does affect the overall performance presented in the following, and we discuss in Section~\ref{sec:Discussions} how they change according to this choice. We note that for real observations, such a purity cut cannot easily be applied. However, calibration of the purity as a function of the catalog S/N limit could be done with help from the \textit{Euclid} deep survey or using external data at various wavelengths. The completeness performance will be affected when changing the S/N threshold of the trimmed catalog, as can be seen in Figure~\ref{fig:roc_curve}.

In the following subsections, we address the performance of the six cluster finder algorithms and compare their behaviors in terms of cluster completeness, purity, dispersion introduced by the detection in their mass proxy, the redshift and centroid recovery, fragmentation and over-merging. The selection of \textit{Euclid} algorithms was done by prioritizing high redshift objects ($z \gtrsim 1$) at high mass ($M \gtrsim 10^{14}$ M$_{\odot}$), those that are expected to carry most of the statistical power in constraining cosmological parameters of interest to \textit{Euclid} \citep[e.g., the dark energy equation of state,][]{Sartoris2016}.

\subsection{Completeness and purity}\label{sec:Completeness_and_purity}
\begin{figure*}[h]
\resizebox{\textwidth}{!} {
\begin{tabular}{lll}
\includegraphics[trim=0cm 1.8cm 3cm 0cm, clip=true, scale=1]{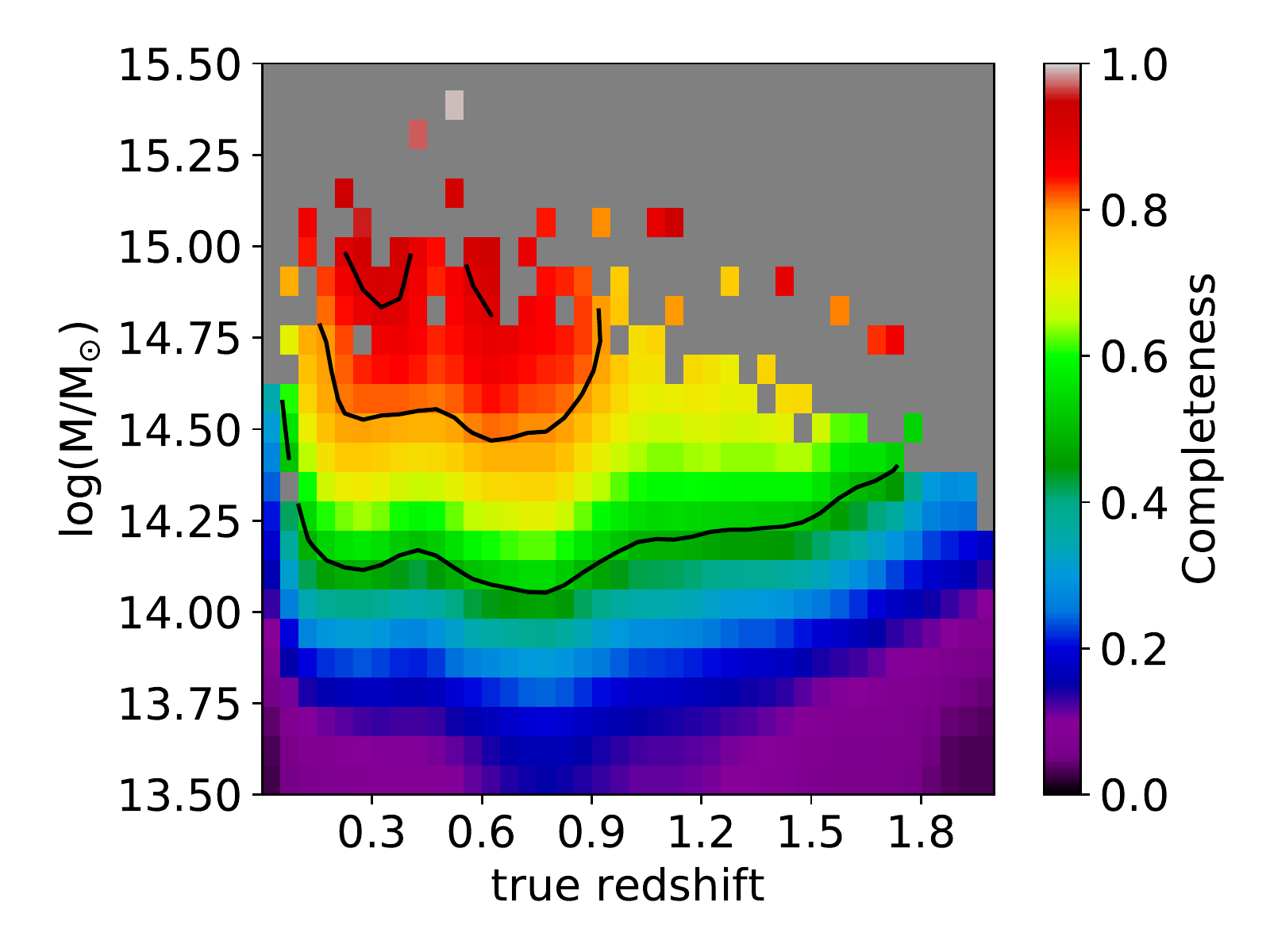} \put(-65,250){\makebox(0,0){\rotatebox{0}{\LARGE \textcolor{white}{AMASCFI}}}} &
\includegraphics[trim=3.2cm 1.8cm 3cm 0cm, clip=true, scale=1]{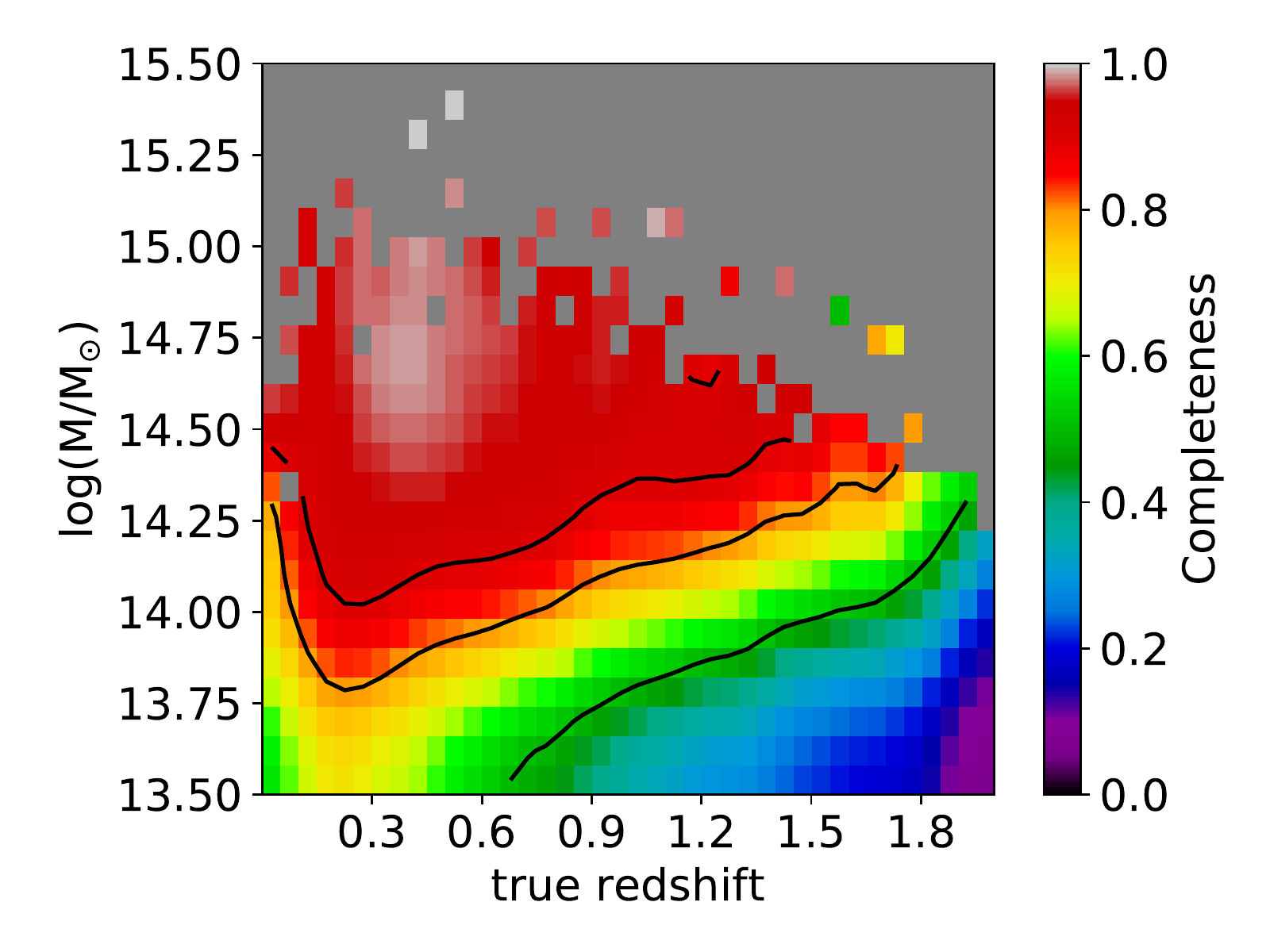} \put(-65,250){\makebox(0,0){\rotatebox{0}{\LARGE \textcolor{white}{AMICO}}}} &
\includegraphics[trim=3.2cm 1.8cm 0cm 0cm, clip=true, scale=1]{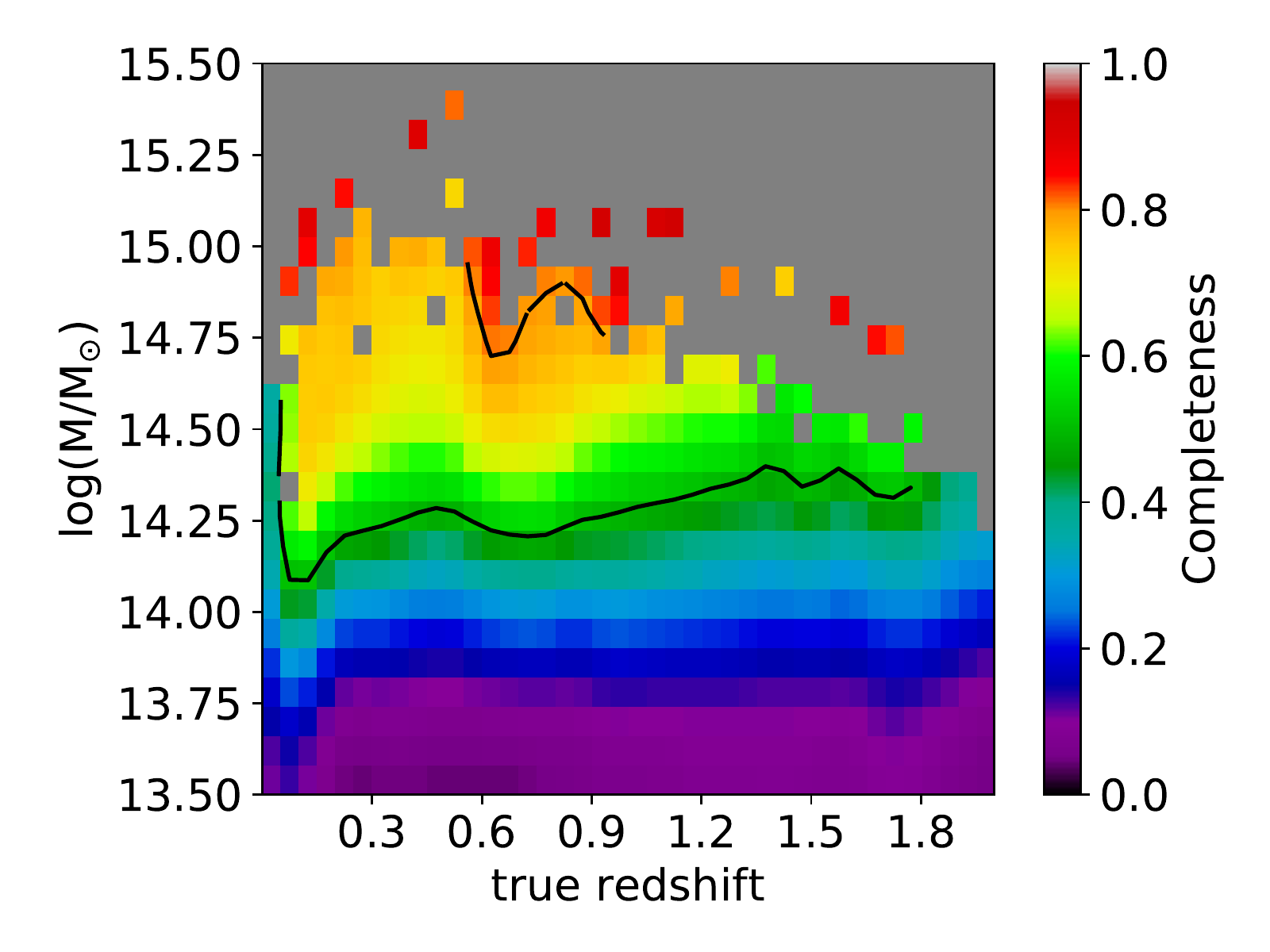} \put(-165,250){\makebox(0,0){\rotatebox{0}{\LARGE \textcolor{white}{HCFA}}}} \\
\includegraphics[trim=0cm 0cm 3cm 0cm, clip=true, scale=1]{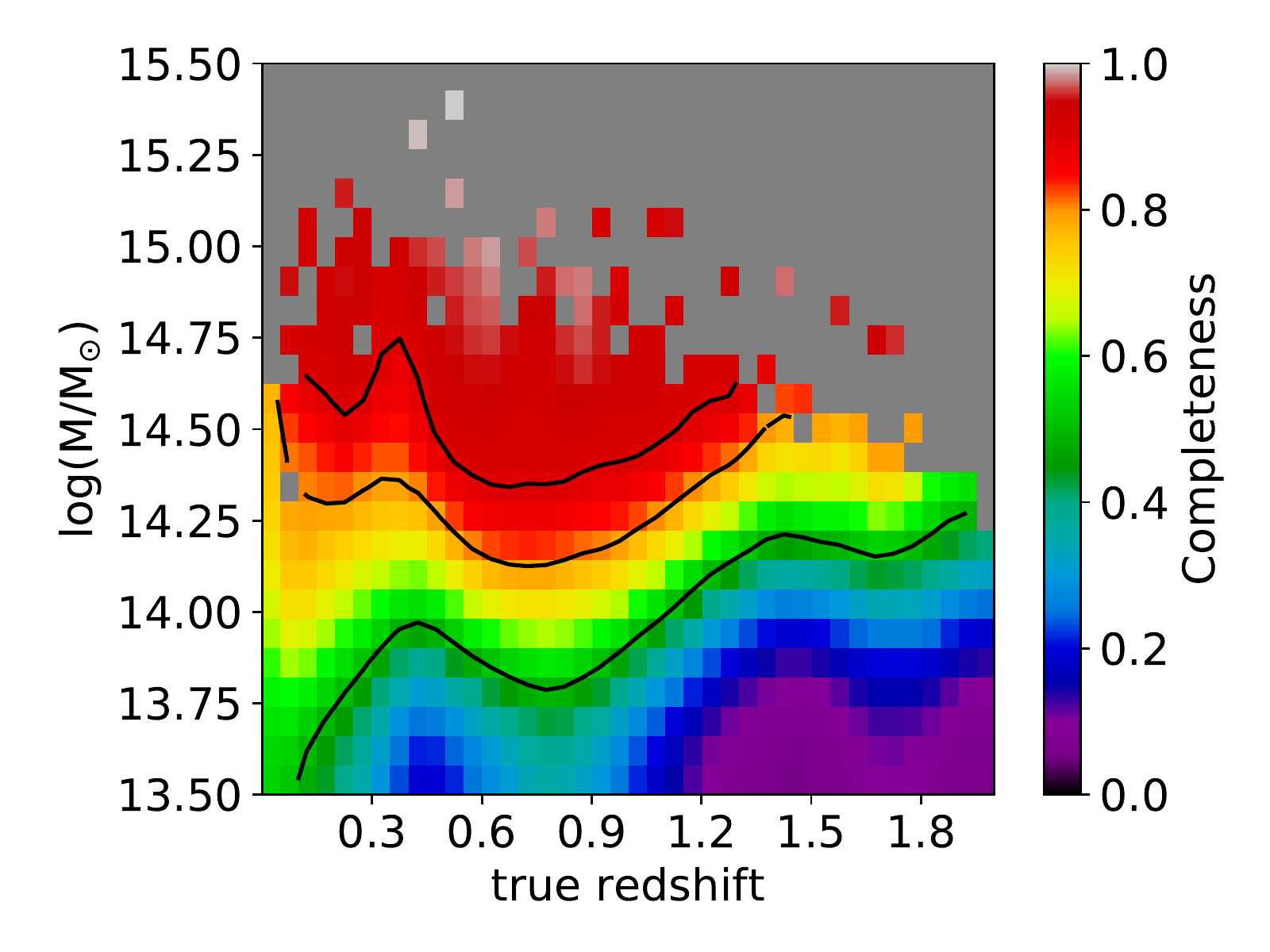} \put(-65,295){\makebox(0,0){\rotatebox{0}{\LARGE \textcolor{white}{PZWav}}}} &
\includegraphics[trim=3.2cm 0cm 3cm 0cm, clip=true, scale=1]{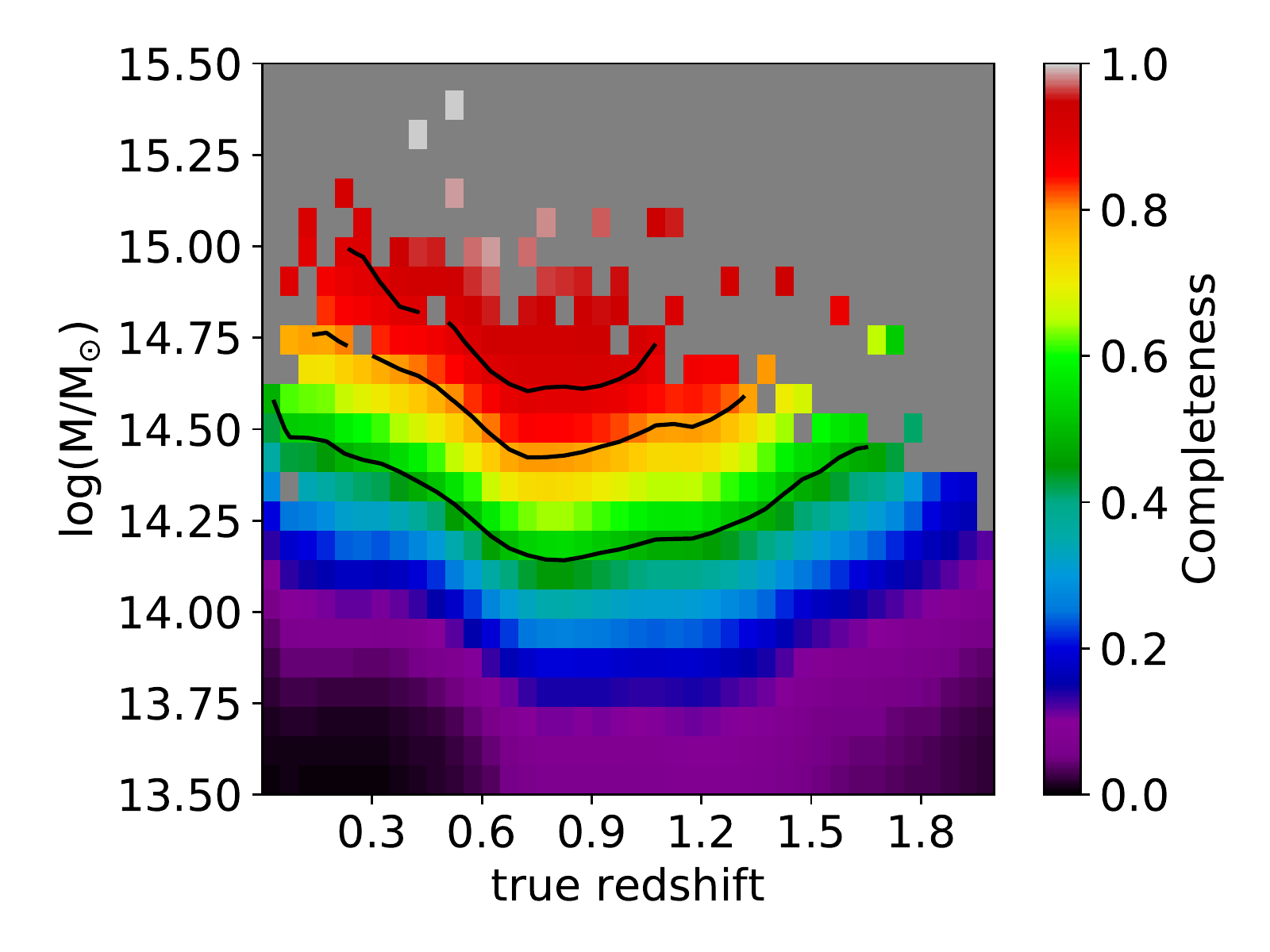} \put(-65,295){\makebox(0,0){\rotatebox{0}{\LARGE \textcolor{white}{sFoF}}}} &
\includegraphics[trim=3.2cm 0cm 0cm 0cm, clip=true, scale=1]{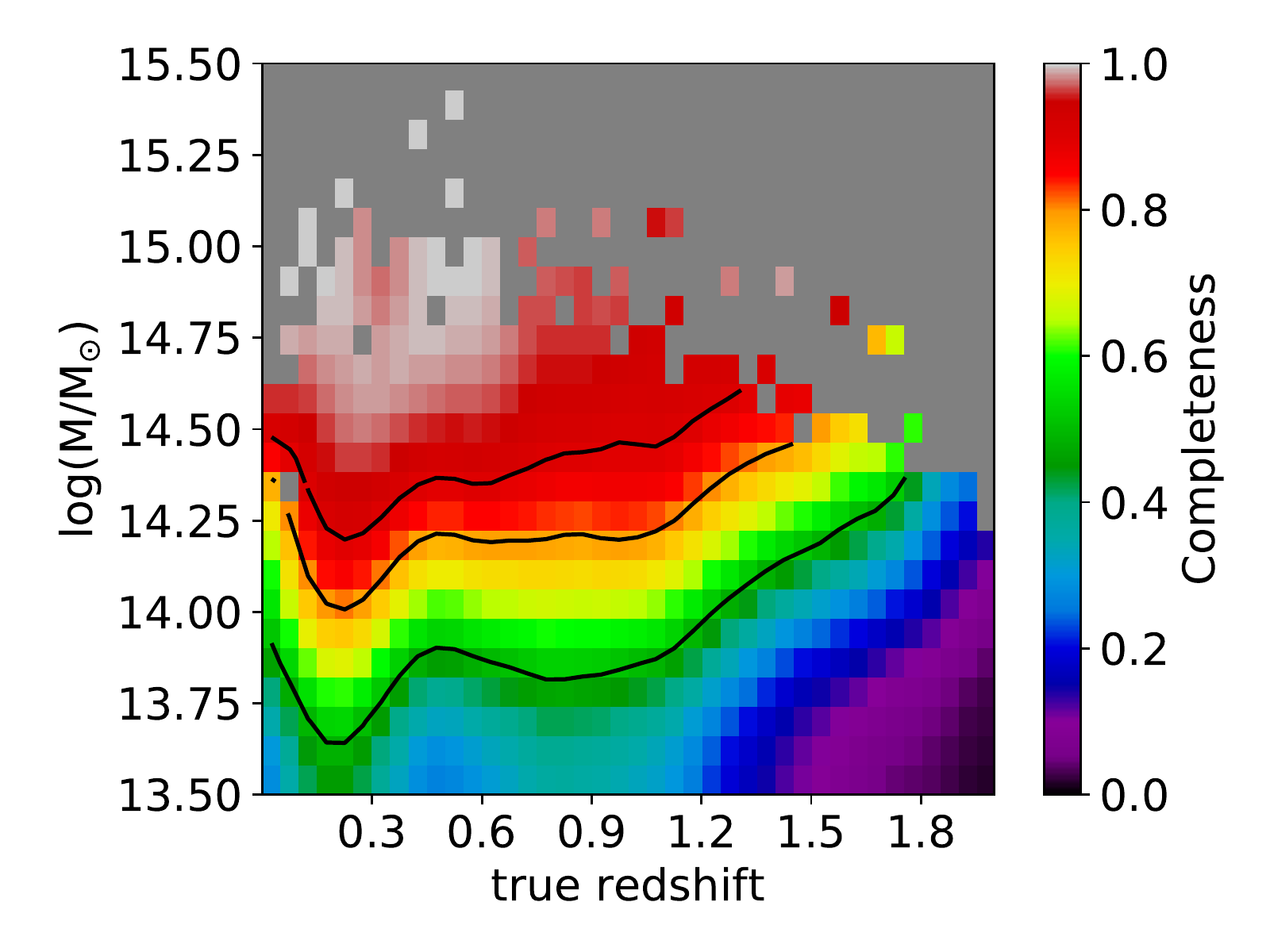} \put(-165,295){\makebox(0,0){\rotatebox{0}{\LARGE \textcolor{white}{WaZP}}}} \\
\end{tabular}}
\caption{\footnotesize{Completeness as a function of mass and redshift for the six final CFC algorithms. The black lines corresponds to completenesses of 50, 80 and 90\%. The pixel size corresponds to about $\Delta {\rm log}\left(M/{\rm M}_{\odot}\right) = 0.08$ and $\Delta z = 0.05$, but the image was smoothed with a Gaussian kernel to a resolution of 0.20 ${\rm log}\left(M/{\rm M}_{\odot}\right)$ $\times$ 0.25 (FWHM) for display purposes. The mask is applied where the number of clusters per pixel is zero before smoothing. Each catalog of detections has been trimmed to the most reliable detections, ensuring a mean purity of 80\% in the range $M>10^{14}$ M$_{\odot}$ and $z_{\rm true}<2$.}}
\label{fig:completeness_m_z}
\end{figure*}

\begin{figure*}[h]
\resizebox{\textwidth}{!} {
\begin{tabular}{lll}
\includegraphics[trim=0cm 1.8cm 0cm 0cm, clip=true, scale=1]{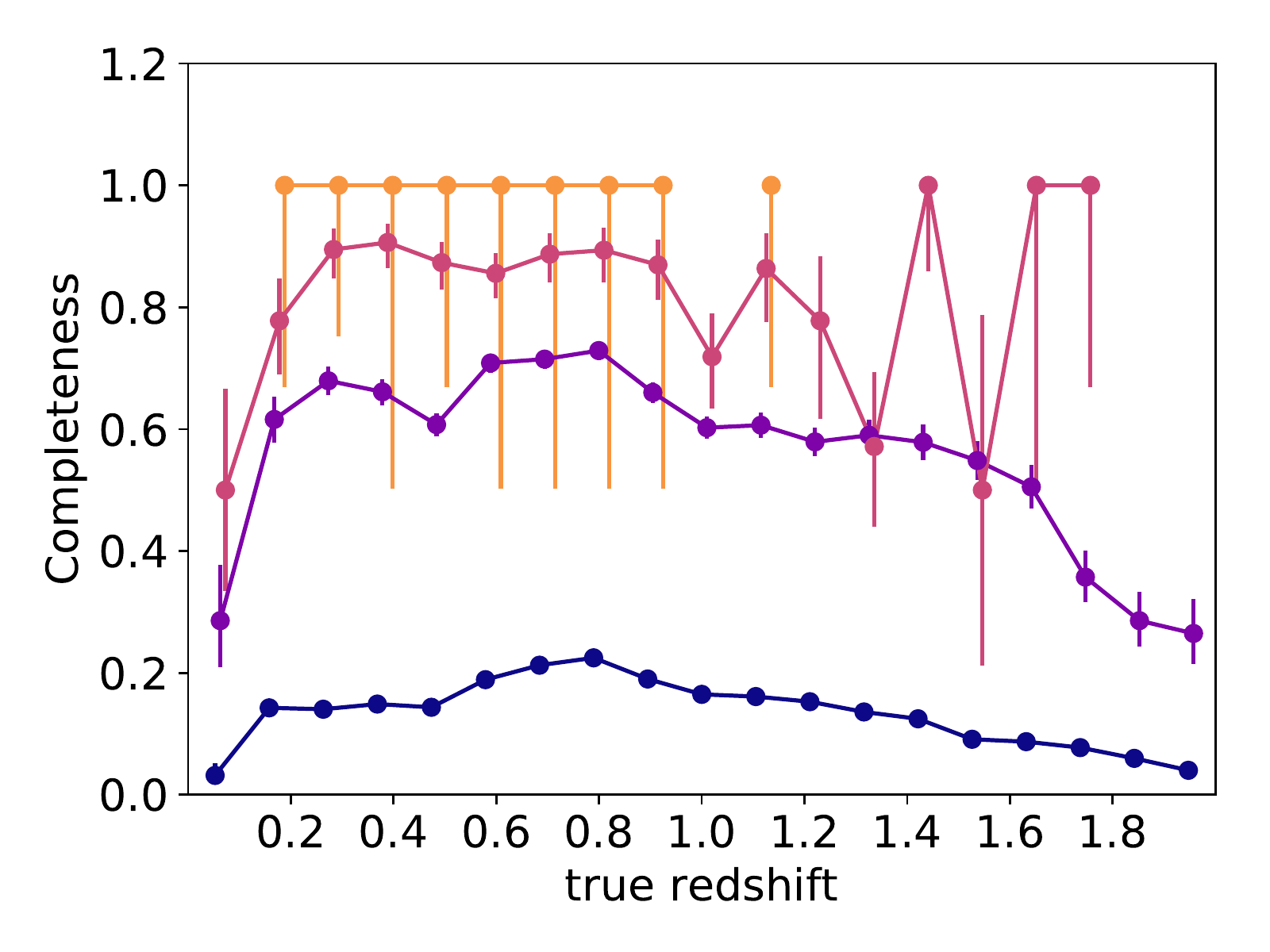} \put(-200,260){\makebox(0,0){\rotatebox{0}{\LARGE AMASCFI}}} &
\includegraphics[trim=2.2cm 1.8cm 0cm 0cm, clip=true, scale=1]{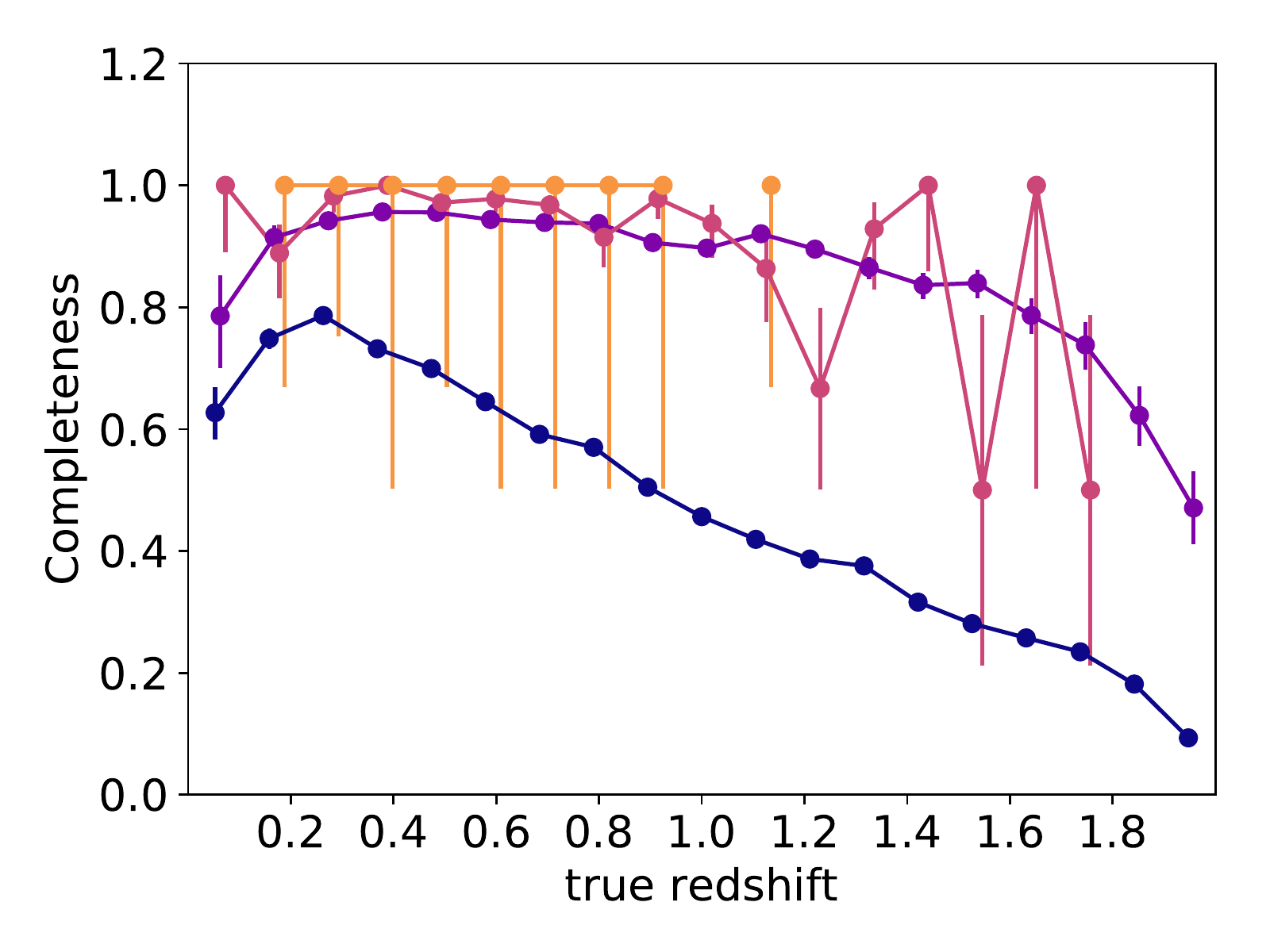} \put(-200,260){\makebox(0,0){\rotatebox{0}{\LARGE AMICO}}} &
\includegraphics[trim=2.2cm 1.8cm 0cm 0cm, clip=true, scale=1]{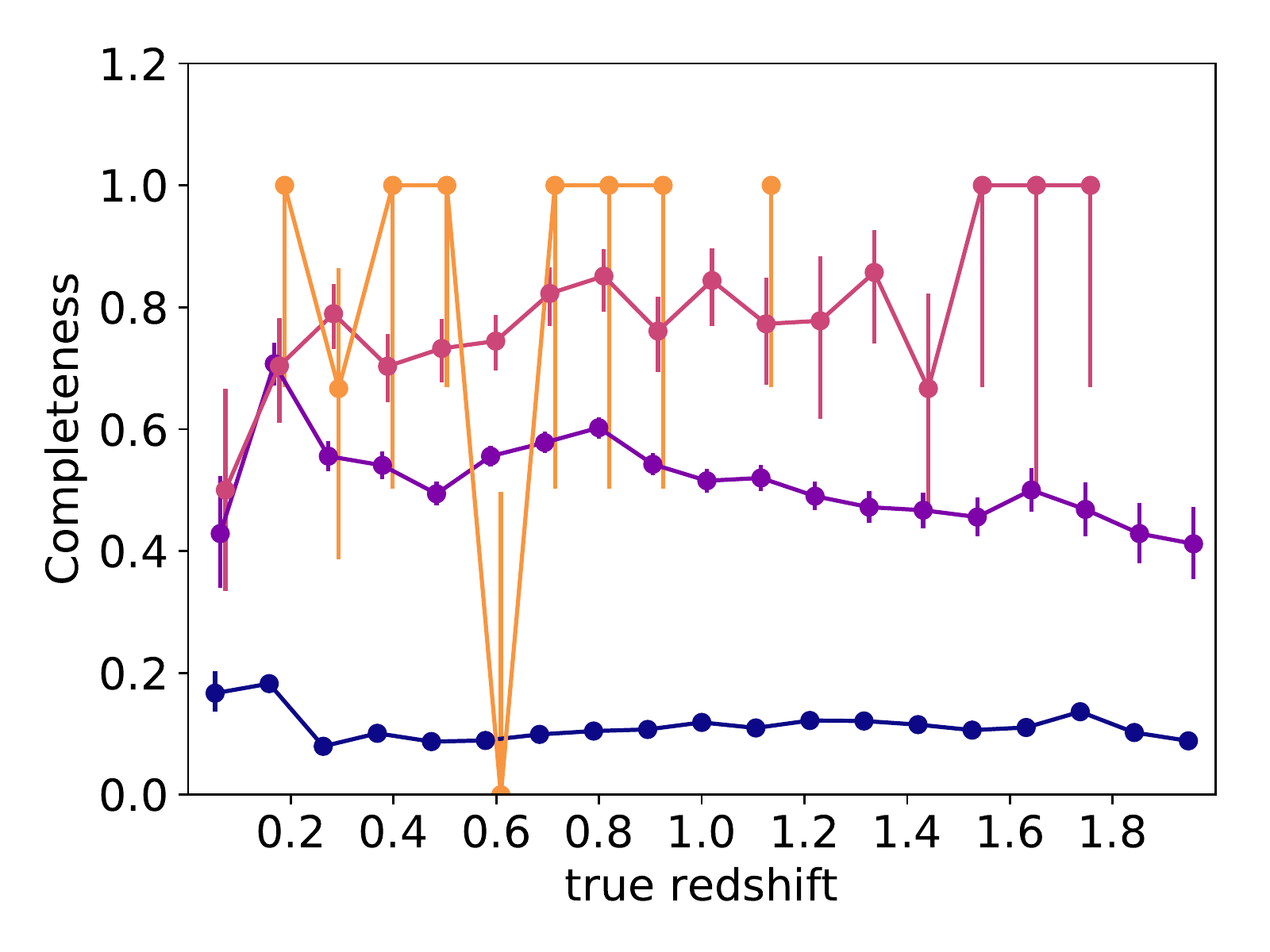} \put(-200,260){\makebox(0,0){\rotatebox{0}{\LARGE HCFA}}}\\
\includegraphics[trim=0cm 0cm 0cm 0cm, clip=true, scale=1]{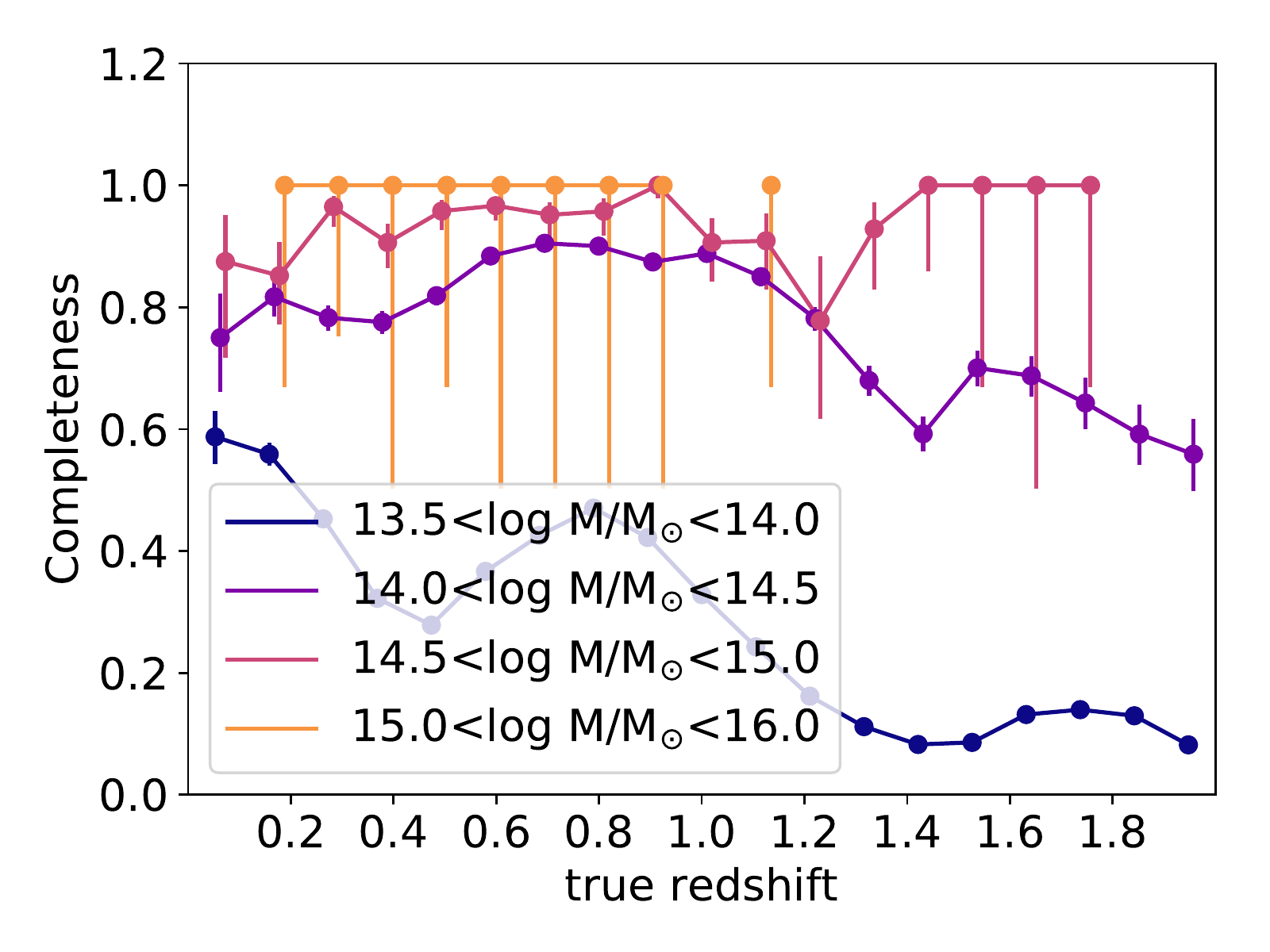} \put(-200,310){\makebox(0,0){\rotatebox{0}{\LARGE PZWav}}} &
\includegraphics[trim=2.2cm 0cm 0cm 0cm, clip=true, scale=1]{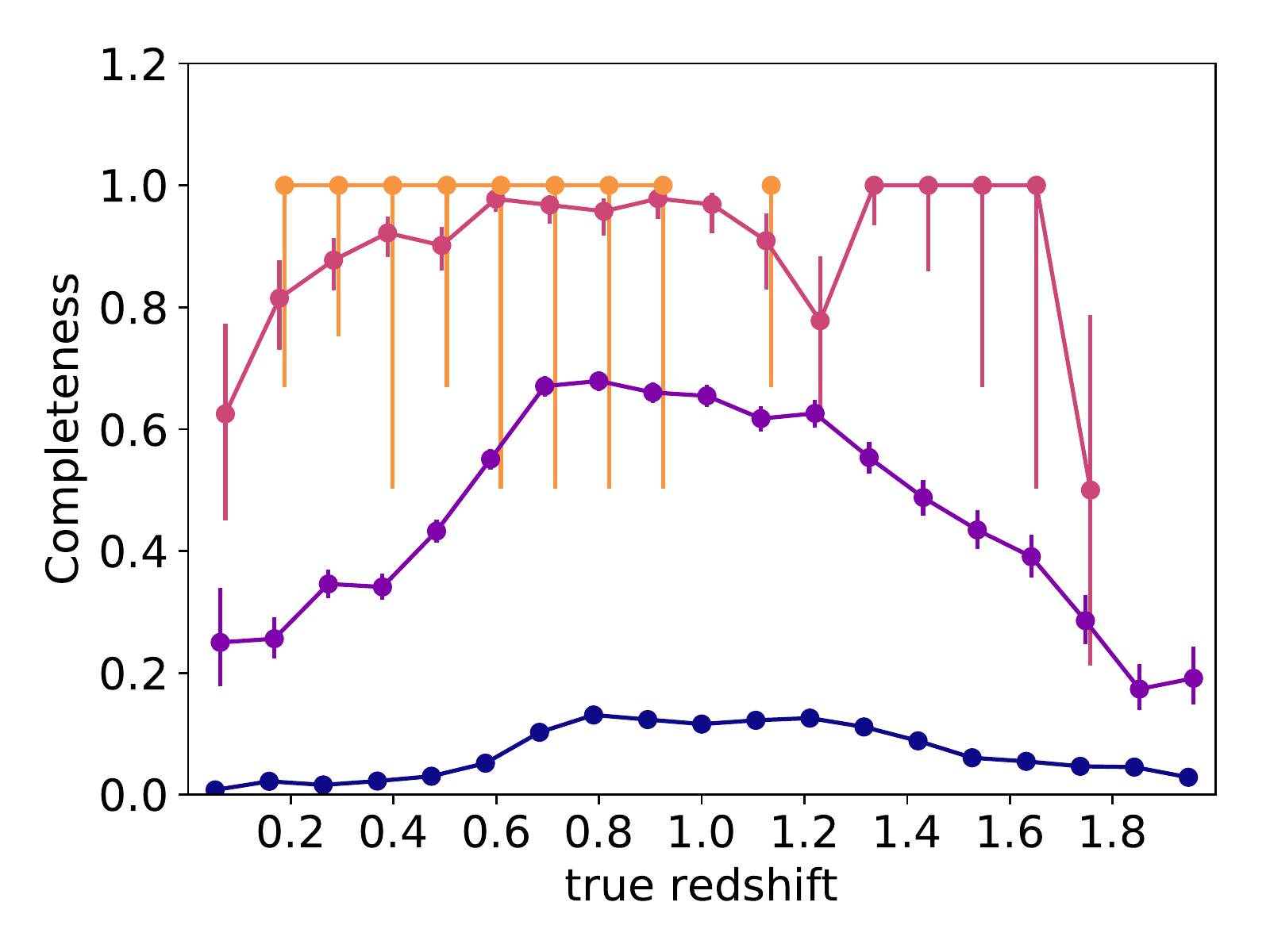} \put(-200,310){\makebox(0,0){\rotatebox{0}{\LARGE sFoF}}} &
\includegraphics[trim=2.2cm 0cm 0cm 0cm, clip=true, scale=1]{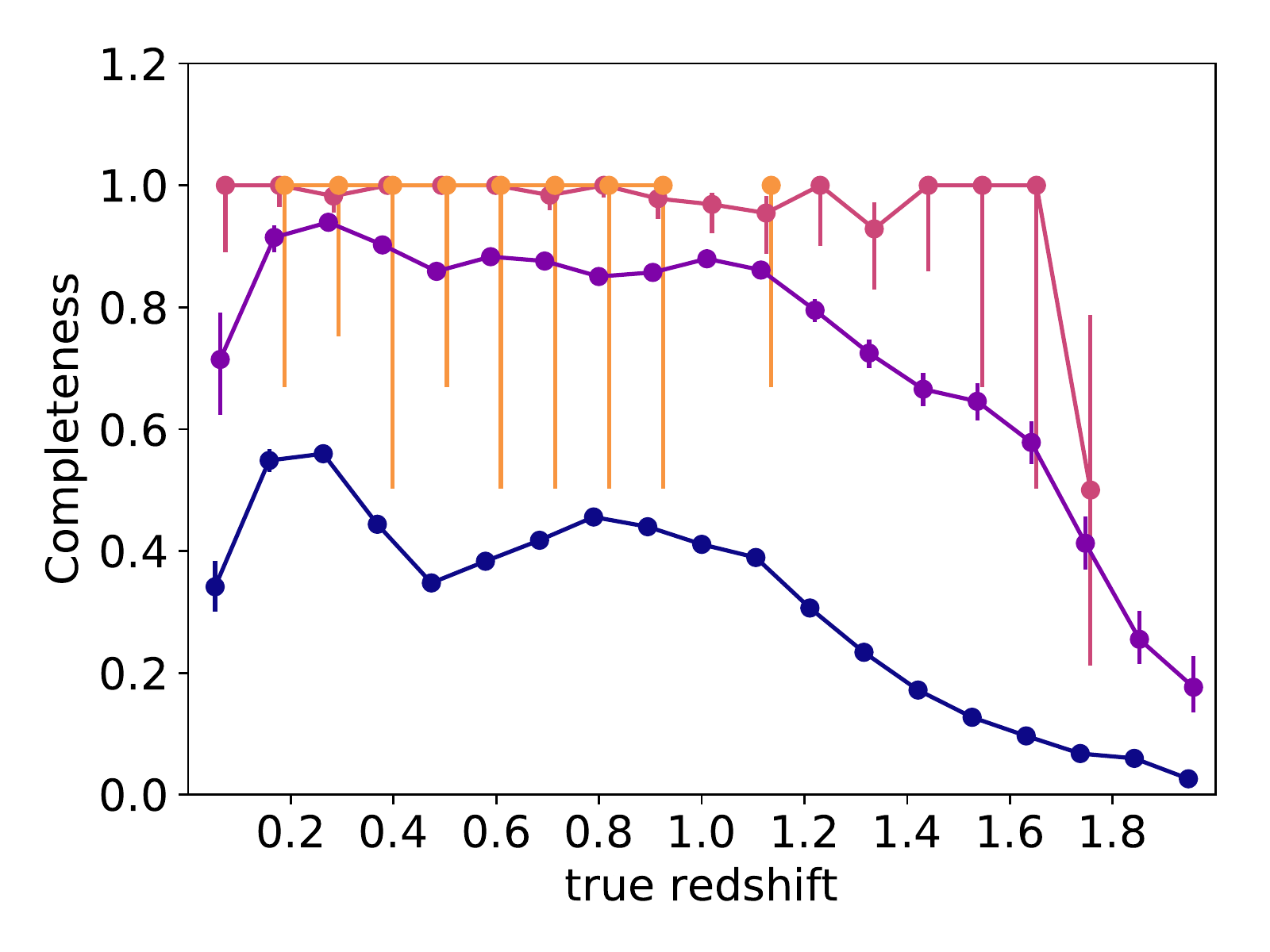} \put(-200,310){\makebox(0,0){\rotatebox{0}{\LARGE WaZP}}}
\end{tabular}}
\caption{\footnotesize{Completeness as a function of redshift in four mass bins, for the six final CFC algorithms. The four colors correspond to different mass bins, respectively $[10^{13.5}, 10^{14}]$, $[10^{14}, 10^{14.5}]$, $[10^{14.5}, 10^{15}]$ and $[10^{15}, 10^{16}]$ M$_{\odot}$. The error bars represent the 68\% confidence interval, following the Wilson score interval approach. Each catalog of detections has been trimmed to the most reliable detections insuring a mean purity of 80\% in the range $M>10^{14}$ M$_{\odot}$ and $z_{\rm true}<2$.}}
\label{fig:completeness_z}
\end{figure*}

The completeness as a function of mass and redshift, is shown in Figure~\ref{fig:completeness_m_z} for all six cluster finders. It is computed after removing the least reliable detections to reach a mean purity of 80\% for all algorithms. As it is not possible to provide error bars in Figure~\ref{fig:completeness_m_z}, we also show the completeness as a function of redshift in different mass bins in Figure~\ref{fig:completeness_z}. The error bars are computed using binomial statistics according to the Wilson score interval approach based on the number of detections in each bin. We stress that the mock catalog is the same for all detection algorithms, therefore these error bars only reflect the absolute statistical uncertainty and should not be considered when comparing the differences between the cluster finders.

All six detection algorithms provide high levels of completeness. As expected, higher mass systems are better detected than lower mass ones, regardless of the cluster finder. Clusters at masses $M>10^{15}$ M$_{\odot}$ are all detected, except for two of them that are missed by HCFA at $z \simeq$ 0.25 and 0.6. In the case of AMICO and PZWav, up to 50\% of the clusters are recovered down to masses of $10^{13.5}$ M$_{\odot}$ at low redshift. We observe a different redshift evolution for the completeness for the various finders given their different sensitivity to the mock properties. Nonetheless, a drop of up to 10\% in cluster completeness is seen at $z\sim0.5$ for most finders (except AMICO), and it could correspond to a feature in the photometric redshifts discussed in Section~\ref{sec:Photometric_redshift_properties}. The redshift evolution of the completeness could also be affected by issues related to the ranking because some codes may have given high rank preferentially in a given redshift bin and could lose some detections at specific redshifts.

\begin{figure}[h]
\includegraphics[width=0.5\textwidth]{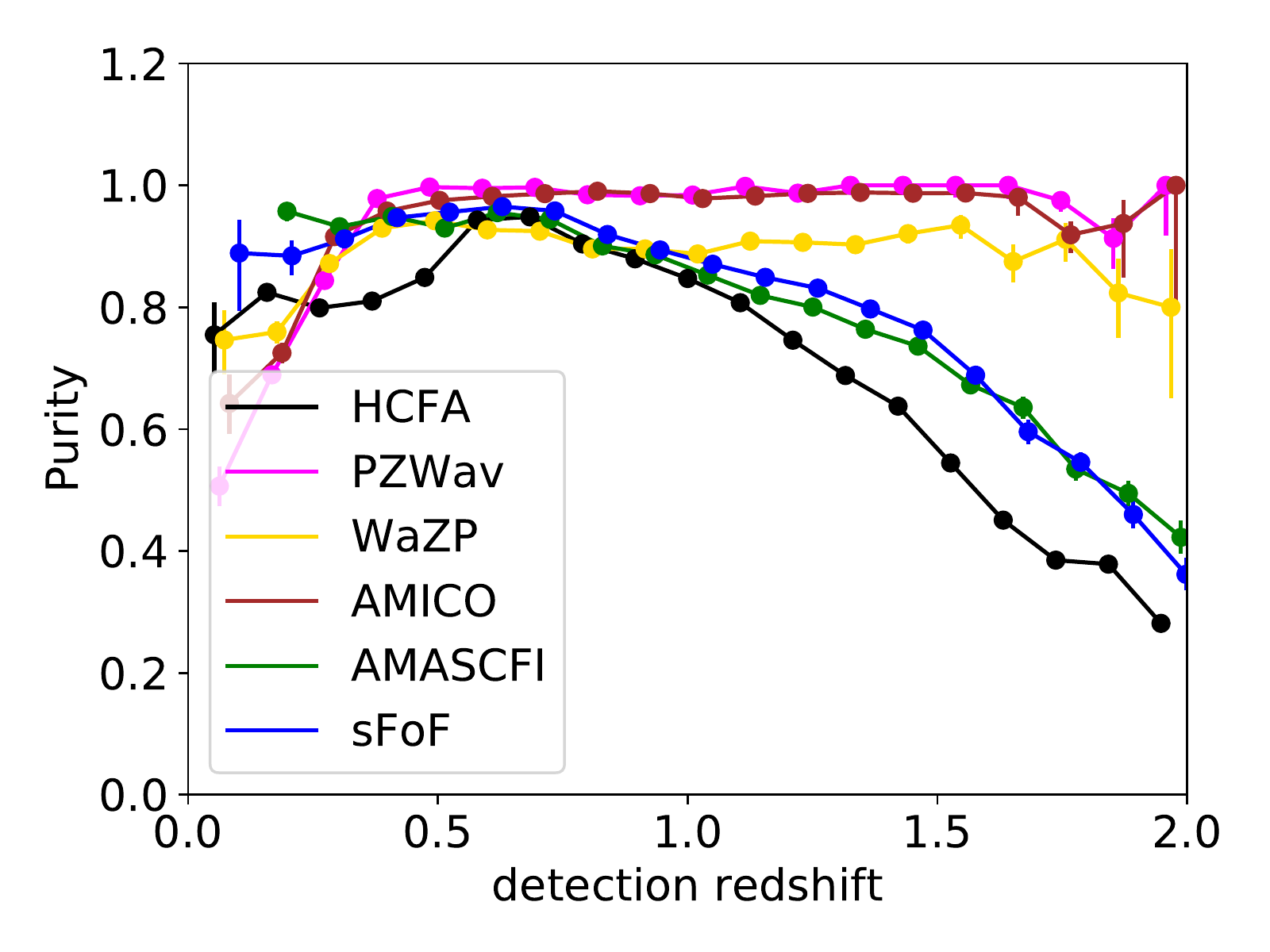}
\includegraphics[width=0.5\textwidth]{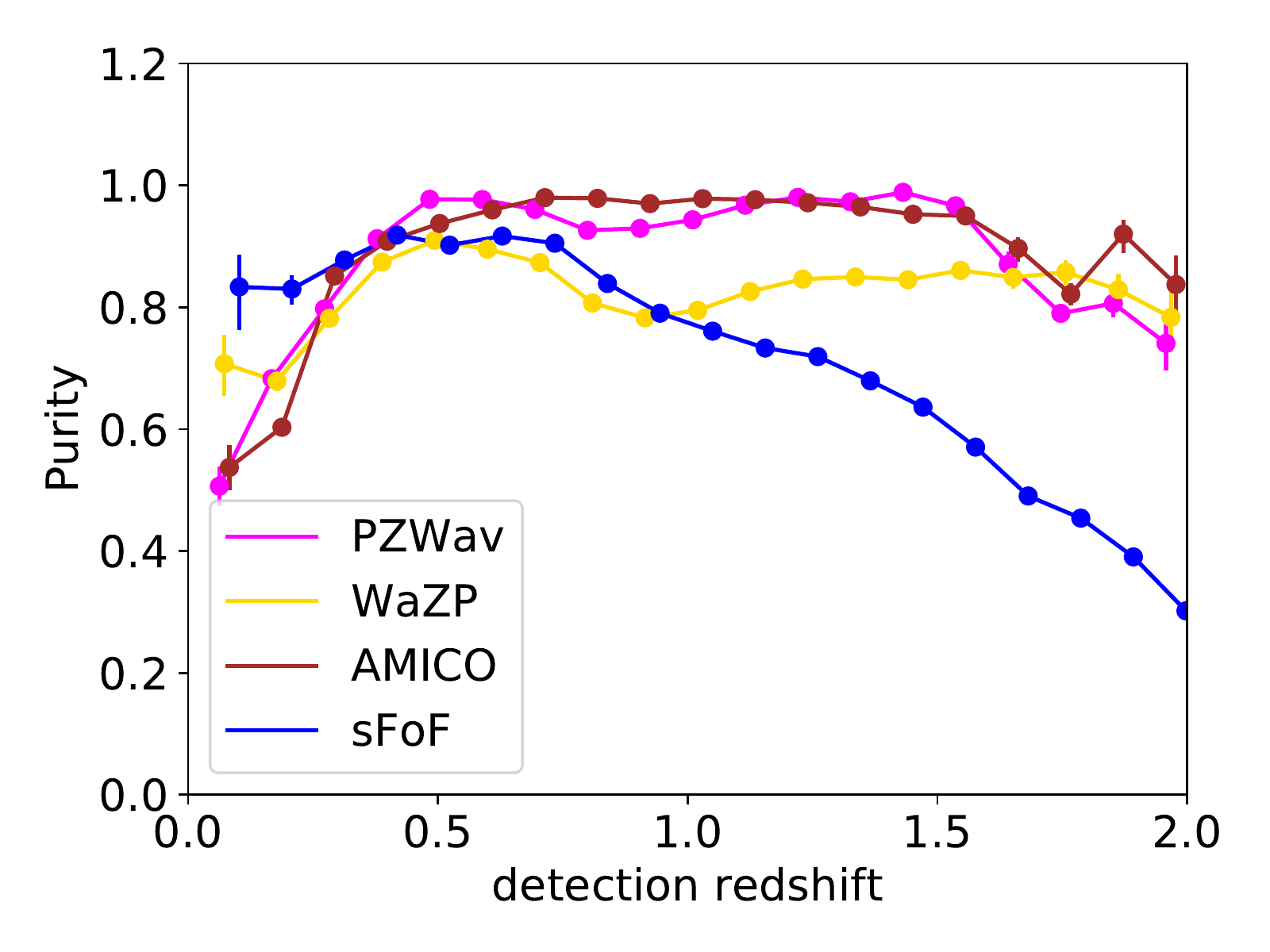}
\caption{\footnotesize{Purity as a function of redshift for a mean completeness of the detection catalogs set to $65\%$ (top panel) and $80\%$ (bottom panel) in the range $0<z_{\rm true}<2$ and for masses $M>10^{14}$ M$_{\odot}$. AMASCFI and HCFA do not reach the $80\%$ completeness level and are thus not represented in the bottom panel. Error bars are computed as in Figure~\ref{fig:completeness_z}.}}
\label{fig:purity_z_comparison}
\end{figure}

In order to assess the quality of the purity and compare it among cluster finders, we restrict the original detection catalogs to a fixed mean completeness for $0 < z_{\rm true} < 2$ and $M>10^{14}$ M$_{\odot}$ (see also the discussion of Section~\ref{sec:Comparison_criteria}, Figure~\ref{tab:perf_summary} and Table~\ref{tab:perf_summary}). We first choose a mean completeness of $65\%$, which is reached by all the codes, and investigate the evolution of purity as a function of redshift (Figure~\ref{fig:purity_z_comparison} top panel). We can see that the purity evolves very differently with redshift for the different algorithms. AMICO, PZWav and WaZP provide pure samples up to high redshift, except for $z \lesssim 0.25$, where the purity drops. This might be due to a strong dependence on the photometric redshift quality of these algorithms, as the outlier fraction in the photometric redshifts strongly increases at $z<0.25$. It could also be caused by the unavoidable fragmentation of very rich clusters, with secondary fragments being counted as impurities. Additionally, in this regime, the mass detection limit of these algorithms (see Figure \ref{fig:completeness_m_z}) may be sufficiently low such that some detected clusters are below the mass threshold of the mock and are thus counted as impurities (see also the discussion in Section \ref{sec:Comparison_criteria}). On the other hand, AMASCFI, HCFA and sFoF catalogs are more pure at low redshifts, but their purity smoothly declines as redshift increases. We also analyzed the redshift evolution of the purity for detection catalogs trimmed to reach $80\%$ completeness. Since HCFA and AMASCFI do not reach this completeness limit, they are not considered here. The comparison of the four resulting algorithms is shown in Figure~\ref{fig:purity_z_comparison} (bottom panel). We can see that the purity is still above about 90\% and relatively flat in the redshift range $[0.25, 2]$ for AMICO and PZWav. The performance of WaZP is slightly lower, but still very good at these redshifts (about 80\%-90\%). The purity of the algorithm sFoF, on the other hand, smoothly declines as redshift increases in this regime. At low redshifts (below 0.25), sFoF remains relatively flat with a purity of about 85\%, while the purity of AMICO, PZWav and WaZP slightly decline, down to about 60\% to 70\%. Nevertheless, we stress that at low redshifts, the mean completeness of these algorithms is much higher than the mean completeness that we impose for this comparison exercise, due to redshift variations. Thus, contamination may arise from objects at the detection limits of the catalog.
 
\subsection{Mass-richness scaling relation}\label{sec:Mass_richness_scaling_relation}
Detections matched to mock clusters were used to investigate the quality of the richness provided by the cluster finders. The aim of this section is to qualitatively compare the goodness of the mass proxy derived from the various algorithms, without going to a detailed characterization of the mass-observable relation, which will be addressed in future work. To do so, we use the full catalogs of true detections (matched clusters), considering all available ranks, because of higher statistics and higher completeness, but the results are not sensitive to this choice because only objects well above the detection limit (see below) are used. The analysis described in Section~\ref{sec:mock_mass_lambda_relation} is reproduced using a mass proxy as provided by the cluster finders. While the richness definition varies from algorithm to algorithm, they are all expected to scale with mass. In order to have comparable numerical values, the richness values provided by the respective algorithms were normalized to the minimum richness. However, this only affects the normalization of the mass-richness scaling, which is not a concern for the present comparison. To mitigate effects induced by Malmquist bias, resulting from the detection limit of the algorithms, we fit the scaling relation with a power law of slope $s$: $\lambda_{\rm det} \propto M^{s}$ in the range where the median richness is above the detection limit by more than three times NMAD.

\begin{figure*}[h]
\resizebox{\textwidth}{!} {
\begin{tabular}{lll}
\includegraphics[trim=0cm 1.8cm 0cm 0cm, clip=true, scale=1]{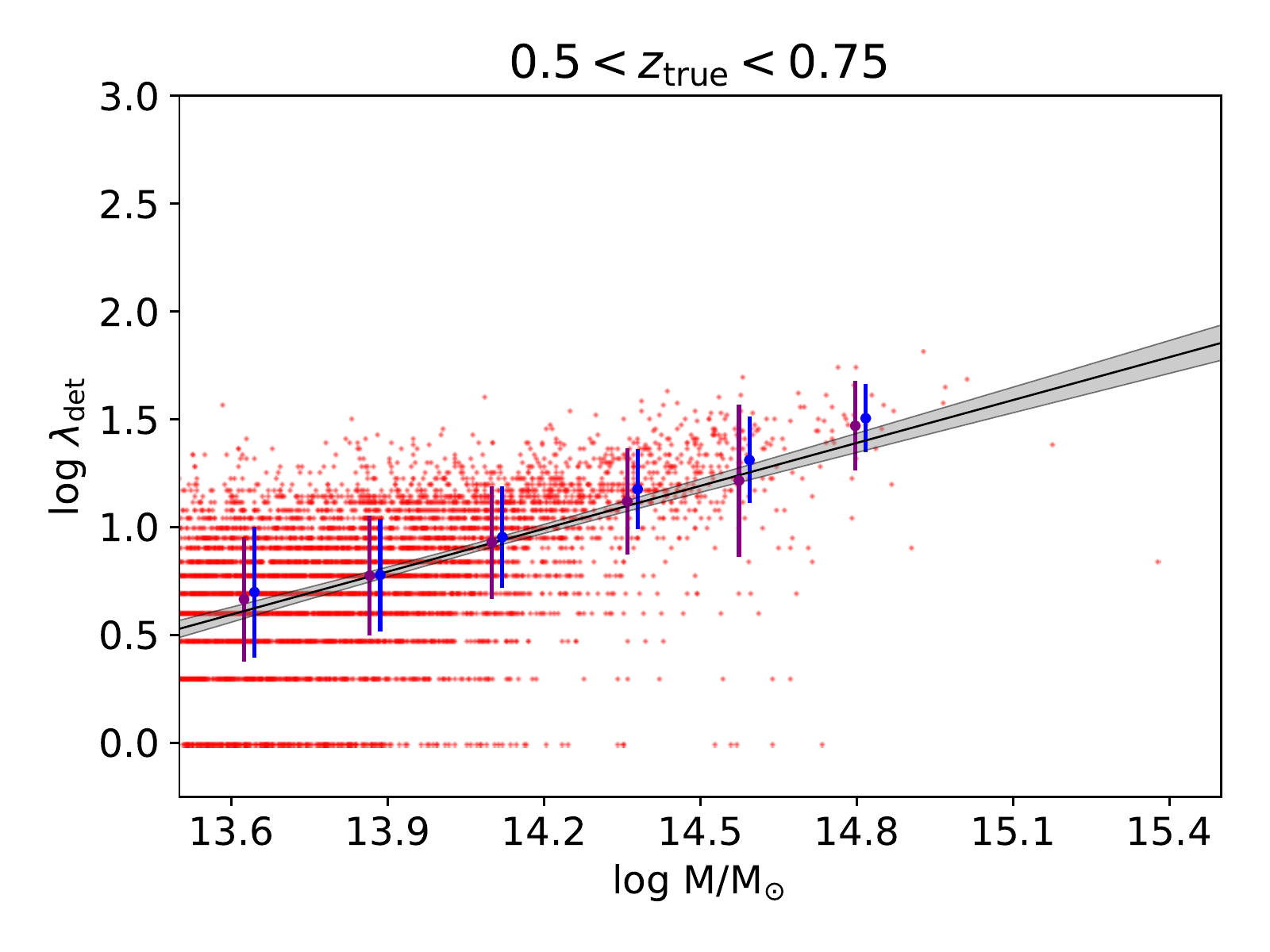} \put(-70,30){\makebox(0,0){\rotatebox{0}{\LARGE AMASCFI}}} &
\includegraphics[trim=2.1cm 1.8cm 0cm 0cm, clip=true, scale=1]{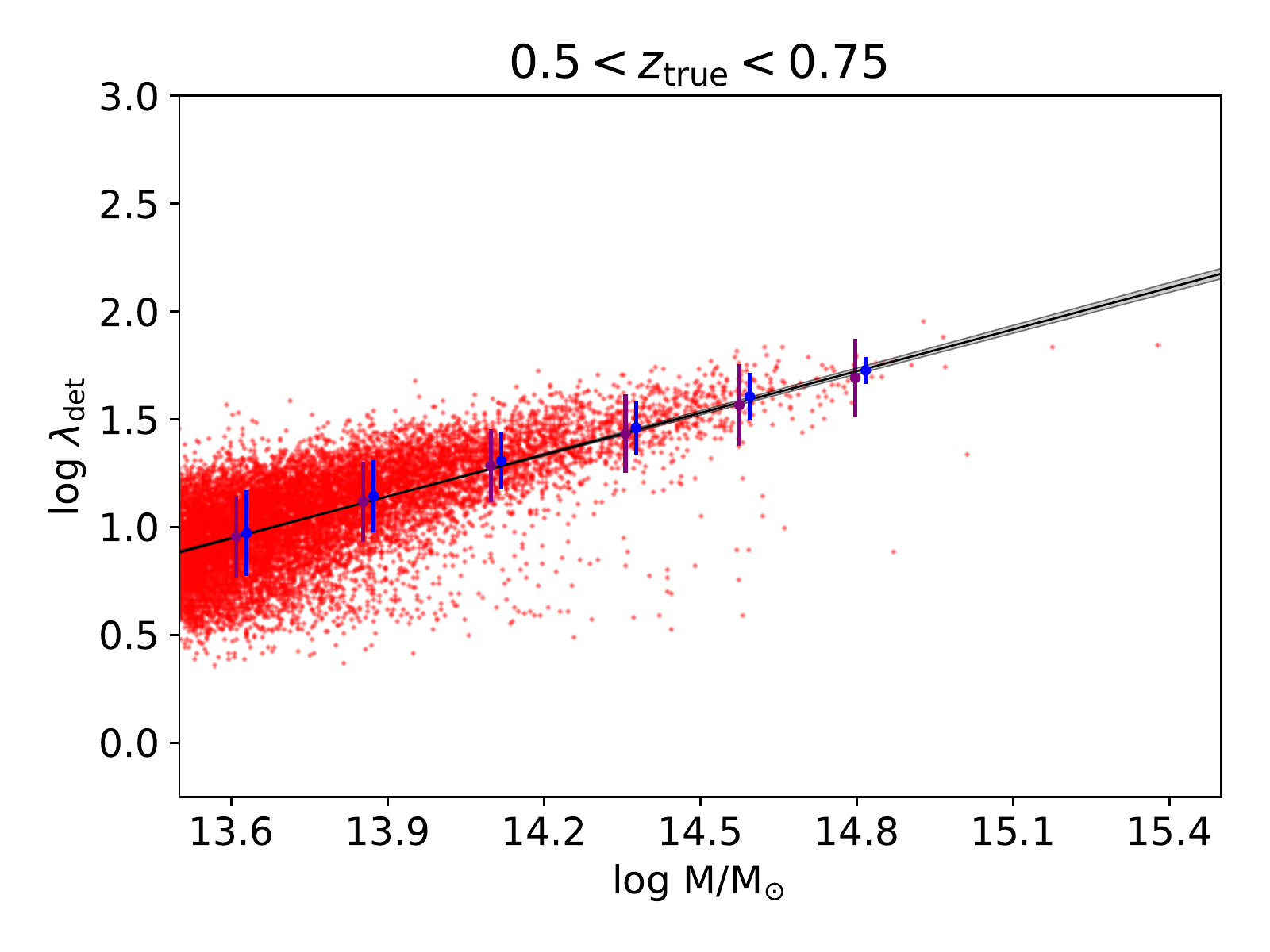} \put(-70,30){\makebox(0,0){\rotatebox{0}{\LARGE AMICO}}} &
\includegraphics[trim=2.1cm 1.8cm 0cm 0cm, clip=true, scale=1]{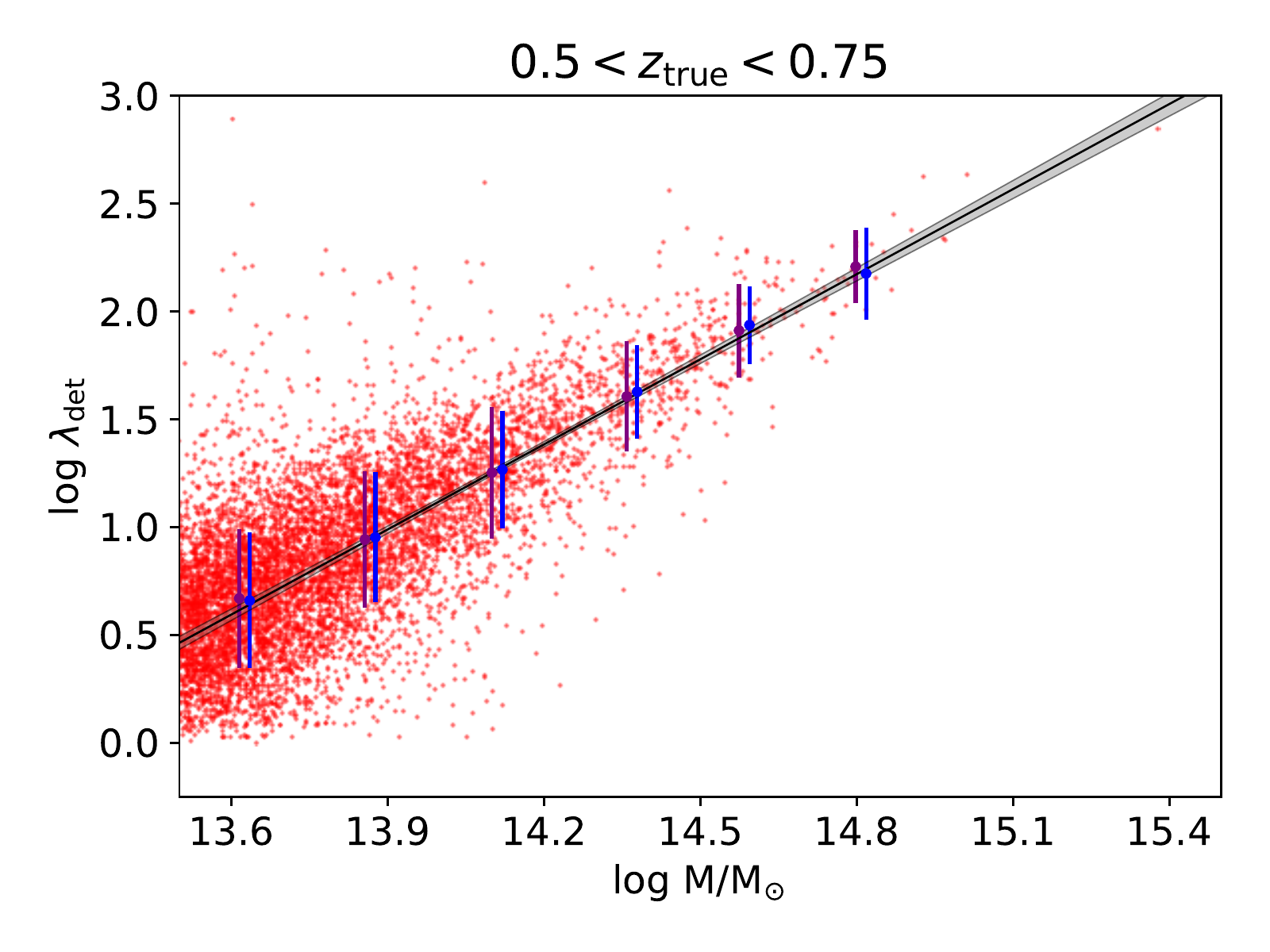} \put(-70,30){\makebox(0,0){\rotatebox{0}{\LARGE HCFA}}}\\
\includegraphics[trim=0cm 0cm 0cm 0cm, clip=true, scale=1]{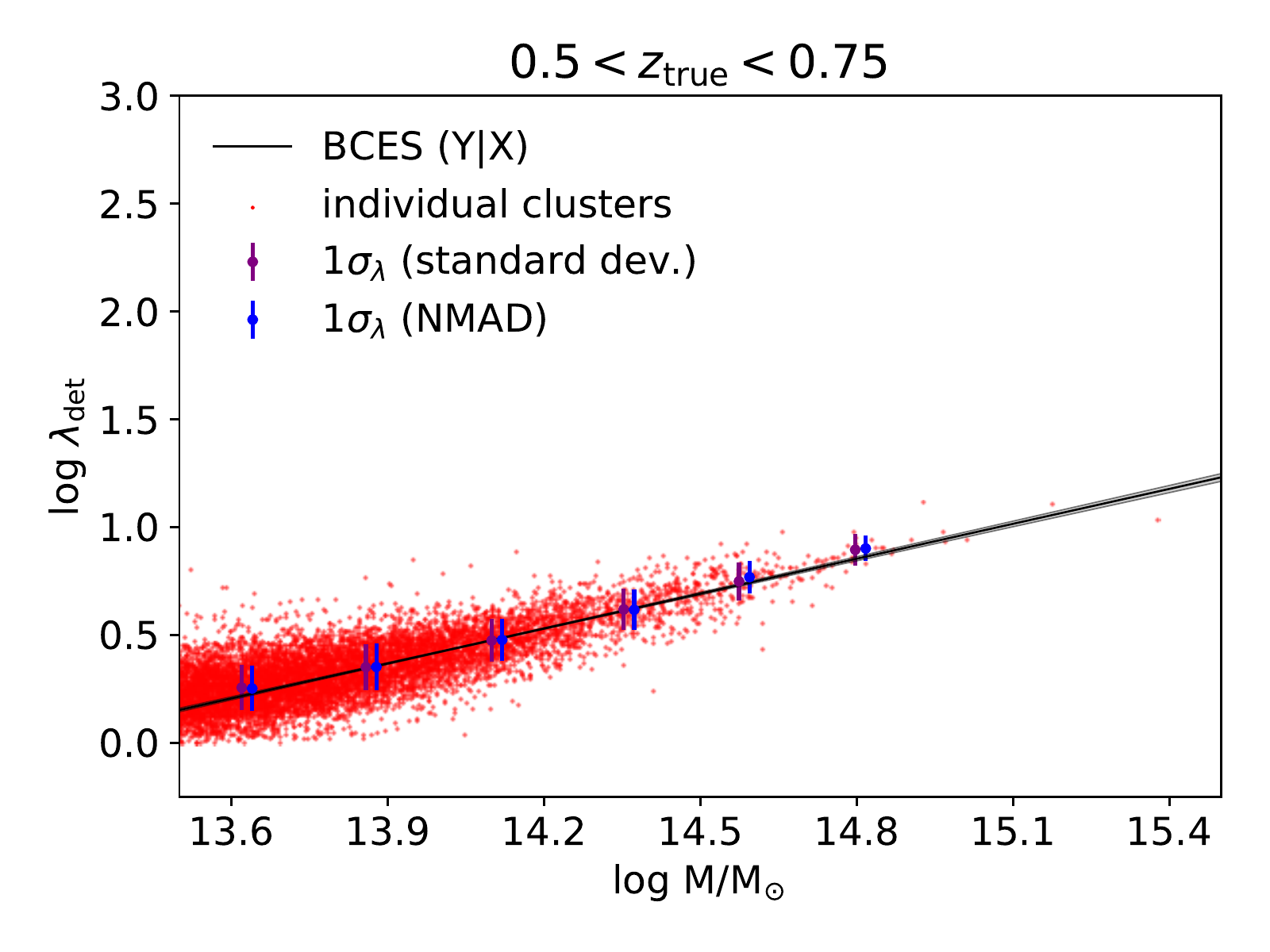} \put(-70,80){\makebox(0,0){\rotatebox{0}{\LARGE PZWav}}} &
\includegraphics[trim=2.1cm 0cm 0cm 0cm, clip=true, scale=1]{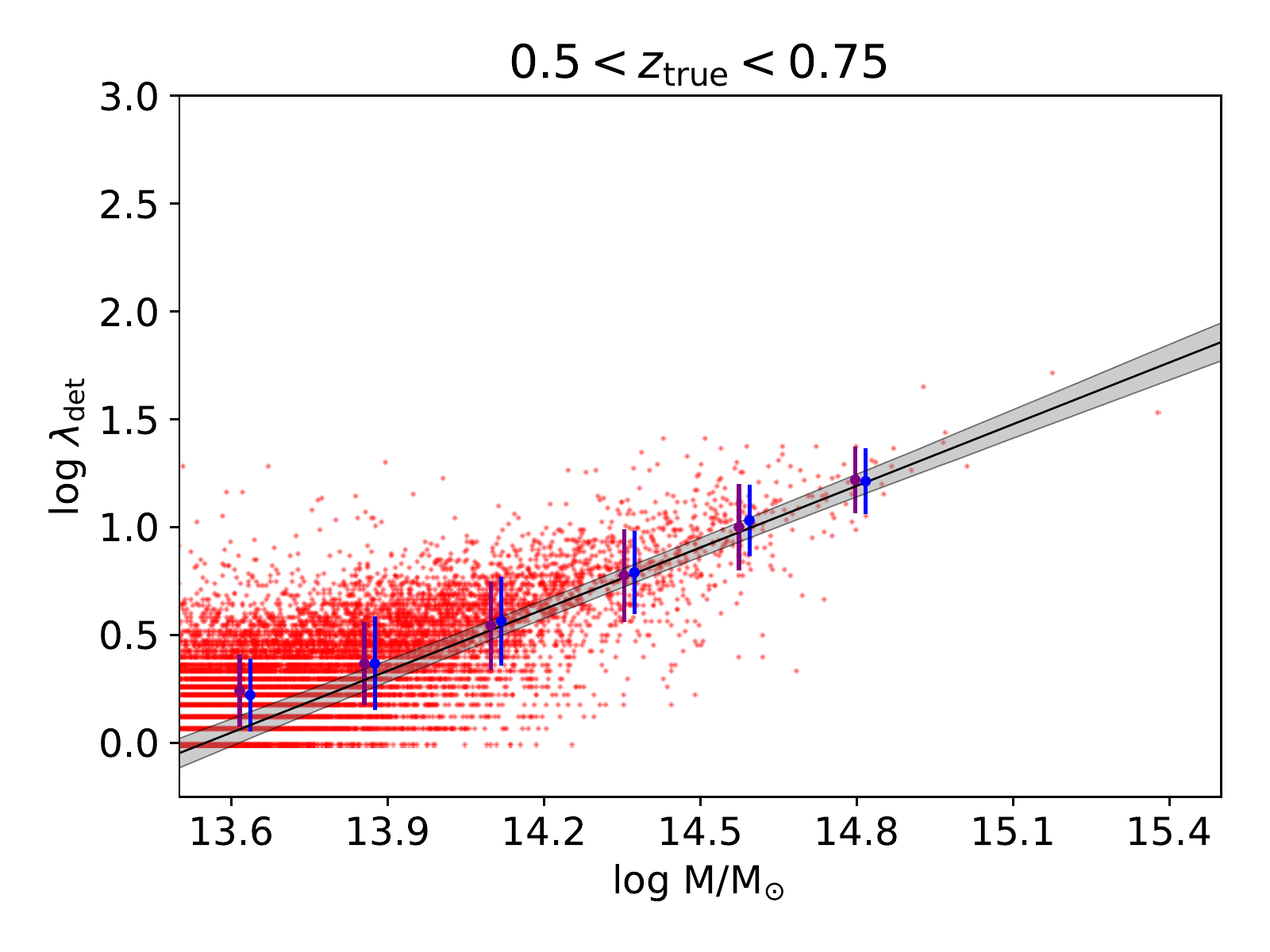} \put(-70,80){\makebox(0,0){\rotatebox{0}{\LARGE sFoF}}} &
\includegraphics[trim=2.1cm 0cm 0cm 0cm, clip=true, scale=1]{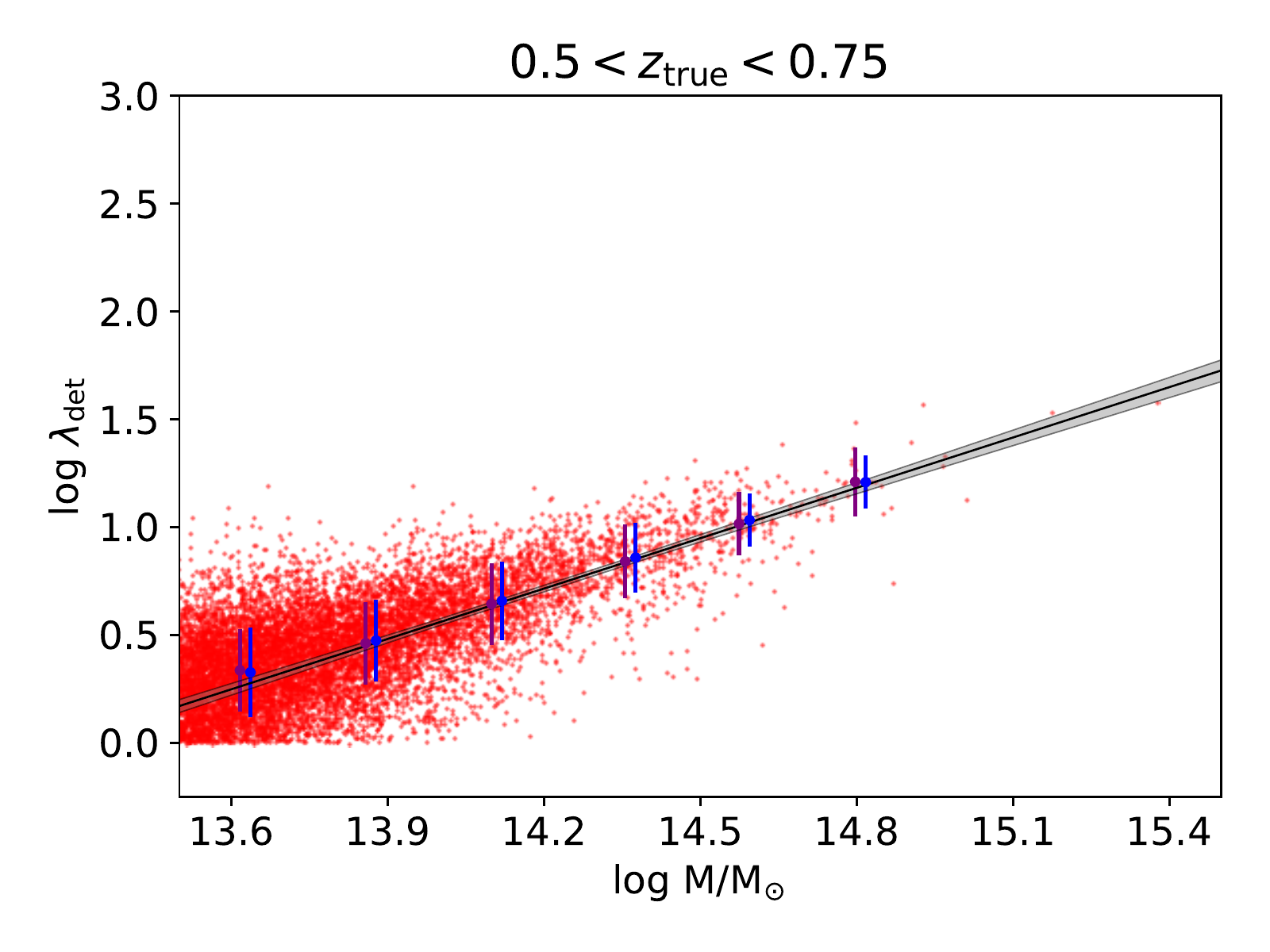} \put(-70,80){\makebox(0,0){\rotatebox{0}{\LARGE WaZP}}}
\end{tabular}}
\caption{\footnotesize{Example of the mass-richness scaling, for redshifts in the range $z_{\rm true} = [0.5, 0.75]$ for all the cluster finders. The legend is the same as in Figure~\ref{fig:mass_lambda_mocks}. To improve statistics, the full detection catalogs, corresponding to true detections, were used to produce the figure. The original richness provided by the finders was renormalized for better comparison. Note that the richness provided by AMASCFI and sFoF are discrete, while it is continuous for the other codes.}}
\label{fig:mass_lambda_algo}
\end{figure*}

Figure~\ref{fig:mass_lambda_algo} illustrates the scaling relation and its best-fit for the six detection algorithms, in the redshift range [0.5, 0.75]. This figure can be directly compared to Figure~\ref{fig:mass_lambda_mocks}. We can see that all cluster finder mass proxies have different definitions, which scale in different ways with mass. For instance, the mass proxy scales with different slopes: ${0.65}$, ${0.61}$, ${1.27}$, ${0.51}$, ${0.96}$ and ${0.78}$ for AMASCFI, AMICO, HCFA, PZWav, sFoF and WaZP, respectively. The slope remains stable independently of the number of matched clusters we consider. The mass proxy is discrete in the case of AMASCFI and sFoF, while it is continuous for the other ones. Even an ideal mass proxy derived from the true galaxy members in mock catalogs would have a significant scatter with respect to the true halo mass, as we have shown in Section~\ref{sec:mock_mass_lambda_relation}. As our aim is to measure the performance of the algorithms, we subtract this intrinsic scatter from the measured one. The statistical scatter induced in the detection process is computed as $\sigma_{\rm det} = \sqrt{\sigma_{\rm meas}^2/s_{\rm meas}^2 - \sigma_{\rm int}^2/s_{\rm int}^2}$, where $\sigma_{\rm meas}$ is the overall scatter of richness at a given mass as measured after the detections are performed and $s_{\rm meas}$ the associated slope, and $\sigma_{\rm int}$ is the intrinsic scatter as discussed in Section~\ref{sec:mock_mass_lambda_relation} with $s_{\rm int}$ the associated slope. We note that the different mass dependence, given by the slope $s$, between all the algorithms, is accounted for when computing the scatter. The scatter $\sigma_{\rm det}$ was measured for different mass bins and redshift bins. We provide its median value for masses above $10^{14}$ M$_{\odot}$ in Table~\ref{tab:perf_summary}. Once normalized by the slope, HCFA presents the smallest scatter, with a median value of 0.15 dex. Most algorithms perform reasonably well, with $\sigma_{\rm det} \simeq 0.18$ dex, going up to 0.27 dex for AMASCFI \citep[see, e.g.,][for comparison with low scatter richness-based mass proxies]{Andreon2015}.

\subsection{Redshift and centroid reconstruction}\label{sec:centroid_and_redshift_differences}
\begin{figure*}[h]
\resizebox{\textwidth}{!} {
\begin{tabular}{lll}
\includegraphics[trim=0cm 1.7cm 0cm 0cm, clip=true, scale=1]{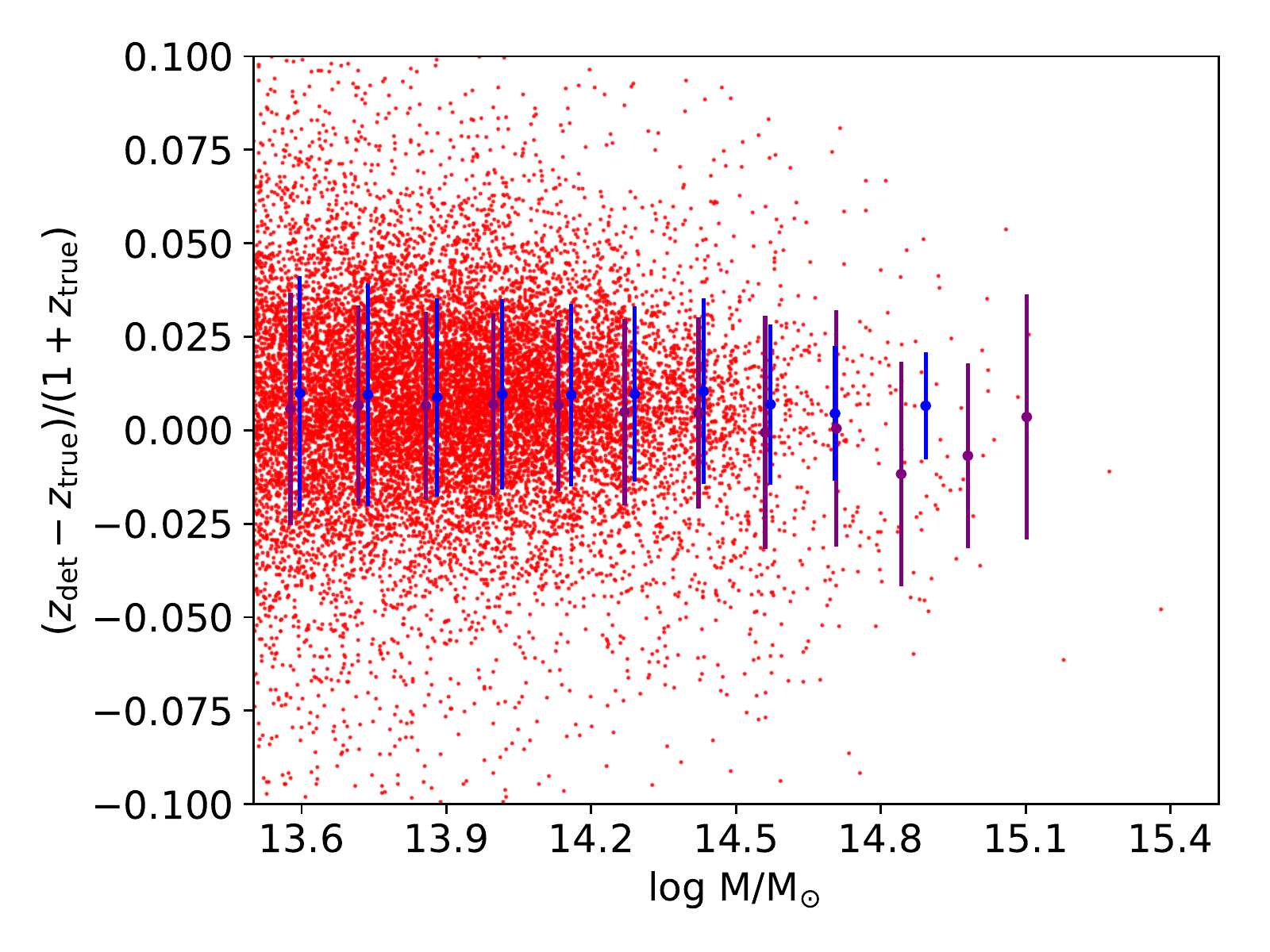} \put(-90,240){\makebox(0,0){\rotatebox{0}{\LARGE AMASCFI}}} &
\includegraphics[trim=3.0cm 1.7cm 0cm 0cm, clip=true, scale=1]{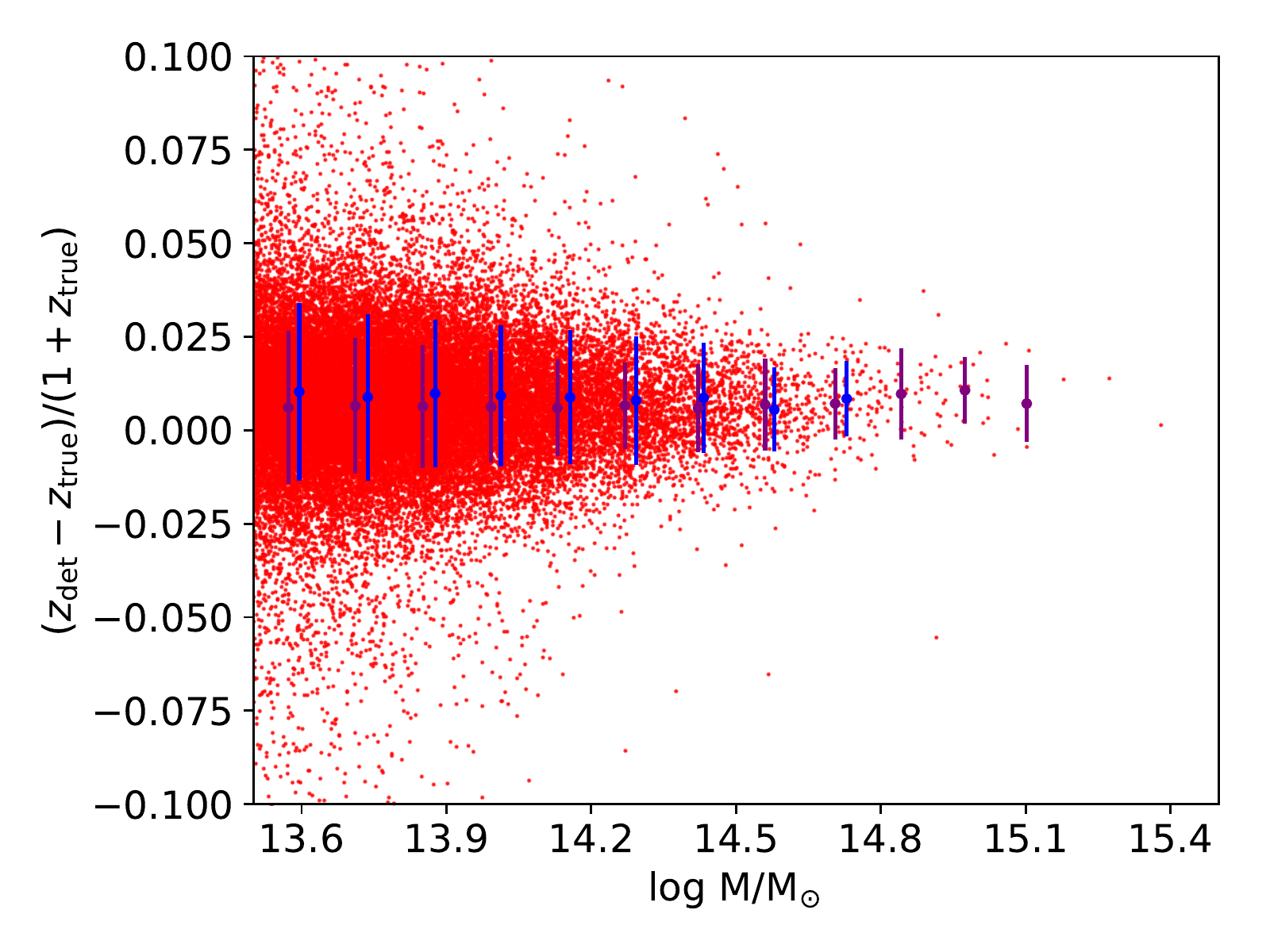} \put(-90,240){\makebox(0,0){\rotatebox{0}{\LARGE AMICO}}} &
\includegraphics[trim=3.0cm 1.7cm 0cm 0cm, clip=true, scale=1]{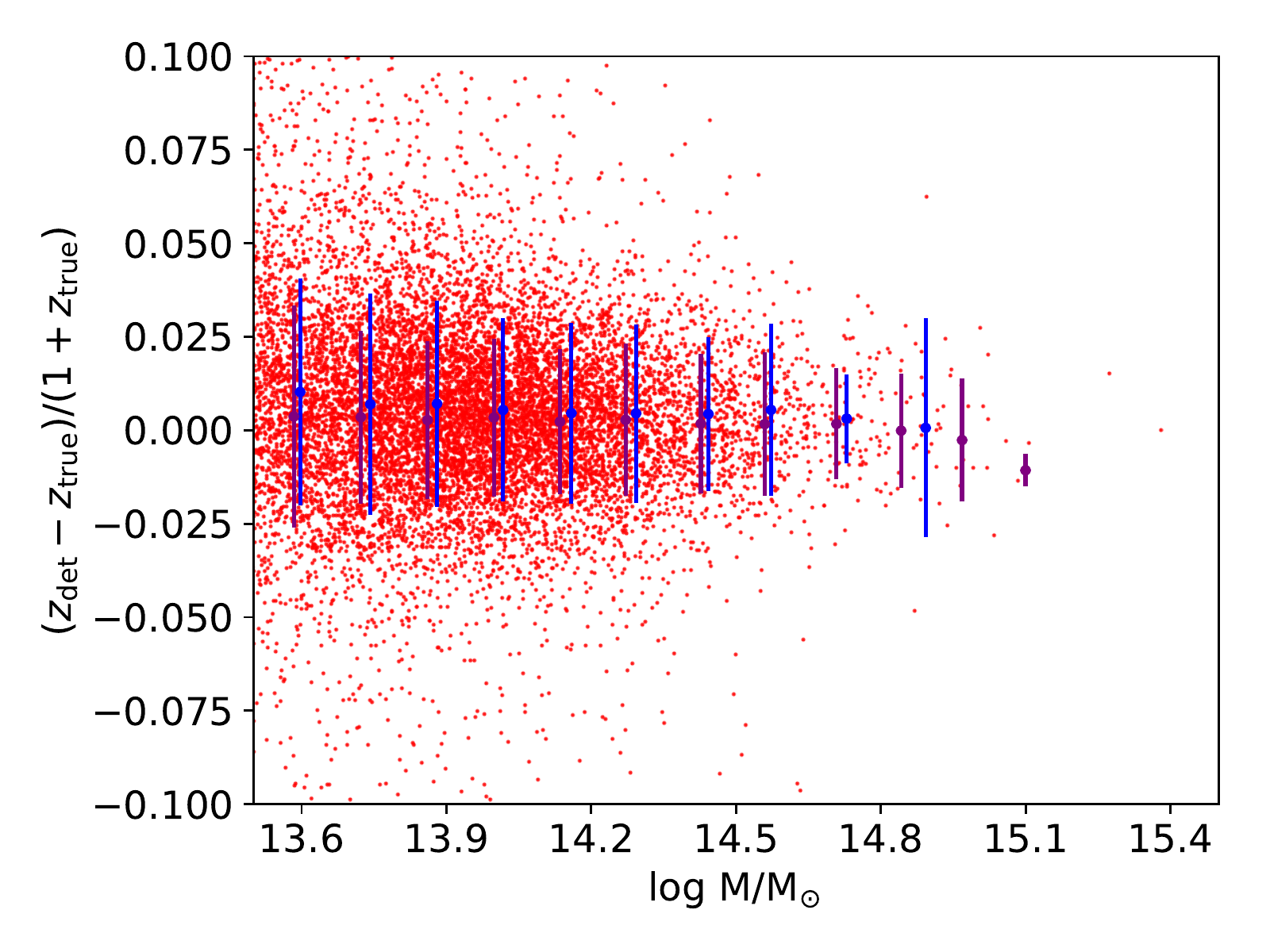} \put(-90,240){\makebox(0,0){\rotatebox{0}{\LARGE HCFA}}}\\
\includegraphics[trim=0cm 0cm 0cm 0cm, clip=true, scale=1]{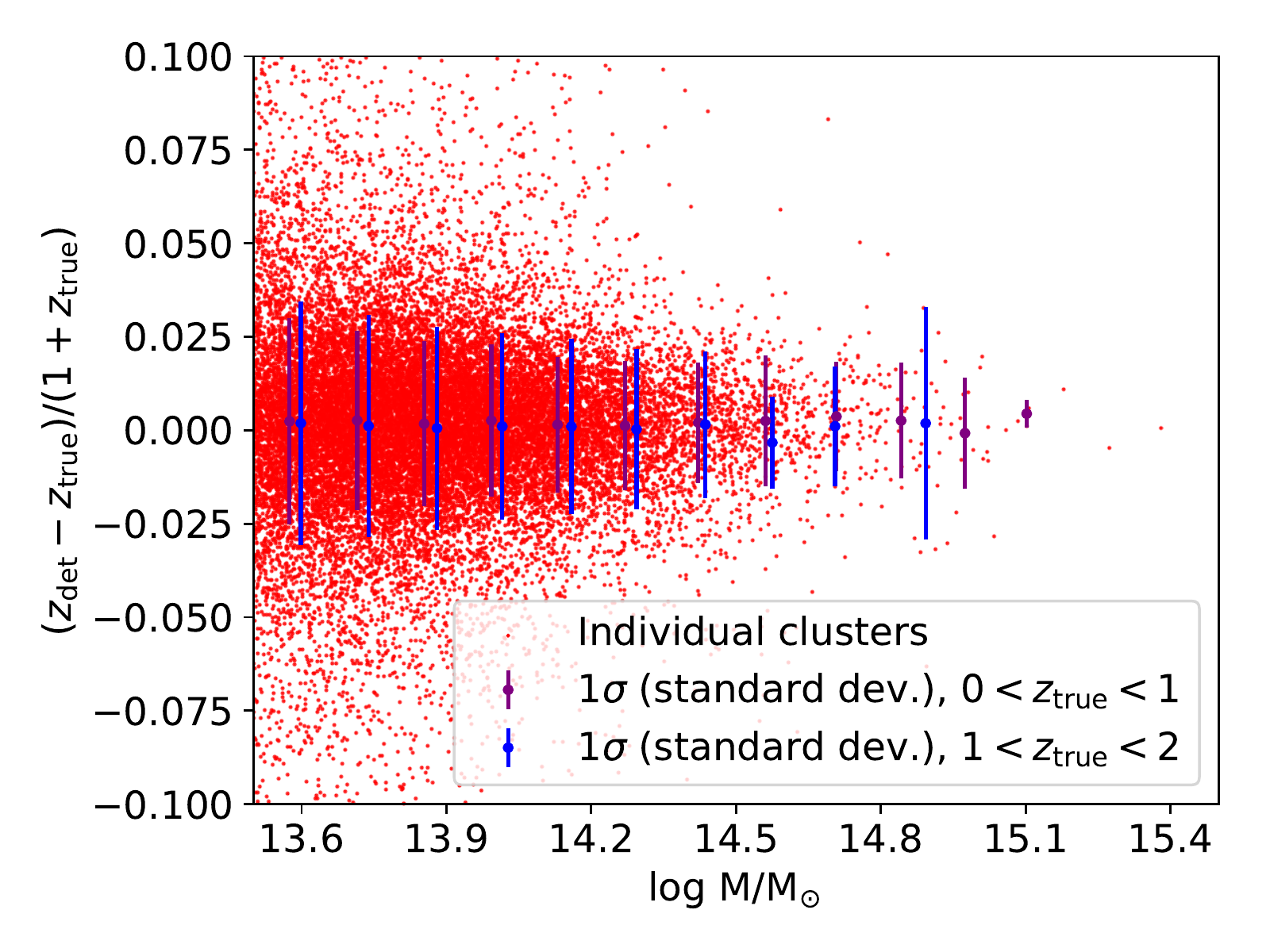} \put(-90,280){\makebox(0,0){\rotatebox{0}{\LARGE PZWav}}} &
\includegraphics[trim=3.0cm 0cm 0cm 0cm, clip=true, scale=1]{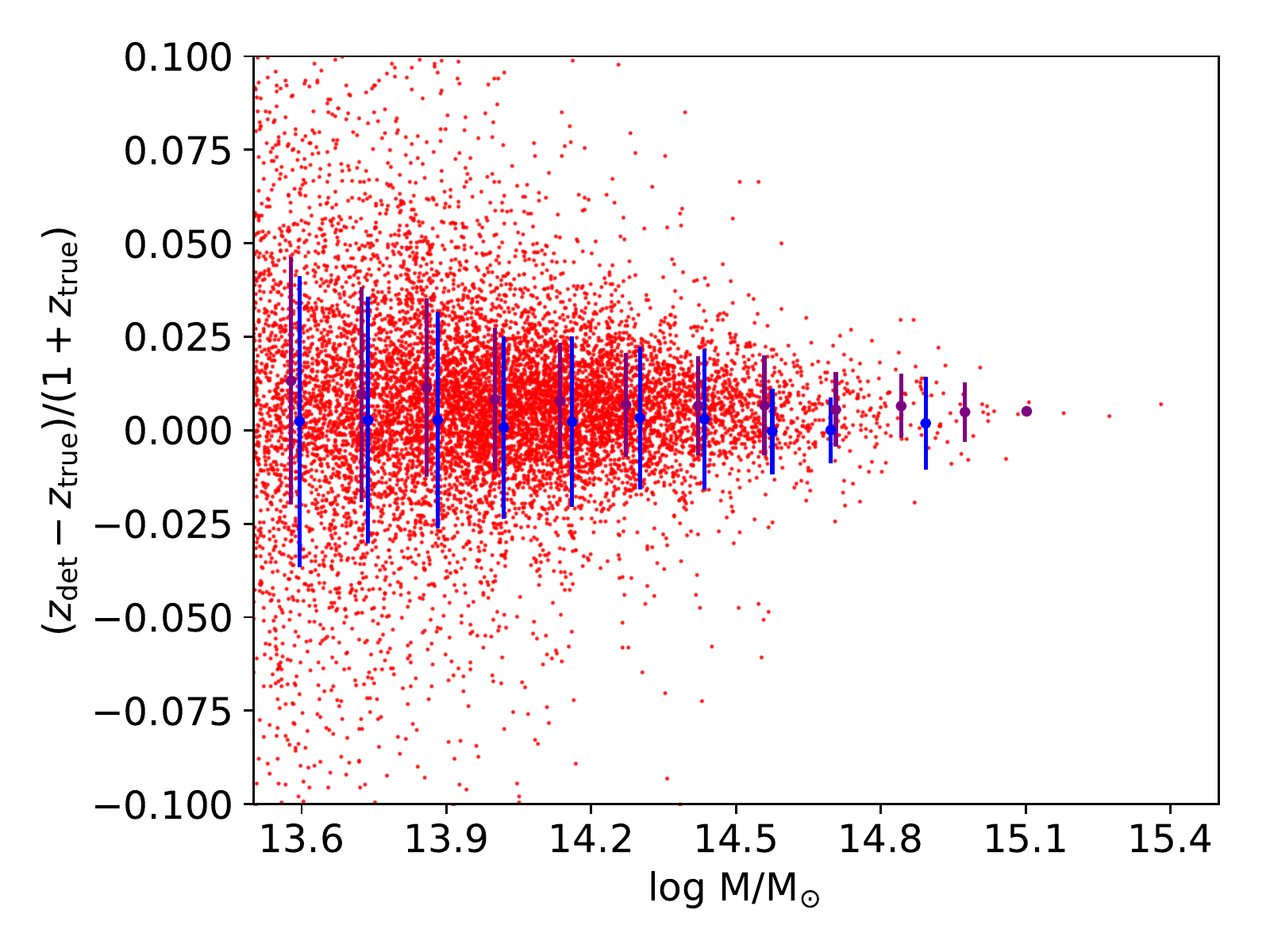} \put(-90,280){\makebox(0,0){\rotatebox{0}{\LARGE sFoF}}} &
\includegraphics[trim=3.0cm 0cm 0cm 0cm, clip=true, scale=1]{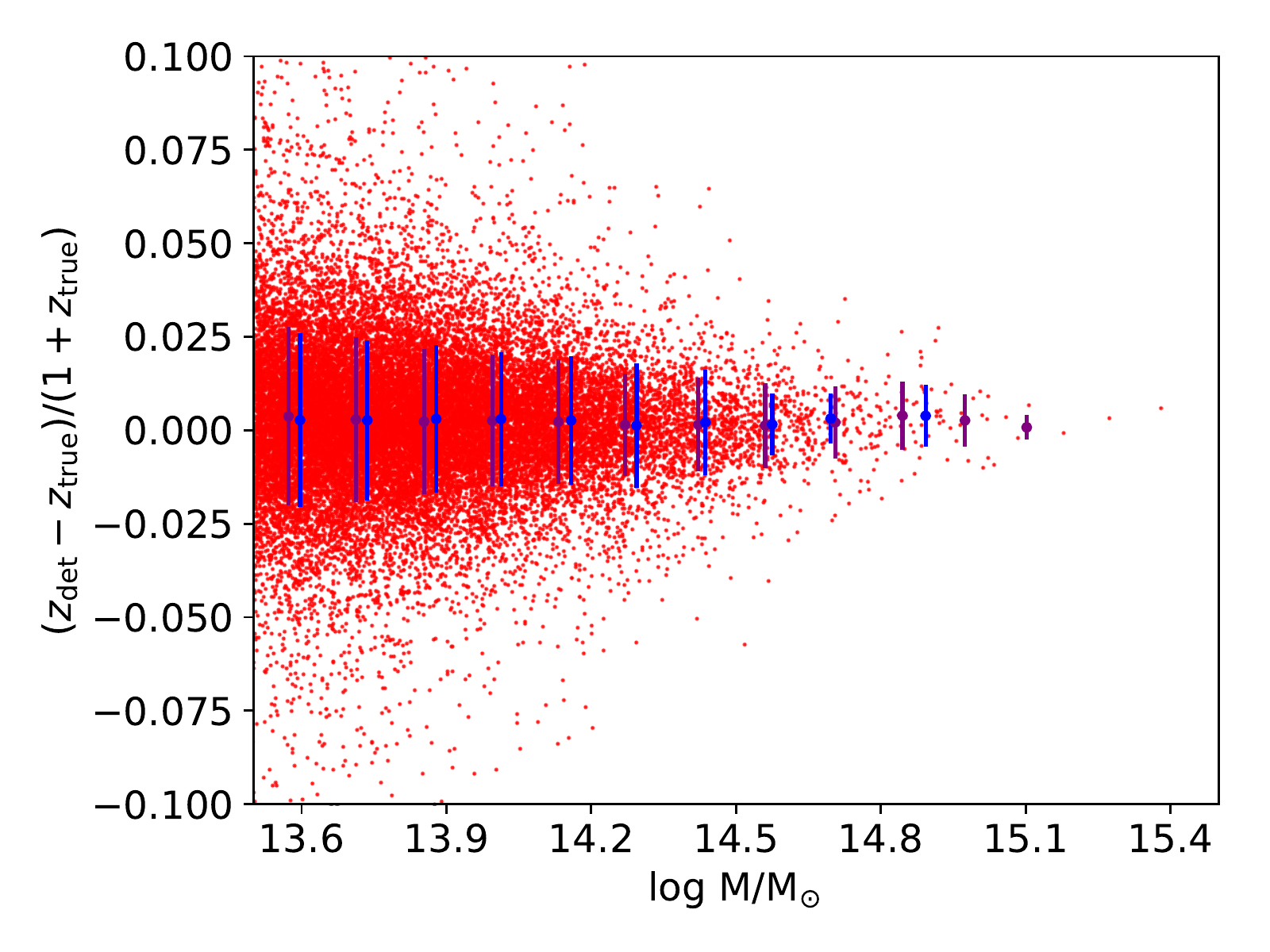} \put(-90,280){\makebox(0,0){\rotatebox{0}{\LARGE WaZP}}}
\end{tabular}}
\caption{\footnotesize{Difference between the redshift associated to a detected cluster and the mock cluster redshift from the simulation, as a function of mock cluster mass, for all cluster finders. The red points show the individual clusters and the error bars provide the standard deviation of the distribution as a function of mass (in bins of 0.14 dex) for two redshift bins. Each catalog of detections has been trimmed to the most reliable detections ensuring a mean purity of 80\% in the range $M>10^{14}$ M$_{\odot}$ and $z_{\rm true}<2$.}}
\label{fig:redshift_recovery}
\end{figure*}

The matched detections are used to compute the differences between the recovered cluster redshifts and the true redshifts. Figure~\ref{fig:redshift_recovery} shows the distribution of this redshift difference for each algorithm and for different mass intervals. We also present the standard deviation of this distribution as a function of mass in two redshift bins ($0<z_{\rm true}<1$ and $1<z_{\rm true}<2$). Because of the matching procedure, this difference is limited to $(1+z_{\rm true}) \sigma_{z_{\rm phot}}$ with a typical standard deviation of $\Delta z / (1+z_{\rm true}) \sim 0.02$ at $M > 10^{14}$ M$_{\odot}$ (compared to 0.2 for the depth of the cylinder along the line of sight used for the matching procedure). We observe that the redshifts of more massive clusters are slightly better recovered than lower mass ones for all detection algorithms (typically an increase in the scatter by a factor of two from $10^{14.5}$ to $10^{13.5}$ M$_{\odot}$), because more galaxies are accessible to perform the redshift estimates. Only a very small redshift dependence is visible once normalized by $\left(1+z_{\rm true}\right)$. While the redshift difference distribution is always compatible with zero, we can observe a small mass-independent redshift bias for all cluster finders, which reflect the photometric redshift bias (see Section~\ref{sec:Photometric_redshift_properties}). The standard deviation of the redshift difference is shown in Table~\ref{tab:perf_summary} for massive clusters ($M > 10^{14}$ M$_{\odot}$). AMICO and WaZP are the ones that perform the best, reaching $\Delta z / (1+z_{\rm true}) = 0.015$, but all the algorithms demonstrate a relatively tight performance, with a maximum of 0.025 for AMASCFI.

\begin{figure*}[h]
\resizebox{\textwidth}{!} {
\begin{tabular}{lll}
\includegraphics[trim=0cm 1.7cm 0cm 0cm, clip=true, scale=1]{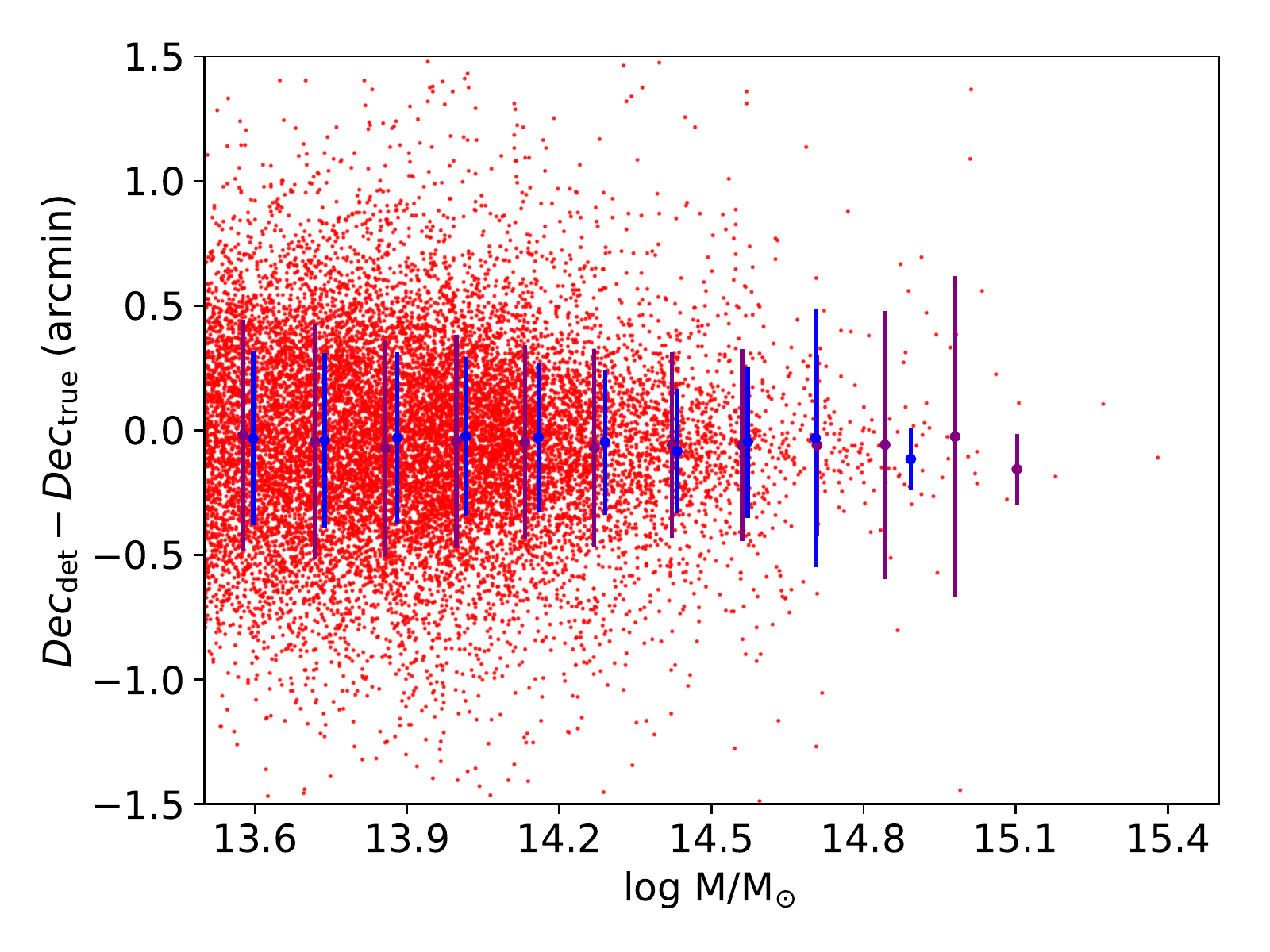} \put(-90,240){\makebox(0,0){\rotatebox{0}{\LARGE AMASCFI}}} &
\includegraphics[trim=2.4cm 1.7cm 0cm 0cm, clip=true, scale=1]{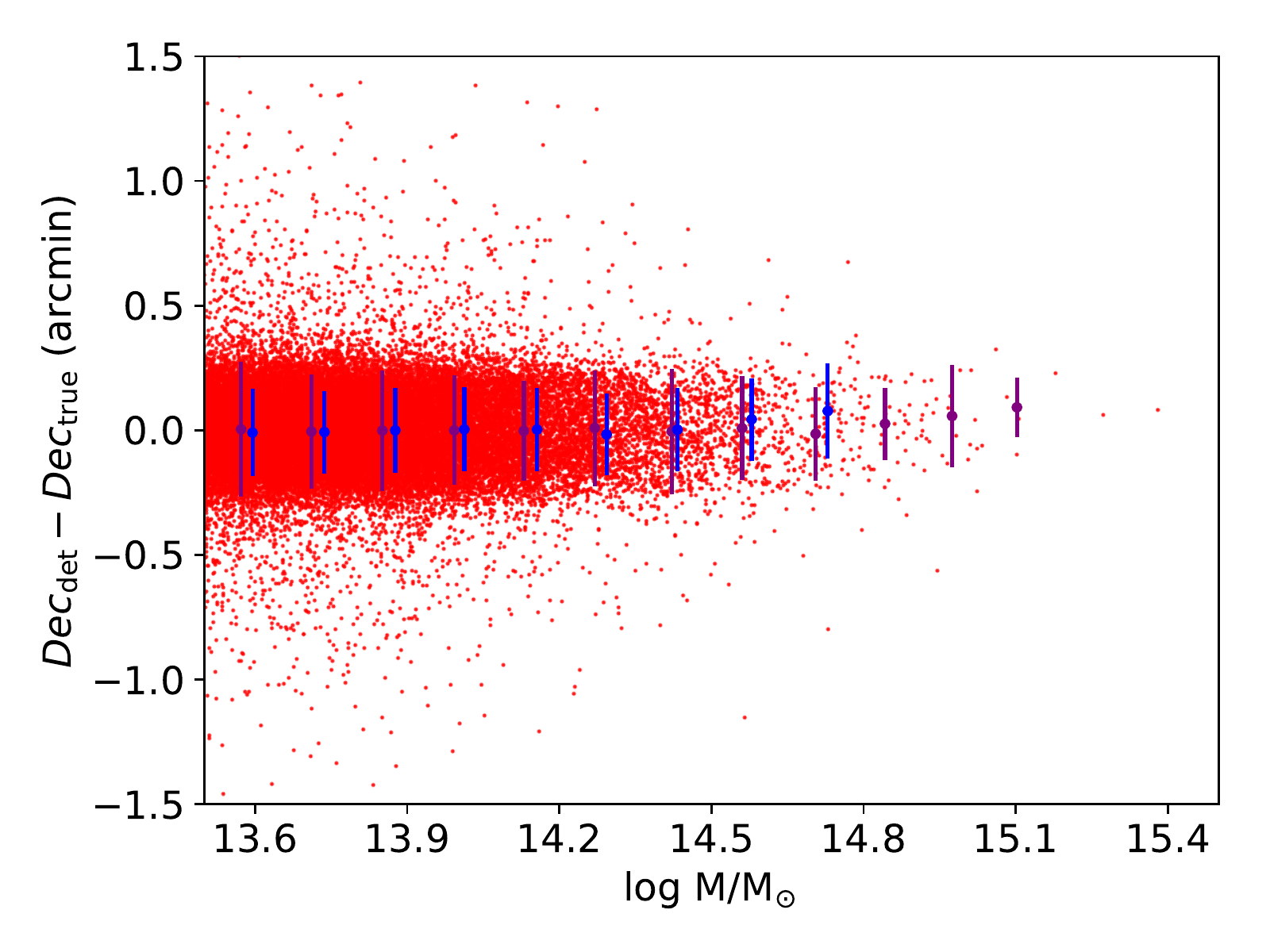} \put(-90,240){\makebox(0,0){\rotatebox{0}{\LARGE AMICO}}} &
\includegraphics[trim=2.4cm 1.7cm 0cm 0cm, clip=true, scale=1]{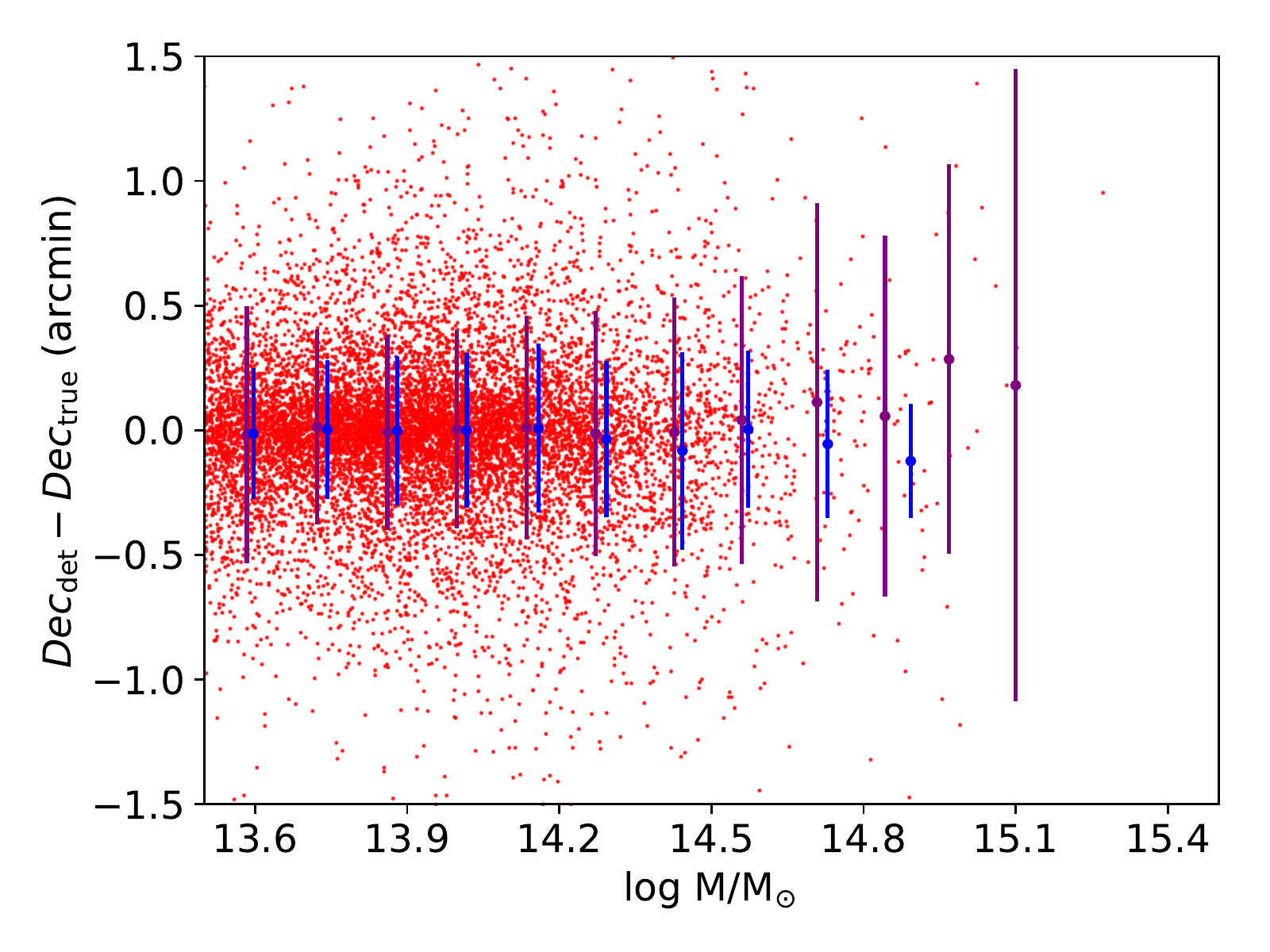} \put(-90,240){\makebox(0,0){\rotatebox{0}{\LARGE HCFA}}}\\
\includegraphics[trim=0cm 0cm 0cm 0cm, clip=true, scale=1]{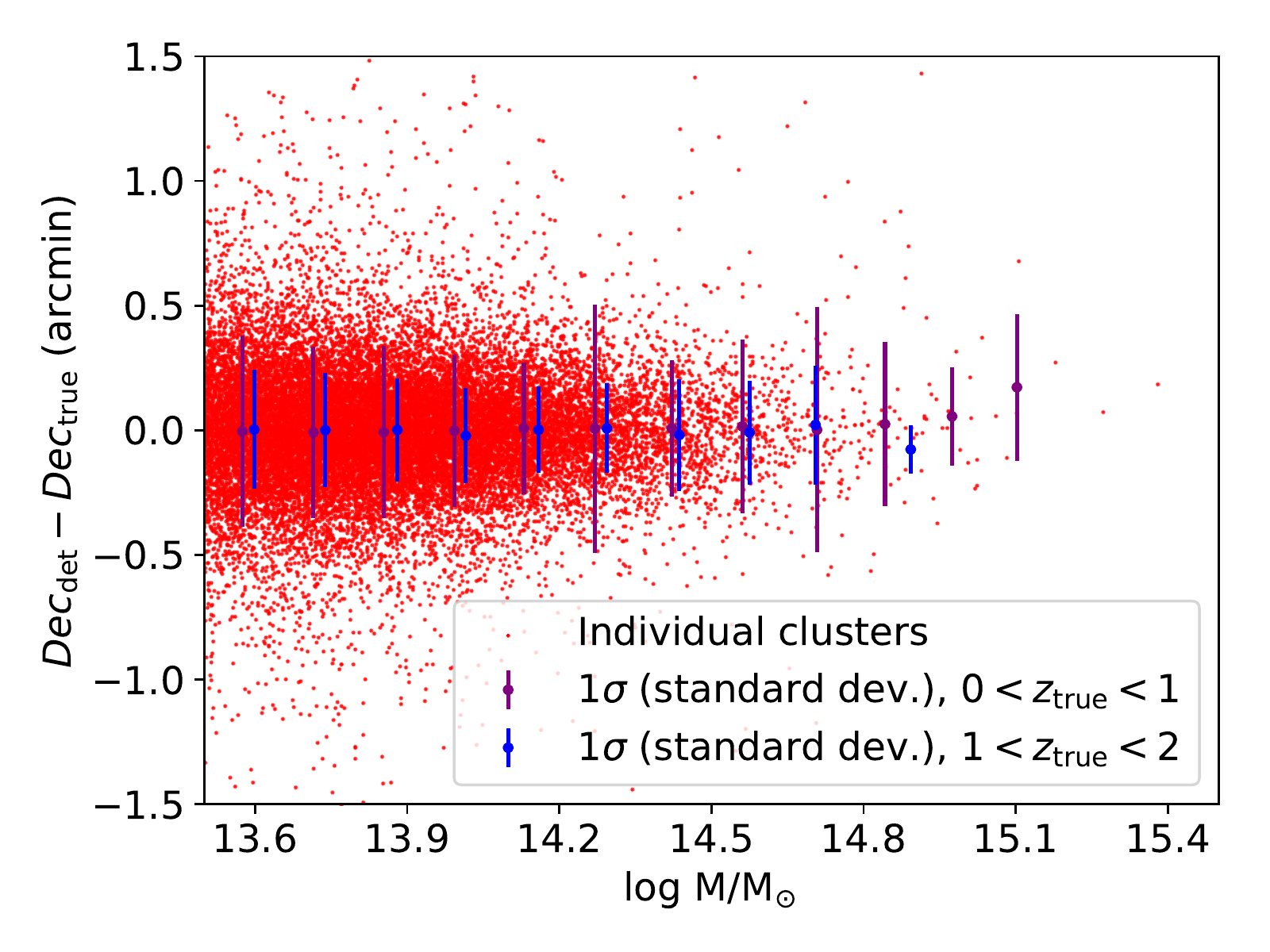} \put(-90,280){\makebox(0,0){\rotatebox{0}{\LARGE PZWav}}} &
\includegraphics[trim=2.4cm 0cm 0cm 0cm, clip=true, scale=1]{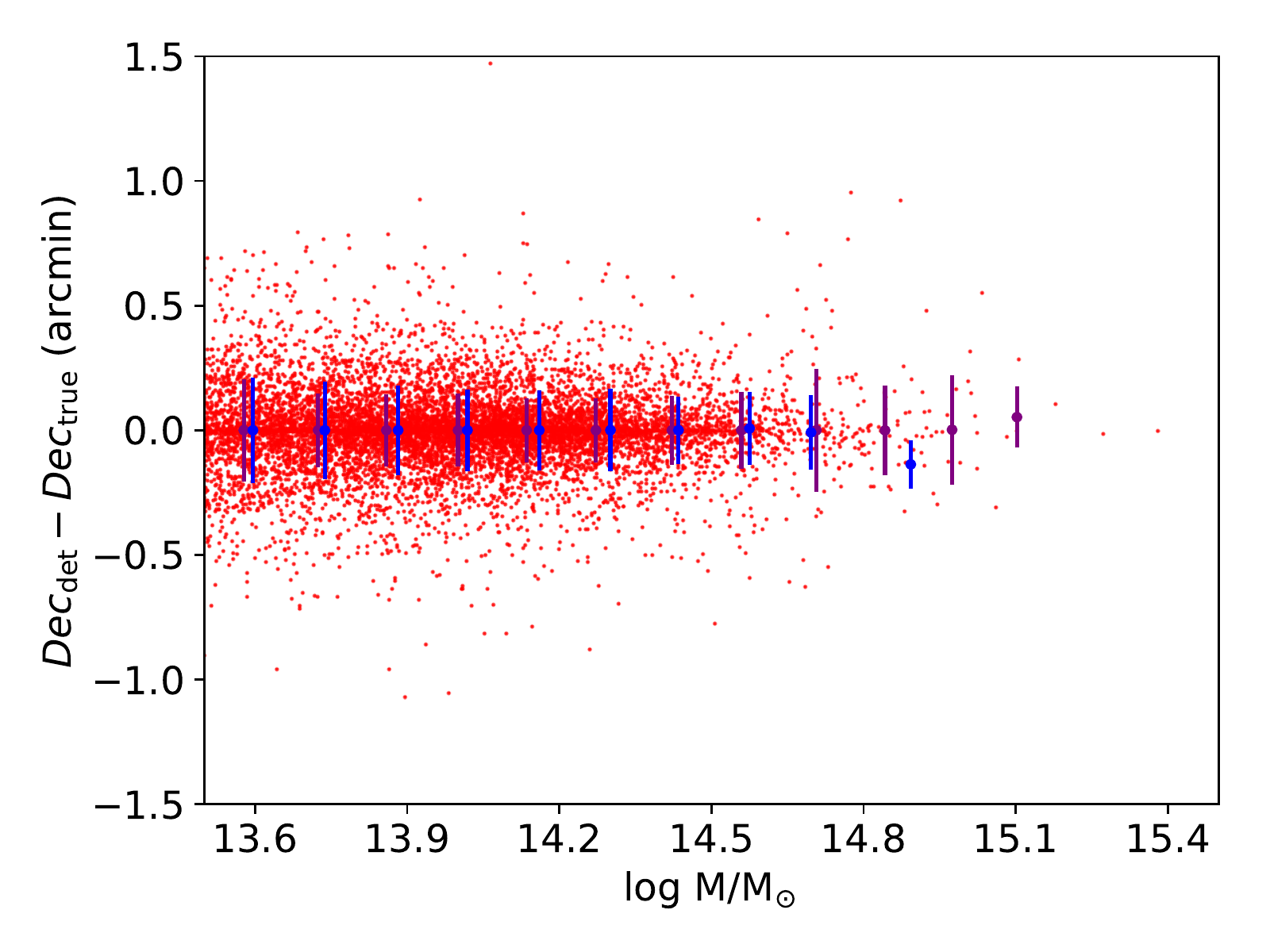} \put(-90,280){\makebox(0,0){\rotatebox{0}{\LARGE sFoF}}} &
\includegraphics[trim=2.4cm 0cm 0cm 0cm, clip=true, scale=1]{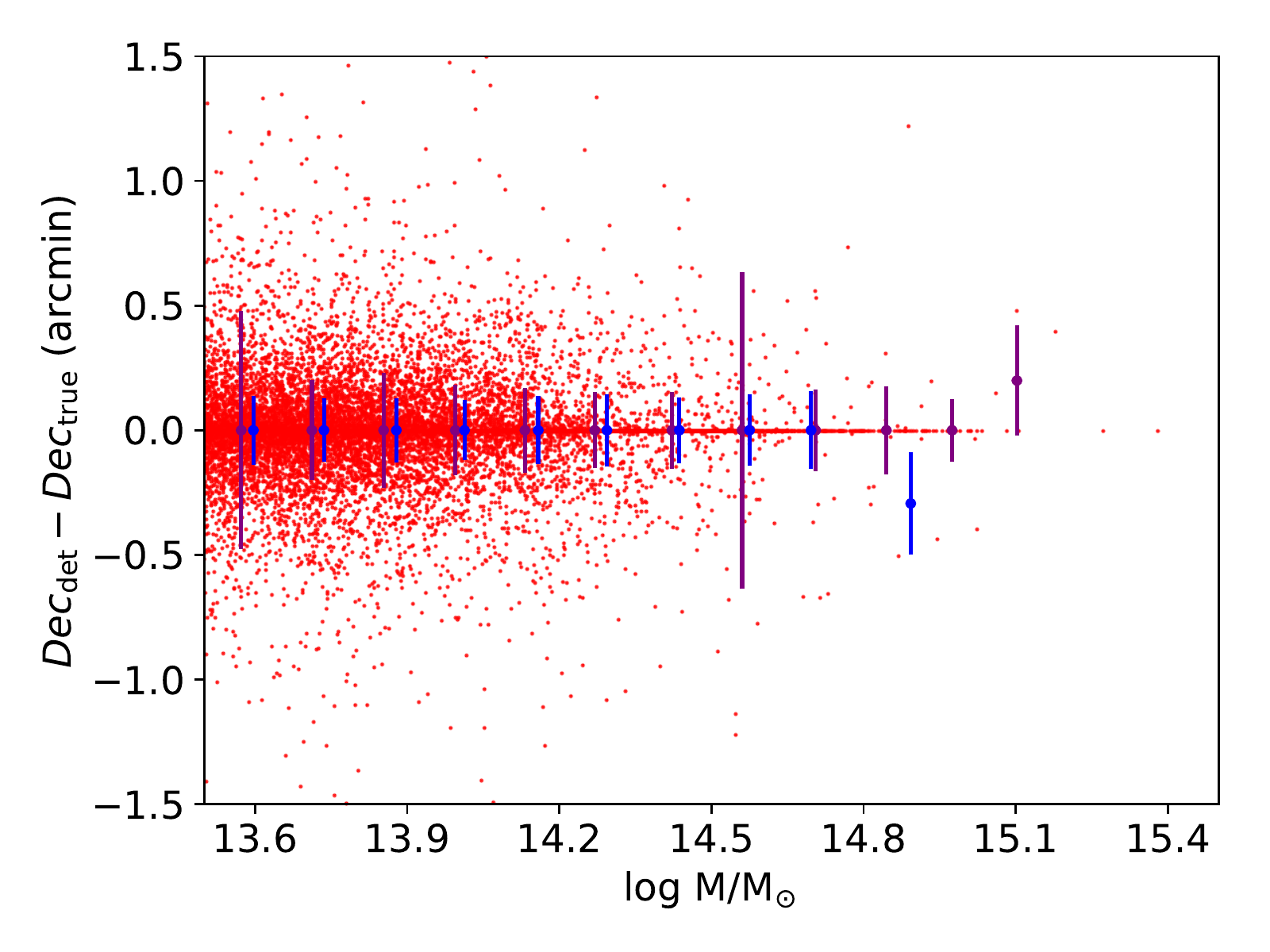} \put(-90,280){\makebox(0,0){\rotatebox{0}{\LARGE WaZP}}}
\end{tabular}}
\caption{\footnotesize{Difference between the declination associated to a detected cluster and the mock cluster declination from the simulation, as a function of mock cluster mass, for all cluster finders. The red points show the individual clusters and the error bars provide the standard deviation of the distribution as a function of mass (in bins of 0.14 dex width) for two redshift bins. Each catalog of detections has been trimmed to the most reliable detections insuring a mean purity of 80\% in the range $M>10^{14}$ M$_{\odot}$ and $z_{\rm true}<2$.}}
\label{fig:centroid_recovery}
\end{figure*}

As for the redshifts, we compute the centroid offset between the detection centers and the mock cluster catalog centers (see Figure~\ref{fig:centroid_recovery}, where the difference is shown in terms of declination offset). We do not observe significant evolution of the distribution with mass or redshift. This is likely due to the fact that more massive clusters are better detected, but also more extended, which compensates the precision in the centroid recovery. Similarly, nearby clusters are more extended, but also better detected than their high redshift counterparts. The mean, median and standard deviation values of the centroid offset distributions are reported in Table~\ref{tab:perf_summary} at $M>10^{14}$ M$_{\odot}$, and corresponds to a typical offset of 0.2 arcmin. Using the mean of the centroid offset distribution as a quality indicator, the algorithms WaZP, sFoF and AMICO present the best coordinate determinations, down to 0.10 arcmin. The highest mean of the angular offset distribution is 0.46 arcmin, for HCFA. We note that in the case of WaZP, a large fraction of clusters are detected with zero offset as WaZP defines the center as the peak of a density map, or as the brightest member if found within 50 kpc of the density peak.

\subsection{Fragmentation and over-merging}\label{sec:fragmentation_and_overmerging}
We compute the fraction of fragmented clusters and the fraction of over-merging according to the definitions given in Section~\ref{sec:Purity_completeness_fragmentation_and_overmerging}: the number of fragments and overmerged structures are defined as the number of possible one-way associations with respect to the mock clusters and the detected clusters, respectively. We note that our definition of fragmentation and overmerging is tailored to the cluster definition in the mocks, and is thus somewhat arbitrary. We compute the over-merging rate only in the case of two or more overmerged structures. In the case of fragmentation, we compute it for 2 fragments or more, but also for 5 fragments or more (this corresponds to the fraction of mock clusters for which more than one, or more than four, associations are possible).

\begin{figure*}[h]
\resizebox{\textwidth}{!} {
\begin{tabular}{lll}
\includegraphics[trim=0cm 1.9cm 0cm 0cm, clip=true, scale=1]{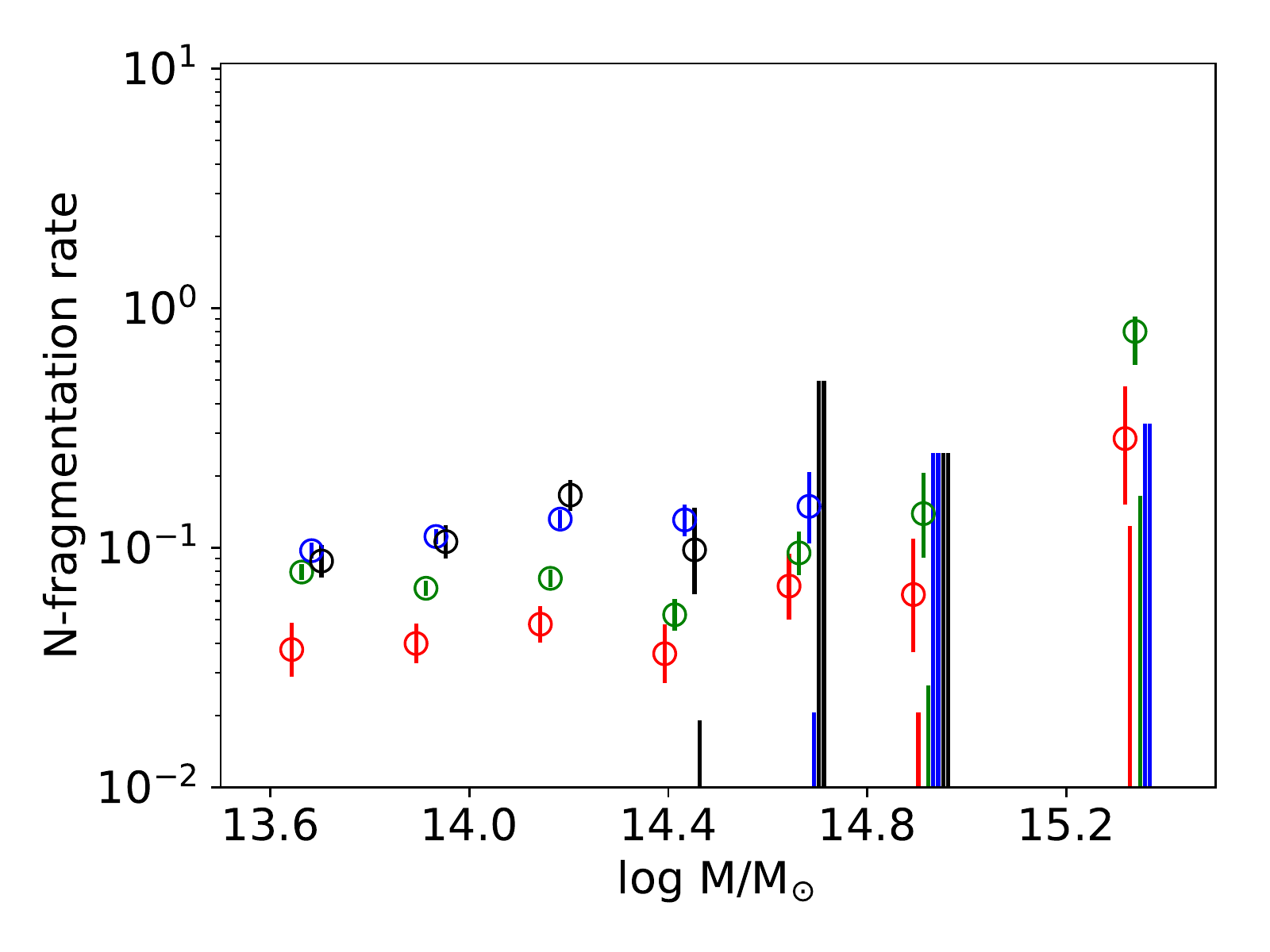} \put(-70,30){\makebox(0,0){\rotatebox{0}{\LARGE AMASCFI}}} &
\includegraphics[trim=2.6cm 1.9cm 0cm 0cm, clip=true, scale=1]{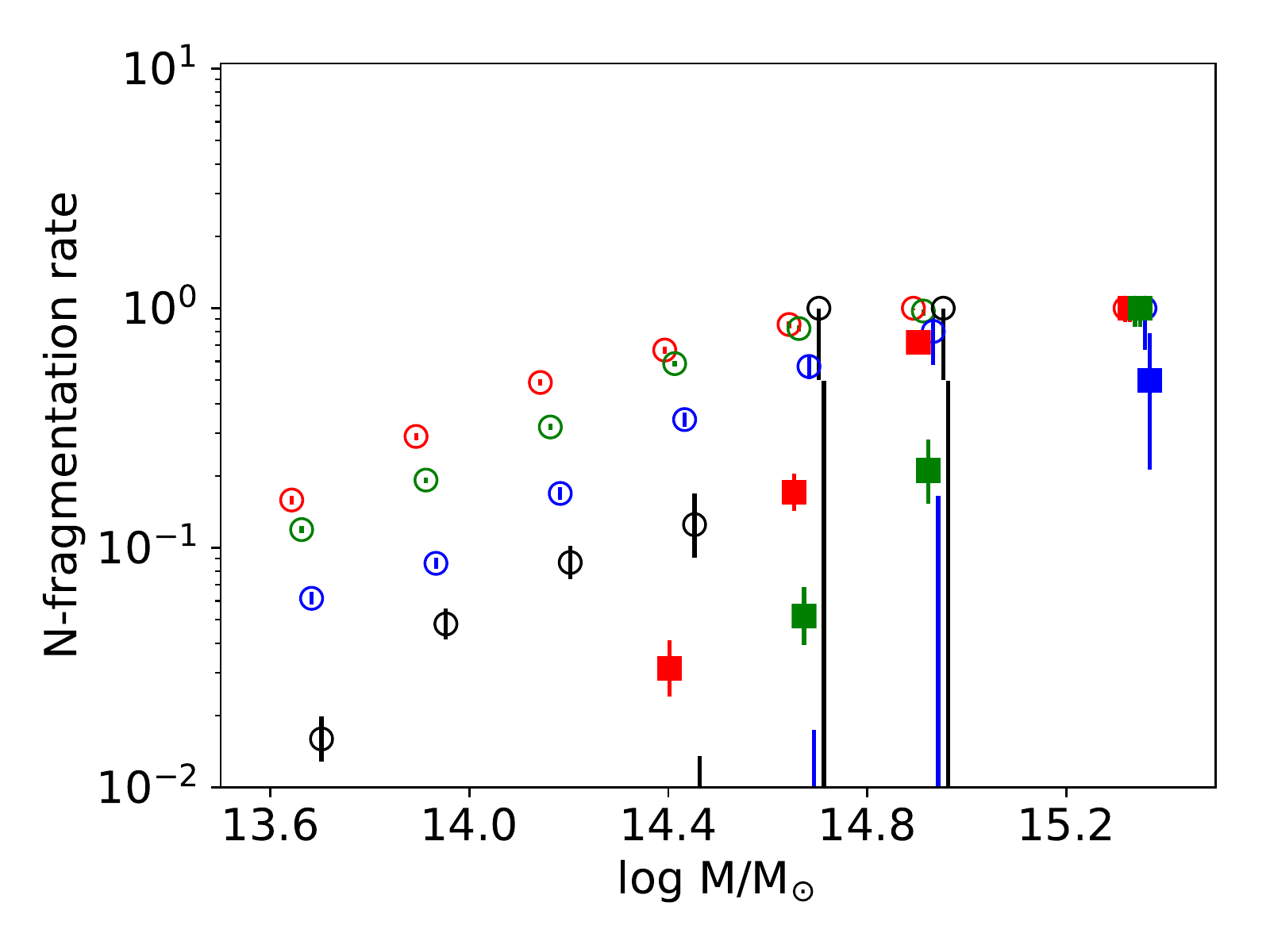} \put(-70,30){\makebox(0,0){\rotatebox{0}{\LARGE AMICO}}} &
\includegraphics[trim=2.6cm 1.9cm 0cm 0cm, clip=true, scale=1]{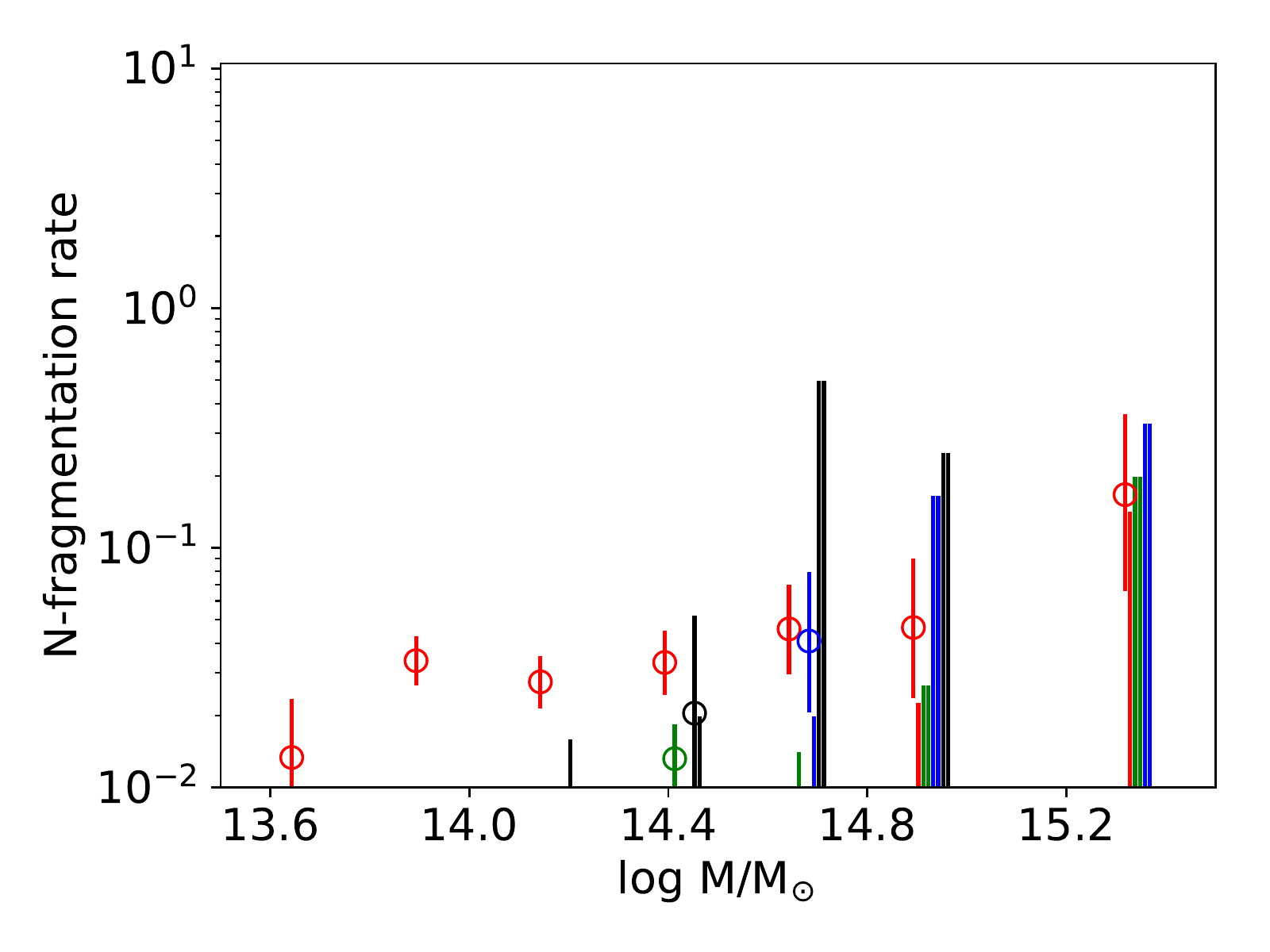} \put(-70,30){\makebox(0,0){\rotatebox{0}{\LARGE HCFA}}}\\
\includegraphics[trim=0cm 0cm 0cm 0cm, clip=true, scale=1]{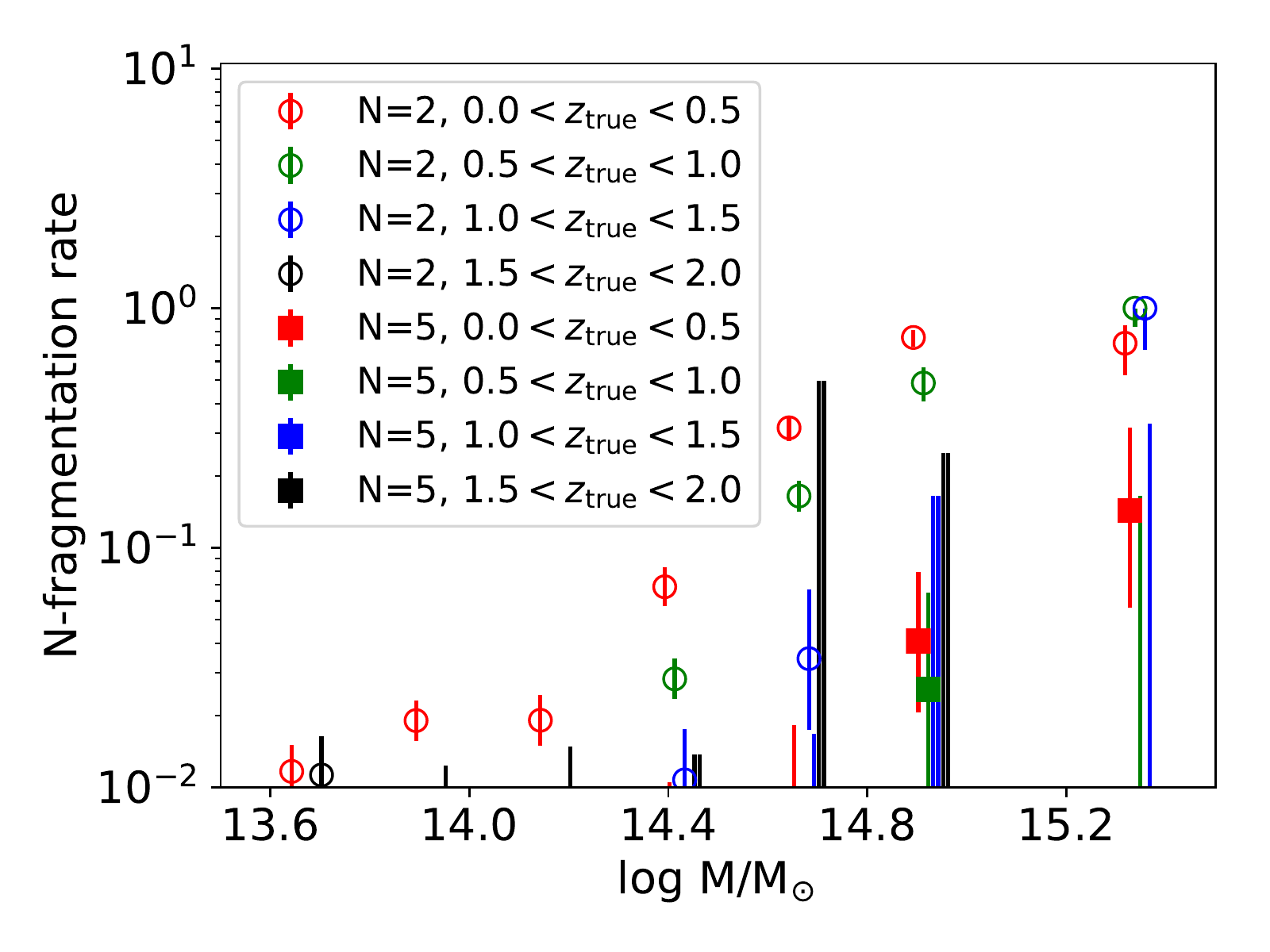} \put(-70,80){\makebox(0,0){\rotatebox{0}{\LARGE PZWav}}} &
\includegraphics[trim=2.6cm 0cm 0cm 0cm, clip=true, scale=1]{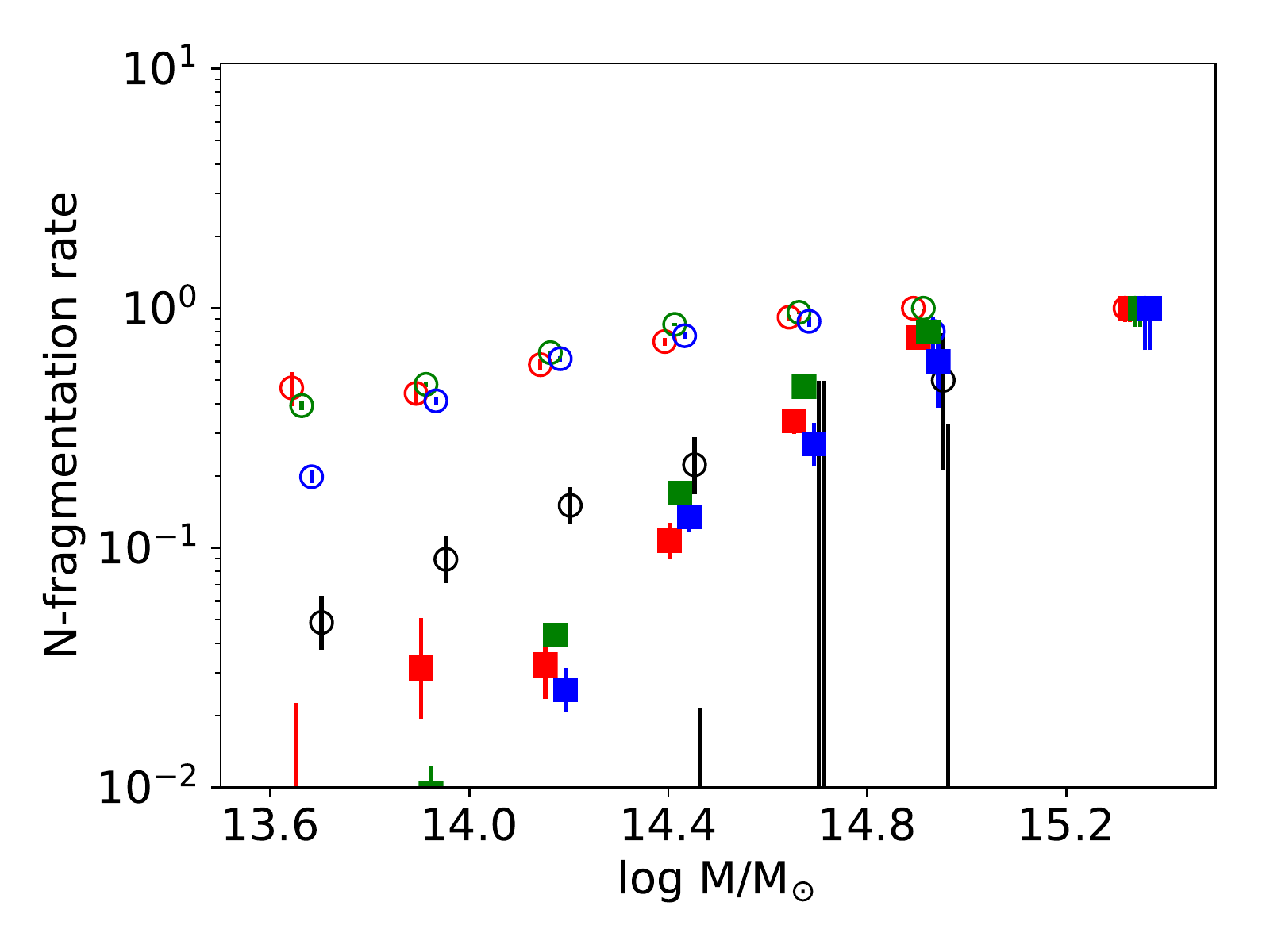} \put(-70,80){\makebox(0,0){\rotatebox{0}{\LARGE sFoF}}} &
\includegraphics[trim=2.6cm 0cm 0cm 0cm, clip=true, scale=1]{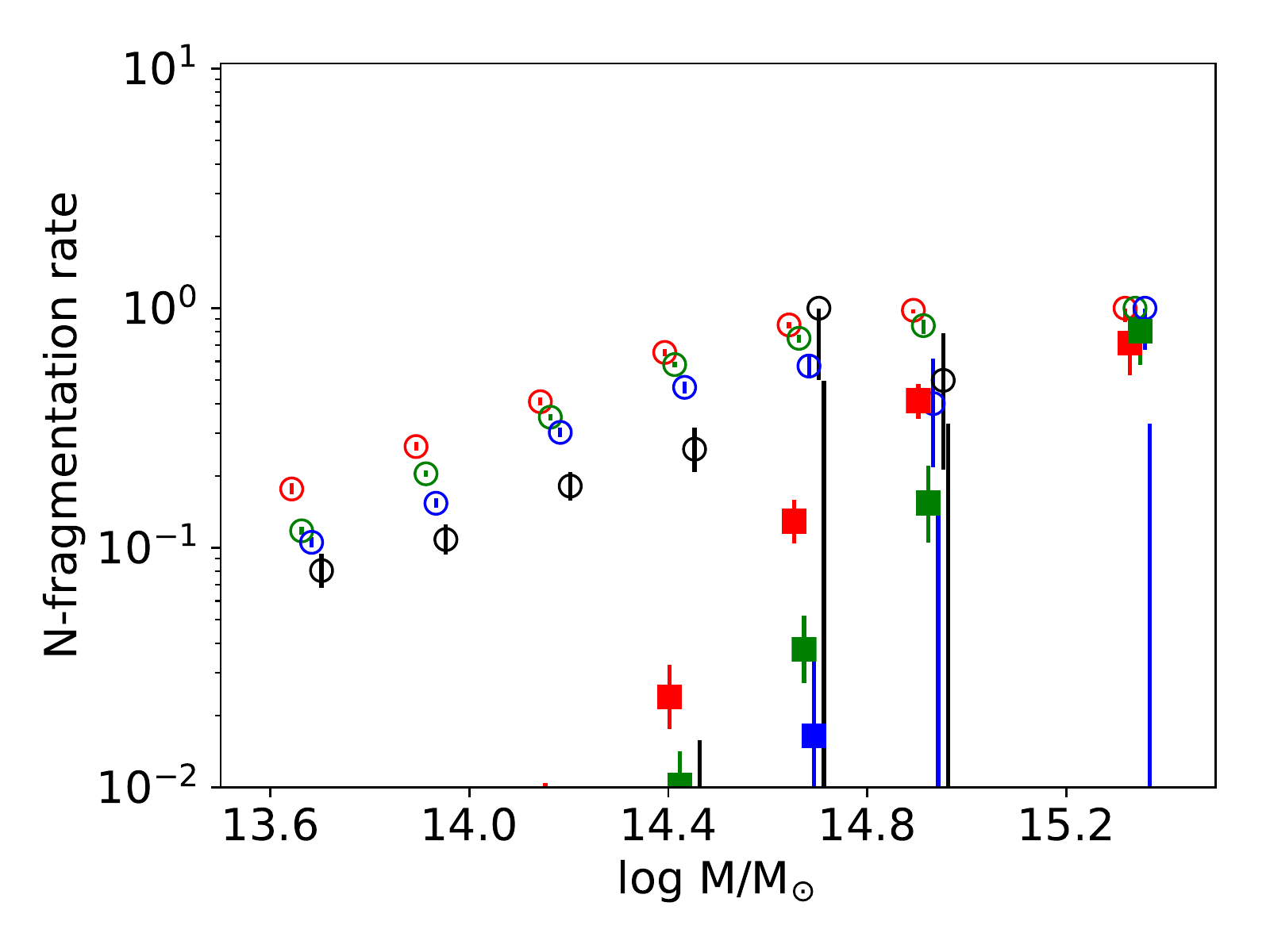} \put(-70,80){\makebox(0,0){\rotatebox{0}{\LARGE WaZP}}}
\end{tabular}}
\caption{\footnotesize{Fraction of fragmented mock clusters as a function of mass, for different redshift bins, for six detection algorithms. The figure shows the fragmentation for a number of fragments larger than one (two or more) and four (five or more), as given in the legend. The symbols that are not visible in the figure correspond to small fragmentation rate, below the figure range. Error bars are computed as in Figure~\ref{fig:completeness_z}. Each catalog of detections has been trimmed to the most reliable detections insuring a mean purity of 80\% in the range $M>10^{14}$ M$_{\odot}$ and $z_{\rm true}<2$.}}
\label{fig:fragmentation}
\end{figure*}

\begin{figure*}[h]
\resizebox{\textwidth}{!} {
\begin{tabular}{lll}
\includegraphics[trim=0cm 1.9cm 0cm 0cm, clip=true, scale=1]{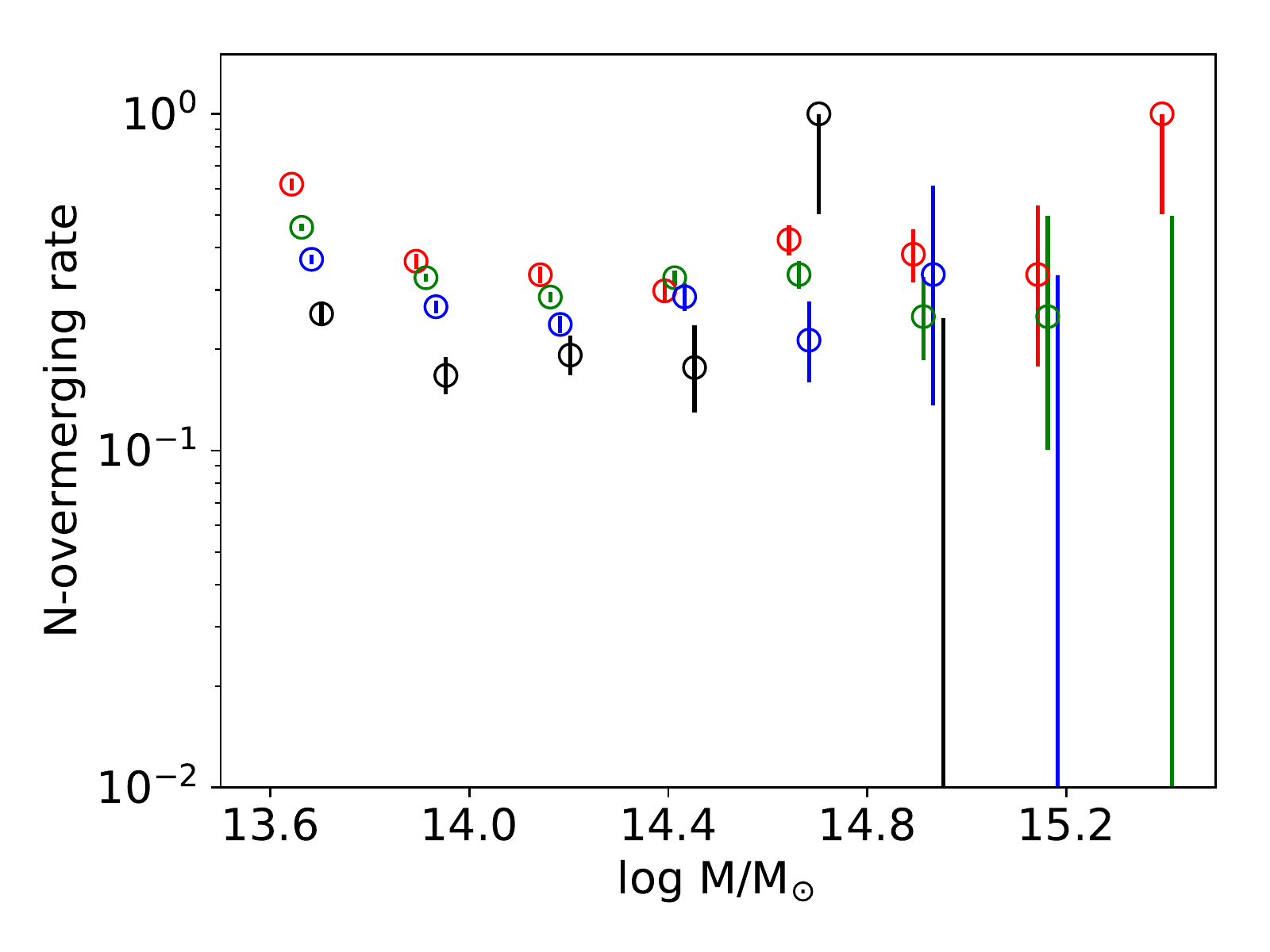} \put(-70,250){\makebox(0,0){\rotatebox{0}{\LARGE AMASCFI}}} &
\includegraphics[trim=2.6cm 1.9cm 0cm 0cm, clip=true, scale=1]{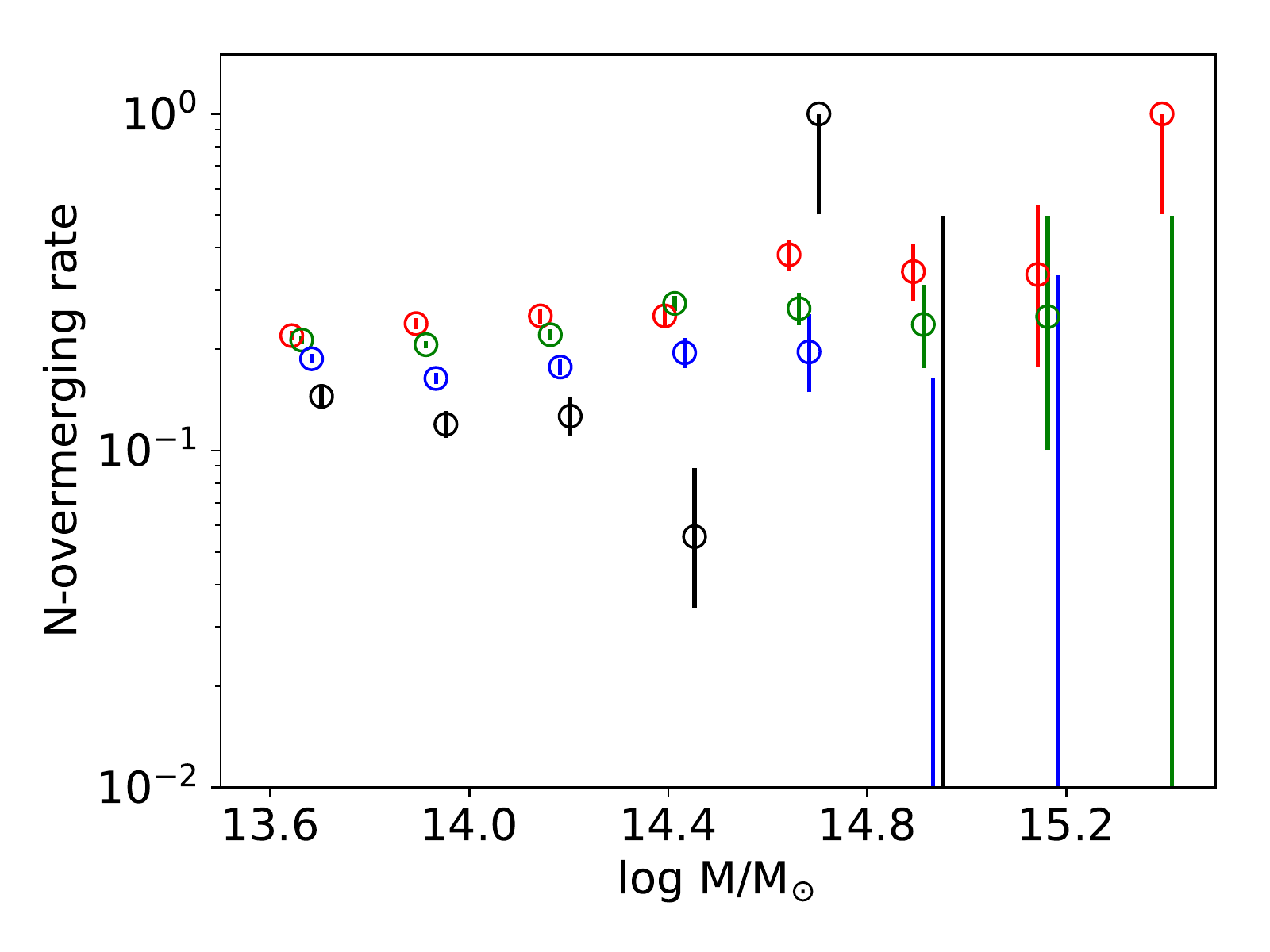} \put(-70,250){\makebox(0,0){\rotatebox{0}{\LARGE AMICO}}} &
\includegraphics[trim=2.6cm 1.9cm 0cm 0cm, clip=true, scale=1]{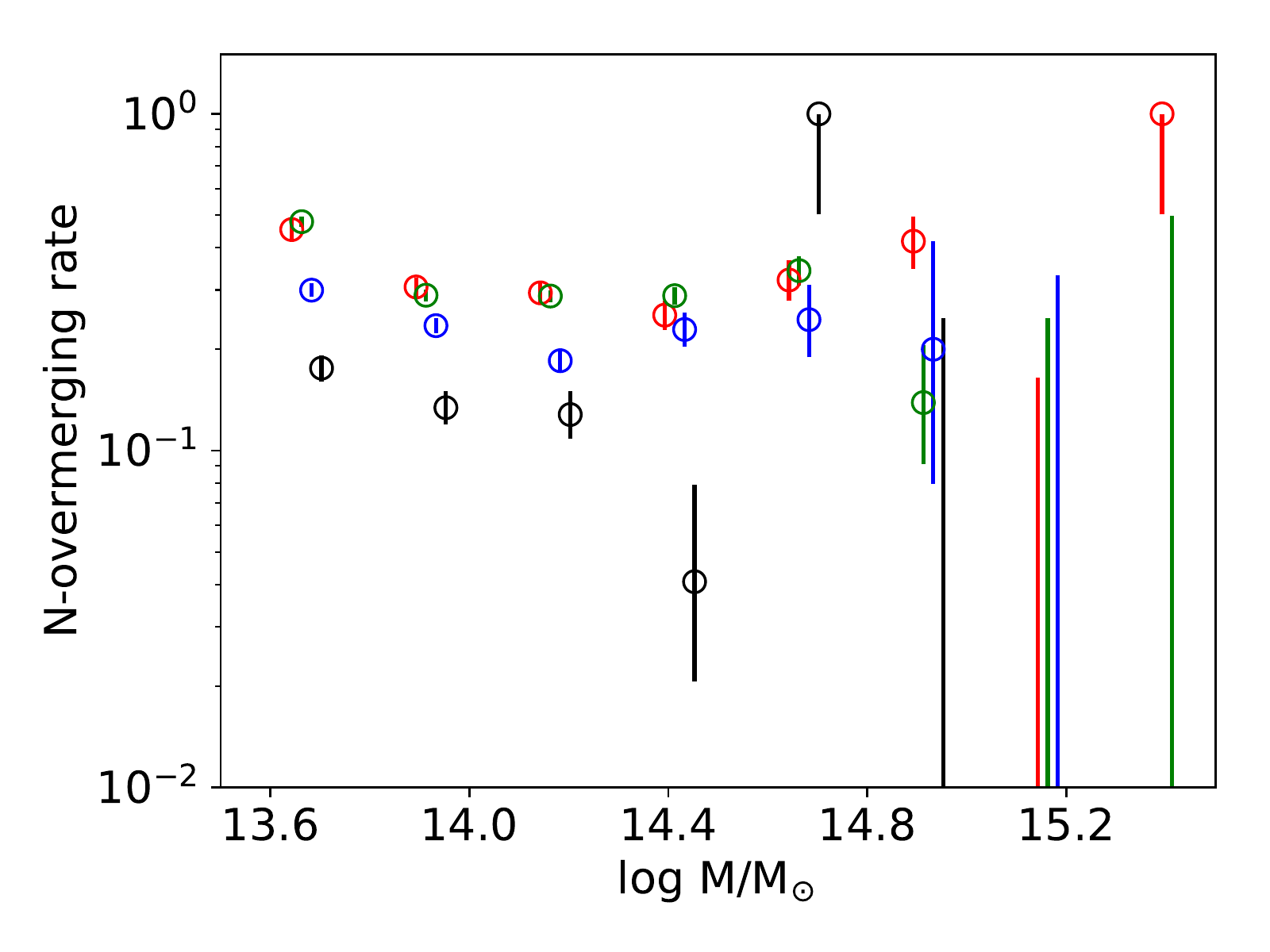} \put(-70,250){\makebox(0,0){\rotatebox{0}{\LARGE HCFA}}}\\
\includegraphics[trim=0cm 0cm 0cm 0cm, clip=true, scale=1]{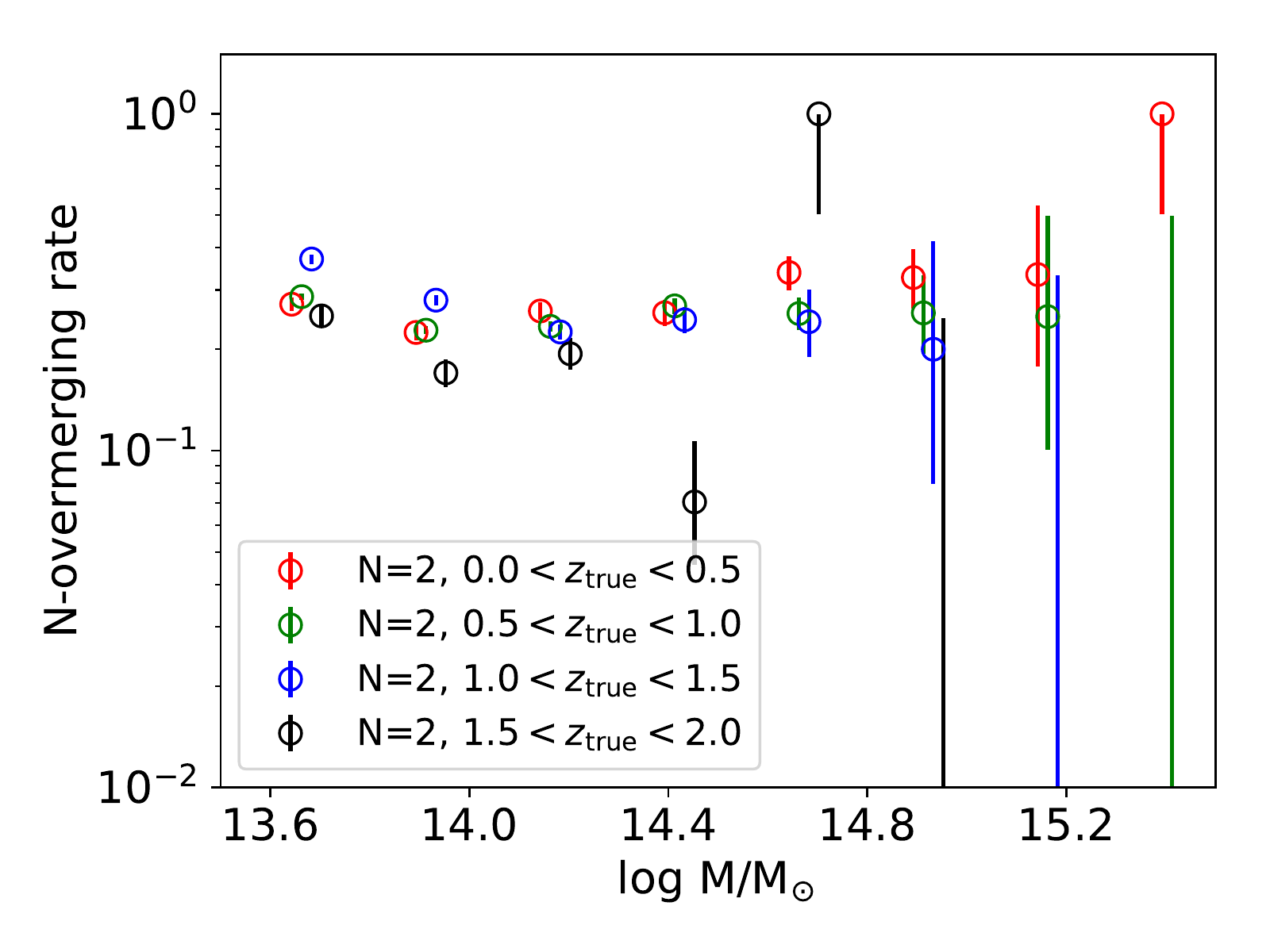} \put(-70,290){\makebox(0,0){\rotatebox{0}{\LARGE PZWav}}} &
\includegraphics[trim=2.6cm 0cm 0cm 0cm, clip=true, scale=1]{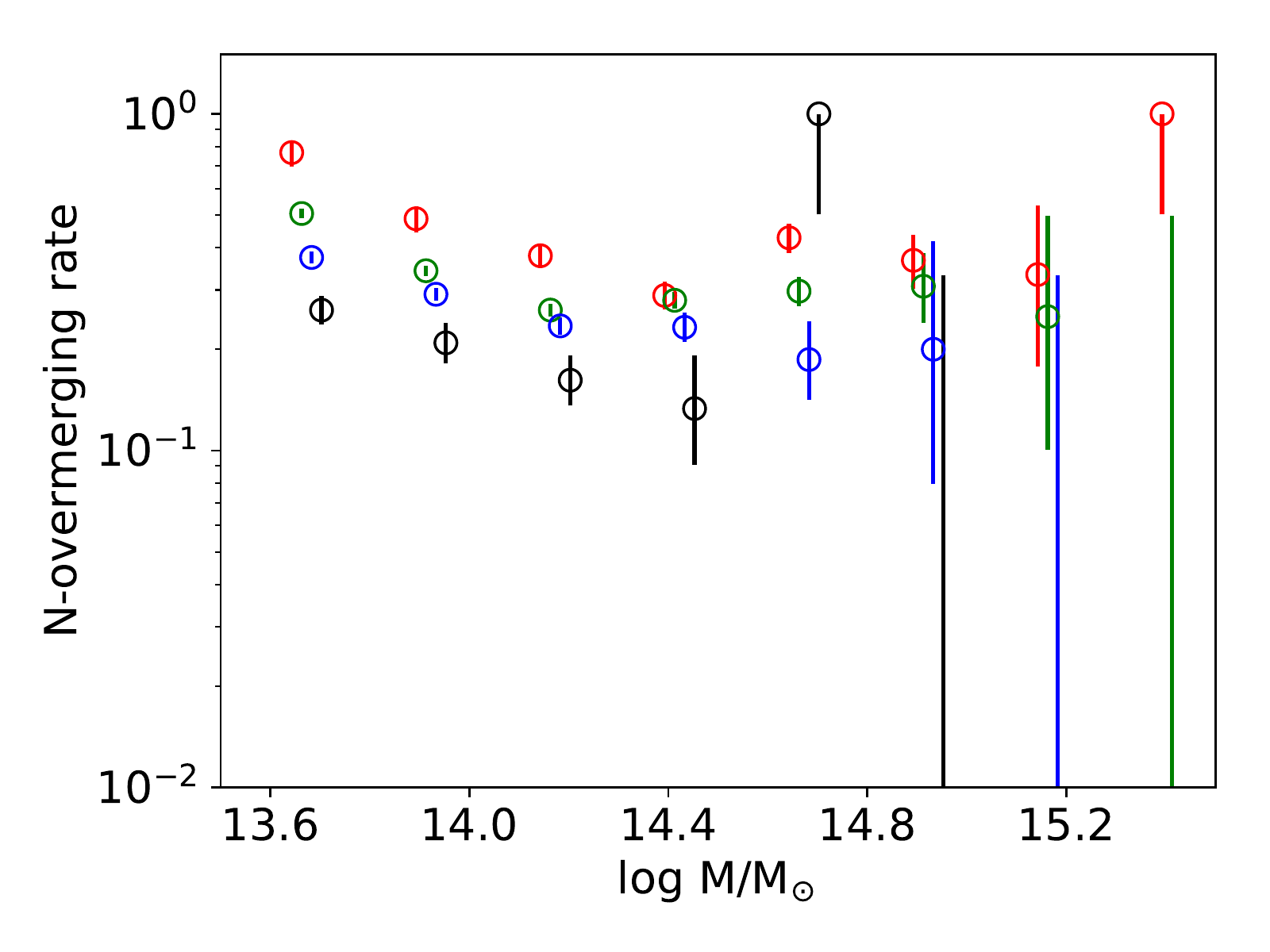} \put(-70,290){\makebox(0,0){\rotatebox{0}{\LARGE sFoF}}} &
\includegraphics[trim=2.6cm 0cm 0cm 0cm, clip=true, scale=1]{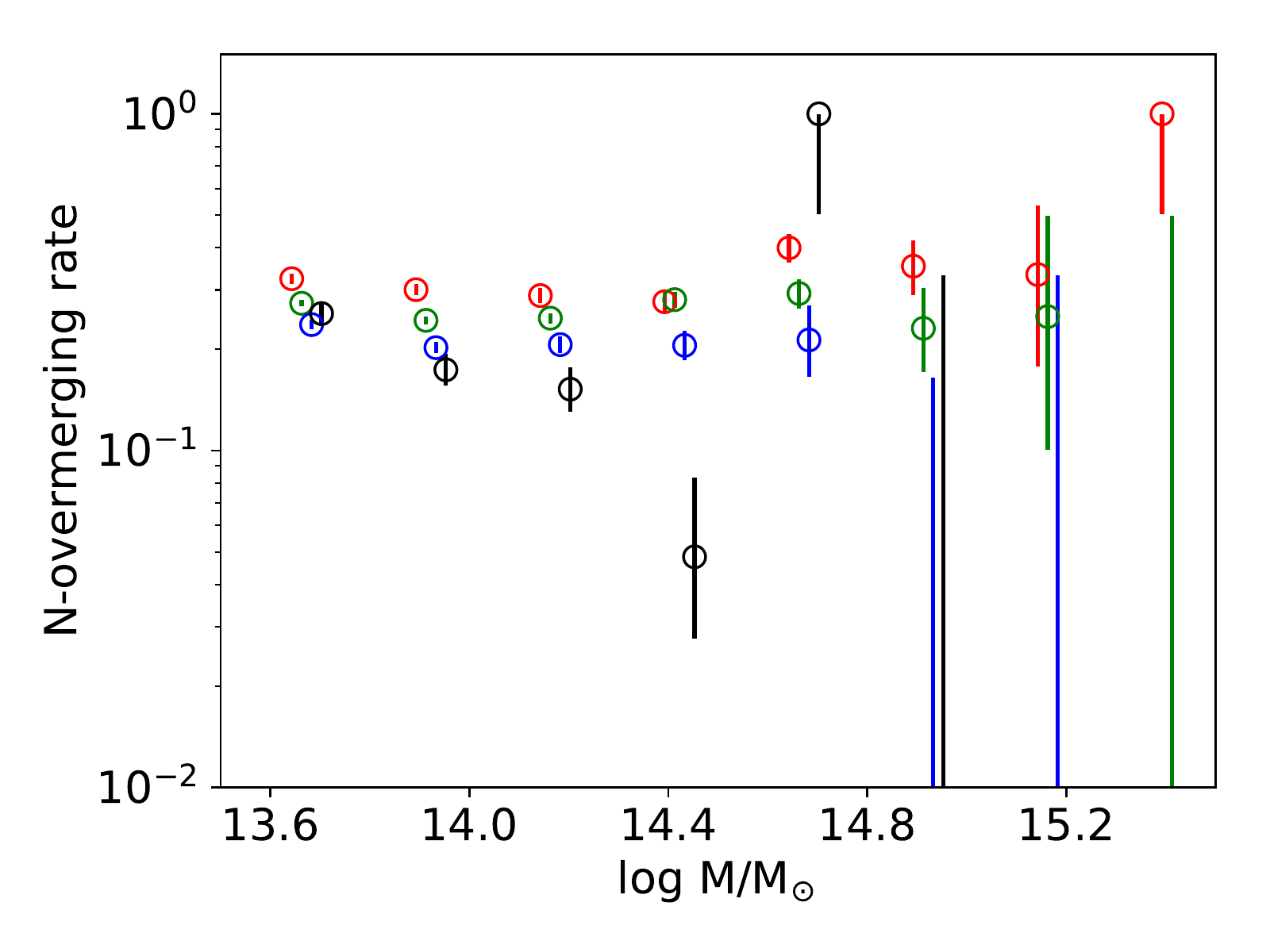} \put(-70,290){\makebox(0,0){\rotatebox{0}{\LARGE WaZP}}}
\end{tabular}}
\caption{\footnotesize{Fraction of overmerged detected clusters (two or more mock clusters associated with a detection), as a function of mass, for different redshift bins, for six detection algorithms. Error bars are computed as in Figure~\ref{fig:completeness_z}. Each catalog of detections has been trimmed to the most reliable detections insuring a mean purity of 80\% in the range $M>10^{14}$ M$_{\odot}$ and $z_{\rm true}<2$.}}
\label{fig:overmerging}
\end{figure*}

The N-fragmentation rate and the N-over-merging rate are given in Figures~\ref{fig:fragmentation}~and~\ref{fig:overmerging} as a function of mass for different redshift bins. All cluster finders present a fragmentation rate that increases with mass and decreases with redshift. This is expected as more substructures become accessible as the mass increases. Similarly, substructures cannot be resolved at higher redshift as fewer galaxies are accessible. Fragmentation could also arise because a code detects lower mass structures that are inside the matching radius of a larger system. In this sense, a code that is more complete at lower masses will automatically have a higher fragmentation rate. We observe significant differences between the different cluster finders. While AMASCFI, HCFA and PZWav present low fragmentation rates, AMICO and WaZP present a higher fragmentation rate, and sFoF tends to fragment clusters up to very low masses and with a large number of fragments (more than 5 fragments are common, as already illustrated in Figure~\ref{fig:matching_illustration} for this cluster finder). The over-merging rate remains at the level of about 10\% and is relatively constant in mass. For most algorithms, we observe an increase in the over-merging rate for lower mass systems, which is likely due to the fact that these objects are more abundant. The over-merging rate is comparable for all detection algorithms, even if AMASCFI, HCFA and sFoF algorithms are slightly more affected than AMICO, PZWav and WaZP. We note that similarly to the fragmentation rate, the overmerging rate is affected by the matching procedure. Indeed, two objects that are correctly detected by a cluster finder, but aligned along the line-of-sight within an overlapping matching cylinder, will appear as overmerged systems. This might be the reason for the two points in Figure~\ref{fig:overmerging} at an overmerging rate of unity (black and red points at $M \sim 10^{14.75}$ and $M \sim 10^{15.25}$ M$_{\odot}$). Both of these points correspond to one single cluster that is overmerged for all detection algorithms.

The fragmentation and over-merging rates are limited in terms of quality assessment because they are largely dependent on the matching procedure. Nevertheless, they provide complementary information and have an impact on the scatter in the scaling relations, depending on how the fragments are distributed in terms of richness. In Figure~\ref{fig:mass_lambda_algo}, we observe outliers that are likely related to over-merging, above the mean relation, and fragmentation, below the mean relation. Fragmentation and over-merging processes have been investigated recently on BAHAMAS hydrodynamical simulations \citep{McCarthy2017} in the context of a friend-of-friend cluster finder algorithm, showing in particular the impact of fragmentation on the scaling relations \citep{Jakobs2018}.

\section{Discussion}\label{sec:Discussions}
\subsection{Global performance, algorithm selection and comparison to previous work}
Among the six codes that have been tested in the final CFC, four of them reach a mean completeness of $80\%$ for masses larger than $10^{14}$ M$_{\odot}$, with high values of purity as shown in Table~\ref{tab:cluster_finder_summary}. AMICO and PZWav demonstrate excellent performance at high redshift in terms of completeness and purity (as a function of ranking), which is of particular importance for cosmological purposes. Based on these results, as detailed in Section~\ref{sec:Performance}, the algorithm presenting the best overall performance, in the context of this work, is the AMICO code. AMICO was therefore selected to be implemented in the \textit{Euclid} pipeline.

Because of the current uncertainties in the physical processes that drive cluster formation at high redshift, which are one of the main targets of \textit{Euclid}, it is important for the detection algorithms to be robust with respect to the underlying cluster properties. In Table~\ref{tab:complementarity_amico}, we provide the complementarity of the cluster finder catalogs with respect to AMICO. This shows the gain achieved on the completeness when using AMICO plus another code. This helps in particular to increase the completeness at the high mass and high redshift. The most efficient codes in terms of complementarity to AMICO are WaZP, sFoF and PZWav, with relative performance varying according to the mass and redshift ranges considered. Among these three codes, PZWav has the best performance in terms of completeness and purity at high redshifts (see Figures \ref{fig:completeness_z} and \ref{fig:purity_z_comparison}). This is the regime where \textit{Euclid} is expected to have the most impact on cluster detection with respect to optical ground based and X-ray surveys. PZWav was thus also selected to be implemented in the \textit{Euclid} cluster detection pipeline. The detailed implementation of the two codes and their joint utilization is under investigation and we leave this point for future publications.

AMICO and PZWav detected around 7700 and 7100 clusters, respectively, in a 300 deg$^2$ region with a purity of 80\% for $M>10^{14}$ M$_{\odot}$ and up to $z_{\rm true}=2$. Assuming that these numbers scale with the \textit{Euclid} survey area of 15000 deg$^2$, we expect a total number of clusters of $\sim 3.5 \times 10^{5}$ in the same mass and redshift range. At redshifts between 1 and 2, this number becomes $\sim 10^{5}$. These numbers of clusters are in reasonable agreement with those obtained by \cite{Sartoris2016} (i.e., $\sim 2 \times 10^5$ at all redshifts and $\sim 4 \times 10^4$ at $z_{\rm true} \geq 1$, for S/N $\gtrsim 5$). At a S/N $\gtrsim 3$, these numbers are larger by a factor of 10\footnote{Here, the quoted S/N is defined as a number of galaxies within $R_{500}$ normalized by the rms of the field counts within the same radius, see \cite{Sartoris2016} for more details.}.

The \textit{Euclid} cluster selection function was also investigated using the same mock as the one used in the present paper \citep{Ascaso2017}. Galaxy clusters were detected using the Bayesian Cluster Finder \citep[BCF,][]{Ascaso2012,Ascaso2014,Ascaso2015}. The results by \cite{Ascaso2017} are in broad agreement with those presented in this paper, but are based on a different methodology. Nonetheless, the exact shape of the selection function is different, as it also depends on the algorithm considered \citep[almost flat at $z_{\rm true}<1$ and increases at higher redshifts in][]{Ascaso2017}.

\begin{table}[]
\caption{\footnotesize{Complementarity of the final CFC detection algorithm with respect to the AMICO results. Two redshift ranges are considered, [0, 1], and [1, 2], both for masses larger than $10^{14}$ M$_{\odot}$. The two numbers correspond to the common fraction of clusters, with respect to AMICO, and the fraction of extra detections, with respect to the total. For example, adding PZWAV to AMICO, in the second redshift bin, would add 5.8\% extra clusters, while they have already 81.6\% clusters in common.}}
\begin{center}
\resizebox{0.5\textwidth}{!} {
\begin{tabular}{c||c|c}
\hline
\hline
Finder & $z = [0, 1]$, $M > 10^{14}$ M$_{\odot}$ &$z = [1, 2]$, $M > 10^{14}$ M$_{\odot}$ \\
-- & \multicolumn{2}{c}{(common fraction, fraction of extra detections)} \\
\hline
\hline
AMASCFI & (71.2, 2.7)\% & (60.7, 4.6)\% \\
HCFA     & (59.0, 2.2)\% & (53.6, 4.5)\% \\
PZWav   & (89.3, 3.2)\% & (81.6, 5.8)\% \\
sFoF       & (60.3, 2.0)\% & (59.5, 3.4)\% \\
WaZP     & (90.1, 4.2)\%  & (78.9, 5.5)\%  \\
\hline
\end{tabular}
}
\end{center}
\label{tab:complementarity_amico}
\end{table}

\subsection{Representativity of the mock and limitations}
The mock used was constructed to be representative of expected \textit{Euclid} data. However, it was originally designed for large-scale galaxy clustering studies, and not for galaxy cluster studies. It is known to present limitations, as discussed in Section~\ref{sec:Simulations}. For instance, clusters are more concentrated than expected from observations and the density profiles of clusters are truncated, which could affect the detection performance. The LF also differs from the expectations of passive evolution, depending on the redshift regime we are interested in. The differences that we observe among the different detection algorithms are based on different information and hypotheses regarding the cluster properties. These differences reflect the intrinsic performance of the codes, their sensitivity to the underlying structural properties of galaxy clusters and their sensitivity to photometric errors. The accuracy to which the \textit{Euclid} selection function will be determined and how systematic effects impact on the derived cosmological constraints will strongly depend on our ability to understand and model the properties of \textit{Euclid} clusters. The work presented in this paper gives the current status of the performance of cluster detection within \textit{Euclid}. This assessment is limited by the simulation used, which is not necessarily fully representative of the true Universe, especially since the properties of distant clusters remain poorly known to date. Nonetheless, despite the fact that the mock may not be fully representative of the true Universe, the relative performance of the algorithms is expected to be fairly stable given their behavior during previous challenges on other mocks (see the Appendix for further details).

In addition to the cluster properties, we note that the cosmological parameters assumed in the simulation differ from current constraints, in particular $\Omega_{\rm m} = 0.25$ and $\sigma_8 = 0.9$ \citep[versus $0.3156 \pm 0.0091$ and $0.831 \pm 0.013$,][]{Planck2015XXIV}. This is expected to lead to fewer projection effects with other clusters along the line-of-sight and to increase the overall number of clusters.

Finally, we note that the photometric redshifts were computed using \textit{Euclid} $YJH$ bands together with $grizY$ bands from assumed ground-based observations. We stress that the addition of $u$-band data, in a more optimistic case, is expected to significantly improve the photometric redshift quality \citep{Ascaso2015}, which in turn can improve the performance of the cluster finders. \textit{Euclid} ground-based complementary observations are currently under way  and, combined with the large surveys expected to be released at the time \textit{Euclid} data will be available, will improve the precision and robustness of photometric redshifts. Furthermore, our analysis is based on the \textit{Euclid} wide photometric survey. Using \textit{Euclid} spectroscopic data will provide detection of the $H\alpha$ line down to a flux limit of $2 \times 10^{-16}$ erg s$^{-1}$ cm$^{-2}$ ($5 \times 10^{-17}$ erg s$^{-1}$ cm$^{-2}$ for the deep survey) over the redshift interval $0.9 < z < 1.8$, so that we expect to further improve the detections of clusters in the high redshift regime (e.g., improve the purity by identifying wrong galaxy associations in redshifts).

In this work, we restricted the mock cluster catalog to masses of $M > 10^{13.25}$ M$_{\odot}$, corresponding on average to $M_{200} > 10^{13.15}$ M$_{\odot}$, and even less for $M_{500}$, depending on the cluster concentration. Given this choice, lower mass objects that can be detected by cluster finders appear as impurities. Therefore, the established performance presented here is conservative. By using different mass cuts for the catalog, we have verified that these objects do not significantly affect our findings.

Given the limited mock area and since clusters are extended objects, clusters for which the central galaxy was out of the footprint were not included in the mock cluster catalog. These objects can also show up as impurities in the detections, depending on how the cluster finders deal with edges. However, by using a catalog made from the barycenter of cluster member galaxies (i.e., including all the galaxies within the footprint), we have observed that edge effects are subdominant.

In this paper, our baseline choice for comparing the performance was to use a purity threshold of 80\% for the different cluster finder catalogs. Changing this limit leads to an overall shift in the accessible mass limit at a given completeness that is similar for all algorithms. For instance, it can be seen on average in Figure~\ref{fig:roc_curve}, for the range $z_{\rm true} = [0, 2]$ and $M > 10^{14}$ M$_{\odot}$, that the completeness is reduced by a factor of typically $\sim 1.4$ when requiring a purity of 90\%. When focussing on the high mass range, $M>10^{14.5}$ M$_{\odot}$, the completeness is much less affected with an overall mean reduction by a factor of $\sim 1.03$. The best tradeoff between purity and completeness and the best range of mass that should be considered will be further addressed in future work.

\section{Summary and conclusions}\label{sec:Conclusions}
In this paper, we have presented the methodology and the results of the final \textit{Euclid} Cluster Finder Challenge. This activity was organized to estimate the performance of galaxy cluster detection algorithms within \textit{Euclid} and to select the cluster finders to be implemented in the \textit{Euclid} pipeline. A total of six algorithms, based on various assumptions and techniques, were considered. They were applied to a galaxy mock that is believed to provide a fairly good representation of expected \textit{Euclid} data, albeit being significantly smaller in terms of sky coverage. We estimated the performance of the different algorithms by matching the detected clusters to the known mock clusters. We observe that the mass-dependent bias on completeness, due to the matching procedure, is estimated to be below 3\% over the mass range considered. If unaccounted for, such biases may affect the derived mass-observable scaling relations, as well as the normalization of cluster counts. Several other methods to estimate the selection function are being developed within the consortium and they should allow us to mitigate such effects.

All six algorithms performed well, three of them reaching a completeness and purity higher than 80\%, down to masses of $10^{14}$ M$_{\odot}$ and up to redshift of 2. Among the competing cluster finders, the AMICO and PZWav codes were selected to be implemented in the \textit{Euclid} cluster pipeline.

This work was based on the application of cluster finders to a galaxy mock, which was characterized, and found to show some limitations in the context of galaxy cluster detection. The physical processes at play in distant clusters \citep[the ones which should provide most of the cosmological power for \textit{Euclid},][]{Sartoris2016} are not yet fully understood (e.g., space densities, structural properties, star formation rate). Thus, improving the robustness of our results would require a better modeling of the structural properties of these objects, to be implemented in the mock. Indeed, any mismodeling in the mock could affect the absolute estimate of performance of a given algorithm. In this context, data from the deep surveys and from other surveys of extremely distant clusters will help to characterize the high-redshift tail of \textit{Euclid} clusters and improve the performance of cluster finder algorithms as well as the precision on its absolute determination.

We note that the performance of the detection codes presented in this paper are those at the time of this selection. The codes continue to be developed and optimized, as well as more representative simulations of the \textit{Euclid} survey become available. Our knowledge of cluster properties at high redshift is expected to significantly improve in the coming years with the advent of new facilities (e.g., \textit{JWST}) and may impact cluster detection methodologies. New methodologies may also become available in coming years. For these reasons, integration of the codes into the \textit{Euclid} pipeline is being configured in a flexible way to allow for the possibility of updating and adding codes, given valid scientific motivations.

The results reported in this paper show that, with current cosmological parameters, \textit{Euclid} has the potential of detecting an unprecedented number of galaxy clusters ($>10^5$), up to redshift 2 and over more than two orders of magnitude in mass (down to $10^{13.5}$ M$_{\odot}$). We note, however, that predictions of number counts are subject to modeling uncertainty; in particular, as we have commented above, our simulations do not represent the currently favored cosmological parameters \citep{Planck2016XIII}, with larger $\sigma_8$ and lower matter density, and do not reproduce certain known cluster properties. Nonetheless, the analytical selection function assumed in \cite{Sartoris2016} and the one obtained in \cite{Ascaso2017} are in line with our findings, such that the accuracy of the \cite{Sartoris2016} forecast is strengthened by the present work. The exquisite leverage provided by \textit{Euclid}, in terms of mass and redshifts, thus enable very competitive constraints on cosmological parameters from cluster number counts. As pointed in \cite{Sartoris2016}, the main challenges will be calibrating the mass-richness relation for the sample and characterizing the selection function in a mass and redshift regime that remains relatively unexplored with current observations. An in-depth investigation of the systematic effects related to the mass determination will be possible thanks to the wide mass and redshift range accessible with \textit{Euclid}, its internal mass calibration available from WL and velocity dispersion, and multi-wavelength synergies available at the time \textit{Euclid} is operating.

While this paper presents an estimate of the expected \textit{Euclid} performance in terms of galaxy cluster detection, many developments are still ongoing within the consortium. In particular, we note that activities dedicated to the assessment of the cluster galaxy membership, the characteristic radius of the detected clusters, and the definition of the richness are being pursued. Improvements in the purity will also be achieved by using spectroscopic data, while an internal calibration of the mass-richness relation will benefit from masses estimated from stacking weak-lensing signal \citep[see][for the assessment of the performance in the \textit{Euclid} context]{Kohlinger2015}. Therefore, significant progress is possible within the coming years. In addition, the work presented here will be reproduced using different simulations where clusters may present different properties, following improvements in our understanding of cluster formation and evolution, particularly at high redshift. Similarly, the choice of the photometric redshift codes may have an impact on the cluster detection, which will be quantified by testing the available codes.

\begin{acknowledgements}
We are thankful to the anonymous referee for useful comments that helped improve the quality of the paper.
This work is part of the ongoing effort dedicated to the scientific preparation of the \textit{Euclid} mission and we are grateful to the \textit{Euclid} consortium. The \textit{Euclid} Consortium acknowledges the European Space Agency and the support of a number of agencies and institutes that have supported the development of \textit{Euclid}. A detailed complete list is available on the \textit{Euclid} web site (\texttt{http://www.euclid-ec.org}). In particular the Academy of Finland, the Agenzia Spaziale Italiana, the Belgian Science Policy, the Canadian \textit{Euclid} Consortium, the Centre National d'Etudes Spatiales, the Deutsches Zentrum f\"ur Luft- and Raumfahrt, the Danish Space Research Institute, the Funda\c{c}\~{a}o para a Ci\^{e}nca e a Tecnologia, the Ministerio de Economia y Competitividad, the National Aeronautics and Space Administration, the Netherlandse Onderzoekschool Voor Astronomie, the Norvegian Space Center, the Romanian Space Agency, the State Secretariat for Education, Research and Innovation (SERI) at the Swiss Space Office (SSO), and the United Kingdom Space Agency.
This work is based on simulation products created for the \textit{Euclid} Consortium. They were created on the DiRAC Data Centric system at Durham University, operated by the Institute for Computational Cosmology on behalf of the STFC DiRAC HPC Facility (www.dirac.ac.uk). This equipment was funded by BIS National E-infrastructure capital grant ST/K00042X/1, STFC capital grant ST/H008519/1 and ST/K00087X/1, and STFC DiRAC Operations grant ST/K003267/1 and Durham University. DiRAC is part of the National E-Infrastructure.
R\'emi Adam acknowledges support from Spanish Ministerio de Econom\'ia and Competitividad (MINECO) through grant number AYA2015-66211-C2-2.
R\'emi Adam acknowledges fundings from the CNES post-doctoral fellowship program.
Christophe Benoist, Alberto Cappi, Sophie Maurogordato, Marina Ricci, Pier-Francesco Rocci and Martin Vannier acknowledge funding from the CNES program (CNES/INSU).
Pier-Francesco Rocci acknowledges funding from a CNES grant.
Fabio Bellagamba thanks the support from the grants ASI n.I/023/12/0 ``Attivit\`a relative alla fase B2/C per la missione \textit{Euclid}''.
Fabio Bellagamba and Stefano Andreon thank the support PRIN MIUR 2015 ``Cosmology and Fundamental Physics: Illuminating the Dark Universe with \textit{Euclid}''.
Matteo Maturi was supported by the SFB-Transregio TR33 'The Dark Universe'.
Anastasio D\'iaz-S\'anchez acknowledges support from project ESP2015-69020-C2-1-R (MINECO).
Anthony Gonzalez acknowledges support from NASA ROSES grant 12-EUCLID12-0004.
Florence Durret acknowledges long term funding from CNES.
This research made use of Astropy, a community-developed core Python package for Astronomy \citep{Astropy2013}, in addition to NumPy \citep{VanDerWalt2011}, SciPy \citep{Jones2001} and Ipython \citep{Perez2007}.
Figures were generated using Matplotlib \citep{Hunter2007}.
\end{acknowledgements}

\bibliography{biblio}

\begin{thebibliography}{166}
\expandafter\ifx\csname natexlab\endcsname\relax\def\natexlab#1{#1}\fi

\bibitem[{{Abbott} {et~al.}(2018){Abbott}, {Abdalla}, {Allam}, {Amara},
  {Annis}, {Asorey}, {Avila}, {Ballester}, {Banerji}, {Barkhouse}, {Baruah},
  {Baumer}, {Bechtol}, {Becker}, {Benoit-L{\'e}vy}, {Bernstein}, {Bertin},
  {Blazek}, {Bocquet}, {Brooks}, {Brout}, {Buckley-Geer}, {Burke}, {Busti},
  {Campisano}, {Cardiel-Sas}, {Carnero Rosell}, {Carrasco Kind}, {Carretero},
  {Castander}, {Cawthon}, {Chang}, {Chen}, {Conselice}, {Costa}, {Crocce},
  {Cunha}, {D'Andrea}, {da Costa}, {Das}, {Daues}, {Davis}, {Davis}, {De
  Vicente}, {DePoy}, {DeRose}, {Desai}, {Diehl}, {Dietrich}, {Dodelson},
  {Doel}, {Drlica-Wagner}, {Eifler}, {Elliott}, {Evrard}, {Farahi}, {Fausti
  Neto}, {Fernandez}, {Finley}, {Flaugher}, {Foley}, {Fosalba}, {Friedel},
  {Frieman}, {Garc{\'{\i}}a-Bellido}, {Gaztanaga}, {Gerdes}, {Giannantonio},
  {Gill}, {Glazebrook}, {Goldstein}, {Gower}, {Gruen}, {Gruendl}, {Gschwend},
  {Gupta}, {Gutierrez}, {Hamilton}, {Hartley}, {Hinton}, {Hislop}, {Hollowood},
  {Honscheid}, {Hoyle}, {Huterer}, {Jain}, {James}, {Jeltema}, {Johnson},
  {Johnson}, {Kacprzak}, {Kent}, {Khullar}, {Klein}, {Kovacs}, {Koziol},
  {Krause}, {Kremin}, {Kron}, {Kuehn}, {Kuhlmann}, {Kuropatkin}, {Lahav},
  {Lasker}, {Li}, {Li}, {Liddle}, {Lima}, {Lin}, {L{\'o}pez-Reyes}, {MacCrann},
  {Maia}, {Maloney}, {Manera}, {March}, {Marriner}, {Marshall}, {Martini},
  {McClintock}, {McKay}, {McMahon}, {Melchior}, {Menanteau}, {Miller},
  {Miquel}, {Mohr}, {Morganson}, {Mould}, {Neilsen}, {Nichol}, {Nogueira},
  {Nord}, {Nugent}, {Nunes}, {Ogando}, {Old}, {Pace}, {Palmese},
  {Paz-Chinch{\'o}n}, {Peiris}, {Percival}, {Petravick}, {Plazas}, {Poh},
  {Pond}, {Porredon}, {Pujol}, {Refregier}, {Reil}, {Ricker}, {Rollins},
  {Romer}, {Roodman}, {Rooney}, {Ross}, {Rykoff}, {Sako}, {Sanchez}, {Sanchez},
  {Santiago}, {Saro}, {Scarpine}, {Scolnic}, {Serrano}, {Sevilla-Noarbe},
  {Sheldon}, {Shipp}, {Silveira}, {Smith}, {Smith}, {Smith}, {Soares-Santos},
  {Sobreira}, {Song}, {Stebbins}, {Suchyta}, {Sullivan}, {Swanson}, {Tarle},
  {Thaler}, {Thomas}, {Thomas}, {Troxel}, {Tucker}, {Vikram}, {Vivas},
  {Walker}, {Wechsler}, {Weller}, {Wester}, {Wolf}, {Wu}, {Yanny}, {Zenteno},
  {Zhang}, {Zuntz}, {DES Collaboration}, {Juneau}, {Fitzpatrick}, {Nikutta},
  {Nidever}, {Olsen}, {Scott}, \& {Data Lab}}]{Abbott2018}
{Abbott}, T.~M.~C. {et~al.} 2018, \apjs, 239, 18

\bibitem[{{Abell}(1958)}]{Abell1958}
{Abell}, G.~O. 1958, \apjs, 3, 211

\bibitem[{{Adam} {et~al.}(2015){Adam}, {Comis}, {Mac{\'{\i}}as-P{\'e}rez},
  {Adane}, {Ade}, {Andr{\'e}}, {Beelen}, {Belier}, {Beno{\^i}t}, {Bideaud},
  {Billot}, {Blanquer}, {Bourrion}, {Calvo}, {Catalano}, {Coiffard},
  {Cruciani}, {D'Addabbo}, {D{\'e}sert}, {Doyle}, {Goupy}, {Kramer},
  {Leclercq}, {Martino}, {Mauskopf}, {Mayet}, {Monfardini}, {Pajot}, {Pascale},
  {Perotto}, {Pointecouteau}, {Ponthieu}, {Rev{\'e}ret}, {Ritacco},
  {Rodriguez}, {Savini}, {Schuster}, {Sievers}, {Tucker}, \&
  {Zylka}}]{Adam2015}
{Adam}, R. {et~al.} 2015, \aap, 576, A12

\bibitem[{{Adami} {et~al.}(2010){Adami}, {Durret}, {Benoist}, {Coupon},
  {Mazure}, {Meneux}, {Ilbert}, {Blaizot}, {Arnouts}, {Cappi}, {Garilli},
  {Guennou}, {Lebrun}, {Lef{\`e}vre}, {Maurogordato}, {McCracken}, {Mellier},
  {Slezak}, {Tresse}, \& {Ulmer}}]{Adami2010}
{Adami}, C. {et~al.} 2010, \aap, 509, A81

\bibitem[{{Adami} \& {Mazure}(1999)}]{Adami1999}
{Adami}, C., \& {Mazure}, A. 1999, \aaps, 134, 393

\bibitem[{{Adami} {et~al.}(2000){Adami}, {Ulmer}, {Romer}, {Nichol}, {Holden},
  \& {Pildis}}]{Adami2000}
{Adami}, C., {Ulmer}, M.~P., {Romer}, A.~K., {Nichol}, R.~C., {Holden}, B.~P.,
  \& {Pildis}, R.~A. 2000, \apjs, 131, 391

\bibitem[{{Akritas} \& {Bershady}(1996)}]{Akritas1996}
{Akritas}, M.~G., \& {Bershady}, M.~A. 1996, \apj, 470, 706

\bibitem[{{Alberts} {et~al.}(2016){Alberts}, {Pope}, {Brodwin}, {Chung},
  {Cybulski}, {Dey}, {Eisenhardt}, {Galametz}, {Gonzalez}, {Jannuzi},
  {Stanford}, {Snyder}, {Stern}, \& {Zeimann}}]{Alberts2016}
{Alberts}, S. {et~al.} 2016, \apj, 825, 72

\bibitem[{{Allen} {et~al.}(2011){Allen}, {Evrard}, \& {Mantz}}]{Allen2011}
{Allen}, S.~W., {Evrard}, A.~E., \& {Mantz}, A.~B. 2011, \araa, 49, 409

\bibitem[{{Amendola} {et~al.}(2013){Amendola}, {Appleby}, {Bacon}, {Baker},
  {Baldi}, {Bartolo}, {Blanchard}, {Bonvin}, {Borgani}, {Branchini}, {Burrage},
  {Camera}, {Carbone}, {Casarini}, {Cropper}, {de Rham}, {Di Porto}, {Ealet},
  {Ferreira}, {Finelli}, {Garc{\'{\i}}a-Bellido}, {Giannantonio}, {Guzzo},
  {Heavens}, {Heisenberg}, {Heymans}, {Hoekstra}, {Hollenstein}, {Holmes},
  {Horst}, {Jahnke}, {Kitching}, {Koivisto}, {Kunz}, {La Vacca}, {March},
  {Majerotto}, {Markovic}, {Marsh}, {Marulli}, {Massey}, {Mellier}, {Mota},
  {Nunes}, {Percival}, {Pettorino}, {Porciani}, {Quercellini}, {Read},
  {Rinaldi}, {Sapone}, {Scaramella}, {Skordis}, {Simpson}, {Taylor}, {Thomas},
  {Trotta}, {Verde}, {Vernizzi}, {Vollmer}, {Wang}, {Weller}, \&
  {Zlosnik}}]{Amendola2013}
{Amendola}, L. {et~al.} 2013, Living Reviews in Relativity, 16, 6

\bibitem[{{Andreon}(2015)}]{Andreon2015}
{Andreon}, S. 2015, \aap, 582, A100

\bibitem[{{Andreon} {et~al.}(2009){Andreon}, {Maughan}, {Trinchieri}, \&
  {Kurk}}]{Andreon2009}
{Andreon}, S., {Maughan}, B., {Trinchieri}, G., \& {Kurk}, J. 2009, \aap, 507,
  147

\bibitem[{{Andreon} {et~al.}(2005){Andreon}, {Punzi}, \& {Grado}}]{Andreon2005}
{Andreon}, S., {Punzi}, G., \& {Grado}, A. 2005, \mnras, 360, 727

\bibitem[{{Annunziatella} {et~al.}(2014){Annunziatella}, {Biviano}, {Mercurio},
  {Nonino}, {Rosati}, {Balestra}, {Presotto}, {Girardi}, {Gobat}, {Grillo},
  {Kelson}, {Medezinski}, {Postman}, {Scodeggio}, {Brescia}, {Demarco},
  {Fritz}, {Koekemoer}, {Lemze}, {Lombardi}, {Sartoris}, {Umetsu}, {Vanzella},
  {Bradley}, {Coe}, {Donahue}, {Infante}, {Kuchner}, {Maier}, {Reg{\H{o}}s},
  {Verdugo}, \& {Ziegler}}]{Annunziatella2014}
{Annunziatella}, M. {et~al.} 2014, \aap, 571, A80

\bibitem[{{Ascaso} {et~al.}(2017){Ascaso}, {Mei}, {Bartlett}, \&
  {Ben{\'{\i}}tez}}]{Ascaso2017}
{Ascaso}, B., {Mei}, S., {Bartlett}, J.~G., \& {Ben{\'{\i}}tez}, N. 2017,
  \mnras, 464, 2270

\bibitem[{{Ascaso} {et~al.}(2015){Ascaso}, {Mei}, \&
  {Ben{\'{\i}}tez}}]{Ascaso2015}
{Ascaso}, B., {Mei}, S., \& {Ben{\'{\i}}tez}, N. 2015, \mnras, 453, 2515

\bibitem[{{Ascaso} {et~al.}(2012){Ascaso}, {Wittman}, \&
  {Ben{\'{\i}}tez}}]{Ascaso2012}
{Ascaso}, B., {Wittman}, D., \& {Ben{\'{\i}}tez}, N. 2012, \mnras, 420, 1167

\bibitem[{{Ascaso} {et~al.}(2014){Ascaso}, {Wittman}, \& {Dawson}}]{Ascaso2014}
{Ascaso}, B., {Wittman}, D., \& {Dawson}, W. 2014, \mnras, 439, 1980

\bibitem[{{Astropy Collaboration} {et~al.}(2013){Astropy Collaboration},
  {Robitaille}, {Tollerud}, {Greenfield}, {Droettboom}, {Bray}, {Aldcroft},
  {Davis}, {Ginsburg}, {Price-Whelan}, {Kerzendorf}, {Conley}, {Crighton},
  {Barbary}, {Muna}, {Ferguson}, {Grollier}, {Parikh}, {Nair}, {Unther},
  {Deil}, {Woillez}, {Conseil}, {Kramer}, {Turner}, {Singer}, {Fox}, {Weaver},
  {Zabalza}, {Edwards}, {Azalee Bostroem}, {Burke}, {Casey}, {Crawford},
  {Dencheva}, {Ely}, {Jenness}, {Labrie}, {Lim}, {Pierfederici}, {Pontzen},
  {Ptak}, {Refsdal}, {Servillat}, \& {Streicher}}]{Astropy2013}
{Astropy Collaboration} {et~al.} 2013, \aap, 558, A33

\bibitem[{Aurenhammer {et~al.}(2013)Aurenhammer, Klein, \&
  Lee}]{Aurenhammer2013}
Aurenhammer, F., Klein, R., \& Lee, D.-T. 2013, Voronoi Diagrams and Delaunay
  Triangulations, 1st edn. (River Edge, NJ, USA: World Scientific Publishing
  Co., Inc.)

\bibitem[{{Barkhouse} {et~al.}(2007){Barkhouse}, {Yee}, \&
  {L{\'o}pez-Cruz}}]{Barkhouse2007}
{Barkhouse}, W.~A., {Yee}, H.~K.~C., \& {L{\'o}pez-Cruz}, O. 2007, \apj, 671,
  1471

\bibitem[{{Bellagamba} {et~al.}(2011){Bellagamba}, {Maturi}, {Hamana},
  {Meneghetti}, {Miyazaki}, \& {Moscardini}}]{Bellagamba2011}
{Bellagamba}, F., {Maturi}, M., {Hamana}, T., {Meneghetti}, M., {Miyazaki}, S.,
  \& {Moscardini}, L. 2011, \mnras, 413, 1145

\bibitem[{{Bellagamba} {et~al.}(2018){Bellagamba}, {Roncarelli}, {Maturi}, \&
  {Moscardini}}]{Bellagamba2018}
{Bellagamba}, F., {Roncarelli}, M., {Maturi}, M., \& {Moscardini}, L. 2018,
  \mnras, 473, 5221

\bibitem[{{Bellagamba} {et~al.}(2019){Bellagamba}, {Sereno}, {Roncarelli},
  {Maturi}, {Radovich}, {Bardelli}, {Puddu}, {Moscardini}, {Getman},
  {Hildebrandt}, \& {Napolitano}}]{Bellagamba2018b}
{Bellagamba}, F. {et~al.} 2019, \mnras, 484, 1598

\bibitem[{{Ben{\'{\i}}tez}(2000)}]{Benitez2000}
{Ben{\'{\i}}tez}, N. 2000, \apj, 536, 571

\bibitem[{{Benitez} {et~al.}(2014){Benitez}, {Dupke}, {Moles}, {Sodre},
  {Cenarro}, {Marin-Franch}, {Taylor}, {Cristobal}, {Fernandez-Soto}, {Mendes
  de Oliveira}, {Cepa-Nogue}, {Abramo}, {Alcaniz}, {Overzier},
  {Hernandez-Monteagudo}, {Alfaro}, {Kanaan}, {Carvano}, {Reis}, {Martinez
  Gonzalez}, {Ascaso}, {Ballesteros}, {Xavier}, {Varela}, {Ederoclite},
  {Vazquez Ramio}, {Broadhurst}, {Cypriano}, {Angulo}, {Diego}, {Zandivarez},
  {Diaz}, {Melchior}, {Umetsu}, {Spinelli}, {Zitrin}, {Coe}, {Yepes}, {Vielva},
  {Sahni}, {Marcos-Caballero}, {Shu Kitaura}, {Maroto}, {Masip}, {Tsujikawa},
  {Carneiro}, {Gonzalez Nuevo}, {Carvalho}, {Reboucas}, {Carvalho}, {Abdalla},
  {Bernui}, {Pigozzo}, {Ferreira}, {Chandrachani Devi}, {Bengaly}, {Campista},
  {Amorim}, {Asari}, {Bongiovanni}, {Bonoli}, {Bruzual}, {Cardiel}, {Cava},
  {Cid Fernandes}, {Coelho}, {Cortesi}, {Delgado}, {Diaz Garcia}, {Espinosa},
  {Galliano}, {Gonzalez-Serrano}, {Falcon-Barroso}, {Fritz}, {Fernandes},
  {Gorgas}, {Hoyos}, {Jimenez-Teja}, {Lopez-Aguerri}, {Lopez-San Juan},
  {Mateus}, {Molino}, {Novais}, {OMill}, {Oteo}, {Perez-Gonzalez}, {Poggianti},
  {Proctor}, {Ricciardelli}, {Sanchez-Blazquez}, {Storchi-Bergmann}, {Telles},
  {Schoennell}, {Trujillo}, {Vazdekis}, {Viironen}, {Daflon},
  {Aparicio-Villegas}, {Rocha}, {Ribeiro}, {Borges}, {Martins}, {Marcolino},
  {Martinez-Delgado}, {Perez-Torres}, {Siffert}, {Calvao}, {Sako}, {Kessler},
  {Alvarez-Candal}, {De Pra}, {Roig}, {Lazzaro}, {Gorosabel}, {Lopes de
  Oliveira}, {Lima-Neto}, {Irwin}, {Liu}, {Alvarez}, {Balmes}, {Chueca},
  {Costa-Duarte}, {da Costa}, {Dantas}, {Diaz}, {Fabregat}, {Ferrari},
  {Gavela}, {Gracia}, {Gruel}, {Gutierrez}, {Guzman}, {Hernandez-Fernandez},
  {Herranz}, {Hurtado-Gil}, {Jablonsky}, {Laporte}, {Le Tiran}, {Licandro},
  {Lima}, {Martin}, {Martinez}, {Montero}, {Penteado}, {Pereira}, {Peris},
  {Quilis}, {Sanchez-Portal}, {Soja}, {Solano}, {Torra}, \&
  {Valdivielso}}]{Benitez2014}
{Benitez}, N. {et~al.} 2014, ArXiv e-prints:1403.5237

\bibitem[{{Ben{\'{\i}}tez} {et~al.}(2004){Ben{\'{\i}}tez}, {Ford}, {Bouwens},
  {Menanteau}, {Blakeslee}, {Gronwall}, {Illingworth}, {Meurer}, {Broadhurst},
  {Clampin}, {Franx}, {Hartig}, {Magee}, {Sirianni}, {Ardila}, {Bartko},
  {Brown}, {Burrows}, {Cheng}, {Cross}, {Feldman}, {Golimowski}, {Infante},
  {Kimble}, {Krist}, {Lesser}, {Levay}, {Martel}, {Miley}, {Postman}, {Rosati},
  {Sparks}, {Tran}, {Tsvetanov}, {White}, \& {Zheng}}]{Benitez2004}
{Ben{\'{\i}}tez}, N. {et~al.} 2004, \apjs, 150, 1

\bibitem[{{Benoist}(2014)}]{Benoist2014}
{Benoist}, C. 2014, in Building the Euclid Cluster Survey - Scientific Program,
  proceedings of a conference held July 6-11 2014 at the Sexten Center for
  Astrophysics., 8

\bibitem[{{Bertin} \& {Arnouts}(1996)}]{Bertin1996}
{Bertin}, E., \& {Arnouts}, S. 1996, \aaps, 117, 393

\bibitem[{{Bleem} {et~al.}(2015){Bleem}, {Stalder}, {de Haan}, {Aird}, {Allen},
  {Applegate}, {Ashby}, {Bautz}, {Bayliss}, {Benson}, {Bocquet}, {Brodwin},
  {Carlstrom}, {Chang}, {Chiu}, {Cho}, {Clocchiatti}, {Crawford}, {Crites},
  {Desai}, {Dietrich}, {Dobbs}, {Foley}, {Forman}, {George}, {Gladders},
  {Gonzalez}, {Halverson}, {Hennig}, {Hoekstra}, {Holder}, {Holzapfel},
  {Hrubes}, {Jones}, {Keisler}, {Knox}, {Lee}, {Leitch}, {Liu}, {Lueker},
  {Luong-Van}, {Mantz}, {Marrone}, {McDonald}, {McMahon}, {Meyer}, {Mocanu},
  {Mohr}, {Murray}, {Padin}, {Pryke}, {Reichardt}, {Rest}, {Ruel}, {Ruhl},
  {Saliwanchik}, {Saro}, {Sayre}, {Schaffer}, {Schrabback}, {Shirokoff},
  {Song}, {Spieler}, {Stanford}, {Staniszewski}, {Stark}, {Story}, {Stubbs},
  {Vanderlinde}, {Vieira}, {Vikhlinin}, {Williamson}, {Zahn}, \&
  {Zenteno}}]{Bleem2014}
{Bleem}, L.~E. {et~al.} 2015, \apjs, 216, 27

\bibitem[{{B{\"o}hringer} {et~al.}(2014){B{\"o}hringer}, {Chon}, \&
  {Collins}}]{Bohringer2014}
{B{\"o}hringer}, H., {Chon}, G., \& {Collins}, C.~A. 2014, \aap, 570, A31

\bibitem[{{B{\"o}hringer} {et~al.}(2001){B{\"o}hringer}, {Schuecker}, {Guzzo},
  {Collins}, {Voges}, {Schindler}, {Neumann}, {Cruddace}, {De Grandi},
  {Chincarini}, {Edge}, {MacGillivray}, \& {Shaver}}]{Bohringer2001}
{B{\"o}hringer}, H. {et~al.} 2001, \aap, 369, 826

\bibitem[{{Bolzonella} {et~al.}(2000){Bolzonella}, {Miralles}, \&
  {Pell{\'o}}}]{Bolzonella2000}
{Bolzonella}, M., {Miralles}, J.-M., \& {Pell{\'o}}, R. 2000, \aap, 363, 476

\bibitem[{{Boselli} {et~al.}(1997){Boselli}, {Lequeux}, {Contursi}, {Gavazzi},
  {Boulade}, {Boulanger}, {Cesarsky}, {Dupraz}, {Madden}, {Sauvage},
  {Viallefond}, \& {Vigroux}}]{Boselli1997}
{Boselli}, A. {et~al.} 1997, \aap, 324, L13

\bibitem[{{Botzler} {et~al.}(2004){Botzler}, {Snigula}, {Bender}, \&
  {Hopp}}]{Botzler2004}
{Botzler}, C.~S., {Snigula}, J., {Bender}, R., \& {Hopp}, U. 2004, \mnras, 349,
  425

\bibitem[{{Brodwin} {et~al.}(2012){Brodwin}, {Gonzalez}, {Stanford}, {Plagge},
  {Marrone}, {Carlstrom}, {Dey}, {Eisenhardt}, {Fedeli}, {Gettings}, {Jannuzi},
  {Joy}, {Leitch}, {Mancone}, {Snyder}, {Stern}, \& {Zeimann}}]{Brodwin2012}
{Brodwin}, M. {et~al.} 2012, \apj, 753, 162

\bibitem[{{Brodwin} {et~al.}(2013){Brodwin}, {Stanford}, {Gonzalez}, {Zeimann},
  {Snyder}, {Mancone}, {Pope}, {Eisenhardt}, {Stern}, {Alberts}, {Ashby},
  {Brown}, {Chary}, {Dey}, {Galametz}, {Gettings}, {Jannuzi}, {Miller},
  {Moustakas}, \& {Moustakas}}]{Brodwin2013}
------. 2013, \apj, 779, 138

\bibitem[{{Budzynski} {et~al.}(2012){Budzynski}, {Koposov}, {McCarthy},
  {McGee}, \& {Belokurov}}]{Budzynski2012}
{Budzynski}, J.~M., {Koposov}, S.~E., {McCarthy}, I.~G., {McGee}, S.~L., \&
  {Belokurov}, V. 2012, \mnras, 423, 104

\bibitem[{{Cannon} {et~al.}(2006){Cannon}, {Drinkwater}, {Edge}, {Eisenstein},
  {Nichol}, {Outram}, {Pimbblet}, {de Propris}, {Roseboom}, {Wake}, {Allen},
  {Bland-Hawthorn}, {Bridges}, {Carson}, {Chiu}, {Colless}, {Couch}, {Croom},
  {Driver}, {Fine}, {Hewett}, {Loveday}, {Ross}, {Sadler}, {Shanks}, {Sharp},
  {Smith}, {Stoughton}, {Weilbacher}, {Brunner}, {Meiksin}, \&
  {Schneider}}]{Cannon2006}
{Cannon}, R. {et~al.} 2006, \mnras, 372, 425

\bibitem[{{Cappellari} \& {Copin}(2003)}]{Cappellari2003}
{Cappellari}, M., \& {Copin}, Y. 2003, \mnras, 342, 345

\bibitem[{{Carlberg} {et~al.}(1997){Carlberg}, {Yee}, {Ellingson}, {Morris},
  {Abraham}, {Gravel}, {Pritchet}, {Smecker-Hane}, {Hartwick}, {Hesser},
  {Hutchings}, \& {Oke}}]{Carlberg1997}
{Carlberg}, R.~G. {et~al.} 1997, \apjl, 485, L13

\bibitem[{{Carretero} {et~al.}(2015){Carretero}, {Castander}, {Gazta{\~n}aga},
  {Crocce}, \& {Fosalba}}]{Carretero2015}
{Carretero}, J., {Castander}, F.~J., {Gazta{\~n}aga}, E., {Crocce}, M., \&
  {Fosalba}, P. 2015, \mnras, 447, 646

\bibitem[{{Castignani} \& {Benoist}(2016)}]{Castignani2016}
{Castignani}, G., \& {Benoist}, C. 2016, \aap, 595, A111

\bibitem[{{Cava} {et~al.}(2017){Cava}, {Biviano}, {Mamon}, {Varela}, {Bettoni},
  {D'Onofrio}, {Fasano}, {Fritz}, {Moles}, {Moretti}, \&
  {Poggianti}}]{Cava2017}
{Cava}, A. {et~al.} 2017, \aap, 606, A108

\bibitem[{{Ciliegi} {et~al.}(2005){Ciliegi}, {Zamorani}, {Bondi}, {Pozzetti},
  {Bolzonella}, {Gregorini}, {Garilli}, {Iovino}, {McCracken}, {Mellier},
  {Radovich}, {de Ruiter}, {Parma}, {Bottini}, {Le Brun}, {Le F{\`e}vre},
  {Maccagni}, {Picat}, {Scaramella}, {Scodeggio}, {Tresse}, {Vettolani},
  {Zanichelli}, {Adami}, {Arnaboldi}, {Arnouts}, {Bardelli}, {Cappi},
  {Charlot}, {Contini}, {Foucaud}, {Franzetti}, {Guzzo}, {Ilbert}, {Marano},
  {Marinoni}, {Mathez}, {Mazure}, {Meneux}, {Merighi}, {Merluzzi}, {Paltani},
  {Pollo}, {Zucca}, {Bongiorno}, {Busarello}, {Gavignaud}, {Pell{\`o}},
  {Ripepi}, \& {Rizzo}}]{Ciliegi2005}
{Ciliegi}, P. {et~al.} 2005, \aap, 441, 879

\bibitem[{{Coe} {et~al.}(2006){Coe}, {Ben{\'{\i}}tez}, {S{\'a}nchez}, {Jee},
  {Bouwens}, \& {Ford}}]{Coe2006}
{Coe}, D., {Ben{\'{\i}}tez}, N., {S{\'a}nchez}, S.~F., {Jee}, M., {Bouwens},
  R., \& {Ford}, H. 2006, \aj, 132, 926

\bibitem[{{Collister} \& {Lahav}(2005)}]{Collister2005}
{Collister}, A.~A., \& {Lahav}, O. 2005, \mnras, 361, 415

\bibitem[{{Costanzi} {et~al.}(2019){Costanzi}, {Rozo}, {Rykoff}, {Farahi},
  {Jeltema}, {Evrard}, {Mantz}, {Gruen}, {Mandelbaum}, {DeRose}, {McClintock},
  {Varga}, {Zhang}, {Weller}, {Wechsler}, \& {Aguena}}]{Costanzi2019}
{Costanzi}, M. {et~al.} 2019, \mnras, 482, 490

\bibitem[{{Cucciati} {et~al.}(2010){Cucciati}, {Marinoni}, {Iovino},
  {Bardelli}, {Adami}, {Mazure}, {Scodeggio}, {Maccagni}, {Temporin}, {Zucca},
  {De Lucia}, {Blaizot}, {Garilli}, {Meneux}, {Zamorani}, {Le F{\`e}vre},
  {Cappi}, {Guzzo}, {Bottini}, {Le Brun}, {Tresse}, {Vettolani}, {Zanichelli},
  {Arnouts}, {Bolzonella}, {Charlot}, {Ciliegi}, {Contini}, {Foucaud},
  {Franzetti}, {Gavignaud}, {Ilbert}, {Lamareille}, {McCracken}, {Marano},
  {Merighi}, {Paltani}, {Pell{\`o}}, {Pollo}, {Pozzetti}, {Vergani}, \&
  {P{\'e}rez-Montero}}]{Cucciati2010}
{Cucciati}, O. {et~al.} 2010, \aap, 520, A42

\bibitem[{{Dalton} {et~al.}(1997){Dalton}, {Maddox}, {Sutherland}, \&
  {Efstathiou}}]{Dalton1997}
{Dalton}, G.~B., {Maddox}, S.~J., {Sutherland}, W.~J., \& {Efstathiou}, G.
  1997, \mnras, 289, 263

\bibitem[{{de Haan} {et~al.}(2016){de Haan}, {Benson}, {Bleem}, {Allen},
  {Applegate}, {Ashby}, {Bautz}, {Bayliss}, {Bocquet}, {Brodwin}, {Carlstrom},
  {Chang}, {Chiu}, {Cho}, {Clocchiatti}, {Crawford}, {Crites}, {Desai},
  {Dietrich}, {Dobbs}, {Doucouliagos}, {Foley}, {Forman}, {Garmire}, {George},
  {Gladders}, {Gonzalez}, {Gupta}, {Halverson}, {Hlavacek-Larrondo},
  {Hoekstra}, {Holder}, {Holzapfel}, {Hou}, {Hrubes}, {Huang}, {Jones},
  {Keisler}, {Knox}, {Lee}, {Leitch}, {von der Linden}, {Luong-Van}, {Mantz},
  {Marrone}, {McDonald}, {McMahon}, {Meyer}, {Mocanu}, {Mohr}, {Murray},
  {Padin}, {Pryke}, {Rapetti}, {Reichardt}, {Rest}, {Ruel}, {Ruhl},
  {Saliwanchik}, {Saro}, {Sayre}, {Schaffer}, {Schrabback}, {Shirokoff},
  {Song}, {Spieler}, {Stalder}, {Stanford}, {Staniszewski}, {Stark}, {Story},
  {Stubbs}, {Vanderlinde}, {Vieira}, {Vikhlinin}, {Williamson}, \&
  {Zenteno}}]{deHaan2016}
{de Haan}, T. {et~al.} 2016, \apj, 832, 95

\bibitem[{{De Lucia} {et~al.}(2012){De Lucia}, {Weinmann}, {Poggianti},
  {Arag{\'o}n-Salamanca}, \& {Zaritsky}}]{DeLucia2012}
{De Lucia}, G., {Weinmann}, S., {Poggianti}, B.~M., {Arag{\'o}n-Salamanca}, A.,
  \& {Zaritsky}, D. 2012, \mnras, 423, 1277

\bibitem[{{De Propris}(2017)}]{DePropris2017}
{De Propris}, R. 2017, \mnras, 465, 4035

\bibitem[{{de Propris} {et~al.}(1998){de Propris}, {Eisenhardt}, {Stanford}, \&
  {Dickinson}}]{dePropris1998}
{de Propris}, R., {Eisenhardt}, P.~R., {Stanford}, S.~A., \& {Dickinson}, M.
  1998, \apjl, 503, L45

\bibitem[{{de Propris} \& {Pritchet}(1998)}]{dePropris1998b}
{de Propris}, R., \& {Pritchet}, C.~J. 1998, \aj, 116, 1118

\bibitem[{{de Propris} {et~al.}(1999){de Propris}, {Stanford}, {Eisenhardt},
  {Dickinson}, \& {Elston}}]{DePropris1999}
{de Propris}, R., {Stanford}, S.~A., {Eisenhardt}, P.~R., {Dickinson}, M., \&
  {Elston}, R. 1999, \aj, 118, 719

\bibitem[{{Diehl} \& {Statler}(2006)}]{Diehl2006}
{Diehl}, S., \& {Statler}, T.~S. 2006, \mnras, 368, 497

\bibitem[{{Dietrich} {et~al.}(2014){Dietrich}, {Zhang}, {Song}, {Davis},
  {McKay}, {Baruah}, {Becker}, {Benoist}, {Busha}, {da Costa}, {Hao}, {Maia},
  {Miller}, {Ogando}, {Romer}, {Rozo}, {Rykoff}, \& {Wechsler}}]{Dietrich2014}
{Dietrich}, J.~P. {et~al.} 2014, \mnras, 443, 1713

\bibitem[{{Driver} {et~al.}(1994){Driver}, {Phillipps}, {Davies}, {Morgan}, \&
  {Disney}}]{Driver1994}
{Driver}, S.~P., {Phillipps}, S., {Davies}, J.~I., {Morgan}, I., \& {Disney},
  M.~J. 1994, \mnras, 268, 393

\bibitem[{{Durret} {et~al.}(2015){Durret}, {Adami}, {Bertin}, {Hao},
  {M{\'a}rquez}, {Martinet}, {Maurogordato}, {Sauvaget}, {Scepi}, {Takey}, \&
  {Ulmer}}]{Durret2015}
{Durret}, F. {et~al.} 2015, \aap, 578, A79

\bibitem[{{Durret} {et~al.}(2011){Durret}, {Adami}, {Cappi}, {Maurogordato},
  {M{\'a}rquez}, {Ilbert}, {Coupon}, {Arnouts}, {Benoist}, {Blaizot}, {Edorh},
  {Garilli}, {Guennou}, {Le Brun}, {Le F{\`e}vre}, {Mazure}, {McCracken},
  {Mellier}, {Mezrag}, {Slezak}, {Tresse}, \& {Ulmer}}]{Durret2011}
------. 2011, \aap, 535, A65

\bibitem[{{Eisenhardt} {et~al.}(2008){Eisenhardt}, {Brodwin}, {Gonzalez},
  {Stanford}, {Stern}, {Barmby}, {Brown}, {Dawson}, {Dey}, {Doi}, {Galametz},
  {Jannuzi}, {Kochanek}, {Meyers}, {Morokuma}, \& {Moustakas}}]{Eisenhardt2008}
{Eisenhardt}, P.~R.~M. {et~al.} 2008, \apj, 684, 905

\bibitem[{{Ellis} \& {Jones}(2004)}]{Ellis2004}
{Ellis}, S.~C., \& {Jones}, L.~R. 2004, \mnras, 348, 165

\bibitem[{{Elston} {et~al.}(2006){Elston}, {Gonzalez}, {McKenzie}, {Brodwin},
  {Brown}, {Cardona}, {Dey}, {Dickinson}, {Eisenhardt}, {Jannuzi}, {Lin},
  {Mohr}, {Raines}, {Stanford}, \& {Stern}}]{Elston2006}
{Elston}, R.~J. {et~al.} 2006, \apj, 639, 816

\bibitem[{{Farrens} {et~al.}(2011){Farrens}, {Abdalla}, {Cypriano}, {Sabiu}, \&
  {Blake}}]{Farrens2011}
{Farrens}, S., {Abdalla}, F.~B., {Cypriano}, E.~S., {Sabiu}, C., \& {Blake}, C.
  2011, \mnras, 417, 1402

\bibitem[{{Fioc} \& {Rocca-Volmerange}(1997)}]{Fioc1997}
{Fioc}, M., \& {Rocca-Volmerange}, B. 1997, \aap, 326, 950

\bibitem[{{Gal}(2006)}]{Gal2006}
{Gal}, R.~R. 2006, ArXiv e-prints:astro-ph/0601195

\bibitem[{{Gal} {et~al.}(2003){Gal}, {de Carvalho}, {Lopes}, {Djorgovski},
  {Brunner}, {Mahabal}, \& {Odewahn}}]{Gal2003}
{Gal}, R.~R., {de Carvalho}, R.~R., {Lopes}, P.~A.~A., {Djorgovski}, S.~G.,
  {Brunner}, R.~J., {Mahabal}, A., \& {Odewahn}, S.~C. 2003, \aj, 125, 2064

\bibitem[{{Gavazzi} \& {Soucail}(2007)}]{Gavazzi2007}
{Gavazzi}, R., \& {Soucail}, G. 2007, \aap, 462, 459

\bibitem[{{George} {et~al.}(2011){George}, {Leauthaud}, {Bundy}, {Finoguenov},
  {Tinker}, {Lin}, {Mei}, {Kneib}, {Aussel}, {Behroozi}, {Busha}, {Capak},
  {Coccato}, {Covone}, {Faure}, {Fiorenza}, {Ilbert}, {Le Floc'h}, {Koekemoer},
  {Tanaka}, {Wechsler}, \& {Wolk}}]{George2011}
{George}, M.~R. {et~al.} 2011, \apj, 742, 125

\bibitem[{{Gilbank} {et~al.}(2011){Gilbank}, {Gladders}, {Yee}, \&
  {Hsieh}}]{Gilbank2011}
{Gilbank}, D.~G., {Gladders}, M.~D., {Yee}, H.~K.~C., \& {Hsieh}, B.~C. 2011,
  \aj, 141, 94

\bibitem[{{Gladders} \& {Yee}(2000)}]{Gladders2000}
{Gladders}, M.~D., \& {Yee}, H.~K.~C. 2000, \aj, 120, 2148

\bibitem[{{Gonzalez}(2014)}]{Gonzalez2014}
{Gonzalez}, A. 2014, in Building the Euclid Cluster Survey - Scientific
  Program, proceedings of a conference held July 6-11 2014 at the Sexten Center
  for Astrophysics., 7

\bibitem[{{Gonzalez-Perez} {et~al.}(2014){Gonzalez-Perez}, {Lacey}, {Baugh},
  {Lagos}, {Helly}, {Campbell}, \& {Mitchell}}]{Gonzalez-Perez2014}
{Gonzalez-Perez}, V., {Lacey}, C.~G., {Baugh}, C.~M., {Lagos}, C.~D.~P.,
  {Helly}, J., {Campbell}, D.~J.~R., \& {Mitchell}, P.~D. 2014, \mnras, 439,
  264

\bibitem[{{Goto} {et~al.}(2002){Goto}, {Sekiguchi}, {Nichol}, {Bahcall}, {Kim},
  {Annis}, {Ivezi{\'c}}, {Brinkmann}, {Hennessy}, {Szokoly}, \&
  {Tucker}}]{Goto2002}
{Goto}, T. {et~al.} 2002, \aj, 123, 1807

\bibitem[{{Hasselfield} {et~al.}(2013){Hasselfield}, {Hilton}, {Marriage},
  {Addison}, {Barrientos}, {Battaglia}, {Battistelli}, {Bond}, {Crichton},
  {Das}, {Devlin}, {Dicker}, {Dunkley}, {D{\"u}nner}, {Fowler}, {Gralla},
  {Hajian}, {Halpern}, {Hincks}, {Hlozek}, {Hughes}, {Infante}, {Irwin},
  {Kosowsky}, {Marsden}, {Menanteau}, {Moodley}, {Niemack}, {Nolta}, {Page},
  {Partridge}, {Reese}, {Schmitt}, {Sehgal}, {Sherwin}, {Sievers}, {Sif{\'o}n},
  {Spergel}, {Staggs}, {Swetz}, {Switzer}, {Thornton}, {Trac}, \&
  {Wollack}}]{Hasselfield2013}
{Hasselfield}, M. {et~al.} 2013, \jcap, 7, 8

\bibitem[{{Henriques} {et~al.}(2012){Henriques}, {White}, {Lemson}, {Thomas},
  {Guo}, {Marleau}, \& {Overzier}}]{Henriques2012}
{Henriques}, B.~M.~B., {White}, S.~D.~M., {Lemson}, G., {Thomas}, P.~A., {Guo},
  Q., {Marleau}, G.-D., \& {Overzier}, R.~A. 2012, \mnras, 421, 2904

\bibitem[{{Huchra} \& {Geller}(1982)}]{Huchra1982}
{Huchra}, J.~P., \& {Geller}, M.~J. 1982, \apj, 257, 423

\bibitem[{Hunter(2007)}]{Hunter2007}
Hunter, J.~D. 2007, Computing In Science \& Engineering, 9, 90

\bibitem[{{Icke} \& {van de Weygaert}(1991)}]{Icke1991}
{Icke}, V., \& {van de Weygaert}, R. 1991, \qjras, 32, 85

\bibitem[{{Ilbert} {et~al.}(2006){Ilbert}, {Arnouts}, {McCracken},
  {Bolzonella}, {Bertin}, {Le F{\`e}vre}, {Mellier}, {Zamorani}, {Pell{\`o}},
  {Iovino}, {Tresse}, {Le Brun}, {Bottini}, {Garilli}, {Maccagni}, {Picat},
  {Scaramella}, {Scodeggio}, {Vettolani}, {Zanichelli}, {Adami}, {Bardelli},
  {Cappi}, {Charlot}, {Ciliegi}, {Contini}, {Cucciati}, {Foucaud}, {Franzetti},
  {Gavignaud}, {Guzzo}, {Marano}, {Marinoni}, {Mazure}, {Meneux}, {Merighi},
  {Paltani}, {Pollo}, {Pozzetti}, {Radovich}, {Zucca}, {Bondi}, {Bongiorno},
  {Busarello}, {de La Torre}, {Gregorini}, {Lamareille}, {Mathez}, {Merluzzi},
  {Ripepi}, {Rizzo}, \& {Vergani}}]{Ilbert2006}
{Ilbert}, O. {et~al.} 2006, \aap, 457, 841

\bibitem[{{Jakobs} {et~al.}(2018){Jakobs}, {Viola}, {McCarthy}, {van Waerbeke},
  {Hoekstra}, {Robotham}, {Hinshaw}, {Hojjati}, {Tanimura}, {Tr{\"o}ster},
  {Baldry}, {Heymans}, {Hildebrandt}, {Kuijken}, {Norberg}, {Schaye},
  {Sif{\'o}n}, {van Uitert}, {Valentijn}, {Verdoes Kleijn}, \&
  {Wang}}]{Jakobs2018}
{Jakobs}, A. {et~al.} 2018, \mnras, 480, 3338

\bibitem[{{Jeffrey} {et~al.}(2018){Jeffrey}, {Abdalla}, {Lahav}, {Lanusse},
  {Starck}, {Leonard}, {Kirk}, {Chang}, {Baxter}, {Kacprzak}, {Seitz},
  {Vikram}, {Whiteway}, {Abbott}, {Allam}, {Avila}, {Bertin}, {Brooks},
  {Carnero Rosell}, {Carrasco Kind}, {Carretero}, {Castander}, {Crocce},
  {Cunha}, {D'Andrea}, {da Costa}, {Davis}, {De Vicente}, {Desai}, {Doel},
  {Eifler}, {Evrard}, {Flaugher}, {Fosalba}, {Frieman},
  {Garc{\'{\i}}a-Bellido}, {Gerdes}, {Gruen}, {Gruendl}, {Gschwend},
  {Gutierrez}, {Hartley}, {Honscheid}, {Hoyle}, {James}, {Jarvis}, {Kuehn},
  {Lima}, {Lin}, {March}, {Melchior}, {Menanteau}, {Miquel}, {Plazas}, {Reil},
  {Roodman}, {Sanchez}, {Scarpine}, {Schubnell}, {Sevilla-Noarbe}, {Smith},
  {Soares-Santos}, {Sobreira}, {Suchyta}, {Swanson}, {Tarle}, {Thomas}, \&
  {Walker}}]{Jeffrey2018}
{Jeffrey}, N. {et~al.} 2018, \mnras, 479, 2871

\bibitem[{{Jiang} {et~al.}(2014){Jiang}, {Helly}, {Cole}, \&
  {Frenk}}]{Jiang2014}
{Jiang}, L., {Helly}, J.~C., {Cole}, S., \& {Frenk}, C.~S. 2014, \mnras, 440,
  2115

\bibitem[{Jones {et~al.}(2001)Jones, Oliphant, Peterson, {et~al.}}]{Jones2001}
Jones, E., Oliphant, T., Peterson, P., {et~al.} 2001, {SciPy}: Open source
  scientific tools for {Python}, http://www.scipy.org/

\bibitem[{{Kepner} {et~al.}(1999){Kepner}, {Fan}, {Bahcall}, {Gunn}, {Lupton},
  \& {Xu}}]{Kepner1999}
{Kepner}, J., {Fan}, X., {Bahcall}, N., {Gunn}, J., {Lupton}, R., \& {Xu}, G.
  1999, \apj, 517, 78

\bibitem[{{Kim} {et~al.}(2002){Kim}, {Kepner}, {Postman}, {Strauss}, {Bahcall},
  {Gunn}, {Lupton}, {Annis}, {Nichol}, {Castander}, {Brinkmann}, {Brunner},
  {Connolly}, {Csabai}, {Hindsley}, {Ivezi{\'c}}, {Vogeley}, \&
  {York}}]{Kim2002}
{Kim}, R.~S.~J. {et~al.} 2002, \aj, 123, 20

\bibitem[{{Knobel} {et~al.}(2009){Knobel}, {Lilly}, {Iovino}, {Porciani},
  {Kova{\v c}}, {Cucciati}, {Finoguenov}, {Kitzbichler}, {Carollo}, {Contini},
  {Kneib}, {Le F{\`e}vre}, {Mainieri}, {Renzini}, {Scodeggio}, {Zamorani},
  {Bardelli}, {Bolzonella}, {Bongiorno}, {Caputi}, {Coppa}, {de la Torre}, {de
  Ravel}, {Franzetti}, {Garilli}, {Kampczyk}, {Lamareille}, {Le Borgne}, {Le
  Brun}, {Maier}, {Mignoli}, {Pello}, {Peng}, {Perez Montero}, {Ricciardelli},
  {Silverman}, {Tanaka}, {Tasca}, {Tresse}, {Vergani}, {Zucca}, {Abbas},
  {Bottini}, {Cappi}, {Cassata}, {Cimatti}, {Fumana}, {Guzzo}, {Koekemoer},
  {Leauthaud}, {Maccagni}, {Marinoni}, {McCracken}, {Memeo}, {Meneux}, {Oesch},
  {Pozzetti}, \& {Scaramella}}]{Knobel2009}
{Knobel}, C. {et~al.} 2009, \apj, 697, 1842

\bibitem[{{Koester} {et~al.}(2007){Koester}, {McKay}, {Annis}, {Wechsler},
  {Evrard}, {Bleem}, {Becker}, {Johnston}, {Sheldon}, {Nichol}, {Miller},
  {Scranton}, {Bahcall}, {Barentine}, {Brewington}, {Brinkmann}, {Harvanek},
  {Kleinman}, {Krzesinski}, {Long}, {Nitta}, {Schneider}, {Sneddin}, {Voges},
  \& {York}}]{Koester2007}
{Koester}, B.~P. {et~al.} 2007, \apj, 660, 239

\bibitem[{{K{\"o}hlinger} {et~al.}(2015){K{\"o}hlinger}, {Hoekstra}, \&
  {Eriksen}}]{Kohlinger2015}
{K{\"o}hlinger}, F., {Hoekstra}, H., \& {Eriksen}, M. 2015, \mnras, 453, 3107

\bibitem[{{Lagos} {et~al.}(2012){Lagos}, {Bayet}, {Baugh}, {Lacey}, {Bell},
  {Fanidakis}, \& {Geach}}]{Lagos2012}
{Lagos}, C.~d.~P., {Bayet}, E., {Baugh}, C.~M., {Lacey}, C.~G., {Bell}, T.~A.,
  {Fanidakis}, N., \& {Geach}, J.~E. 2012, \mnras, 426, 2142

\bibitem[{{Laureijs} {et~al.}(2011){Laureijs}, {Amiaux}, {Arduini},
  {Augu{\`e}res}, {Brinchmann}, {Cole}, {Cropper}, {Dabin}, {Duvet}, {Ealet},
  \& et~al.}]{Laureijs2011}
{Laureijs}, R. {et~al.} 2011, ArXiv e-prints:1110.3193

\bibitem[{{Licitra} {et~al.}(2016{\natexlab{a}}){Licitra}, {Mei}, {Raichoor},
  {Erben}, \& {Hildebrandt}}]{Licitra2016a}
{Licitra}, R., {Mei}, S., {Raichoor}, A., {Erben}, T., \& {Hildebrandt}, H.
  2016{\natexlab{a}}, \mnras, 455, 3020

\bibitem[{{Licitra} {et~al.}(2016{\natexlab{b}}){Licitra}, {Mei}, {Raichoor},
  {Erben}, {Hildebrandt}, {Mu{\~n}oz}, {Van Waerbeke}, {C{\^o}t{\'e}},
  {Cuillandre}, {Duc}, {Ferrarese}, {Gwyn}, {Huertas-Company}, {Lan{\c c}on},
  {Parroni}, \& {Puzia}}]{Licitra2016b}
{Licitra}, R. {et~al.} 2016{\natexlab{b}}, \apj, 829, 44

\bibitem[{{Lin} {et~al.}(2006){Lin}, {Mohr}, {Gonzalez}, \&
  {Stanford}}]{Lin2006}
{Lin}, Y.-T., {Mohr}, J.~J., {Gonzalez}, A.~H., \& {Stanford}, S.~A. 2006,
  \apjl, 650, L99

\bibitem[{{Lin} {et~al.}(2004){Lin}, {Mohr}, \& {Stanford}}]{Lin2004}
{Lin}, Y.-T., {Mohr}, J.~J., \& {Stanford}, S.~A. 2004, \apj, 610, 745

\bibitem[{{LSST Science Collaboration} {et~al.}(2009){LSST Science
  Collaboration}, {Abell}, {Allison}, {Anderson}, {Andrew}, {Angel}, {Armus},
  {Arnett}, {Asztalos}, {Axelrod}, \& et~al.}]{LSST2009}
{LSST Science Collaboration} {et~al.} 2009, ArXiv e-prints:0912.0201

\bibitem[{{Majumdar} \& {Mohr}(2004)}]{Majumdar2004}
{Majumdar}, S., \& {Mohr}, J.~J. 2004, \apj, 613, 41

\bibitem[{{Mamon} {et~al.}(2010){Mamon}, {Biviano}, \& {Murante}}]{Mamon2010}
{Mamon}, G.~A., {Biviano}, A., \& {Murante}, G. 2010, \aap, 520, A30

\bibitem[{{Mana} {et~al.}(2013){Mana}, {Giannantonio}, {Weller}, {Hoyle},
  {H{\"u}tsi}, \& {Sartoris}}]{Mana2013}
{Mana}, A., {Giannantonio}, T., {Weller}, J., {Hoyle}, B., {H{\"u}tsi}, G., \&
  {Sartoris}, B. 2013, \mnras, 434, 684

\bibitem[{{Mannucci} {et~al.}(2001){Mannucci}, {Basile}, {Poggianti},
  {Cimatti}, {Daddi}, {Pozzetti}, \& {Vanzi}}]{Mannucci2001}
{Mannucci}, F., {Basile}, F., {Poggianti}, B.~M., {Cimatti}, A., {Daddi}, E.,
  {Pozzetti}, L., \& {Vanzi}, L. 2001, \mnras, 326, 745

\bibitem[{{Mantz} {et~al.}(2015){Mantz}, {von der Linden}, {Allen},
  {Applegate}, {Kelly}, {Morris}, {Rapetti}, {Schmidt}, {Adhikari}, {Allen},
  {Burchat}, {Burke}, {Cataneo}, {Donovan}, {Ebeling}, {Shandera}, \&
  {Wright}}]{Mantz2015}
{Mantz}, A.~B. {et~al.} 2015, \mnras, 446, 2205

\bibitem[{{Marinoni} {et~al.}(2002){Marinoni}, {Davis}, {Newman}, \&
  {Coil}}]{Marinoni2002}
{Marinoni}, C., {Davis}, M., {Newman}, J.~A., \& {Coil}, A.~L. 2002, \apj, 580,
  122

\bibitem[{{Maturi} {et~al.}(2019){Maturi}, {Bellagamba}, {Radovich},
  {Roncarelli}, {Sereno}, {Moscardini}, {Bardelli}, \& {Puddu}}]{Maturi2018}
{Maturi}, M., {Bellagamba}, F., {Radovich}, M., {Roncarelli}, M., {Sereno}, M.,
  {Moscardini}, L., {Bardelli}, S., \& {Puddu}, E. 2019, \mnras, 485, 498

\bibitem[{{Maturi} {et~al.}(2005){Maturi}, {Meneghetti}, {Bartelmann}, {Dolag},
  \& {Moscardini}}]{Maturi2005}
{Maturi}, M., {Meneghetti}, M., {Bartelmann}, M., {Dolag}, K., \& {Moscardini},
  L. 2005, \aap, 442, 851

\bibitem[{{Mazure} {et~al.}(2007){Mazure}, {Adami}, {Pierre}, {Le F{\`e}vre},
  {Arnouts}, {Duc}, {Ilbert}, {Lebrun}, {Meneux}, {Pacaud}, {Surdej}, \&
  {Valtchanov}}]{Mazure2007}
{Mazure}, A. {et~al.} 2007, \aap, 467, 49

\bibitem[{{McCarthy} {et~al.}(2017){McCarthy}, {Schaye}, {Bird}, \& {Le
  Brun}}]{McCarthy2017}
{McCarthy}, I.~G., {Schaye}, J., {Bird}, S., \& {Le Brun}, A.~M.~C. 2017,
  \mnras, 465, 2936

\bibitem[{{Mei} {et~al.}(2009){Mei}, {Holden}, {Blakeslee}, {Ford}, {Franx},
  {Homeier}, {Illingworth}, {Jee}, {Overzier}, {Postman}, {Rosati}, {Van der
  Wel}, \& {Bartlett}}]{Mei2009}
{Mei}, S. {et~al.} 2009, \apj, 690, 42

\bibitem[{{Mei} {et~al.}(2015){Mei}, {Scarlata}, {Pentericci}, {Newman},
  {Weiner}, {Ashby}, {Castellano}, {Conselice}, {Finkelstein}, {Galametz},
  {Grogin}, {Koekemoer}, {Huertas-Company}, {Lani}, {Lucas}, {Papovich},
  {Rafelski}, \& {Teplitz}}]{Mei2015}
------. 2015, \apj, 804, 117

\bibitem[{{Merson} {et~al.}(2013){Merson}, {Baugh}, {Helly}, {Gonzalez-Perez},
  {Cole}, {Bielby}, {Norberg}, {Frenk}, {Benson}, {Bower}, {Lacey}, \&
  {Lagos}}]{Merson2013}
{Merson}, A.~I. {et~al.} 2013, \mnras, 429, 556

\bibitem[{{Mohr} {et~al.}(2012){Mohr}, {Armstrong}, {Bertin}, {Daues}, {Desai},
  {Gower}, {Gruendl}, {Hanlon}, {Kuropatkin}, {Lin}, {Marriner}, {Petravic},
  {Sevilla}, {Swanson}, {Tomashek}, {Tucker}, \& {Yanny}}]{Mohr2012}
{Mohr}, J.~J. {et~al.} 2012, in \procspie, Vol. 8451, Software and
  Cyberinfrastructure for Astronomy II, 84510D

\bibitem[{{Molino} {et~al.}(2017){Molino}, {Ben{\'{\i}}tez}, {Ascaso}, {Coe},
  {Postman}, {Jouvel}, {Host}, {Lahav}, {Seitz}, {Medezinski}, {Rosati},
  {Schoenell}, {Koekemoer}, {Jimenez-Teja}, {Broadhurst}, {Melchior},
  {Balestra}, {Bartelmann}, {Bouwens}, {Bradley}, {Czakon}, {Donahue}, {Ford},
  {Graur}, {Graves}, {Grillo}, {Infante}, {Jha}, {Kelson}, {Lazkoz}, {Lemze},
  {Maoz}, {Mercurio}, {Meneghetti}, {Merten}, {Moustakas}, {Nonino}, {Orgaz},
  {Riess}, {Rodney}, {Sayers}, {Umetsu}, {Zheng}, \& {Zitrin}}]{Molino2017}
{Molino}, A. {et~al.} 2017, \mnras, 470, 95

\bibitem[{{Muzzin} {et~al.}(2007){Muzzin}, {Yee}, {Hall}, {Ellingson}, \&
  {Lin}}]{Muzzin2007}
{Muzzin}, A., {Yee}, H.~K.~C., {Hall}, P.~B., {Ellingson}, E., \& {Lin}, H.
  2007, \apj, 659, 1106

\bibitem[{{Nakata} {et~al.}(2001){Nakata}, {Kajisawa}, {Yamada}, {Kodama},
  {Shimasaku}, {Tanaka}, {Doi}, {Furusawa}, {Hamabe}, {Iye}, {Kimura},
  {Komiyama}, {Miyazaki}, {Okamura}, {Ouchi}, {Sasaki}, {Sekiguchi}, {Yagi}, \&
  {Yasuda}}]{Nakata2001}
{Nakata}, F. {et~al.} 2001, \pasj, 53, 1139

\bibitem[{{Navarro} {et~al.}(1996){Navarro}, {Frenk}, \& {White}}]{Navarro1996}
{Navarro}, J.~F., {Frenk}, C.~S., \& {White}, S.~D.~M. 1996, \apj, 462, 563

\bibitem[{{Neyrinck}(2008)}]{Neyrinck2008}
{Neyrinck}, M.~C. 2008, \mnras, 386, 2101

\bibitem[{{Noirot} {et~al.}(2016){Noirot}, {Vernet}, {De Breuck}, {Wylezalek},
  {Galametz}, {Stern}, {Mei}, {Brodwin}, {Cooke}, {Gonzalez}, {Hatch},
  {Rettura}, \& {Stanford}}]{Noirot2016}
{Noirot}, G. {et~al.} 2016, \apj, 830, 90

\bibitem[{Okabe {et~al.}(2000)Okabe, Boots, Sugihara, \& Chiu}]{Okabe2000}
Okabe, A., Boots, B., Sugihara, K., \& Chiu, S.~N. 2000, Spatial Tessellations:
  Concepts and Applications of {V}oronoi Diagrams, 2nd edn., Series in
  Probability and Statistics (John Wiley and Sons, Inc.)

\bibitem[{{Old} {et~al.}(2015){Old}, {Wojtak}, {Mamon}, {Skibba}, {Pearce},
  {Croton}, {Bamford}, {Behroozi}, {de Carvalho}, {Mu{\~n}oz-Cuartas},
  {Gifford}, {Gray}, {von der Linden}, {Merrifield}, {Muldrew}, {M{\"u}ller},
  {Pearson}, {Ponman}, {Rozo}, {Rykoff}, {Saro}, {Sepp}, {Sif{\'o}n}, \&
  {Tempel}}]{Old2015}
{Old}, L. {et~al.} 2015, \mnras, 449, 1897

\bibitem[{{Olsen} {et~al.}(2007){Olsen}, {Benoist}, {Cappi}, {Maurogordato},
  {Mazure}, {Slezak}, {Adami}, {Ferrari}, \& {Martel}}]{Olsen2007}
{Olsen}, L.~F. {et~al.} 2007, \aap, 461, 81

\bibitem[{{Pacaud} {et~al.}(2016){Pacaud}, {Clerc}, {Giles}, {Adami},
  {Sadibekova}, {Pierre}, {Maughan}, {Lieu}, {Le F{\`e}vre}, {Alis}, {Altieri},
  {Ardila}, {Baldry}, {Benoist}, {Birkinshaw}, {Chiappetti},
  {D{\'e}mocl{\`e}s}, {Eckert}, {Evrard}, {Faccioli}, {Gastaldello}, {Guennou},
  {Horellou}, {Iovino}, {Koulouridis}, {Le Brun}, {Lidman}, {Liske},
  {Maurogordato}, {Menanteau}, {Owers}, {Poggianti}, {Pomar{\`e}de}, {Pompei},
  {Ponman}, {Rapetti}, {Reiprich}, {Smith}, {Tuffs}, {Valageas}, {Valtchanov},
  {Willis}, \& {Ziparo}}]{Pacaud2016}
{Pacaud}, F. {et~al.} 2016, \aap, 592, A2

\bibitem[{{Parroni} {et~al.}(2017){Parroni}, {Mei}, {Erben}, {Van Waerbeke},
  {Raichoor}, {Ford}, {Licitra}, {Meneghetti}, {Hildebrandt}, {Miller},
  {C{\^o}t{\'e}}, {Covone}, {Cuillandre}, {Duc}, {Ferrarese}, {Gwyn}, \&
  {Puzia}}]{Parroni2017}
{Parroni}, C. {et~al.} 2017, \apj, 848, 114

\bibitem[{P\'erez \& Granger(2007)}]{Perez2007}
P\'erez, F., \& Granger, B.~E. 2007, Computing in Science and Engineering, 9,
  21

\bibitem[{{Pierre}(1990)}]{Pierre1990}
{Pierre}, M. 1990, \aap, 229, 7

\bibitem[{{Pierre} {et~al.}(2016){Pierre}, {Pacaud}, {Adami}, {Alis},
  {Altieri}, {Baran}, {Benoist}, {Birkinshaw}, {Bongiorno}, {Bremer}, {Brusa},
  {Butler}, {Ciliegi}, {Chiappetti}, {Clerc}, {Corasaniti}, {Coupon}, {De
  Breuck}, {Democles}, {Desai}, {Delhaize}, {Devriendt}, {Dubois}, {Eckert},
  {Elyiv}, {Ettori}, {Evrard}, {Faccioli}, {Farahi}, {Ferrari}, {Finet},
  {Fotopoulou}, {Fourmanoit}, {Gandhi}, {Gastaldello}, {Gastaud},
  {Georgantopoulos}, {Giles}, {Guennou}, {Guglielmo}, {Horellou}, {Husband},
  {Huynh}, {Iovino}, {Kilbinger}, {Koulouridis}, {Lavoie}, {Le Brun}, {Le
  Fevre}, {Lidman}, {Lieu}, {Lin}, {Mantz}, {Maughan}, {Maurogordato},
  {McCarthy}, {McGee}, {Melin}, {Melnyk}, {Menanteau}, {Novak}, {Paltani},
  {Plionis}, {Poggianti}, {Pomarede}, {Pompei}, {Ponman}, {Ramos-Ceja},
  {Ranalli}, {Rapetti}, {Raychaudury}, {Reiprich}, {Rottgering}, {Rozo},
  {Rykoff}, {Sadibekova}, {Santos}, {Sauvageot}, {Schimd}, {Sereno}, {Smith},
  {Smol{\v c}i{\'c}}, {Snowden}, {Spergel}, {Stanford}, {Surdej}, {Valageas},
  {Valotti}, {Valtchanov}, {Vignali}, {Willis}, \& {Ziparo}}]{Pierre2016}
{Pierre}, M. {et~al.} 2016, \aap, 592, A1

\bibitem[{{Planck Collaboration} {et~al.}(2014){Planck Collaboration}, {Ade},
  {Aghanim}, {Armitage-Caplan}, {Arnaud}, {Ashdown}, {Atrio-Barandela},
  {Aumont}, {Baccigalupi}, {Banday}, \& et~al.}]{Planck2013XX}
{Planck Collaboration} {et~al.} 2014, \aap, 571, A20

\bibitem[{{Planck Collaboration} {et~al.}(2016{\natexlab{a}}){Planck
  Collaboration}, {Ade}, {Aghanim}, {Arnaud}, {Ashdown}, {Aumont},
  {Baccigalupi}, {Banday}, {Barreiro}, {Barrena}, \& et~al.}]{Planck2015XXVII}
------. 2016{\natexlab{a}}, \aap, 594, A27

\bibitem[{{Planck Collaboration} {et~al.}(2016{\natexlab{b}}){Planck
  Collaboration}, {Ade}, {Aghanim}, {Arnaud}, {Ashdown}, {Aumont},
  {Baccigalupi}, {Banday}, {Barreiro}, {Bartlett}, \& et~al.}]{Planck2016XIII}
------. 2016{\natexlab{b}}, \aap, 594, A13

\bibitem[{{Planck Collaboration} {et~al.}(2016{\natexlab{c}}){Planck
  Collaboration}, {Ade}, {Aghanim}, {Arnaud}, {Ashdown}, {Aumont},
  {Baccigalupi}, {Banday}, {Barreiro}, {Bartlett}, \& et~al.}]{Planck2015XXIV}
------. 2016{\natexlab{c}}, \aap, 594, A24

\bibitem[{{Platen} {et~al.}(2007){Platen}, {van de Weygaert}, \&
  {Jones}}]{Platen2007}
{Platen}, E., {van de Weygaert}, R., \& {Jones}, B.~J.~T. 2007, \mnras, 380,
  551

\bibitem[{{Popesso} {et~al.}(2005){Popesso}, {B{\"o}hringer}, {Romaniello}, \&
  {Voges}}]{Popesso2005}
{Popesso}, P., {B{\"o}hringer}, H., {Romaniello}, M., \& {Voges}, W. 2005,
  \aap, 433, 415

\bibitem[{{Postman} {et~al.}(1996){Postman}, {Lubin}, {Gunn}, {Oke}, {Hoessel},
  {Schneider}, \& {Christensen}}]{Postman1996}
{Postman}, M., {Lubin}, L.~M., {Gunn}, J.~E., {Oke}, J.~B., {Hoessel}, J.~G.,
  {Schneider}, D.~P., \& {Christensen}, J.~A. 1996, \aj, 111, 615

\bibitem[{{Radovich} {et~al.}(2017){Radovich}, {Puddu}, {Bellagamba},
  {Roncarelli}, {Moscardini}, {Bardelli}, {Grado}, {Getman}, {Maturi}, {Huang},
  {Napolitano}, {McFarland}, {Valentijn}, \& {Bilicki}}]{Radovich2017}
{Radovich}, M. {et~al.} 2017, \aap, 598, A107

\bibitem[{{Raichoor} \& {Andreon}(2012)}]{Raichoor2012}
{Raichoor}, A., \& {Andreon}, S. 2012, \aap, 543, A19

\bibitem[{{Ramella} {et~al.}(2001){Ramella}, {Boschin}, {Fadda}, \&
  {Nonino}}]{Ramella2001}
{Ramella}, M., {Boschin}, W., {Fadda}, D., \& {Nonino}, M. 2001, \aap, 368, 776

\bibitem[{{Ramella} {et~al.}(2004){Ramella}, {Boschin}, {Geller}, {Mahdavi}, \&
  {Rines}}]{Ramella2004}
{Ramella}, M., {Boschin}, W., {Geller}, M.~J., {Mahdavi}, A., \& {Rines}, K.
  2004, \aj, 128, 2022

\bibitem[{{Rettura} {et~al.}(2014){Rettura}, {Martinez-Manso}, {Stern}, {Mei},
  {Ashby}, {Brodwin}, {Gettings}, {Gonzalez}, {Stanford}, \&
  {Bartlett}}]{Rettura2014}
{Rettura}, A. {et~al.} 2014, \apj, 797, 109

\bibitem[{{Ricci} {et~al.}(2018){Ricci}, {Benoist}, {Maurogordato}, {Adami},
  {Chiappetti}, {Gastaldello}, {Guglielmo}, {Poggianti}, {Sereno}, {Adam},
  {Arnouts}, {Cappi}, {Koulouridis}, {Pacaud}, {Pierre}, \&
  {Ramos-Ceja}}]{Ricci2018}
{Ricci}, M. {et~al.} 2018, \aap, 620, A13

\bibitem[{{Rozo} {et~al.}(2010){Rozo}, {Wechsler}, {Rykoff}, {Annis}, {Becker},
  {Evrard}, {Frieman}, {Hansen}, {Hao}, {Johnston}, {Koester}, {McKay},
  {Sheldon}, \& {Weinberg}}]{Rozo2010}
{Rozo}, E. {et~al.} 2010, \apj, 708, 645

\bibitem[{{Rykoff} {et~al.}(2014){Rykoff}, {Rozo}, {Busha}, {Cunha},
  {Finoguenov}, {Evrard}, {Hao}, {Koester}, {Leauthaud}, {Nord}, {Pierre},
  {Reddick}, {Sadibekova}, {Sheldon}, \& {Wechsler}}]{Rykoff2014}
{Rykoff}, E.~S. {et~al.} 2014, \apj, 785, 104

\bibitem[{{Saro} {et~al.}(2015){Saro}, {Bocquet}, {Rozo}, {Benson}, {Mohr},
  {Rykoff}, {Soares-Santos}, {Bleem}, {Dodelson}, {Melchior}, {Sobreira},
  {Upadhyay}, {Weller}, {Abbott}, {Abdalla}, {Allam}, {Armstrong}, {Banerji},
  {Bauer}, {Bayliss}, {Benoit-L{\'e}vy}, {Bernstein}, {Bertin}, {Brodwin},
  {Brooks}, {Buckley-Geer}, {Burke}, {Carlstrom}, {Capasso}, {Capozzi},
  {Carnero Rosell}, {Carrasco Kind}, {Chiu}, {Covarrubias}, {Crawford},
  {Crocce}, {D'Andrea}, {da Costa}, {DePoy}, {Desai}, {de Haan}, {Diehl},
  {Dietrich}, {Doel}, {Cunha}, {Eifler}, {Evrard}, {Fausti Neto}, {Fernandez},
  {Flaugher}, {Fosalba}, {Frieman}, {Gangkofner}, {Gaztanaga}, {Gerdes},
  {Gruen}, {Gruendl}, {Gupta}, {Hennig}, {Holzapfel}, {Honscheid}, {Jain},
  {James}, {Kuehn}, {Kuropatkin}, {Lahav}, {Li}, {Lin}, {Maia}, {March},
  {Marshall}, {Martini}, {McDonald}, {Miller}, {Miquel}, {Nord}, {Ogando},
  {Plazas}, {Reichardt}, {Romer}, {Roodman}, {Sako}, {Sanchez}, {Schubnell},
  {Sevilla}, {Smith}, {Stalder}, {Stark}, {Strazzullo}, {Suchyta}, {Swanson},
  {Tarle}, {Thaler}, {Thomas}, {Tucker}, {Vikram}, {von der Linden}, {Walker},
  {Wechsler}, {Wester}, {Zenteno}, \& {Ziegler}}]{Saro2015}
{Saro}, A. {et~al.} 2015, \mnras, 454, 2305

\bibitem[{{Sarron} {et~al.}(2018){Sarron}, {Martinet}, {Durret}, \&
  {Adami}}]{Sarron2018}
{Sarron}, F., {Martinet}, N., {Durret}, F., \& {Adami}, C. 2018, \aap, 613, A67

\bibitem[{{Sartoris} {et~al.}(2016){Sartoris}, {Biviano}, {Fedeli}, {Bartlett},
  {Borgani}, {Costanzi}, {Giocoli}, {Moscardini}, {Weller}, {Ascaso},
  {Bardelli}, {Maurogordato}, \& {Viana}}]{Sartoris2016}
{Sartoris}, B. {et~al.} 2016, \mnras, 459, 1764

\bibitem[{{Schechter}(1976)}]{Schechter1976}
{Schechter}, P. 1976, \apj, 203, 297

\bibitem[{{Shan} {et~al.}(2012){Shan}, {Kneib}, {Tao}, {Fan}, {Jauzac},
  {Limousin}, {Massey}, {Rhodes}, {Thanjavur}, \& {McCracken}}]{Shan2012}
{Shan}, H. {et~al.} 2012, \apj, 748, 56

\bibitem[{{Spergel} {et~al.}(2015){Spergel}, {Gehrels}, {Baltay}, {Bennett},
  {Breckinridge}, {Donahue}, {Dressler}, {Gaudi}, {Greene}, {Guyon}, {Hirata},
  {Kalirai}, {Kasdin}, {Macintosh}, {Moos}, {Perlmutter}, {Postman},
  {Rauscher}, {Rhodes}, {Wang}, {Weinberg}, {Benford}, {Hudson}, {Jeong},
  {Mellier}, {Traub}, {Yamada}, {Capak}, {Colbert}, {Masters}, {Penny},
  {Savransky}, {Stern}, {Zimmerman}, {Barry}, {Bartusek}, {Carpenter}, {Cheng},
  {Content}, {Dekens}, {Demers}, {Grady}, {Jackson}, {Kuan}, {Kruk}, {Melton},
  {Nemati}, {Parvin}, {Poberezhskiy}, {Peddie}, {Ruffa}, {Wallace}, {Whipple},
  {Wollack}, \& {Zhao}}]{Spergel2015}
{Spergel}, D. {et~al.} 2015, ArXiv e-prints:1503.03757

\bibitem[{{Springel} {et~al.}(2005){Springel}, {White}, {Jenkins}, {Frenk},
  {Yoshida}, {Gao}, {Navarro}, {Thacker}, {Croton}, {Helly}, {Peacock}, {Cole},
  {Thomas}, {Couchman}, {Evrard}, {Colberg}, \& {Pearce}}]{Springel2005}
{Springel}, V. {et~al.} 2005, \nat, 435, 629

\bibitem[{{Sridhar} {et~al.}(2017){Sridhar}, {Maurogordato}, {Benoist},
  {Cappi}, \& {Marulli}}]{Sridhar2017}
{Sridhar}, S., {Maurogordato}, S., {Benoist}, C., {Cappi}, A., \& {Marulli}, F.
  2017, \aap, 600, A32

\bibitem[{{Stanford} {et~al.}(2012){Stanford}, {Brodwin}, {Gonzalez},
  {Zeimann}, {Stern}, {Dey}, {Eisenhardt}, {Snyder}, \&
  {Mancone}}]{Stanford2012}
{Stanford}, S.~A. {et~al.} 2012, \apj, 753, 164

\bibitem[{{Starck} {et~al.}(1998){Starck}, {Murtagh}, \&
  {Bijaoui}}]{Starck1998}
{Starck}, J.-L., {Murtagh}, F.~D., \& {Bijaoui}, A. 1998, {Image Processing and
  Data Analysis}, 297

\bibitem[{{Strazzullo} {et~al.}(2016){Strazzullo}, {Daddi}, {Gobat},
  {Valentino}, {Pannella}, {Dickinson}, {Renzini}, {Brammer}, {Onodera},
  {Finoguenov}, {Cimatti}, {Carollo}, \& {Arimoto}}]{Strazzullo2016}
{Strazzullo}, V. {et~al.} 2016, \apjl, 833, L20

\bibitem[{{Strazzullo} {et~al.}(2010){Strazzullo}, {Rosati}, {Pannella},
  {Gobat}, {Santos}, {Nonino}, {Demarco}, {Lidman}, {Tanaka}, {Mullis},
  {Nu{\~n}ez}, {Rettura}, {Jee}, {B{\"o}hringer}, {Bender}, {Bouwens},
  {Dawson}, {Fassbender}, {Franx}, {Perlmutter}, \& {Postman}}]{Strazzullo2010}
------. 2010, \aap, 524, A17

\bibitem[{{Strazzullo} {et~al.}(2006){Strazzullo}, {Rosati}, {Stanford},
  {Lidman}, {Nonino}, {Demarco}, {Eisenhardt}, {Ettori}, {Mainieri}, \&
  {Toft}}]{Strazzullo2006}
------. 2006, \aap, 450, 909

\bibitem[{{Sunyaev} \& {Zel'dovich}(1972)}]{Sunyaev1972}
{Sunyaev}, R.~A., \& {Zel'dovich}, Y.~B. 1972, \apspr, 4, 173

\bibitem[{{Toft} {et~al.}(2004){Toft}, {Mainieri}, {Rosati}, {Lidman},
  {Demarco}, {Nonino}, \& {Stanford}}]{Toft2004}
{Toft}, S., {Mainieri}, V., {Rosati}, P., {Lidman}, C., {Demarco}, R.,
  {Nonino}, M., \& {Stanford}, S.~A. 2004, \aap, 422, 29

\bibitem[{{Trevisan} \& {Mamon}(2017)}]{Trevisan2017b}
{Trevisan}, M., \& {Mamon}, G.~A. 2017, \mnras, 471, 2022

\bibitem[{{Trevisan} {et~al.}(2017){Trevisan}, {Mamon}, \&
  {Stalder}}]{Trevisan2017}
{Trevisan}, M., {Mamon}, G.~A., \& {Stalder}, D.~H. 2017, \mnras, 471, L47

\bibitem[{{van der Burg} {et~al.}(2015){van der Burg}, {Hoekstra}, {Muzzin},
  {Sif{\'o}n}, {Balogh}, \& {McGee}}]{vdb2015}
{van der Burg}, R.~F.~J., {Hoekstra}, H., {Muzzin}, A., {Sif{\'o}n}, C.,
  {Balogh}, M.~L., \& {McGee}, S.~L. 2015, \aap, 577, A19

\bibitem[{{van der Burg} {et~al.}(2014){van der Burg}, {Muzzin}, {Hoekstra},
  {Wilson}, {Lidman}, \& {Yee}}]{vdb2014}
{van der Burg}, R.~F.~J., {Muzzin}, A., {Hoekstra}, H., {Wilson}, G., {Lidman},
  C., \& {Yee}, H.~K.~C. 2014, \aap, 561, A79

\bibitem[{{van der Walt} {et~al.}(2011){van der Walt}, {Colbert}, \&
  {Varoquaux}}]{VanDerWalt2011}
{van der Walt}, S., {Colbert}, S.~C., \& {Varoquaux}, G. 2011, Computing in
  Science and Engineering, 13, 22

\bibitem[{{Veropalumbo} {et~al.}(2014){Veropalumbo}, {Marulli}, {Moscardini},
  {Moresco}, \& {Cimatti}}]{Veropalumbo2014}
{Veropalumbo}, A., {Marulli}, F., {Moscardini}, L., {Moresco}, M., \&
  {Cimatti}, A. 2014, \mnras, 442, 3275

\bibitem[{{Vikhlinin} {et~al.}(2009){Vikhlinin}, {Kravtsov}, {Burenin},
  {Ebeling}, {Forman}, {Hornstrup}, {Jones}, {Murray}, {Nagai}, {Quintana}, \&
  {Voevodkin}}]{Vikhlinin2009}
{Vikhlinin}, A. {et~al.} 2009, \apj, 692, 1060

\bibitem[{{Wen} {et~al.}(2012){Wen}, {Han}, \& {Liu}}]{Wen2012}
{Wen}, Z.~L., {Han}, J.~L., \& {Liu}, F.~S. 2012, \apjs, 199, 34

\bibitem[{{Wetzel} {et~al.}(2013){Wetzel}, {Tinker}, {Conroy}, \& {van den
  Bosch}}]{Wetzel2013}
{Wetzel}, A.~R., {Tinker}, J.~L., {Conroy}, C., \& {van den Bosch}, F.~C. 2013,
  \mnras, 432, 336

\bibitem[{{Wylezalek} {et~al.}(2013){Wylezalek}, {Galametz}, {Stern}, {Vernet},
  {De Breuck}, {Seymour}, {Brodwin}, {Eisenhardt}, {Gonzalez}, {Hatch},
  {Jarvis}, {Rettura}, {Stanford}, \& {Stevens}}]{Wylezalek2013}
{Wylezalek}, D. {et~al.} 2013, \apj, 769, 79

\bibitem[{{Yang} {et~al.}(2008){Yang}, {Mo}, \& {van den Bosch}}]{Yang2008}
{Yang}, X., {Mo}, H.~J., \& {van den Bosch}, F.~C. 2008, \apj, 676, 248

\end{thebibliography}

\begin{appendix}
\section{Summary of the previous challenges}
In preparation for the final cluster finder challenge described in this paper, three other challenges took place between 2013 and 2017, organized within the Work-Package Clusters of Galaxies (Implementation) of the \textit{Euclid} Science Ground Segment Organisation Unit Level3. The aim of these preparatory challenges was to set up and refine the procedure by which to assess the relative performance of the different cluster finder algorithms. Eight different codes were tested on two \textit{Euclid} survey-like mock catalogs, where the positions of the massive halos to be detected by the codes were unknown to the participants to the challenge.

The analysis of these two challenges raised several difficulties that had to be overcome. In particular, the definition of S/N was not common to all cluster finder algorithms, so to compare the relative performance of the algorithms in terms of completeness, we ranked the detections in S/N and selected the first N detections in order of decreasing S/N, separately for each algorithm. To compare the relative performance in terms of purity, we translated the different richness estimates provided by the different algorithms into masses, by fitting a mass-richness relation independently for each algorithm.

To test whether the relative performance of the different codes was dependent on the mock catalog itself, two mocks were considered, one based on semianalytic modeling and one based on halo occupation distribution (HOD). However, we noticed when performing tests on the properties of galaxies in clusters in the different mocks that the cluster number density concentration in the mocks based on HOD simulations was very high ($\sim 8$ times the typical values). This led us to use a mock based on a SAM for the final challenge. Both mocks reproducing the wide and the deep \textit{Euclid} surveys were considered. Initially, mock galaxies were assigned photometric redshifts from a random Gaussian distribution centered on the real galaxy redshift with a dispersion equal to the \textit{Euclid} survey requirement, $\sigma_z =0.03 (1+z)$ or $0.05 (1+z)$, and with an outlier fraction of 5\% or 10\%. At a later stage, photometric redshifts were computed from fits to the galaxy spectral energy distributions (using BPZ by \cite{Benitez2000}, \cite{Coe2006}, and {\tt Hyperz} by \cite{Bolzonella2000}) resulting from the galaxy mock magnitudes with added noise. Galaxy magnitudes in the mocks were rendered closer to real galaxy magnitudes by the use of the {\tt PhotReal} \citep{Ascaso2015} algorithm. We found that modifications to the photometric redshift assignment method had little to no impact on the performance of the detection algorithms.

At the end of the third cluster finder challenge, we considered the whole methodology and analysis pipeline to be sufficiently mature to make a final assessment of the relative performance of the different codes. While eight cluster finder codes in total were tested in the three preliminary challenges, only six of them took part in the final challenge described in this paper.

We hereafter give the description of the two codes (RedGOLD and Voronoi) that did not participate in the final challenge for reasons not related to their performance in the earlier CFCs.

\subsection{RedGOLD}
The RedGOLD cluster detection algorithm \citep{Licitra2016a,Licitra2016b} was developed as a modified version of color-based detection algorithms such as RedMaPPer \citep{Rykoff2014}, taking into account galaxy morphology and color cuts performed on clusters at high redshifts \citep{Mei2009,Mei2015}. The algorithm requires as input a catalog with galaxy positions, photometry and photometric redshifts. It selects overdensities of galaxies in a color-color plane, and it is adapted to select both red passive and blue star-forming galaxies. At redshift $z\lesssim 1.5$, where clusters are mostly dominated by a passive galaxy population and show a tight red sequence, the algorithm selects overdensities of red passive galaxies while minimizing contamination from dusty star-forming galaxies. It imposes an NFW profile and calculates cluster detection significance and richness, which is tightly correlated to weak lensing masses \citep{Parroni2017}. In RedGOLD, the S/N is defined as the significance of detections with respect to the background galaxy density, as described in \cite{Licitra2016a}. In \cite{Licitra2016a,Licitra2016b}, cluster candidate catalogs were obtained for the CFHT-LS and NGVS surveys. When compared to X-ray detected cluster catalogs on these two surveys and the \cite{Henriques2012} mock galaxy catalogs from the Millenium simulation \citep{Springel2005}, RedGOLD was demonstrated to be $80\%$ pure up to $z \sim 1$, and $\sim 100\%$ ($\sim 70\%$) complete at $z \leq 0.6$ ($z \leq1$) for galaxy clusters with $M > 10^{14}$ M$_{\odot}$. RedGOLD participated in the \textit{Euclid} CFC I and II.

\subsection{Voronoi}
The Voronoi diagram and its dual, the Delaunay triangulation, have long been known as a very useful and versatile mathematical techniques in a variety of topics - including 3D reconstruction and modeling of objects, visualization of medical datasets, shape analysis and pattern recognition, computer animation, to mention a few \citep[][for wide a review of their applications in computational geometry]{Okabe2000,Aurenhammer2013}. In short, a Voronoi tessellation on a two-dimensional (tri-dimensional) distribution of objects is a unique plane (volume) partition into convex cells (polytopes), each of them containing one, and only one, such object. These are the set of points which are closer to that object than to any other. Its main advantage consists in its being completely non-parametric and adaptive, thus very useful in all those cases when one does not desire to define a priori a specific spatial scale for the analysis. The first order Delaunay neighbors of a chosen object are then those objects in the centers of the cells (polytopes) sharing a wall with the cell (polytope) of the chosen object. Second order Delaunay neighbors are obtained extending one step further out, that is considering also objects that are first order neighbors of objects in the previous list. The Voronoi tessellation has received attention in astrophysics, and among its many usages we may quote those as a tool to reproduce the foamy distribution of galaxies in space \citep[see][]{Pierre1990,Icke1991}, to find voids in galaxy spatial distribution \citep[see][]{Platen2007,Neyrinck2008}, to bin data to a constant S/N per bin in integral field spectroscopy \citep{Cappellari2003} or X-ray imaging data \citep{Diehl2006}, and to detect clusters both in 2D projected photometric galaxy catalogs \citep{Ramella2001} and in 3D galaxy redshift catalogs \citep[see][]{Marinoni2002,Cucciati2010}.

The Voronoi algorithm written for \textit{Euclid} CFC was tailored for cluster detection in photometric redshift space, with a typical error in photometric redshift equal to $\sigma_{z_{\rm phot}}$. The algorithm works in two steps. 

In the first step the sample of galaxies is split into a series of partially overlapping redshift slices. Each slice is defined with thickness equal to 1.5 $\sigma_{z_{\rm phot}}$ and is separated from the adjacent slices by photometric redshift steps of 0.5 $\sigma_{z_{\rm phot}}$ (a galaxy will then usually be present in more than one redshift slice). In each redshift slice the algorithm starts by building the Voronoi tessellation of the RA-Dec galaxy distribution and then computes, for each galaxy, the area covered by including both first and second order Voronoi-Delaunay neighbors, a choice that is driven by the need to minimize the noise in the area estimate. In each redshift slice, all the areas thus estimated are sorted, a fit to their distribution is computed and only galaxies whose area is below 1.5 $\sigma$ the mean value (thus galaxies located in higher than the mean density regions) are kept as cluster seeds. Around each of these seeds we grow outwards, and keep adding first order neighbors of peripheral galaxies. The growth continues as long as the added members satisfy a request on their second order Voronoi-Delauney neighbors area (should be smaller than a predefined cutoff) and a cutoff on the growth ratio (at least 10 new members added in each subsequent growth step). This procedure is run on all the redshift slices defined in the galaxy catalog. 

The second step of the algorithm then merges together (using RA, Dec and $z_{\rm phot}$ information) results from different redshift slices, thus building the final cluster catalog. Each cluster is defined by a center: median of its putative member galaxies RA, Dec and $z_{\rm phot}$ values. The algorithm computes for each cluster its S/N using expected values for a distribution of points with the same surface density as on the actual $z_{\rm phot}$ shell that contains most of its members - if needed during this step further trimming or enlarging of the outer regions can be done, while also a very bland compactness limit ($n90/n50 < 2$) is applied, where $n50/n90$ is the number of galaxies included within $50\%/90\%$ of the group radius.

The algorithm computes also cluster areas, by summing Voronoi-Delaunay areas of connected galaxies, and observed richness of clusters, after statistical subtraction of expected background galaxy density within the cluster area. This code participated in the \textit{Euclid} CFC I.

\end{appendix}

\end{document}